\documentclass[structabstract]{aa}  

\usepackage{graphicx}
\usepackage{txfonts}
\usepackage{color} 
\usepackage{natbib}
\usepackage{lscape}
\usepackage{rotating}

\def\Rsolar{$R_{\odot}$}
\def\Msolar{$M_{\odot}$}
\newcommand{\Mo}{M_{\odot}}
\newcommand{\Ro}{R_{\odot}}
\newcommand{\peryr}{\,{\rm yr^{-1}}}

\begin{document}

	\title{PopCORN: Hunting down the differences between binary population synthesis codes}
   \author{S.~Toonen \inst{1}
          \and
          J.~S.~W.~Claeys \inst{1}
          \and
          N.~Mennekens \inst{2}
          \and         
          A.~J.~Ruiter \inst{3} 
          }

          \institute{
			Department of Astrophysics, Radboud University Nijmegen,
            P.O. Box 9010, 6500 GL Nijmegen, The Netherlands\\
            \email{silviato@astro.ru.nl} 
            \and Astrophysical Institute, Vrije Universiteit Brussel,
            Pleinlaan 2, 1050 Brussels, Belgium 
            \and Max Planck Institute for Astrophysics,
            Karl-Schwarzschild-Str. 1, 85741 Garching, Germany 
          }

   \date{Received March 27, 2013; accepted November 11, 2013}

  \abstract
   {Binary population synthesis (BPS) modelling is a very effective
     tool to study the evolution and properties of various types of
     close binary systems. The uncertainty in the parameters of the
     model and their effect on a population can be tested in a
     statistical way, which then leads to a deeper understanding of
       the underlying (sometimes poorly understood) physical processes involved. 
 Several BPS codes exist that have been developed with different philosophies and aims. Although BPS has been very successful for studies of many populations of binary stars, in the particular case of the study of the progenitors of supernovae Type Ia, the predicted rates and ZAMS progenitors vary substantially between different BPS codes.}
   {To understand the predictive power of BPS codes, we study the similarities and differences in the predictions of four different BPS codes for low- and intermediate-mass binaries. We investigate the differences in the characteristics of the predicted populations, and whether they are caused by different assumptions made in the BPS codes or by numerical effects, e.g. a lack of accuracy in BPS codes.}  
   {We compare a large number of evolutionary sequences for binary stars, starting with the same initial conditions following the evolution until
     the first (and when applicable, the second) white dwarf (WD) is formed. To simplify the complex problem of comparing BPS codes that are based on many (often different) assumptions, we equalise the assumptions as much as possible to examine the inherent differences of the four BPS codes. 
  }
   {We find that the simulated populations are similar between the
     codes. Regarding the population of binaries with
       one WD, there is very good agreement between the
       physical characteristics, the evolutionary channels that lead to
       the birth of these systems, and their birthrates. 
   Regarding the double WD population, there is a good agreement on
   which evolutionary channels exist to create double WDs and a rough
   agreement on the characteristics of the double WD population. 
Regarding which progenitor systems lead to a single and double WD system and which systems do not, the four codes agree well. Most importantly, we find that for these two populations, the differences in the predictions from the four codes are not due to numerical differences, but because of different inherent assumptions. We identify critical assumptions for BPS studies that need to be studied in more detail.}
   {}

   \maketitle
%
\section{Introduction}
Binary population synthesis codes (hereafter BPS codes) enable the rapid calculation of the evolution of a large number of binary stars over the course of the binary lifetime. With such models, we can study the diverse properties  of binary populations, 
e.g. the chemical enrichment of a region, or the frequency
of an astrophysical event \citep[for a review, see][]{Han01}. We can learn about and study the formation and evolution of stellar systems that are important for a wide range of astronomical topics:\
novae, X-ray binaries, symbiotics, subdwarf B stars, gamma ray bursts, R Coronae Borealis stars, AM CVn stars, Type Ia and Type Ib/c supernovae, runaway stars, binary pulsars, blue stragglers, etc.

To carefully study binary populations, in principle it is necessary to follow the
evolution of every binary system in detail. 
However, it is not feasible to evolve a population of binary stars from the zero-age main-sequence (ZAMS) to remnant formation with a detailed stellar evolution code. Such a task
is computationally expensive as there are many physical
processes which must be taken into account over large physical and
temporal scales, such as tidal evolution, Roche lobe Overflow (RLOF), mass transfer. Moreover not all processes can be modelled with detailed codes or are quite uncertain (or both), e.g. common envelope evolution, contact phases. 
Therefore, simplifying assumptions are made about the binary evolution process and many of its facets are modelled by the use of parameters. This process is generally known as binary population synthesis. Examples of such parametrisation are 
straightforward prescriptions for the stability and rate of mass transfer.
To some degree, 
the effects that are most important for the problem being studied will be more elaborately included in the corresponding BPS codes.
For the evolution of an individual system, the above can of course be an oversimplification. However, for the treatment of the general characteristics of a large population of binaries this process works very well \citep[e.g.][]{Egg89}.

Recently, several BPS codes have been used to study the progenitors of Type Ia supernovae \citep[e.g.][]{Yun94, Han95,Jor97,Yun00,Nel01,Han04,DeD04,Yun05,Lip09,Rui09,Rui11,Men10,Wan10,Men11,Bog09,Bog11,ruiter2013a,Too12,Men13, Cla13}. From these recent studies, it has become evident that the various codes show different results in terms of the SNe Ia rate \citep{Nel13}, in particular for the single degenerate channel in which binary systems can produce a SNe Ia by accretion from a non-degenerate companion to a white dwarf (WD). 
The differences in the predicted SNIa rate are largely, but not completely, due to differences in the assumed retention efficiency of the accretion onto the WD \citep{Bou13}. While it has long been expected by groups working on population synthesis that the differences in the BPS results were the result of different assumptions being made in these various studies rather than numerical in nature, it became ever more clear that a quantitative study of the nature and causes of these differences is necessary. 

This paper aims to do this by clarifying, for four different BPS codes, the respective ingredients and assumptions included in the population codes and comparing models of several simulated populations for which all assumptions have been made the same as much as possible.
We discuss the similarities and differences in the predicted populations and examine the causes for the differences that remain. 
The causes for differences are valuable information for interpreting binary population synthesis results, and as input for the astronomical community to increase our understanding of binary evolution. The project is known as PopCORN - Population synthesis of Compact Objects Research Network. It is not the purpose of this paper to discuss the advantages or shortcomings of the respective methods used in BPS codes, nor to judge
which assumptions made for binary evolutionary aspects are the most desirable.

The paper focuses on low and intermediate mass close binaries, i.e. those with
initial stellar masses below 10\Msolar. The reason for this is
twofold: firstly, as the project originates from differences in the
predictions of SNe Ia rates, the systems that produce WDs are the main focus. Secondly, since the evolution of massive stars is even less straightforward, and its modelling includes even more uncertainties, comparing massive star population synthesis will be a whole new project. 

In Sect.\,\ref{sec:BinEvol} we give an overview of the relevant processes for the evolution of low- and intermediate-mass binaries. Sect.\,\ref{sec:BPS} describes the codes involved in this project. The method we use to conduct the BPS comparison is described in Sect.\,\ref{sec:method}. We compare the simulated populations of systems containing one WD in Sect.\,\ref{sec:swd} and two WDs in Sect.\,\ref{sec:dwd}. 
A more detailed comparison for the most important evolutionary paths is given in Appendix\,\ref{sec:ev_path}. 
In Sect.\,\ref{sec:sum} we summarise and discuss the causes for differences that were found in Sect.\,\ref{sec:Results}. Our conclusions are given in Sect.\,\ref{sec:concl}. An overview of the inherent and typical assumptions of each code can be found in Appendix\,\ref{sec:TNS_inherent}~and~\ref{sec:TNS_equalized} respectively.  

\section{Binary evolution\label{sec:BinEvol}}
In this section we will give a rough outline of binary evolution and the most important processes that take place in low and intermediate mass binaries. The actual implementation in the four BPS codes under consideration in this study is described in Appendix\,\ref{sec:TNS_inherent}~and~\ref{sec:TNS_equalized}.

Low- and intermediate-mass systems with initial periods less than
approximately 10 years and primary masses above approximately 0.8\Msolar, will come into Roche lobe contact within a Hubble time. The stars in a binary system evolve effectively as single stars, slowly increasing in radius and luminosity, until one or both of the stars fills its Roche lobe. At this point mass from the outer layers of the star can flow through the first Lagrangian point leaving the donor star.

Depending on the reaction of the star upon mass loss and the reaction of the Roche lobe upon the rearrangement of mass and angular momentum in the system, mass transfer can be stable or unstable. When mass transfer becomes unstable, the loss of mass from the donor star will cause it to overfill its Roche lobe further. In turn this increases the mass loss rate leading to a runaway process. In comparison, when mass transfer is stable, the donor star will stay approximately within the Roche lobe. Mass transfer is maintained by the expansion of the donor star, or the contraction of the Roche lobe from the rearrangement of mass and angular momentum in the binary system. 

RLOF influences the evolution of the donor star by the decrease in mass. The evolution of the companion star is affected too if some or all of the mass lost by the donor is accreted. This is particularly true if some of the accreted (hydrogen-rich) matter makes its way to the core through internal mixing, where it will thus lead to replenishment of hydrogen, a process known as rejuvenation \citep[see e.g.][]{Van94}. 

Orbits of close binaries are affected by angular momentum loss (AML) from gravitational wave emission \citep[e.g.][]{Pet64}, possibly magnetic braking \citep {Ver81, Kni11} and tidal interaction. 
Magnetic braking extracts angular momentum from a rotating star by a stellar wind that is magnetically coupled to the star. If the star is in corotation with the orbit, angular momentum is essentially also removed from the binary orbit.
Tidal interaction plays a crucial role in circularising
binaries and will strive to synchronise the rotational period of
each star with the orbital period. While it is known that tidal effects will eventually achieve tidal locking of
both components, the strength of tidal effects is still subject to debate \citep[see e.g.][]{Zah77,Hut81}.

\subsection{Stable mass transfer}
\label{sec:mt_stable}
In the case of conservative RLOF the variation in the orbital separation $a$ during the mass transfer phase is dictated solely by the masses. If the gainer star accretes mass non-conservatively, there is a loss of matter and angular momentum from the system. 
We define the accretion efficiency: 
\begin{equation}
\beta = \left| \dot M_{\rm a} / \dot M_{\rm d} \right|,
\label{eq:beta}
\end{equation}
where $M_{\rm d}$ is the mass of the donor star and $M_{\rm a}$ is the mass of the accreting companion. If $\beta <
1$, it is also necessary to make an assumption about how much angular
momentum $J$ is carried away with it. We define this with a parameter $\eta$ such that:

\begin{equation}
\frac{\dot J}{J} = \eta \frac{\dot M}{M_{\rm d} + M_{\rm a}}(1-\beta). 
\label{eq:Jloss}
\end{equation}

Several prescriptions for AML exist (Appendix\,\ref{sec:AMLwind}) and the amount of angular momentum that is lost from the system due to mass loss has a strong influence on the evolution of the binary.

Matter and angular momentum can also be lost through stellar winds. 
As these are usually assumed to be
spherically symmetric, they will extract the specific orbital angular
momentum of the donor star, and result in an increase in the orbital period. If, however, the
wind is allowed to interact with the orbit of the binary, the result
is entirely dependent on this interaction.

\subsection{Unstable mass transfer}
\label{sec:mt_unstable}
During unstable mass transfer, the envelope of the donor star engulfs the companion star. Therefore this phase is often called the common envelope (CE) phase \citep{Pac76}. A merger of the companion and the core of the donor star can be avoided, if the gaseous envelope surrounding them is expelled e.g. by viscous friction that heats the envelope. Because of the loss of significant amounts of mass and angular momentum the CE-phase can have a very strong effect on the binary orbit. In particular it plays an essential role in the formation of short period systems containing at least one compact object. Despite this, the phenomenon is not yet well understood, see \citet{Iva13} for an overview. 

There are several formalisms available to treat the orbital evolution during CE-evolution. The most popular ones are the $\alpha$-formalism \citep{Tut79} and the $\gamma$-formalism \citep{Nel00}. The first considers the energy budget of the initial and final configuration, while the latter is based on the angular momentum balance. Both prescriptions include a parameter after which they are named, which determines the efficiency to remove the envelope. Because such an unstable mass transfer phase occurs on a short timescale, it is often assumed that the gainer does not have the time to gain an appreciable amount of mass during a CE-phase.

The $\alpha$-parameter describes the efficiency of which orbital energy is consumed to unbind the CE according to:
\begin{equation}
E_{\rm gr} = \alpha_{\rm ce} (E_{\rm orb,i}-E_{\rm orb,f}),
\label{eq:ce_a}
\end{equation}
where $E_{\rm orb}$ is the orbital energy, $E_{\rm gr}$ is the binding energy of the envelope and $\alpha_{\rm ce}$ is the efficiency of the energy conversion. The subscript i and f represent the parameter before and after the CE-phase respectively. 
Several prescriptions for the quantities $E_{\rm orb,i}$ and $E_{\rm gr}$ have been proposed \citep{Web84, Ibe93, HTP02} resulting in de facto different $\alpha$-formalisms. We assume $E_{\rm orb,i}$ and $E_{\rm gr}$ as given in the $\alpha$-formalism of \citet{Web84}, such that
\begin{equation}
E_{\rm orb,i} = \frac{GM_{\rm d}M_{\rm a}}{2a_{\rm i}},
\label{eq:Eorbi_web}
\end{equation}
and
\begin{equation}
E_{\rm gr} = \frac{GM_{\rm d} M_{\rm d,env}}{\lambda_{\rm ce} R},
\label{eq:Egr_web}
\end{equation} 
where $R$ is the radius of the donor star, $M_{\rm d,env}$ is the envelope mass of the donor and $\lambda_{\rm ce}$ depends on the structure of the donor \citep{deK87, Dew00, Xu10, Lov11}.

In the case of mass transfer between two giants with loosely bound envelopes, both envelopes can be lost simultaneously. This process is considered by binary\_c/nucsyn, SeBa and StarTrack.
The envelopes are expelled according to 
\begin{equation}
E_{\rm gr,d1} + E_{\rm gr, d2} = \alpha (E_{\rm orb,i}-E_{\rm orb,f}),
\label{eq:ce_dspi}
\end{equation} 
analogous to eq.\,\ref{eq:ce_a}, where  $E_{\rm gr, d1}$ and $E_{\rm gr, d2}$ represents the binding energy of the envelope of the two donor stars. This mechanism is termed a double CE-phase \citep{Bro95}. 

\section{Binary population synthesis codes}
\label{sec:BPS}
In this paper we compare the results of the simulations of four different BPS codes. These codes have been developed throughout the years with different scientific aims and philosophies, which has resulted in different numerical treatments and assumptions to describe binary evolution. An overview of the methods that are inherent to and the typical assumptions in the four BPS codes can be found in Appendix\,\ref{sec:TNS_inherent}~and~\ref{sec:TNS_equalized}. Below a short description is given of each code in alphabetical order:

\subsection{binary\_c/nucsyn\label{sec:binaryC}}
Binary\_c/nucsyn (binary\_c for future reference) is a rapid single star and binary population synthesis code with binary evolution based on \citet{HPT00,HTP02}. Updates and relevant additions are continuously made \citep{Izzard04,Izzard06,Izzard09, Cla13} to improve the code and to compare the effects of different prescriptions for ill-constrained physical processes. 
The most recent updates \citep{Cla13} that are relevant for this paper are a new formulation to determine the mass transfer rate, the accretion efficiency of WDs and the stability criteria for helium star donors and accreting WDs. The code uses analytical formulae based on detailed single star tracks at different metallicities \cite[based on ][]{Pols98,Karakas02}, with integration of different binary features \cite[based on BSE,][]{HTP02}. In addition, the code includes nucleosynthesis to follow the chemical evolution of binary systems and their output to the environment \citep{Izzard04,Izzard06,Izzard09}.

The code is used for different purposes, from the evolution of low-mass stars to high-mass stars. This includes the study of carbon- or nitrogen-enhanced metal-poor stars \cite[CEMP/NEMP-stars,][]{Izzard09,Pols12,Abate13}, the evolution of Barium stars \cite[]{Bona06,Izzard10}, progenitor studies of SNe Ia \citep{Cla13}, the study of rotation of massive stars \cite[]{Mink12} and recently the evolution of triple systems \cite[]{Hamers13}. Although the code has different purposes, the main strength of the code is the combination of a binary evolution code with nucleosynthesis which enables the study of not only the binary effects on populations, but also the chemical evolution of populations and its output to the environment.

\subsection{The Brussels code\label{sec:brussels}}
The Brussels binary evolution population number synthesis code has been under development for the better part of two decades, primarily to study the influence of binary star evolution on the chemical evolution of galaxies. A thorough review of the Brussels PNS code is given by \cite{DeD04}.

The population code uses actual binary evolution calculations (not analytical formulae) performed with the Paczy\'nski-based Brussels binary evolution code, developed over more than three decades at the Astrophysical Institute of the Vrije Universiteit Brussel. An important feature is that the effects of accretion on the further evolution of the secondary star are taken into account. The population code interpolates between the results of several thousands of actual binary evolution models, calculated under the assumption of the ``snowfall model'' by \cite{Neo77} in the case of direct impact, and assuming accretion induced full mixing \cite[see][]{Van94} if accretion occurs through a disk. The actual evolution models have been published by \cite{Van98}. The research done with the Brussels code mainly focuses on the chemical enrichment of galaxies caused by intermediate mass and massive binaries. Therefore the interpolations contained in the population code do not allow for the detailed evolution of stars with initial masses below 3 \Msolar.

In recent years, the code was mainly used to study the progenitors of Type Ia supernovae \citep{Men10,Men13}, the contribution of binaries to the chemical evolution of globular clusters \citep{Van12} and the influence of merging massive close binaries on Type II supernova progenitors \citep{Van13}.

\subsection{SeBa\label{sec:seba}}
SeBa is a fast binary population synthesis code that is originally
developed by \citet{Por96} with substantial updates from \citet{Nel01}, \citet{Too12} and \citet{Too13}. Recent updates include the metallicity dependent
single stellar evolution tracks of \citet{HPT00} for non-degenerate
stars, updated wind mass loss prescriptions and improved
prescriptions for hydrogen and helium accretion, and the stability of
mass transfer.

The philosophy of SeBa is to not a priori define evolution of the binary, but rather to determine this at runtime depending on
the parameters of the stellar system. When more sophisticated
models become available of processes that influence stellar evolution,
these can be included, and the effect can be studied without
altering the formalism of binary interactions. An example of this is the
stability criterion of mass transfer and the mass accretion 
efficiency. 

SeBa has been used to study a large range of stellar populations:
high mass binaries \citep{Por96}, double neutron stars \citep{Por98},
gravitational wave sources \citep{Por96gw, Nel01gw}, double white
dwarfs \citep{Nel01}, AM CVn systems \citep{Nel01b}, sdB stars \citep{Nel10}, SNIa progenitors
\citep{Too12, Bou13}, post-CE binaries \citet{Too13} and ultracompact X-ray binaries \citet{VanH13}.

As part of the software package Starlab, it has been used to simulate
the evolution of dense stellar systems \citep{Por01, Por04}. Recently,
SeBa is incorporated in the Astrophysics Multipurpose Software
Environment, or AMUSE. This is a component library with a homogeneous
interface structure, and can be downloaded for free at {\tt
amusecode.org} \citep{Por09}.

\subsection{StarTrack\label{sec:startrack}}
StarTrack is a Monte Carlo-based single and binary star rapid
evolution code. Stars are evolved at a given metallicity (range: $Z =
0.0001 - 0.03$) by adopting analytical fitting formulae from
evolutionary tracks of detailed single stellar models \citep{HPT00},
and modified over the years in order to incorporate the most important
physics for binary evolution. The orbital parameters (separation,
eccentricity and stellar spins) $a, e, \omega_{1}$ and $\omega_{2}$
are solved numerically as the system evolves, and re-distribution of
angular momentum determines how the orbit behaves. As physical
insights  regarding various aspects of stellar and binary  evolution
become available in the literature, new input physics can be
implemented into the code, and thus the code is continuously being
updated and improved.

The StarTrack code was originally used to predict physical properties of compact objects such and single and double black holes and neutron stars, 
as well as gamma ray bursts and compact object mergers in context of gravitational wave detection with {\em LIGO} \citep{Bel02a,Bel02b,Abb04}. In more recent years, studies with the code have grown to include compact binaries in globular clusters \citep{ivanova2005a}, X-ray binary populations \citep{belczynski2004a,RBH06}, sources of gravitational wave radiation for ground-based and space-based gravitational wave detectors \citep{ruiter2009a,ruiter2010a,BBB10,Bel10b}, gamma ray bursts \citep{belczynski2007a,osh2008,Bel08b}, Type Ia supernovae progenitors \citep{BBR05,Rui09,Rui11,ruiter2013a} and core-collapse supernova explosion mechanisms \citep{belczynski2012a}. The most comprehensive description of the code to date can be found in \citet{Bel08a}, with some updates described in \citet{Rui09} (SNe~Ia), \citet{belczynski2010c} (stellar winds), and \citet{dominik2012a} (wind mass-loss rates, CE). 

\section{Method}
\label{sec:method}
To examine the inherent differences between four BPS codes, we compare the results of a simulation made by these codes in which the assumptions are equalised as far as possible (Sect.\,\ref{sec:standass}). We consider two populations of binaries:

\begin{itemize}
\item Single WDs with a non-degenerate companion (hydrogen-rich or helium-rich star) (SWDs)
\item Double WD systems (DWDs)
\end{itemize}
Of both populations we investigate the initial distributions and the distributions at the moment that the SWD or DWD system forms. We establish the similarities between the results of the different BPS codes. If we notice differences between the results, we analyse these in greater detail by comparing e.g. the evolutionary paths or individual systems. 
A more detailed comparison of the populations of the most important evolutionary paths is given in Appendix\,\ref{sec:ev_path}. 

In the simulation, we assume an initial primary mass $M_{\rm 1, zams}$ between $M_{\rm 1, zams, min}=0.8$\Msolar~and
$M_{\rm 1, zams, max}=10$\Msolar, an initial mass ratio $q_{\rm zams} = M_{\rm 2,zams}/M_{\rm 1, zams}$ between $q_{\rm zams, min} = 0.1$\Msolar$/M_{\rm 1, zams}$ and $q_{\rm zams, max}=1$ and an initial semi-major axis $a_{\rm zams}$ between $a_{\rm zams, min}=5$\Rsolar~and $a_{\rm zams, max}=10^4$\Rsolar~(Table\,\ref{tbl:eq_assump}). Furthermore we assume an initial eccentricity $e_{\rm zams}$ of zero. We consider SWDs and DWDs that are formed within a Hubble time, more specifically 13.7 Gyr. The initial distribution of the primary masses follows
\cite{KTG93}, the initial mass ratio distribution is flat\footnote{Note that the initially imposed constraint on the mass ratio (i.e. 
 $q_{\rm zams,min} = 0.1$\Msolar$/M_{\rm 1, zams}$) affects the overall shape of 
 the resulting $q_{\rm zams}$-distribution. Even though the probability of drawing a mass ratio 
 anywhere is equal, this is strictly only true between $q_{\rm zams}\approx 0.1-1$. 
 Mass ratios lower than approximately 0.1 are drawn less often, since the primary masses 
 cluster around 1\Msolar~due to the IMF, and the lower mass limit of the secondary is assumed to be 0.1\Msolar.}, and the initial distribution of the semi-major
axis is flat in a logarithmic scale.

Not every BPS research group focuses on the full range of stellar masses. Consequently in their codes there are no (valid) prescriptions available for all stellar masses. The research group that uses the Brussels code, mainly focuses on the chemical enrichment of galaxies and therefore is not interested in the evolution of stars with a mass lower than 3\Msolar~(Sect.\,\ref{sec:brussels}).
Consequently, in order to make the comparison with the results of the Brussels code we only compare with a subset of the SWD and DWD populations. We define this subset as the `intermediate mass range', while the entire populations is considered as the `full mass range'.
The `intermediate mass range' is defined in the two populations as follows:
\begin{itemize}
\item for the \emph{SWD population} we only consider WDs originating from initial primary masses higher than 3\Msolar.
\item for the \emph{DWD population} we only consider WDs originating from initial primary and secondary masses both higher than 3\Msolar.
\end{itemize}

In addition, we refer to the 'low mass range' or 'low mass primaries' which encompasses the systems with an initial primary mass lower than 3\Msolar.

BPS codes are ideal to investigate the effect of different assumptions on populations, since a different assumption can cause a shift in e.g. the mass or separation of the population under investigation. We do not have to agree on the exact evolution of individual systems. As long as the shift is small the characteristics of the population do not change. Keeping this in mind when comparing the results of the different BPS codes we define them to agree when simulated populations (of similar evolutionary paths) are recovered at the same regions in the mass and separation space. 

\subsection{Assumptions for this project\label{sec:standass}}
\begin{table}
\footnotesize
\caption{Equalised initial distribution and range of binary parameters}\label{tbl:eq_assump}
\begin{center}
\begin{tabular}{|l|c|}
\hline
Parameter & Initial distribution \\
\hline \hline
$M_{\rm 1,zams}$ (\Msolar) & KTG93 \\
$a_{\rm zams}$ (\Rsolar)& $\propto a^{-1}$ (A83)\\
$q_{\rm zams}$ &  Flat \\
\hline

Parameter & Value\\
\hline \hline
$M_{\rm 1,zams,\rm min}$ & 0.8 (0.1)$^{(1)}$ \\
$M_{\rm 1, zams,\rm max}$ & 10 (100)$^{(1)}$  \\
$a_{\rm zams,\rm min}$ & 5 \\
$a_{\rm zams,\rm max}$ & 1e4 (1e6)$^{(1)}$\\
$q_{\rm zams, \rm min}$ & 0.1/M$_{\rm 1, zams}$\\
$q_{\rm zams, \rm max}$ & 1\\
$e_{\rm zams}$ & 0 \\
Max time (Gyr) & 13.7\\
Binary fraction (\%) &  100 \\
$\beta$ (RLOF) &  1\\
$\alpha_{\rm ce}\lambda_{\rm ce}$ &  1\\
\hline
Physics & Assumption \\
\hline \hline
CE & $\alpha^{(2)}$\\
Wind accretion & No\\
Tides & No\\
Magn. braking & No\\

\hline
\end{tabular}
\tablefoot{ 
\\References in the table: KTG93 = \cite{KTG93}, A83 = \cite{Abt83}.
\\(1) The values outside and inside the brackets represent the values for the simulated and entire stellar population, respectively. 
\\(2) The prescription is based on \cite{Web84}.}
\end{center}
\end{table}

In order to compare the codes we make the most simple assumptions. These are not necessarily believed to be realistic, but are taken to make the comparison feasible. The assumptions for this project are discussed below and shown in Table\,\ref{tbl:eq_assump}.
The typical assumptions taken by the authors in the corresponding BPS codes in their previous research projects are summarised in Table\,\ref{tbl:TS_all} in Appendix\,\ref{sec:TNS_equalized}. For simplicity and brevity, we do not study the effect of these assumptions on the characteristics of SWD and DWD populations in this project. 

\begin{itemize}
\item Mass transfer is assumed to be conservative ($\beta$ = 1) during stable RLOF towards all types of objects. We emphasise that this is not a realistic assumption, especially in the case of a WD accretor. During the CE-phase no material is assumed to be accreted by the companion star ($\beta$ = 0). 

In the Brussels code a constant accretion efficiency of a WD-accretor cannot be implemented and therefore for this study mass transfer to all compact objects is assumed to be unstable and evolve into a CE-phase in this code.

\item As no mass nor angular momentum is lost from RLOF, we do not require an assumption for the specific angular momentum loss of the material. During wind mass loss, we assume the wind matter leaves the system with the specific angular momentum of the donor star. 
However, this assumptions is not possible in the Brussels code (for an overview of the assumptions see Sect.\,\ref{sec:AMLwind}.

\item We use the $\alpha$-prescription of \cite{Web84} to describe the CE-phase (eq.\,\ref{eq:ce_a},~\ref{eq:Eorbi_web}~and~\ref{eq:Egr_web}). We assume that the parameters $\alpha_{ce}$ and $\lambda_{ce}$ are equal to one, mainly for simplicity, but also because the prevalence of this choice in the literature allows for comparison between this and other studies.

\item We assume that matter lost through winds cannot be accreted by the companion star. 

\item Due to the diversity of the prescriptions for magnetic braking
  and tides, we do not consider these effects and they are turned off
  for this paper. However, in StarTrack spin-orbit coupling is still taken into account, as it is firmly integrated with the binary evolution equations.
  
\end{itemize}

\subsection{Normalisation\label{sec:grid}}
When calculating birthrates of evolutionary channels, the simulation has to be normalised to an entire stellar population (Table\,\ref{tbl:eq_assump}). For this work the initial distribution and ranges of $M_{\rm 1, zams}$, $q_{\rm zams}$ and $a_{\rm zams}$ are as discussed in Sect.\,\ref{sec:method} with the exception of the initial primary masses of a stellar population to vary between 0.1 and 100\Msolar, and the semi-major axis between 5 and 10$^{6}$\Rsolar. We assume a binary fraction of 100\%. 

If the star formation rate $S$ in \Msolar$\peryr$ is independent of time, the birthrate of a specific binary type~X (e.g. systems evolving through a specific evolutionary channel) is given by:
\begin{equation}
{\rm Birthrate(X)} = S \frac{\phi(\rm X)}{M_{\rm tot}},
\label{eq:birthrate}
\end{equation} 
with $\phi(\rm X)$ the total number of systems of binary type~X in the simulation, and $M_{\rm tot}$ the total mass of all stellar systems in the entire stellar population. More specifically,

\begin{equation}
\phi({\rm X}) = \int_{0.1}^{100} \int_{0.1/M_{\rm 1, zams}}^{1} \int_{5}^{1e6} x \Psi dM_{\rm 1, zams} dq  da,
\label{eq::intrate}
\end{equation}

with $x=x(M_{\rm 1,zams}, q, a)$ equals 1 for binary systems of type~X, and zero otherwise and $\Psi$ is the initial distribution function of $M_ {\rm 1,zams}$, $q_ {\rm zams}$ and $a_{\rm zams}$. Note that in this project we assume that the initial distribution for $M_{\rm 1,zams}$, $q_{\rm zams}$ and $a_{\rm zams}$ are independent (Table\,\ref{tbl:eq_assump}), such that $\Psi$ is separable:
\begin{equation}
\Psi(M_{\rm 1, zams}, q_{\rm zams}, a_{\rm zams}) = \psi(M_{\rm 1,zams})\varphi(q_{\rm zams})\chi(a_{\rm zams}).
\label{eq:psi}
\end{equation}

The total mass of all stellar systems assuming a 100\% binary fraction is:
\begin{equation}
M_{\rm tot} = \int_{0.1}^{100} \int_{0.1/M_{\rm 1,zams}}^{1} \int_{5}^{1e6} M_{\rm t,zams} \Psi dM_{\rm 1,zams} dq da,
\label{eq:Mbin}
\end{equation} 

where $M_{\rm t, zams} = M_{\rm 1,zams}+M_{\rm 2, zams}$.

For this project a constant star formation rate of 1\Msolar$\peryr$ is assumed. This simple star formation rate is chosen to make the comparison with other codes easier.
\begin{table*}
\caption{Birthrates in $\peryr$ for single and double white
  dwarf systems for the three BPS codes for the full mass range and
  the intermediate mass range. }
\begin{tabular}{|l||c|c|c||c|c|c|c|}\hline
 & \multicolumn{3}{|c||}{Full mass range}  & \multicolumn{4}{|c|}{Intermediate mass range} \\
\hline
 & binary\_c & SeBa & StarTrack & binary\_c & Brussels code & SeBa & StarTrack\\
 \hline  \hline 
SWD systems & 0.048 & 0.052 & 0.048 &   5.1e-3 &  7.8e-3 & 5.2e-3 & 4.4e-3\\
DWD systems & 0.012 & 0.014 & 0.015 &   8.4e-4   &  1.1e-3 & 8.7e-4  &  6.6e-4\\
 \hline 
 \end{tabular}
\label{tbl:birthrates}
\end{table*}

\section{Comparison}
\label{sec:Results}
\subsection{Single white dwarf systems}
\label{sec:swd}


    \begin{figure*}[tbh]
    \centering
    \setlength\tabcolsep{0pt}
    \begin{tabular}{ccc}
	\includegraphics[height=4.6cm, clip=true, trim =8mm 0mm 48.5mm 5mm]{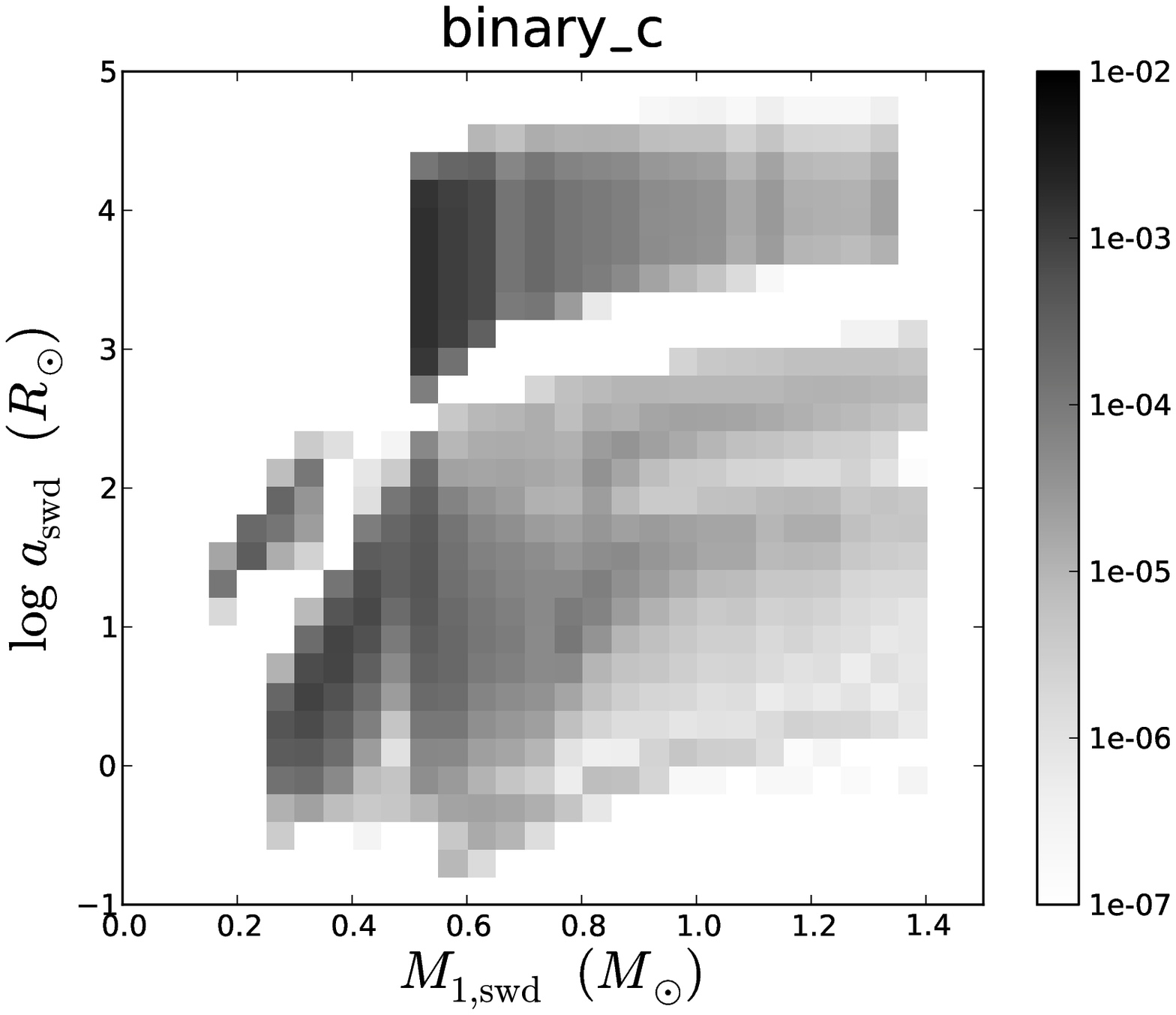} &
	\includegraphics[height=4.6cm, clip=true, trim =20mm 0mm 48.5mm 5mm]{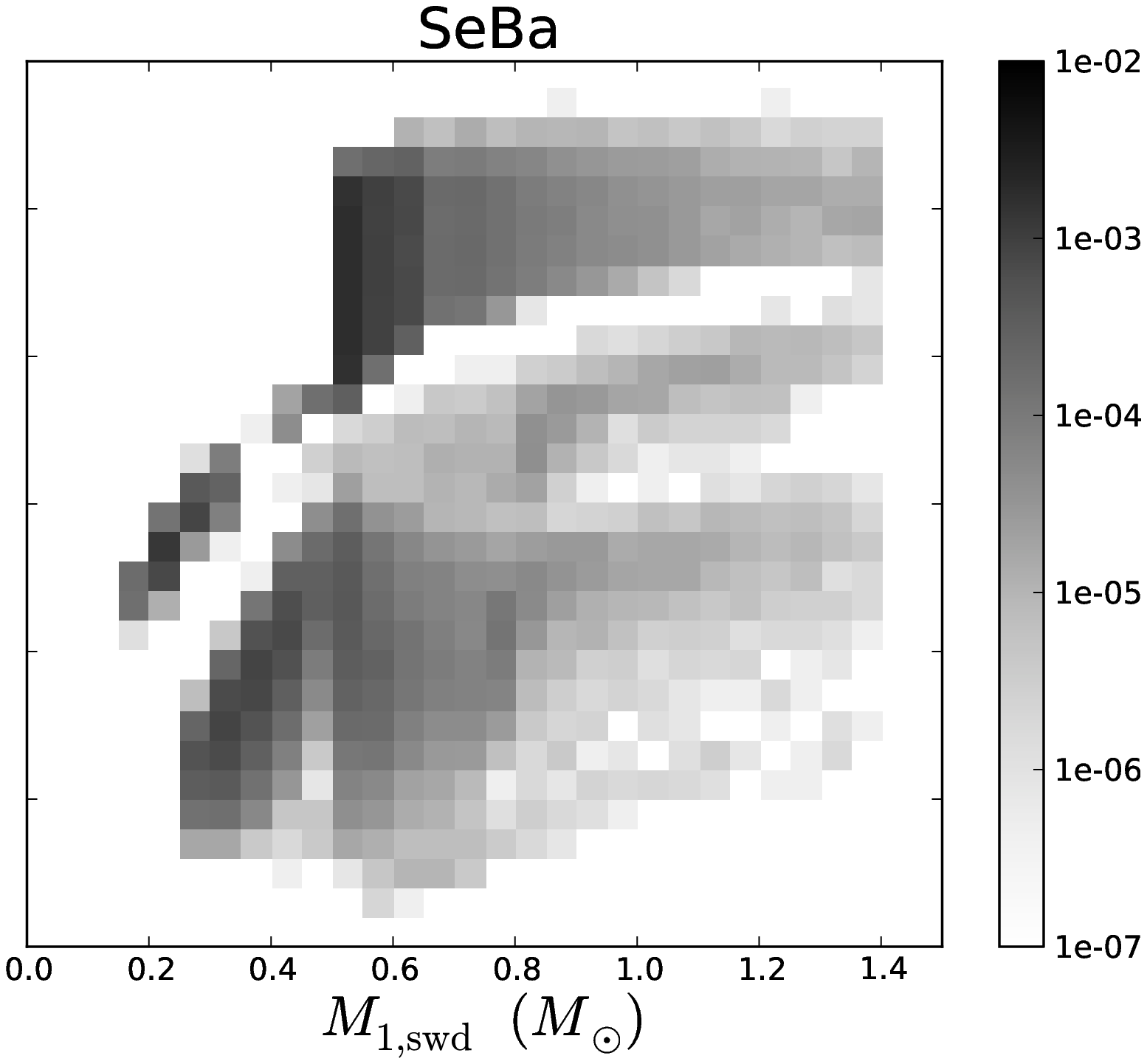} & 
	\includegraphics[height=4.6cm, clip=true, trim =20mm 0mm 23mm 5mm]{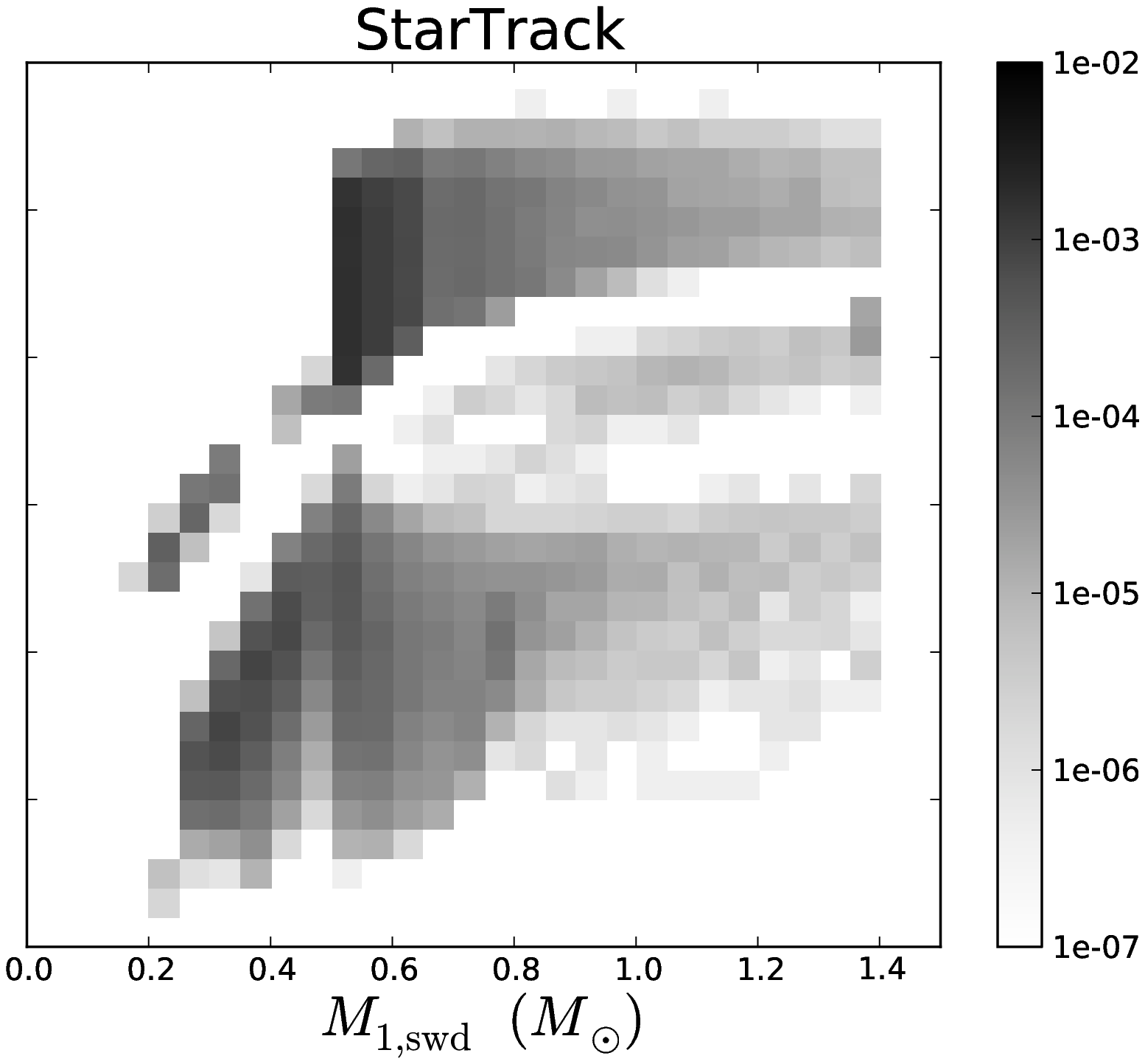} \\
	\end{tabular}
    \caption{Orbital separation versus WD mass for all SWDs in the full mass range at the time of SWD formation. } 
    \label{fig:swd_final_a_sin}
    \end{figure*}

    \begin{figure*}[tbh]
    \centering
    \setlength\tabcolsep{0pt}
    \begin{tabular}{cccc}
	\includegraphics[height=4.6cm, clip=true, trim =8mm 0mm 48.5mm 5mm]{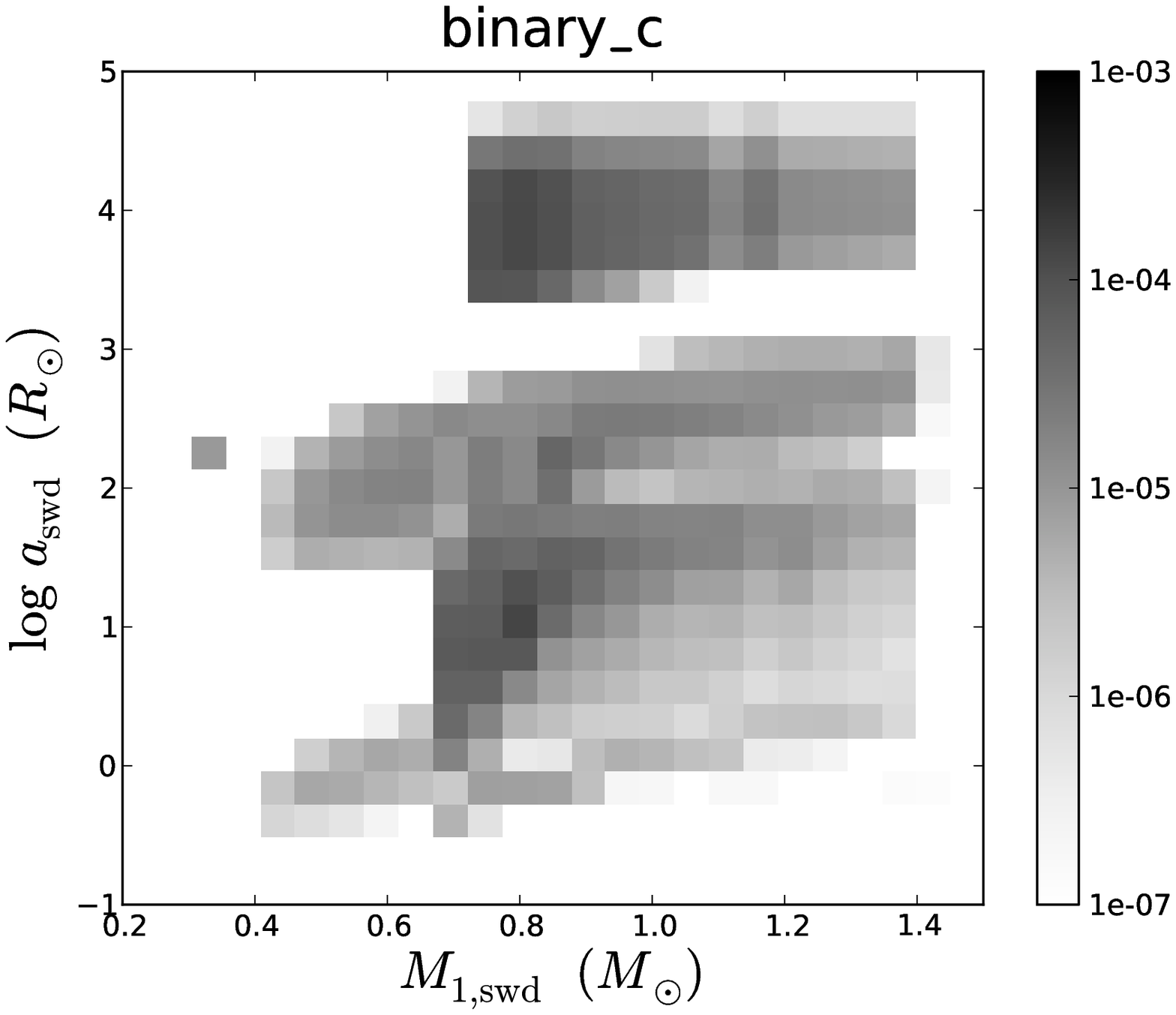} &
	\includegraphics[height=4.6cm, clip=true, trim =20mm 0mm 48.5mm 5mm]{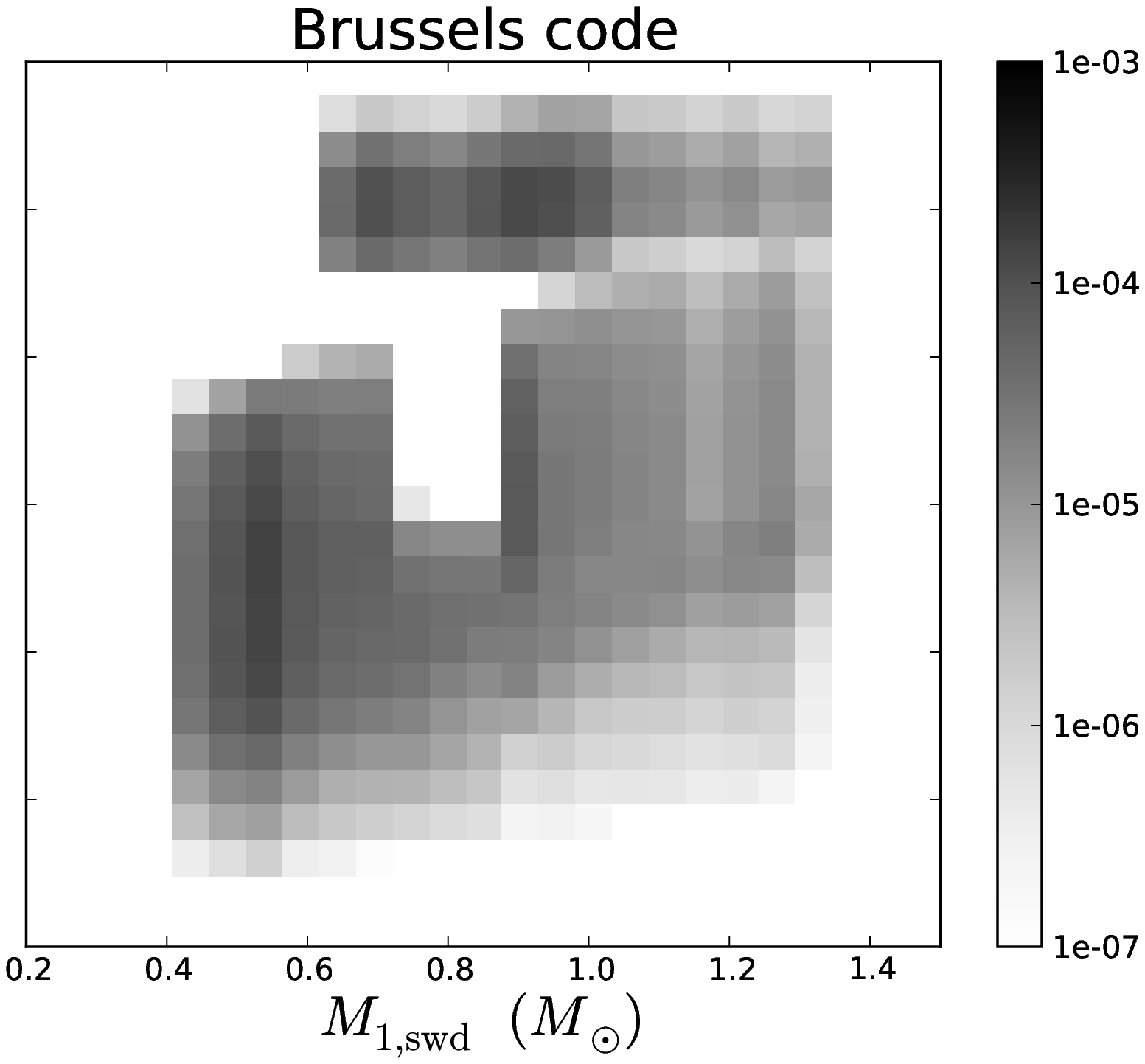} &
	\includegraphics[height=4.6cm, clip=true, trim =20mm 0mm 48.5mm 5mm]{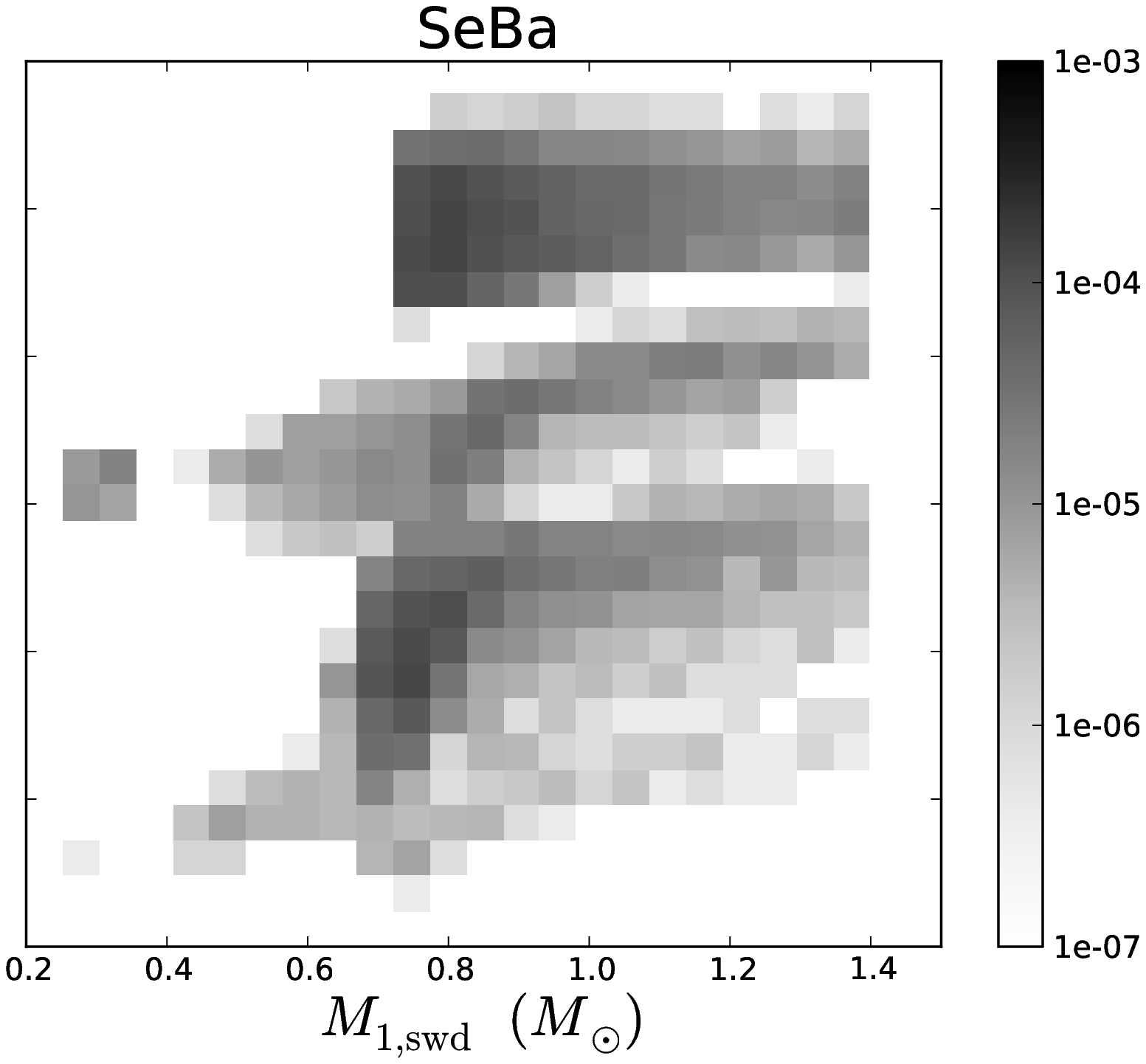} &
	\includegraphics[height=4.6cm, clip=true, trim =20mm 0mm 23mm 5mm]{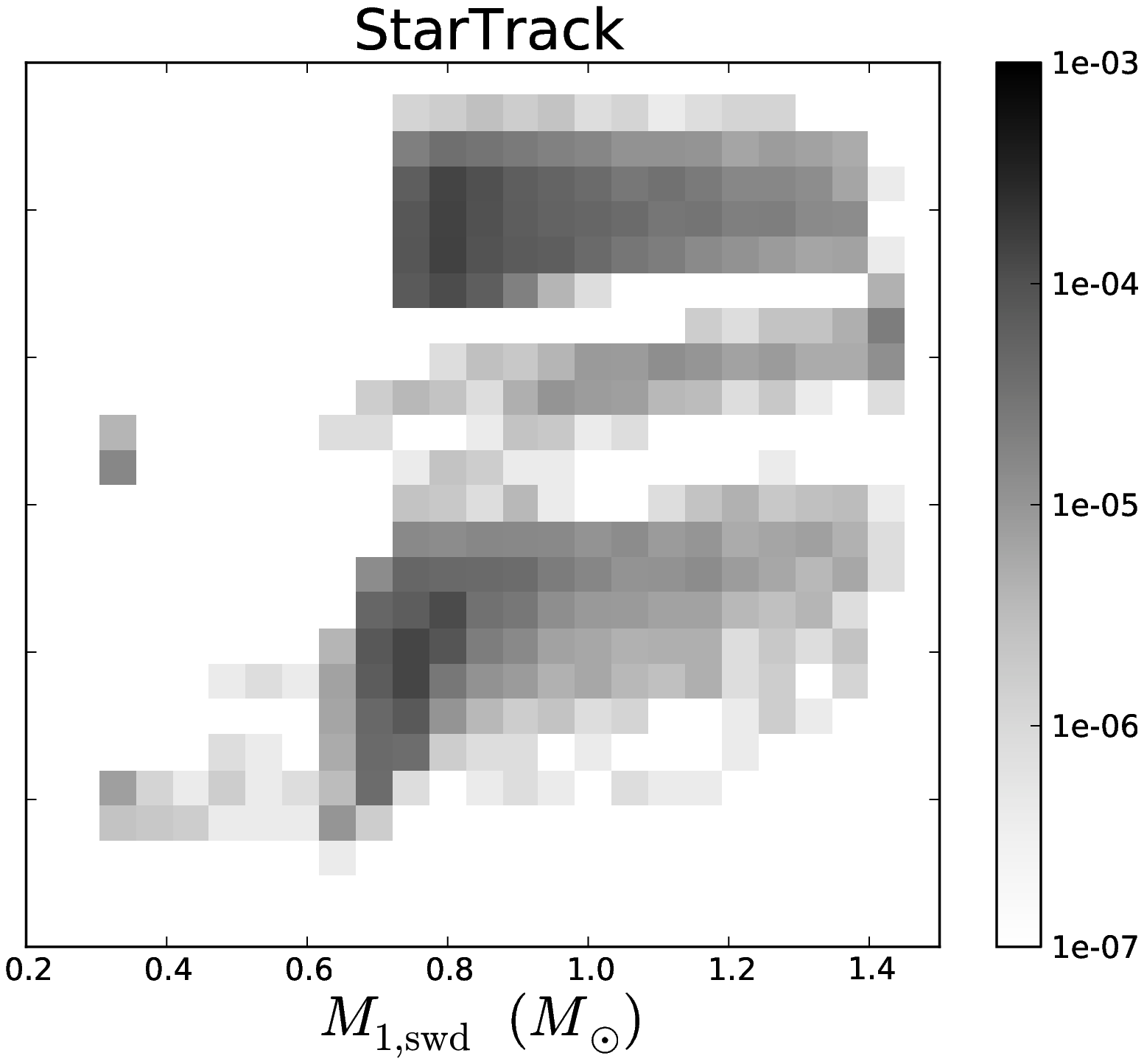} \\
	\end{tabular}
    \caption{Orbital separation versus WD mass for all SWDs in the intermediate mass range at the time of SWD formation. }
    \label{fig:swd_final_a_IM_sin}
    \end{figure*}

     \begin{figure*}[tbh]
    \centering
    \setlength\tabcolsep{0pt}
    \begin{tabular}{ccc}
	\includegraphics[height=4.6cm, clip=true, trim =8mm 0mm 48.5mm 5mm]{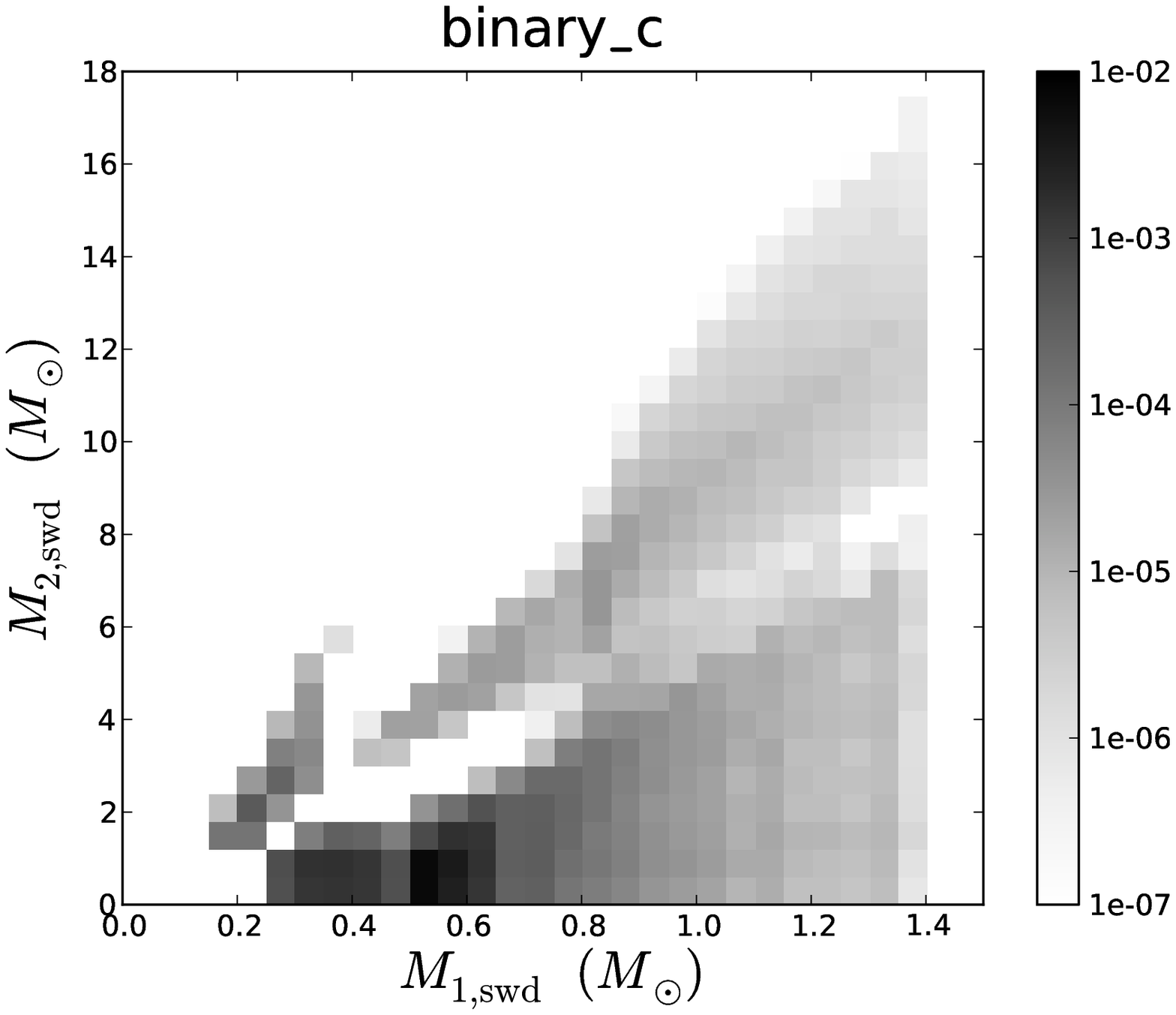} &
	\includegraphics[height=4.6cm, clip=true, trim =20mm 0mm 48.5mm 5mm]{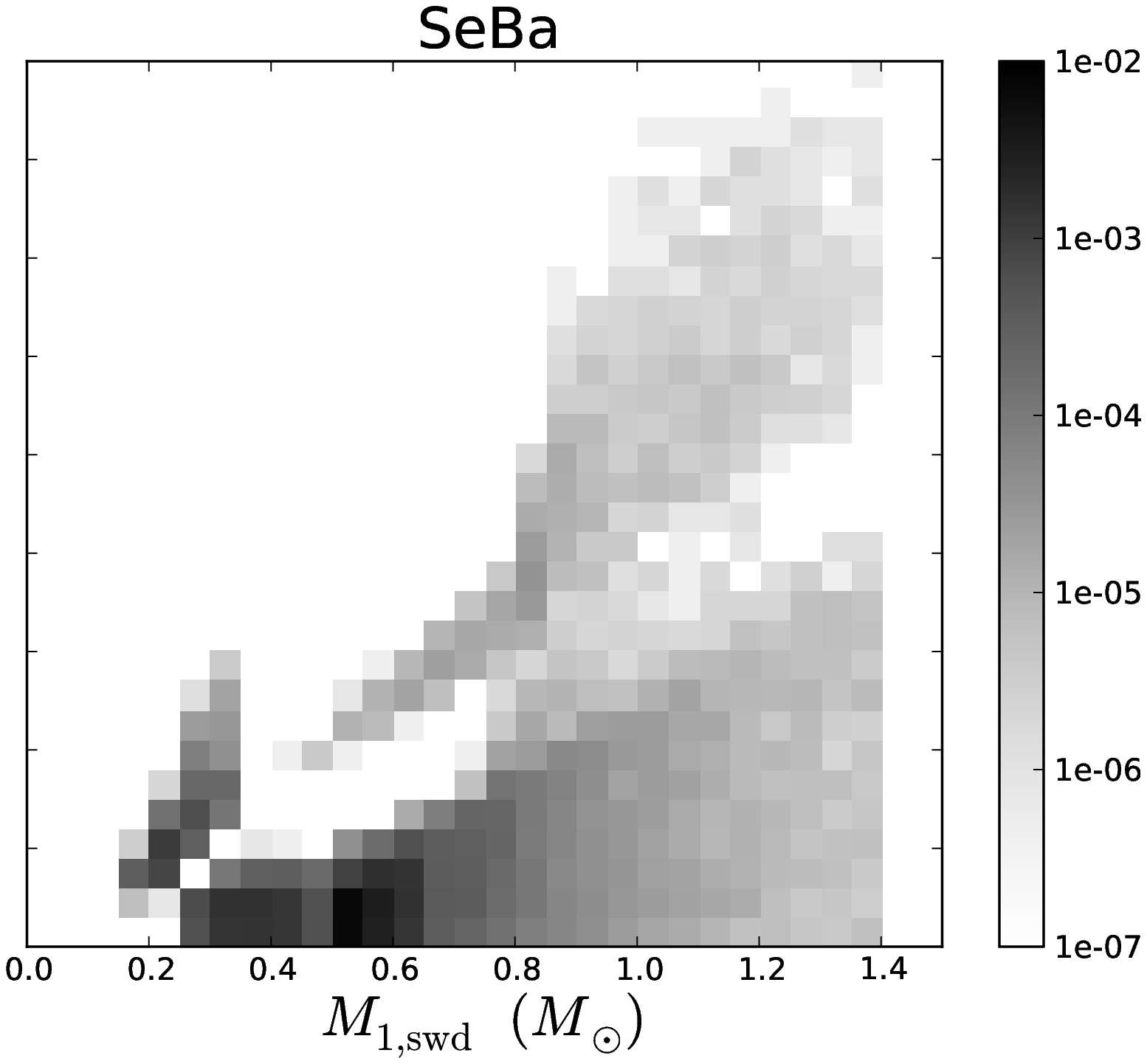} &
	\includegraphics[height=4.6cm, clip=true, trim =20mm 0mm 23mm 5mm]{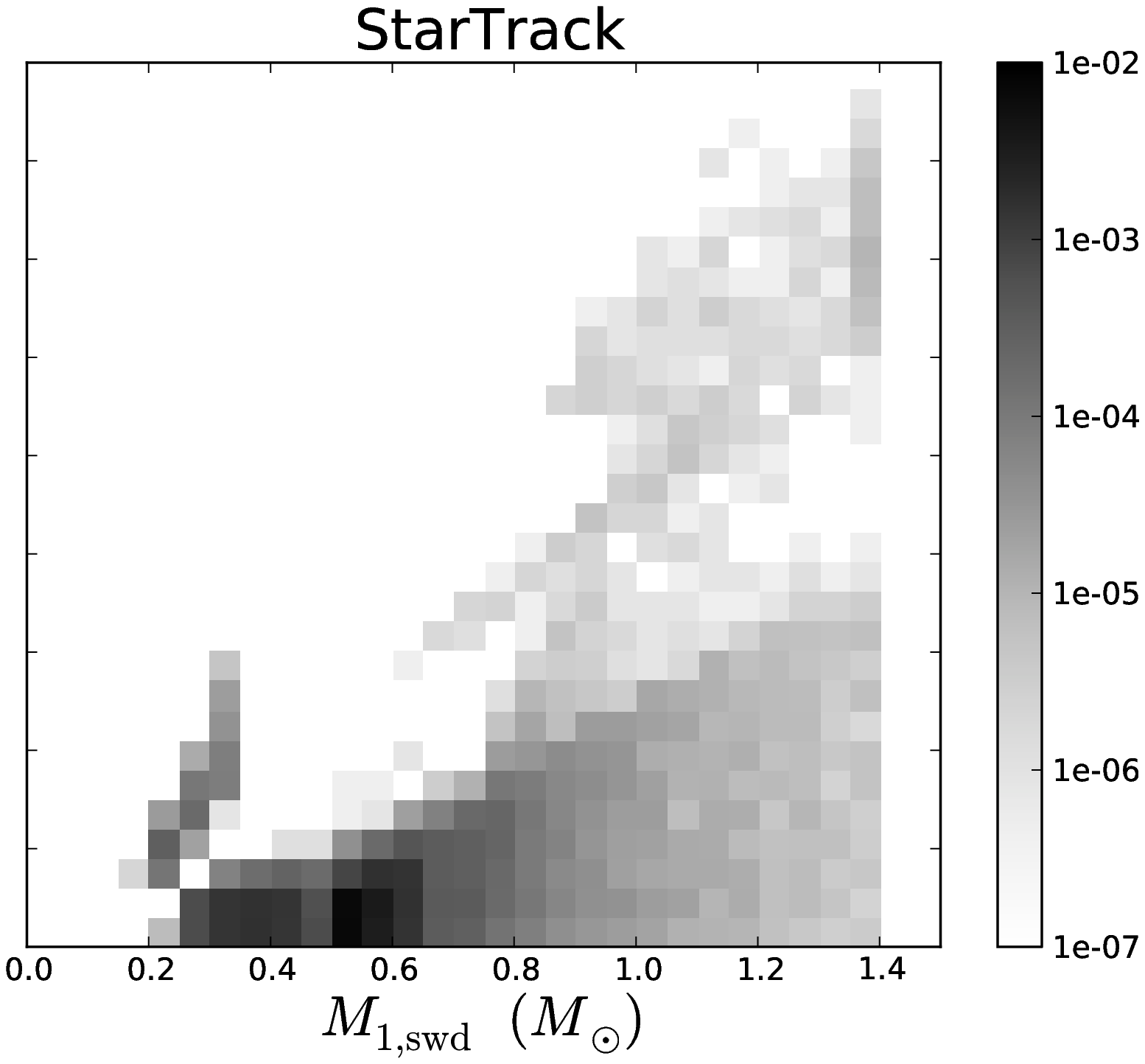} \\
	\end{tabular}
    \caption{Secondary mass versus WD mass for all SWDs in the full mass range at the time of SWD formation. }
    \label{fig:swd_final_m2_sin}
    \end{figure*}

     \begin{figure*}[tbh]
    \centering
    \setlength\tabcolsep{0pt}
    \begin{tabular}{cccc}
	\includegraphics[height=4.6cm, clip=true, trim =8mm 0mm 48.5mm 5mm]{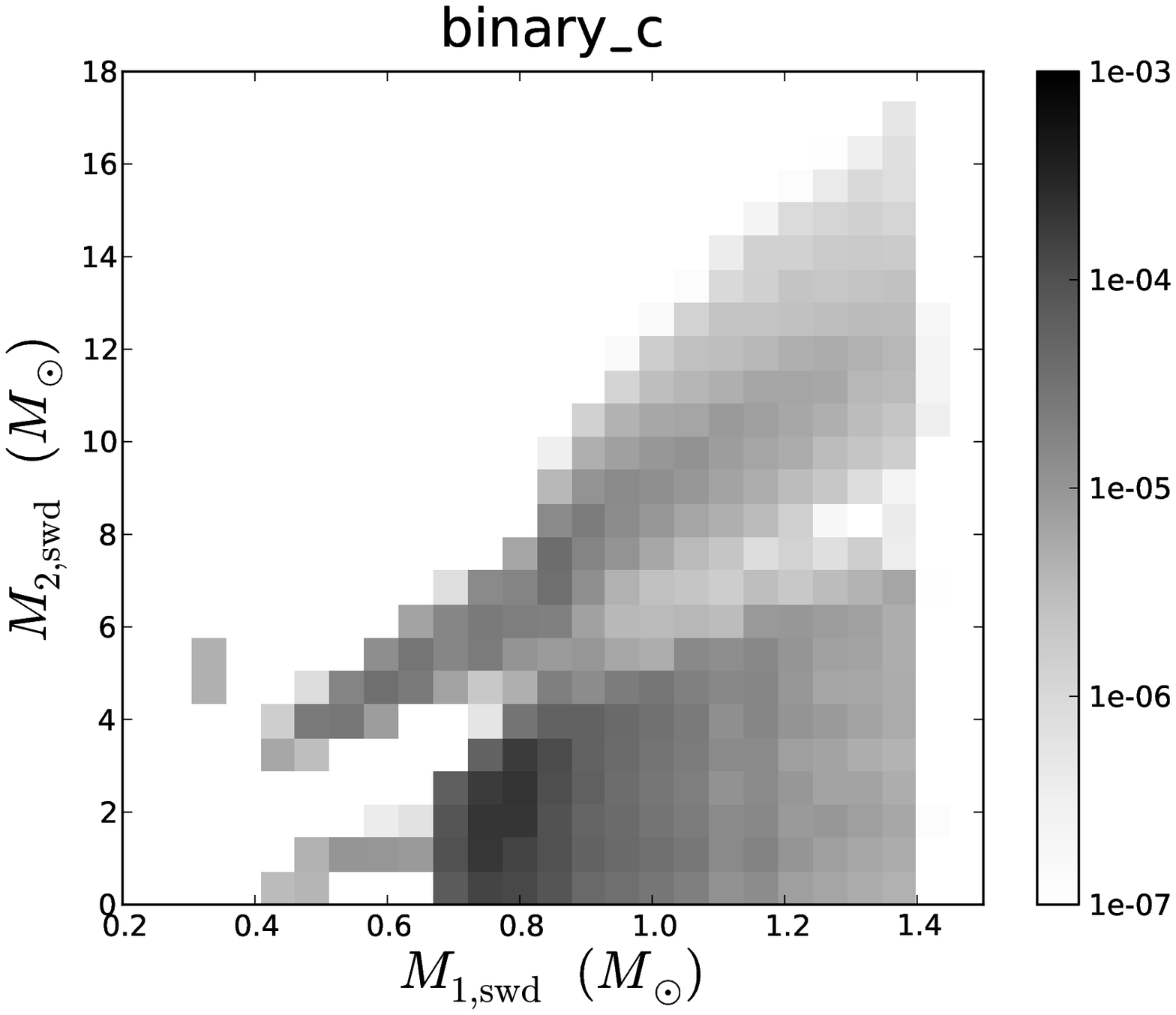} &
	\includegraphics[height=4.6cm, clip=true, trim =20mm 0mm 48.5mm 5mm]{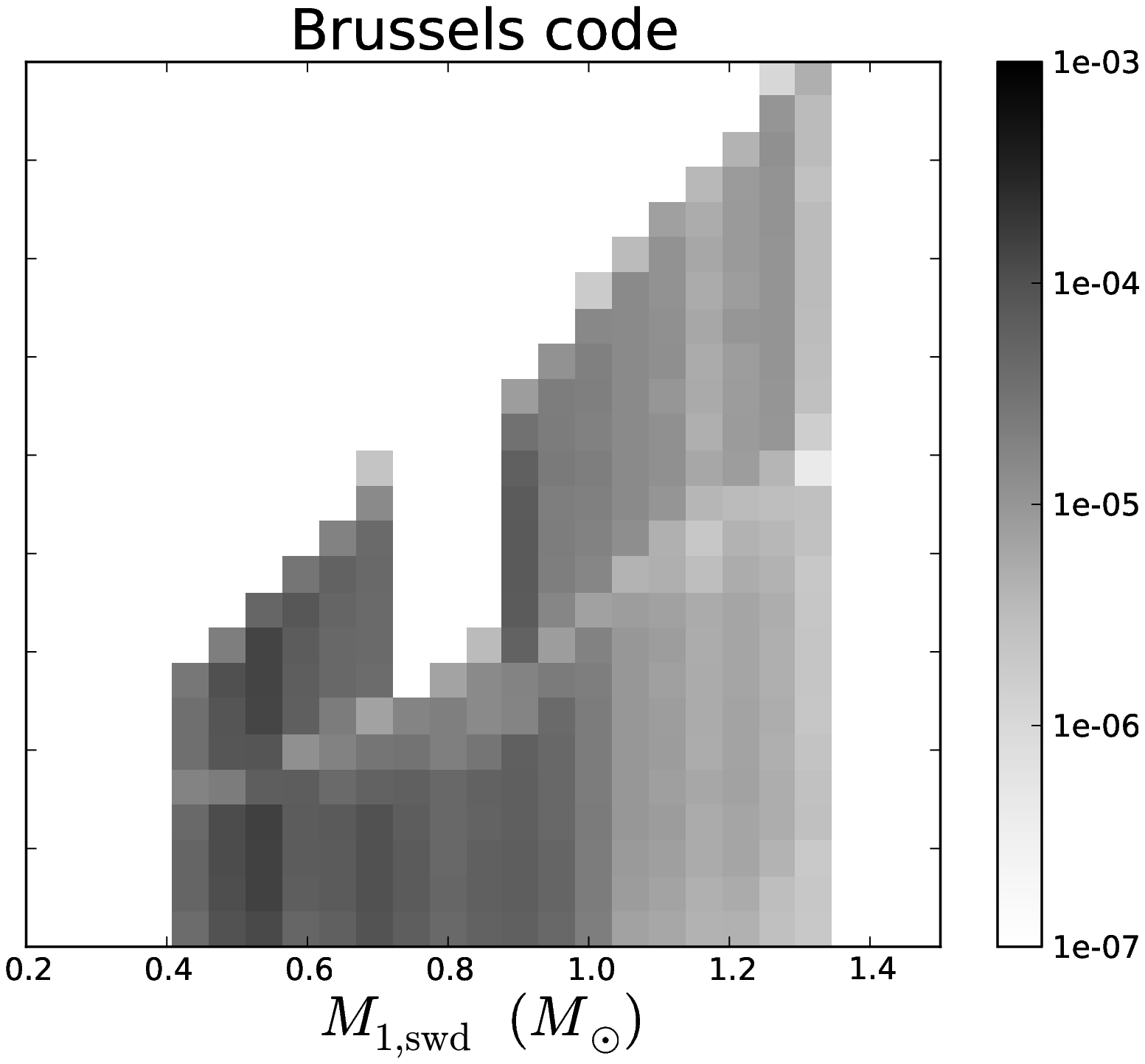} &
	\includegraphics[height=4.6cm, clip=true, trim =20mm 0mm 48.5mm 5mm]{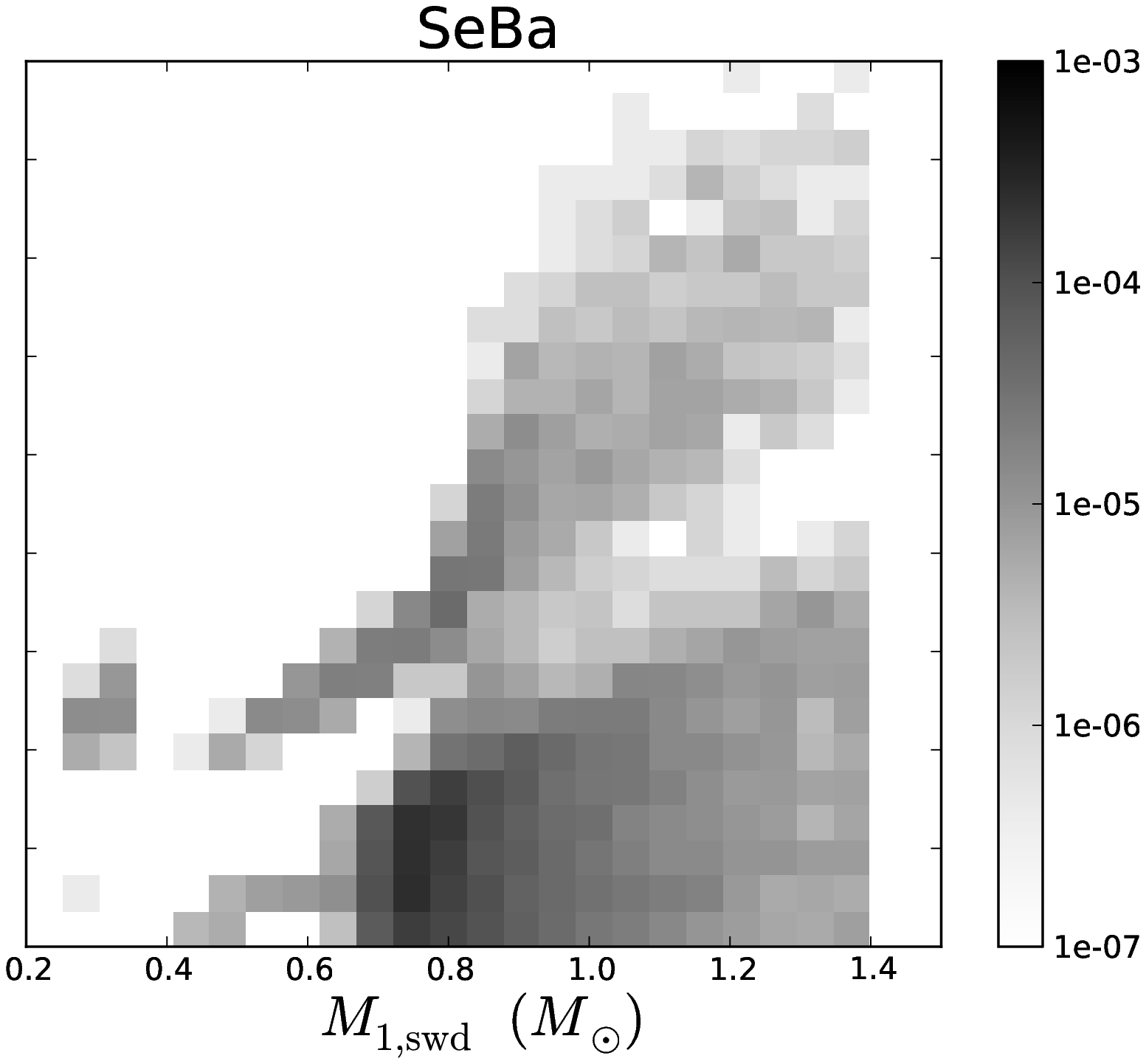} &
	\includegraphics[height=4.6cm, clip=true, trim =20mm 0mm 23mm 5mm]{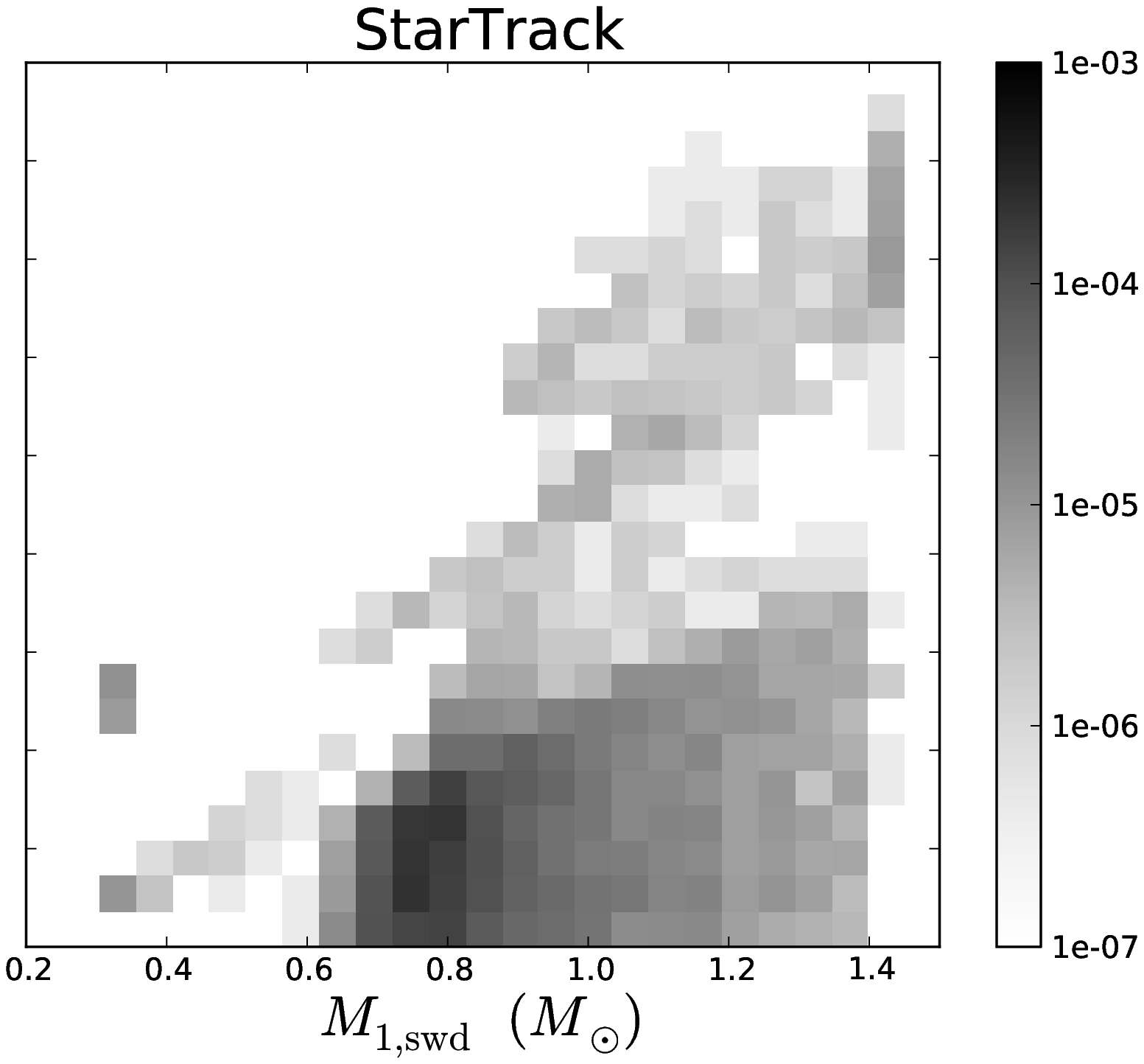} \\	
	\end{tabular}
    \caption{Secondary mass versus WD mass for all SWDs in the intermediate mass range at the time of SWD formation. }
    \label{fig:swd_final_m2_IM_sin}
    \end{figure*}

    \begin{figure*}[tbh]
    \centering
    \setlength\tabcolsep{0pt}
    \begin{tabular}{ccc}
	\includegraphics[height=4.6cm, clip=true, trim =8mm 0mm 48.5mm 5mm]{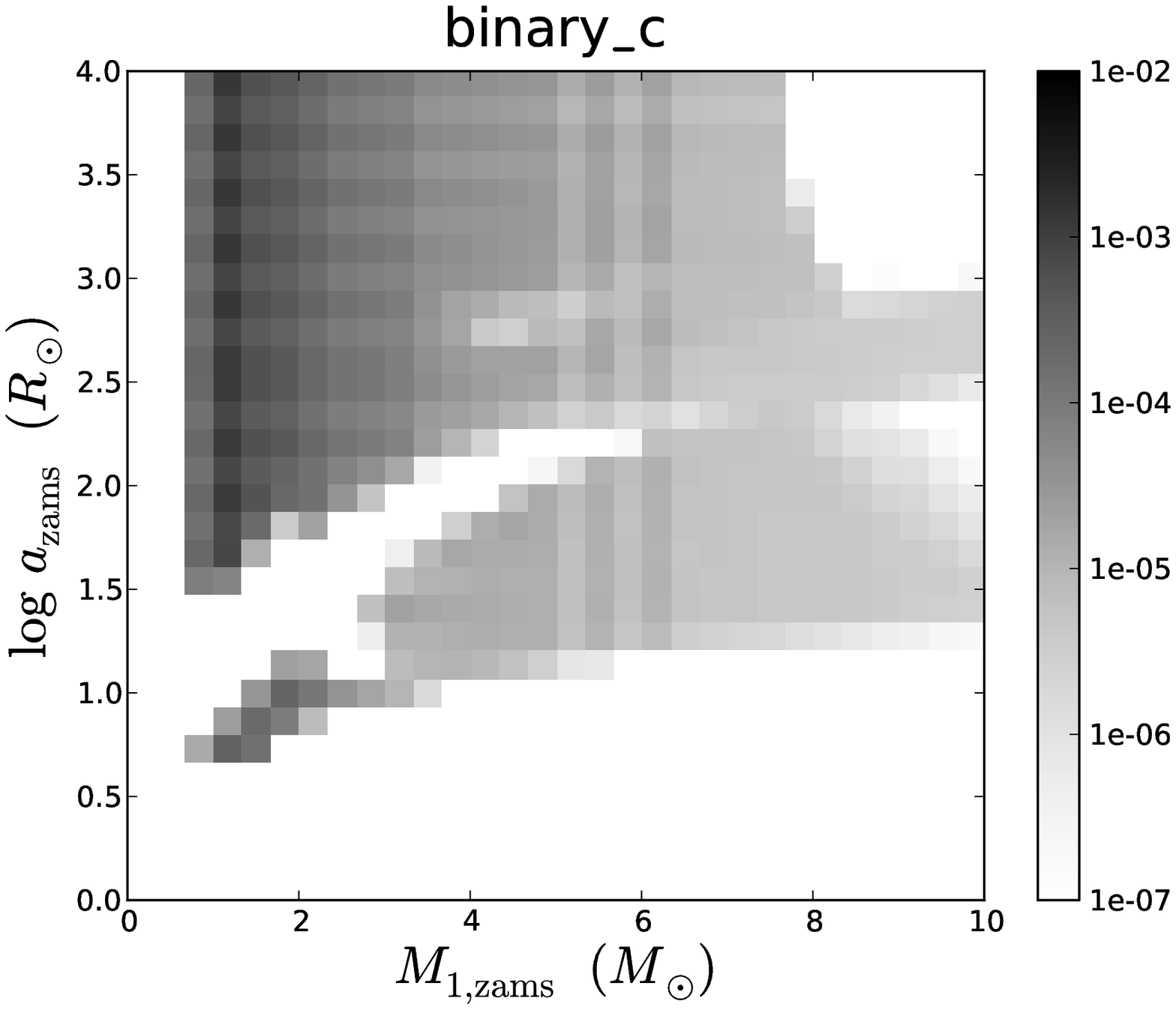} &
	\includegraphics[height=4.6cm, clip=true, trim =20mm 0mm 48.5mm 5mm]{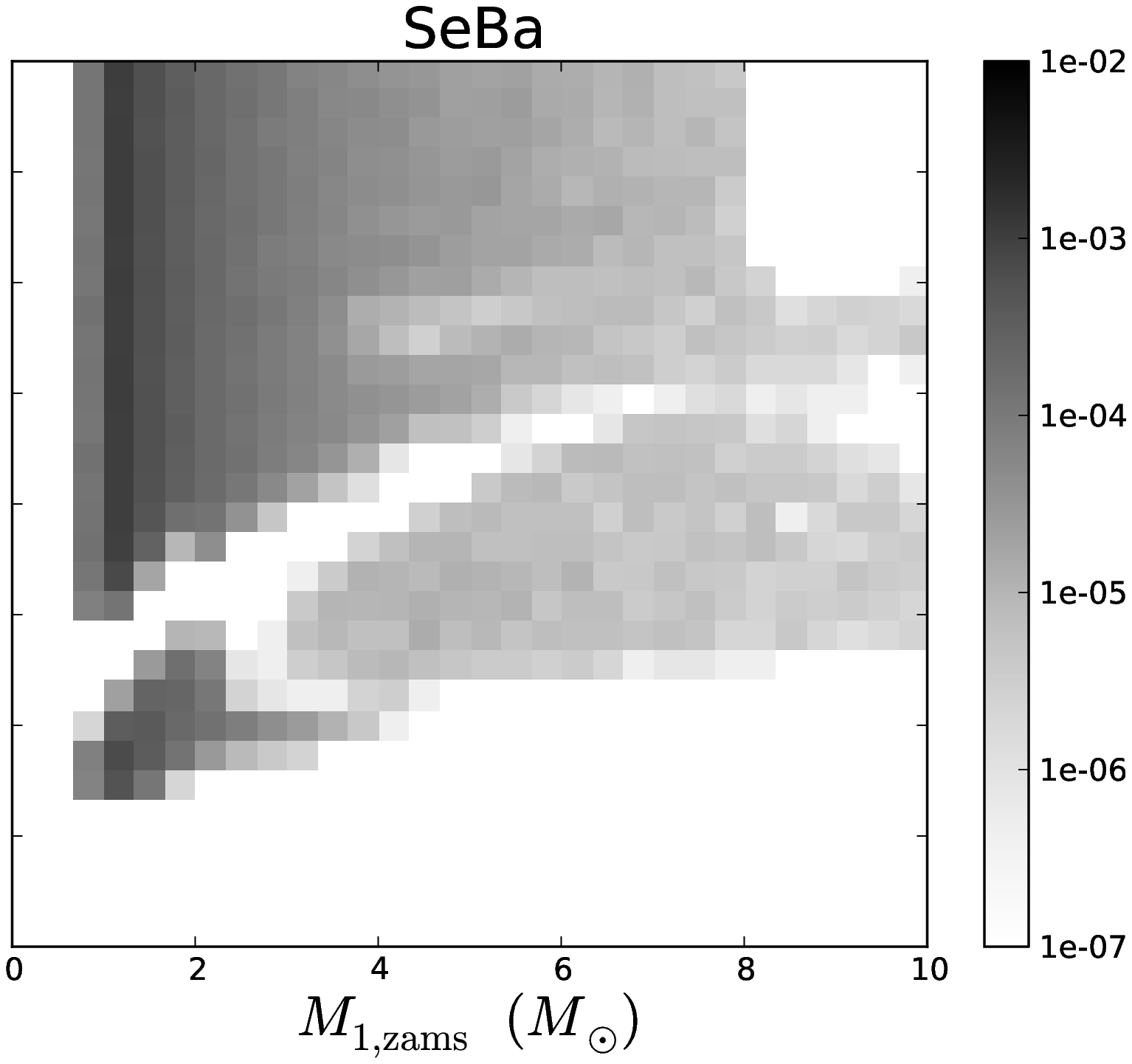} & 
	\includegraphics[height=4.6cm, clip=true, trim =20mm 0mm 23mm 5mm]{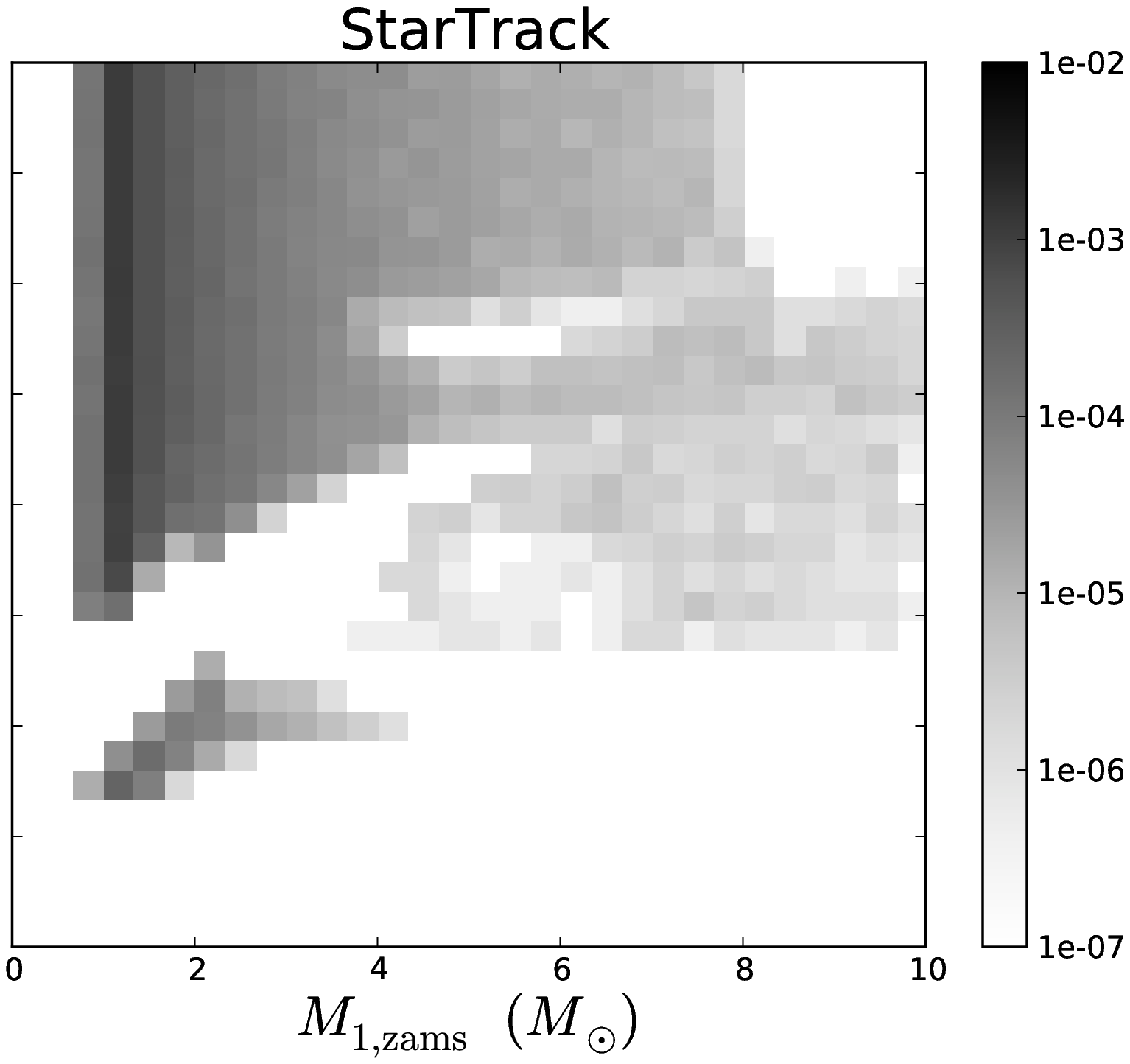} \\
	\end{tabular}
    \caption{Initial orbital separation versus initial primary mass for all SWDs in the full mass range. } 
    \label{fig:swd_zams_a_sin}
    \end{figure*}

    \begin{figure*}[tbh]
    \centering
    \setlength\tabcolsep{0pt}
    \begin{tabular}{cccc}
	\includegraphics[height=4.6cm, clip=true, trim =8mm 0mm 48.5mm 5mm]{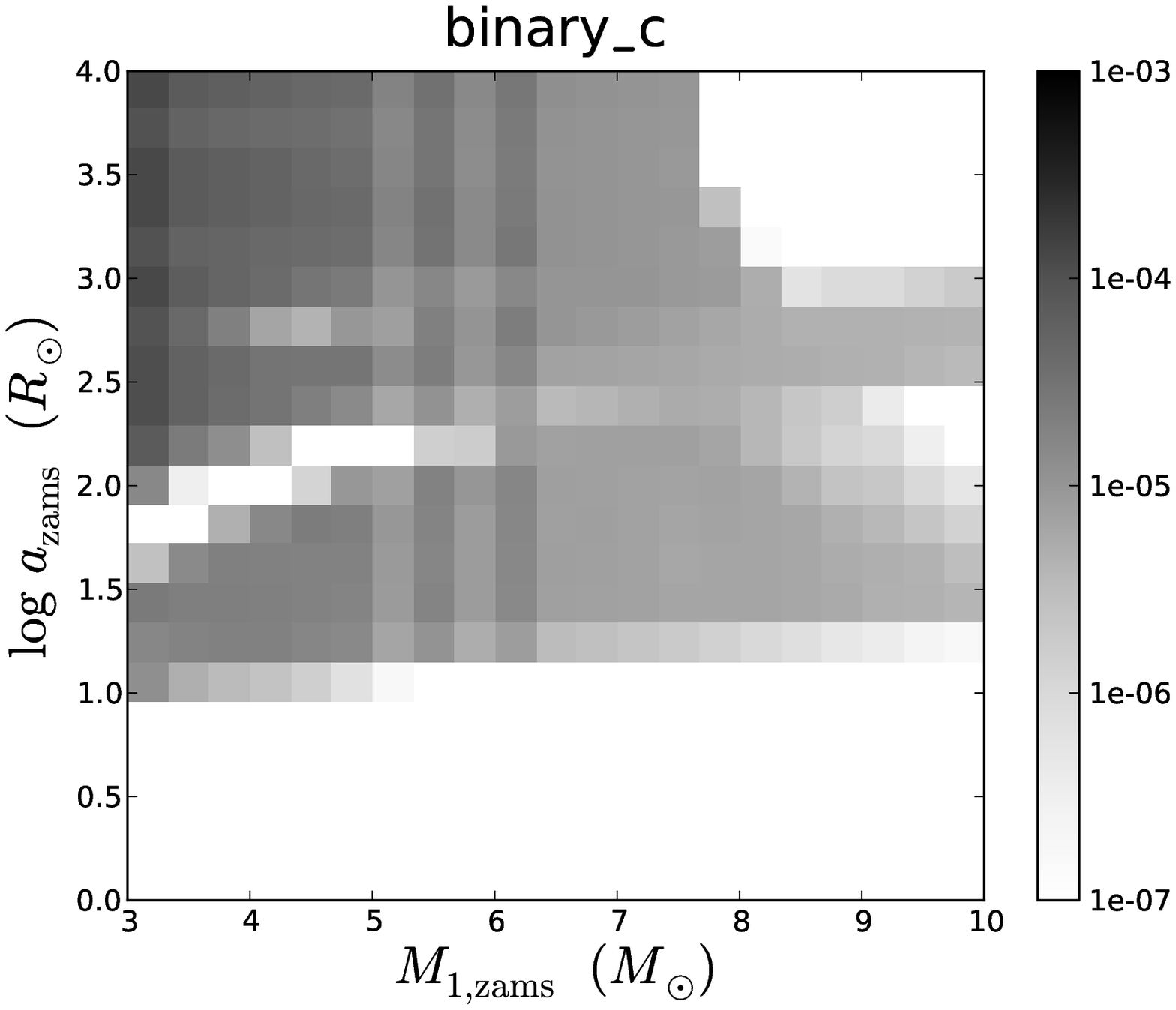} &
	\includegraphics[height=4.6cm, clip=true, trim =20mm 0mm 48.5mm 5mm]{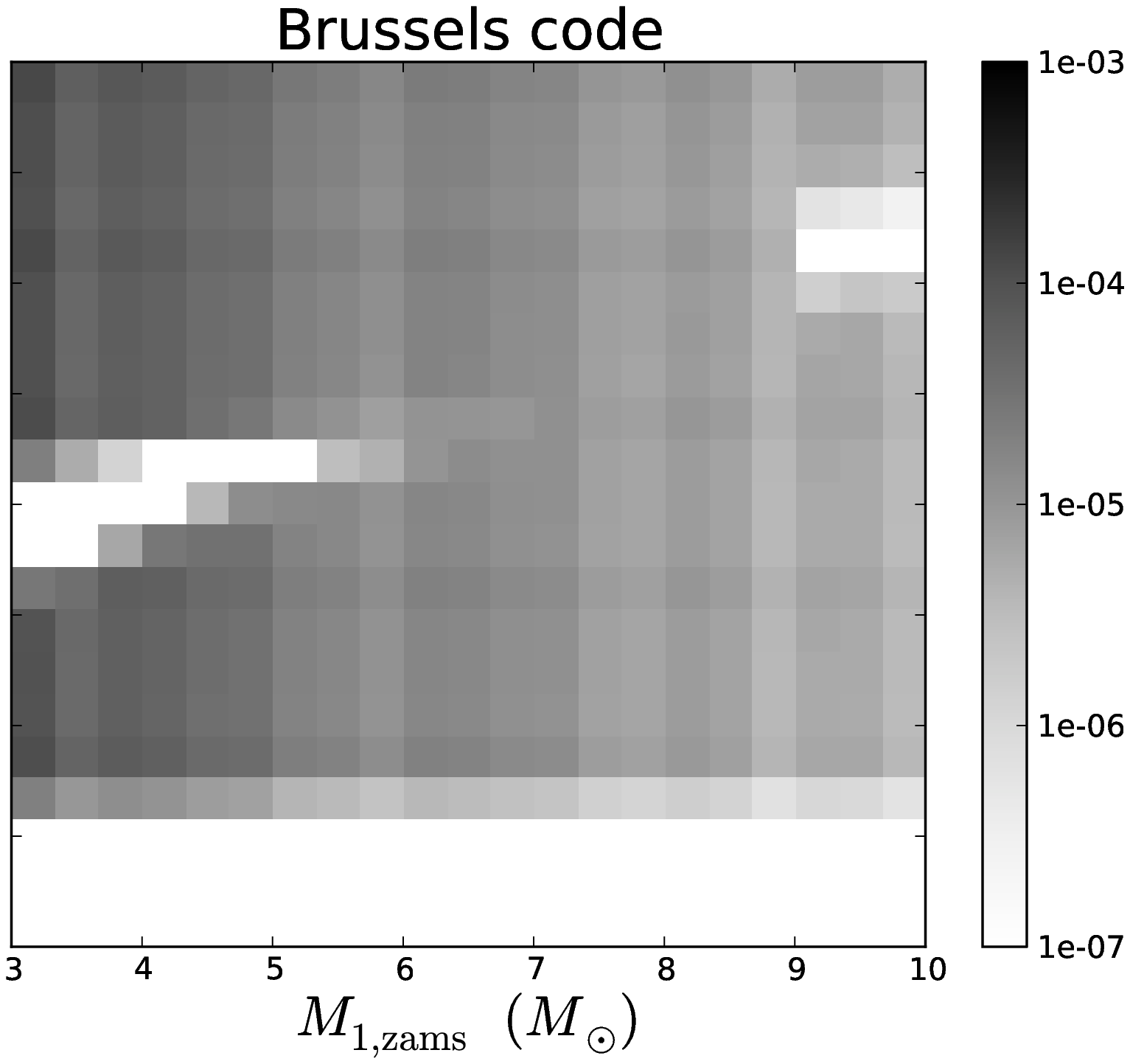} &
	\includegraphics[height=4.6cm, clip=true, trim =20mm 0mm 48.5mm 5mm]{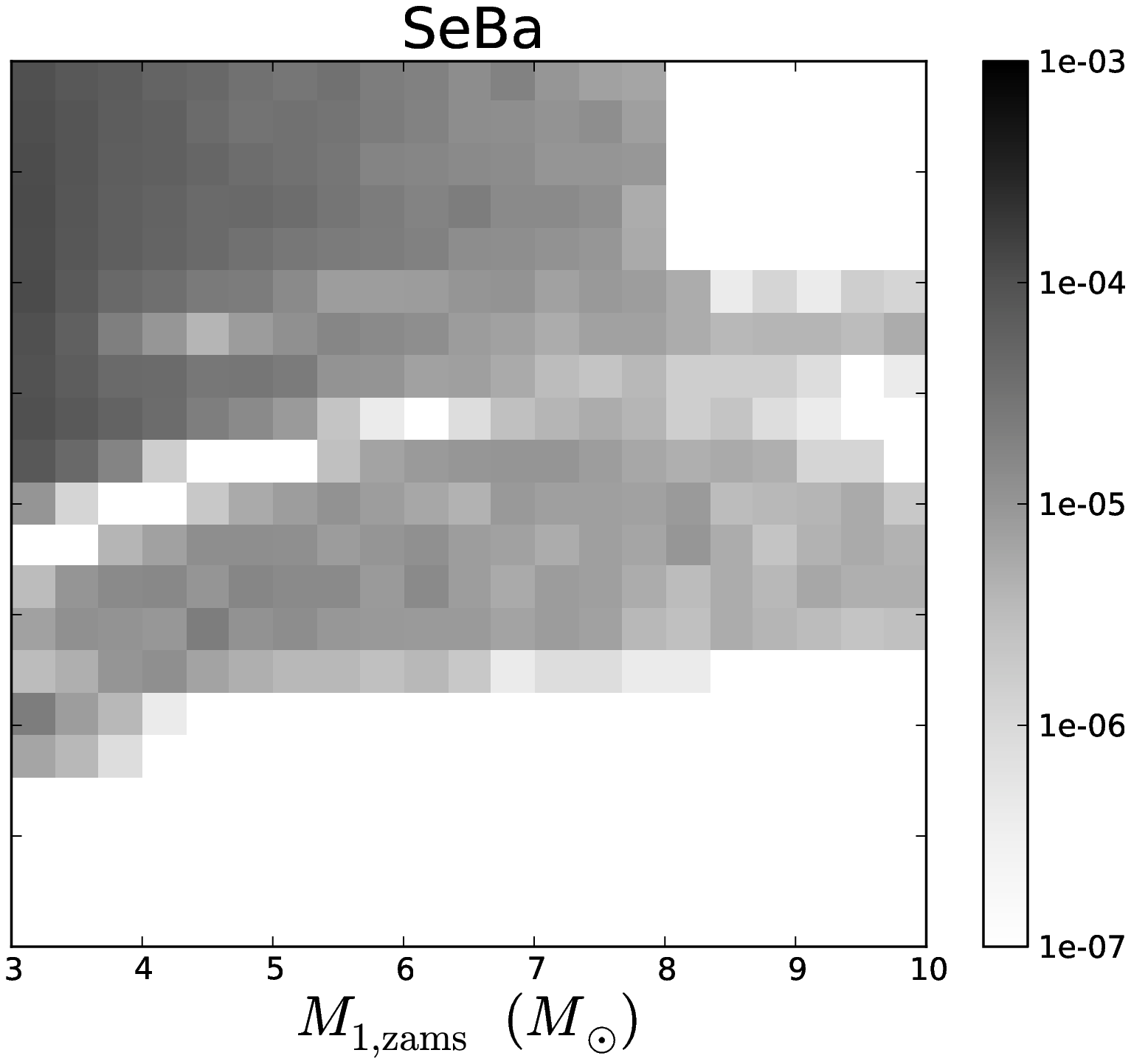} & 
	\includegraphics[height=4.6cm, clip=true, trim =20mm 0mm 23mm 5mm]{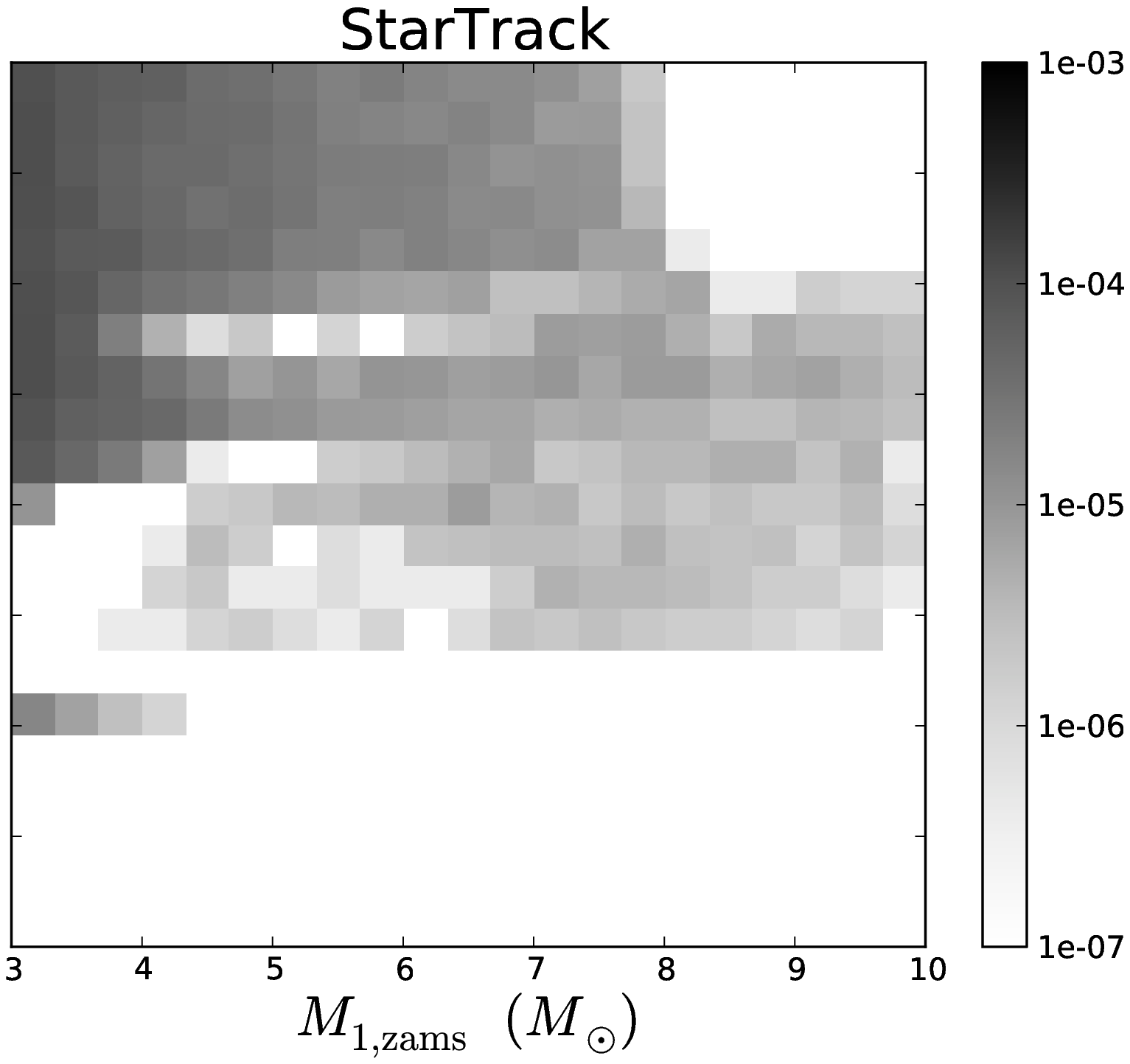} \\
	\end{tabular}
    \caption{Initial orbital separation versus initial primary mass for all SWDs in the intermediate mass range. } 
    \label{fig:swd_zams_a_IM_sin}
    \end{figure*}

Systems containing a WD and a non-degenerate companion have typically undergone a one-directional mass transfer event i.e. one star has lost mass and possibly the other gained mass. The mass transfer event may consist of one or two episodes, either of which may have been stable or unstable. 
The characteristics of the population of SWD systems show the imprint of the mass transfer episodes. Figure~\ref{fig:swd_final_a_sin}~and~\ref{fig:swd_final_a_IM_sin} show the orbital separation $a_{\rm swd}$ as a function of primary mass $M_{\rm 1, swd}$ at the moment of WD formation for the full and intermediate mass range respectively. Likewise Fig.~\ref{fig:swd_final_m2_sin}~and~\ref{fig:swd_final_m2_IM_sin} show the secondary mass $M_{\rm 2, swd}$ as a function of primary mass at WD formation for the full and intermediate mass range. These figures show that in general the codes find very similar SWD systems. 

In more detail, at large separations ($a_{\rm swd} \gtrsim 500$\Rsolar~for the full mass range, and $a_{\rm swd} \gtrsim 2000$\Rsolar~for the intermediate mass range) all codes find systems in which the stars do not interact. 
The population of SWDs with WD masses in the low mass range is very comparable in orbital separation, primary and secondary mass between the codes binary\_c, SeBa and StarTrack. Intermediate mass systems can be divided in two groups, either in separation and/or in secondary mass. According to all codes, intermediate mass systems that undergo a CE-phase (for the first mass transfer episode) are compact with $a_{\rm swd} \lesssim 200$\Rsolar~and have secondary masses up to 10\Msolar. 
Furthermore, the codes agree that in the intermediate mass range, systems for which the first phase of mass transfer is stable are in general more compact than non-interacting systems and less compact than the systems undergoing a CE-phase. The secondary mass is between 3 and 18\Msolar~as it accretes conservatively during stable mass transfer.

The ZAMS configurations for progenitors of SWDs are shown in
Fig.\,\ref{fig:swd_zams_a_sin}~and~\ref{fig:swd_zams_a_IM_sin} with the
separation $a_{\rm zams}$ versus primary mass $M_{\rm
  1,zams}$. 
There is a general agreement between the codes about which progenitor
systems lead to a SWD system and which systems do not. According to
all codes, compact progenitor systems ($a_{\rm zams} \lesssim
400$\Rsolar~for the intermediate mass range, while $a_{\rm zams}
\lesssim 30$\Rsolar~for the low mass range) undergo stable mass transfer for the initial mass transfer episode. Furthermore the codes agree
that for most progenitor systems with orbital separations in the range
$a_{\rm zams} \approx (0.1-3)\cdot 10^3$\Rsolar~the first phase of
mass transfer is unstable. 
Systems with orbital separations that lie between the ranges described above lead to a merging event, thus no SWD system is formed. 
Progenitor systems with $a_{\rm zams} \gtrsim 700$\Rsolar~for the intermediate mass range ($a_{\rm zams}\gtrsim 250$\Rsolar~for the low mass range) are too wide for the primary star to fill its Roche lobe.

\begin{figure}[tbh]
\centering
\includegraphics[width = 8cm]{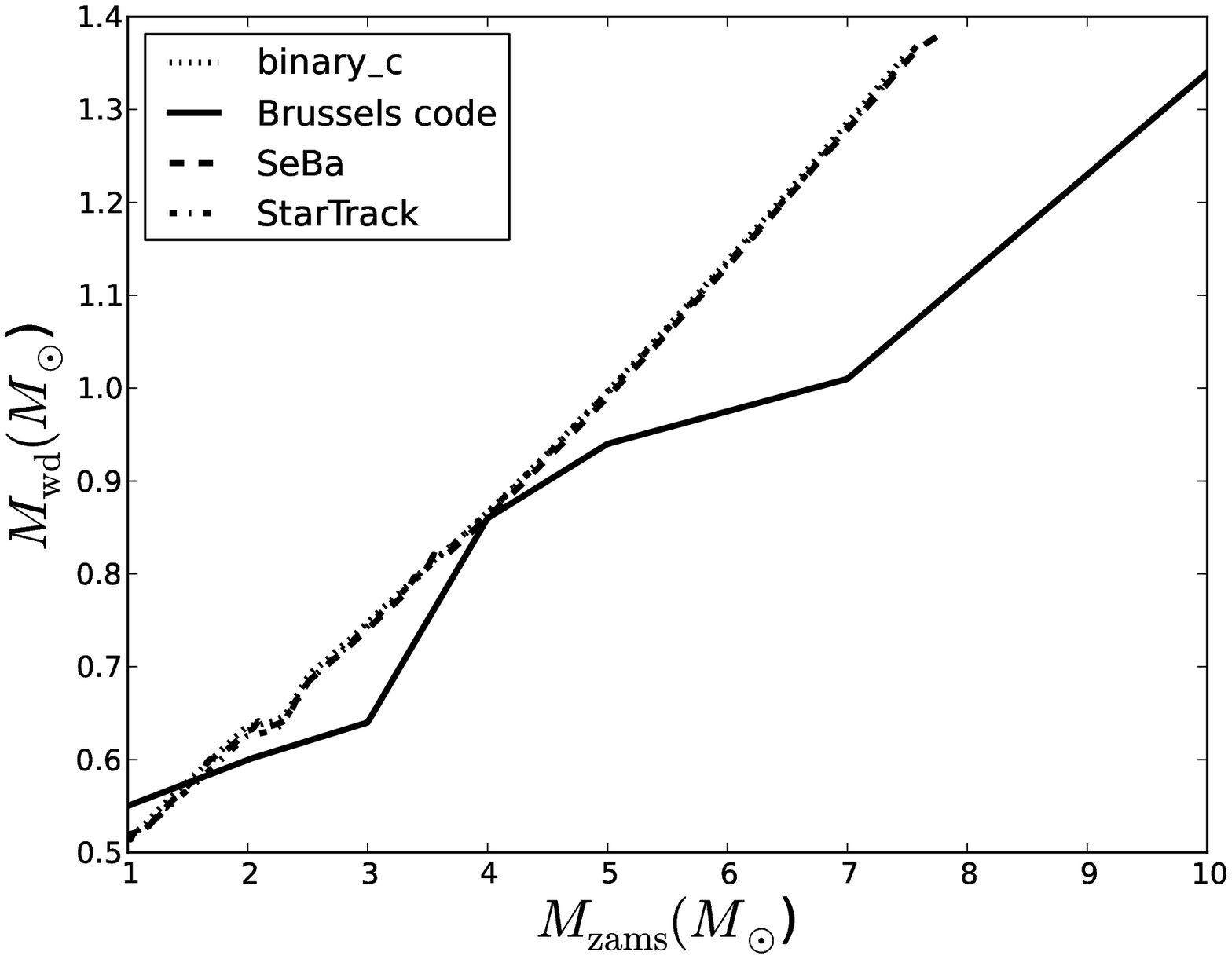} 
\caption{Initial-final mass relation of single stars that become WDs for the
  different groups, dotted line shows the results of binary$\_$c,
  solid line the results of the Brussels code, the dashed line the
  results of SeBa, and the dash-dotted line the results of
  StarTrack.} 
\label{fig:ifm}
\end{figure}

Overall the simulations of the four codes show a good agreement on the characteristics of the population of SWDs in orbital parameters and birthrates (Table\,\ref{tbl:birthrates}), however, differences can be noted. The most important causes are the relation between the initial and WD mass, the stability of mass transfer and the modelling of the stable mass transfer phase. 
The initial-final mass (MiMf)-relation of single stars (Fig.\,\ref{fig:ifm}) is very similar between binary\_c, SeBa and StarTrack, but different than the one from the Brussels code due to different single star prescriptions that are used in the latter code (see also Appendix\,\ref{sec:channel1} for a discussion). The effect on the population of SWD progenitors can be seen in Fig.\,\ref{fig:swd_zams_a_IM_sin} in the maximum mass of the primary stars which is extended from about 8\Msolar~in binary\_c, SeBa and StarTrack to about 10\Msolar~in the Brussels code.  
For binary stars the relation between WD mass and the initial mass is hereafter called the initial-WD mass (MiMwd)-relation (Appendix\,\ref{sec:channel2}). Differences in the MiMwd-relation lead to an increase of systems at small WD masses $\lesssim 0.64$\Msolar~in Fig.\,\ref{fig:swd_final_a_IM_sin} in the Brussels code compared to the other codes. 
The gap in WD masses between 0.7-0.9\Msolar~in the Brussels data in Fig.\,\ref{fig:swd_final_a_IM_sin} is a result of a discontinuity in the MiMwd-relation between the WD masses of primaries that fill their Roche a second time, and those that do not. In the other codes, the primary WD masses of binaries that evolve through these two evolutionary channels are overlapping.
Differences in the stability criteria of mass transfer can be seen in Fig.\,\ref{fig:swd_final_a_IM_sin}~and~\ref{fig:swd_final_m2_IM_sin}, where the StarTrack code shows a decrease of systems that underwent stable mass transfer (Appendix\,\ref{sec:channel3}).
Mass transfer is modelled differently in the codes (Sect.\,\ref{sec:TNS_inherent}) leading to an extension to small separations in the Brussels data compared to the other codes (Fig.\,\ref{fig:swd_zams_a_IM_sin}), and an increase in systems that underwent stable mass transfer at $a_{\rm zams} \approx 10 $\Rsolar~ for $M_{\rm 1, zams}\gtrsim 4$\Msolar~in Fig.\,\ref{fig:swd_final_m2_IM_sin} (Appendix\,\ref{sec:channel5}).

For a more detailed comparison of the SWD population in the full and intermediate mass range, see Appendix\,\ref{sec:ev_path_swd}. 

\subsection{Double white dwarfs}
\label{sec:dwd}


    \begin{figure*}[htb]
    \centering
    \setlength\tabcolsep{0pt}
    \begin{tabular}{ccc}
	\includegraphics[height=4.6cm, clip=true, trim =8mm 0mm 48.5mm 5mm]{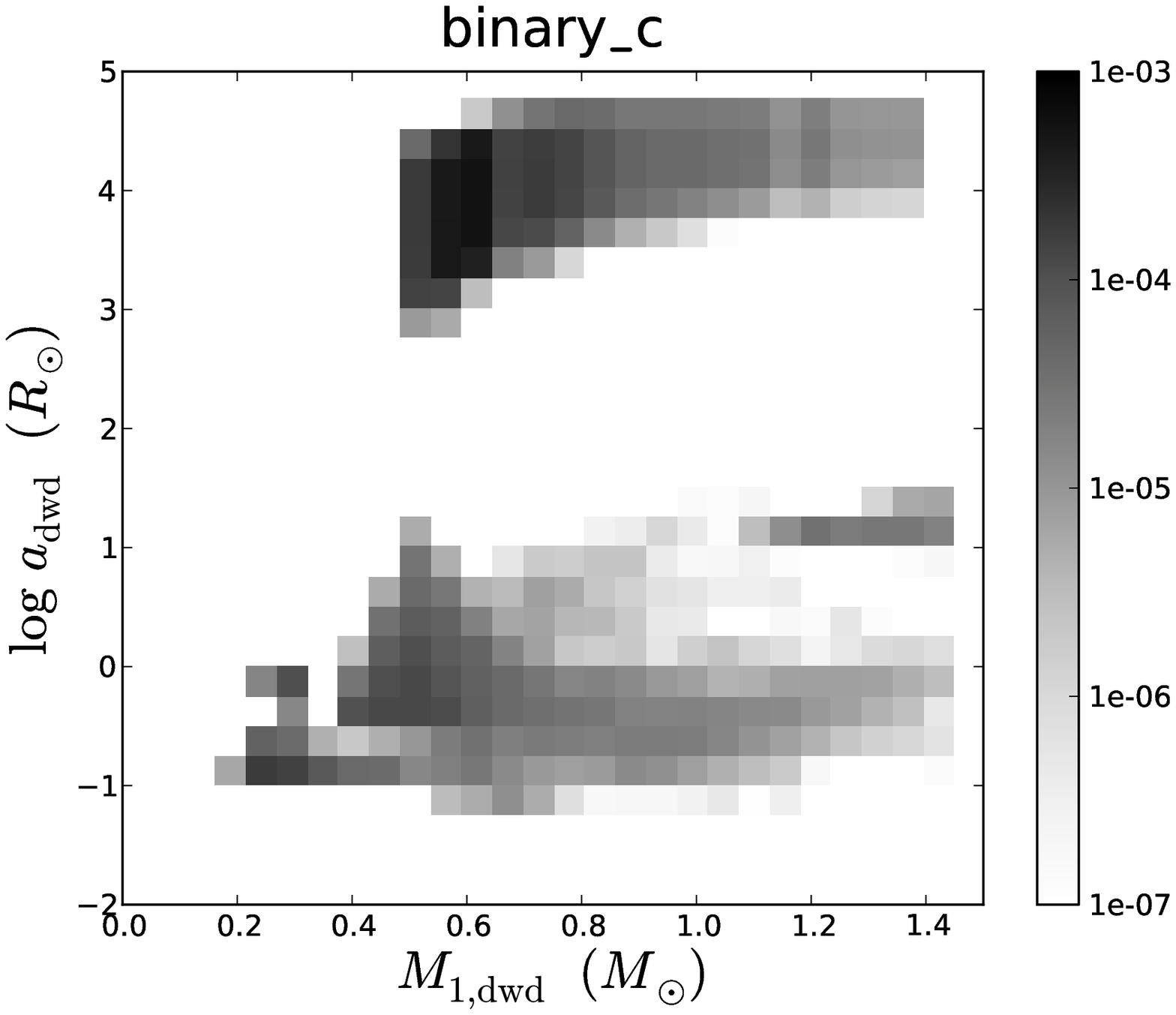} &
	\includegraphics[height=4.6cm, clip=true, trim =20mm 0mm 48.5mm 5mm]{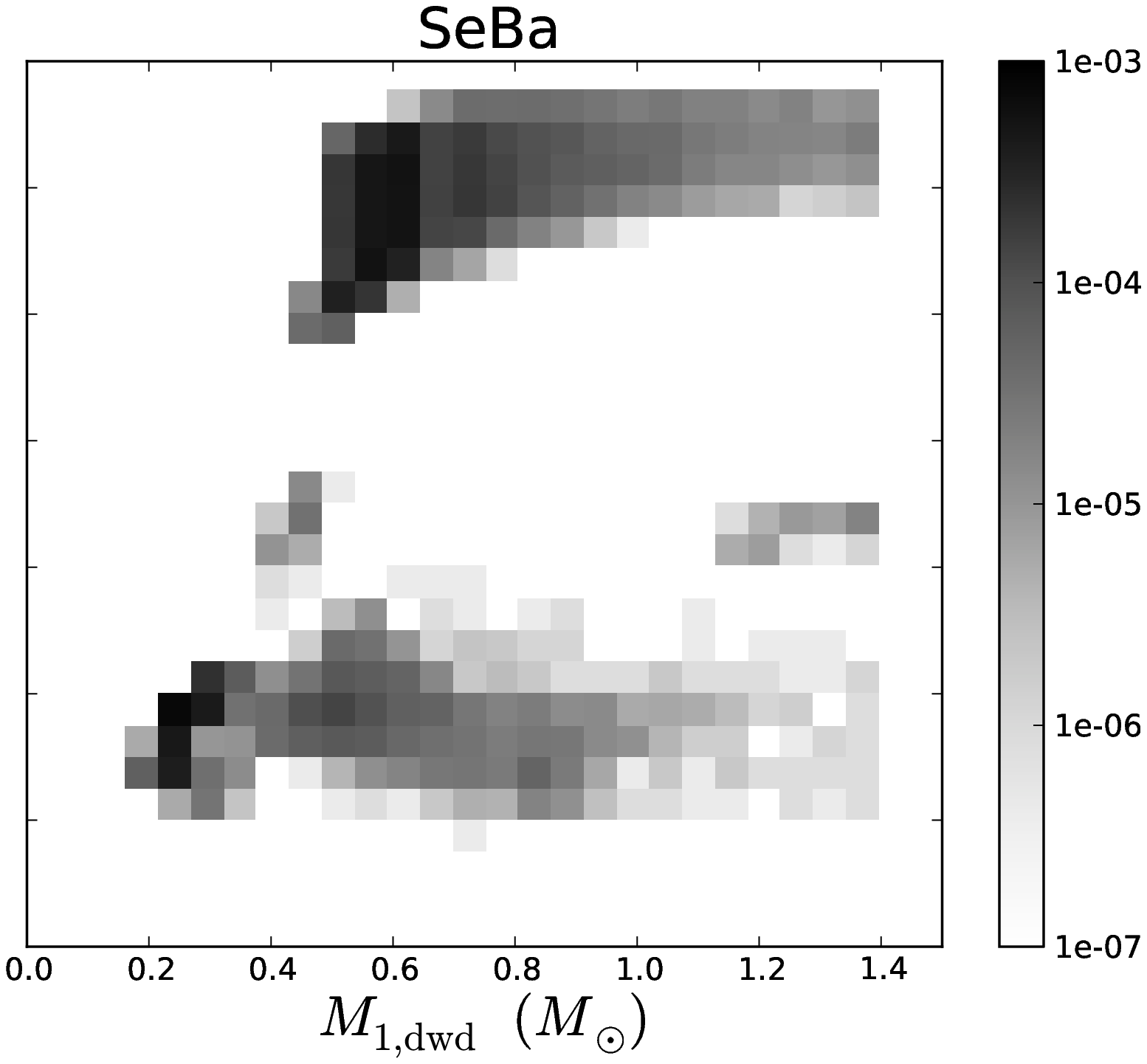}&
	\includegraphics[height=4.6cm, clip=true, trim =20mm 0mm 23mm 5mm]{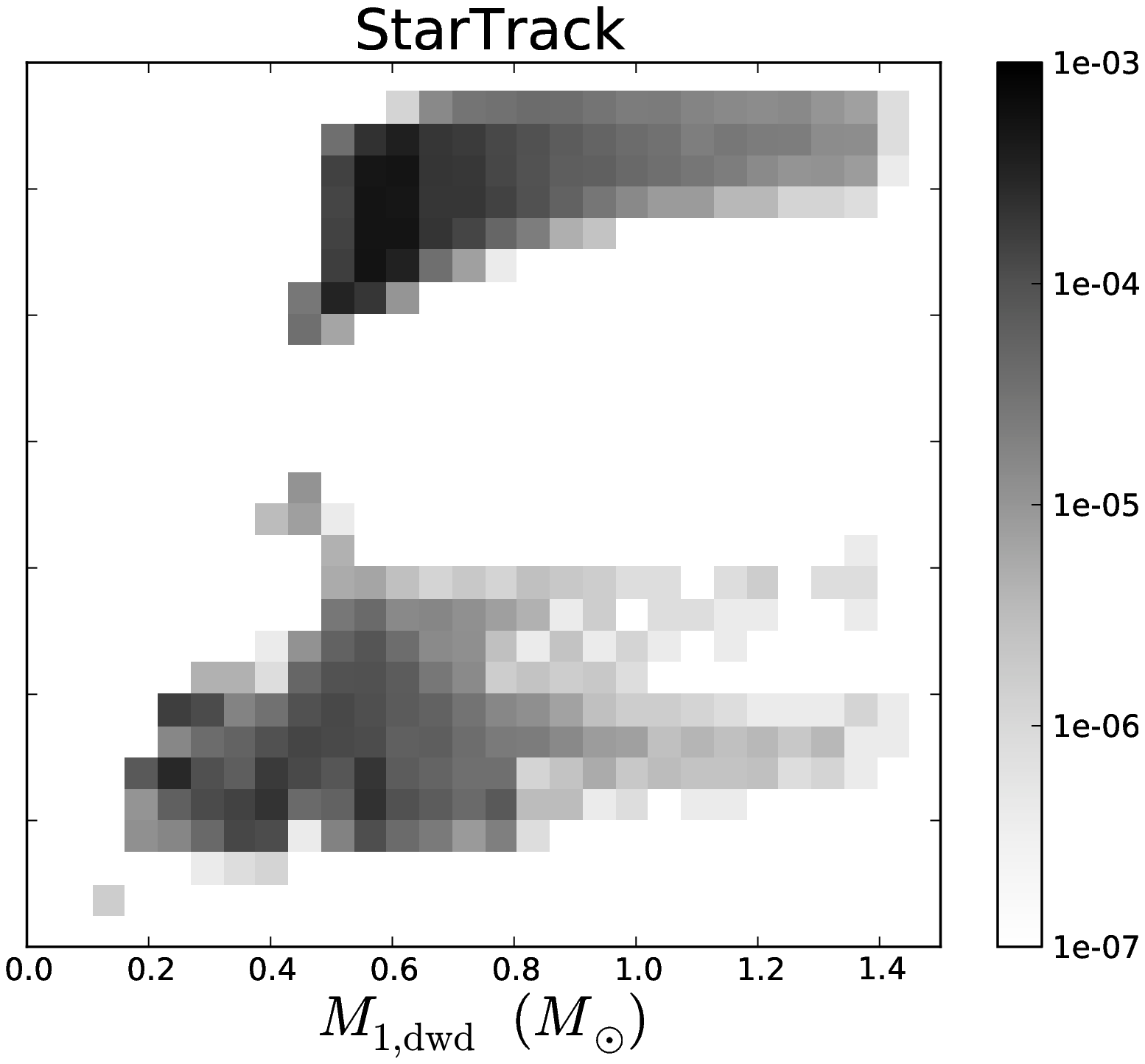} \\	
	\end{tabular}
    \caption{Orbital separation versus primary WD mass for all DWDs in the full mass range at the time of DWD formation. }
    \label{fig:dwd_a_sin}
    \end{figure*}

    \begin{figure*}[htb]
    \centering
    \setlength\tabcolsep{0pt}
    \begin{tabular}{cccc}
	\includegraphics[height=4.6cm, clip=true, trim =8mm 0mm 48.5mm 5mm]{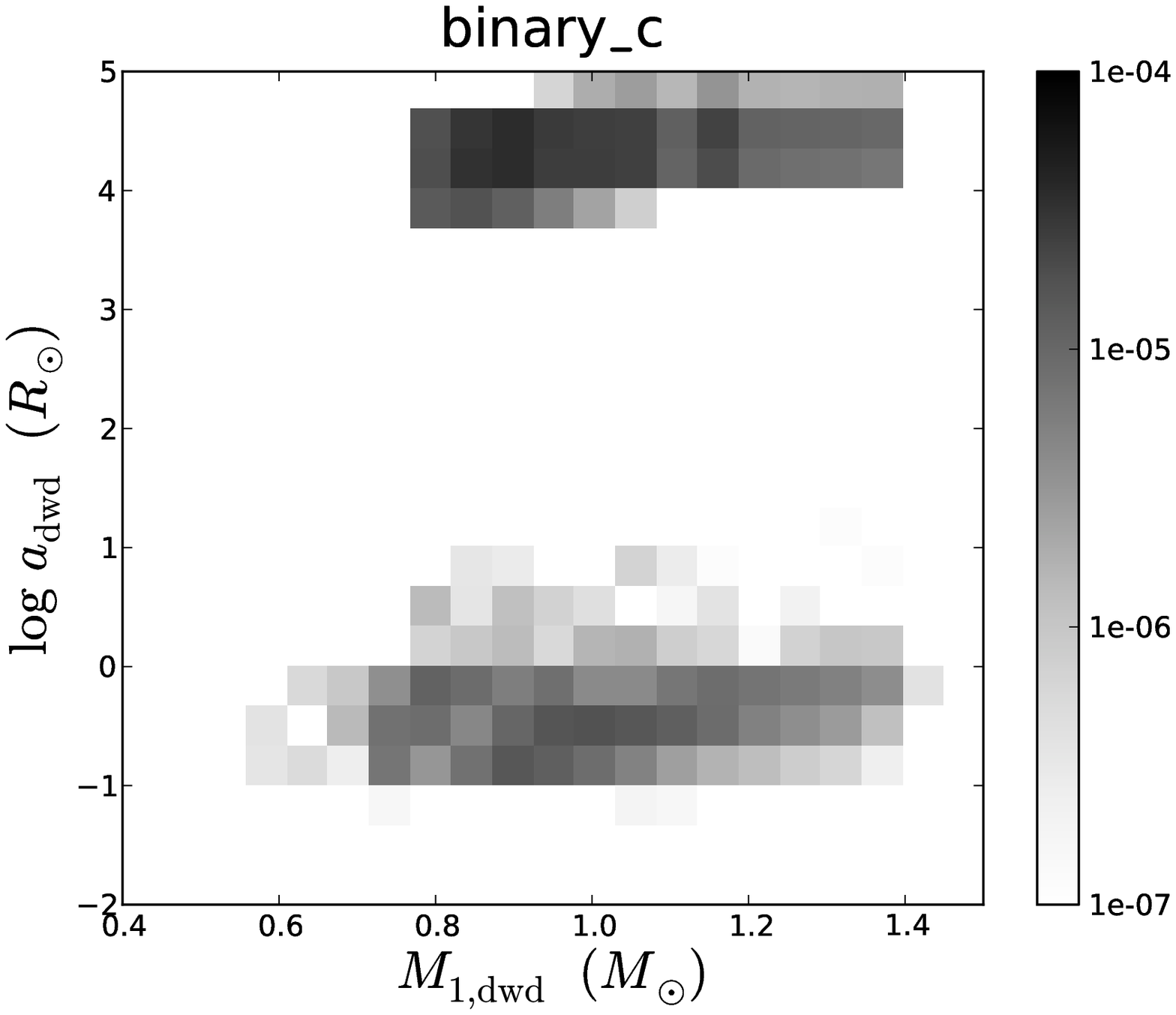} &
	\includegraphics[height=4.6cm, clip=true, trim =20mm 0mm 48.5mm 5mm]{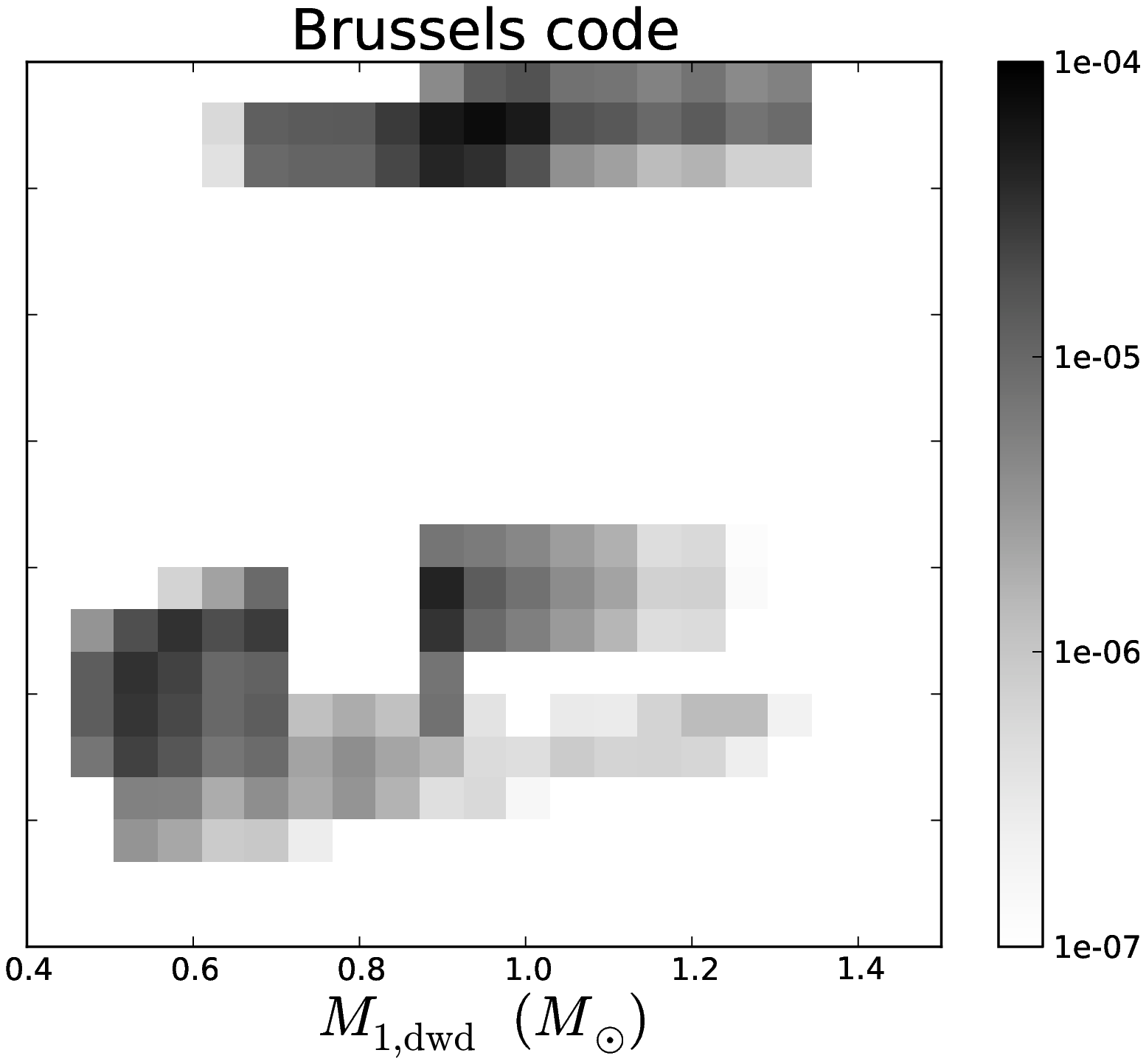} & 
	\includegraphics[height=4.6cm, clip=true, trim =20mm 0mm 48.5mm 5mm]{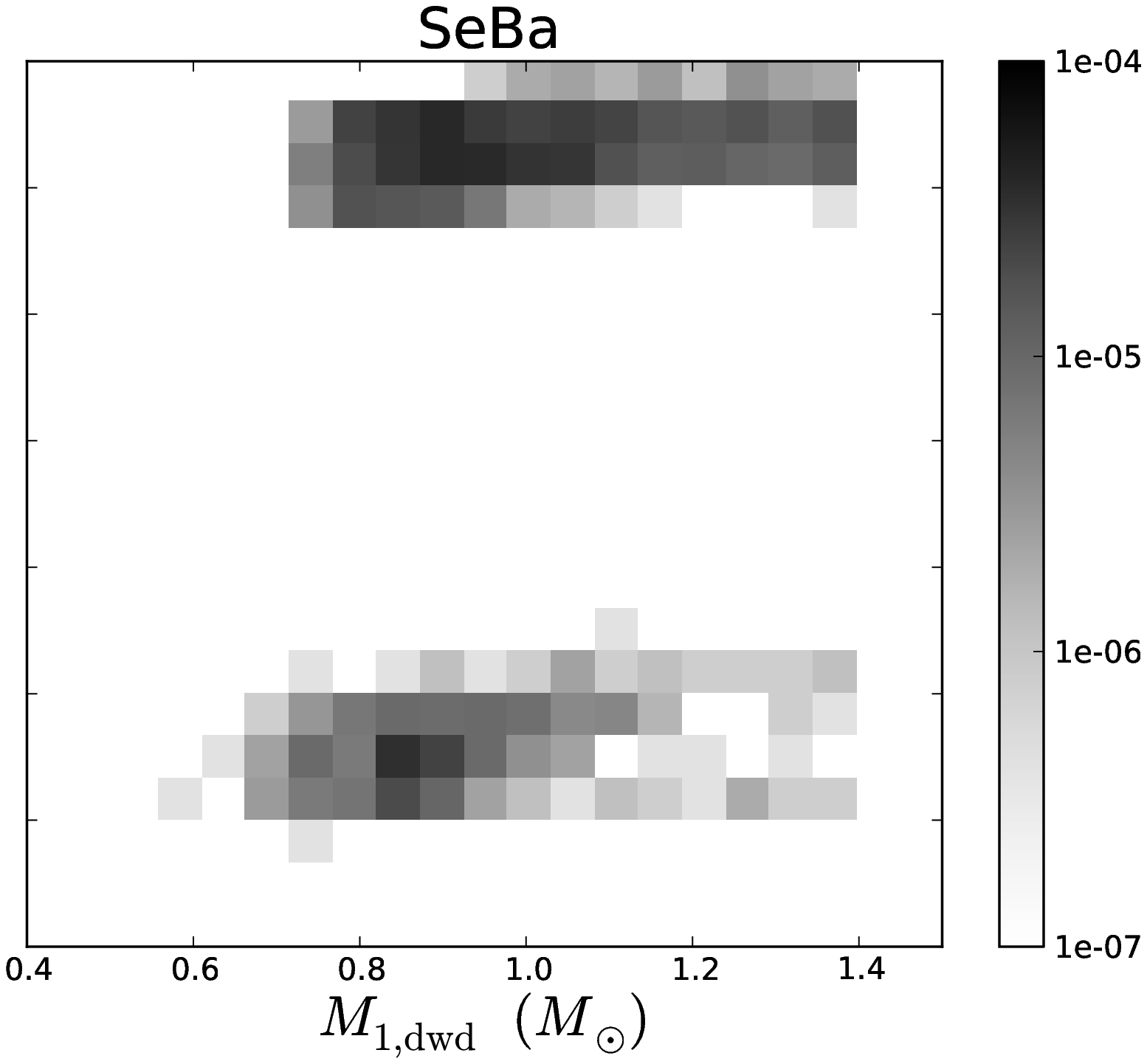} &
	\includegraphics[height=4.6cm, clip=true, trim =20mm 0mm 23mm 5mm]{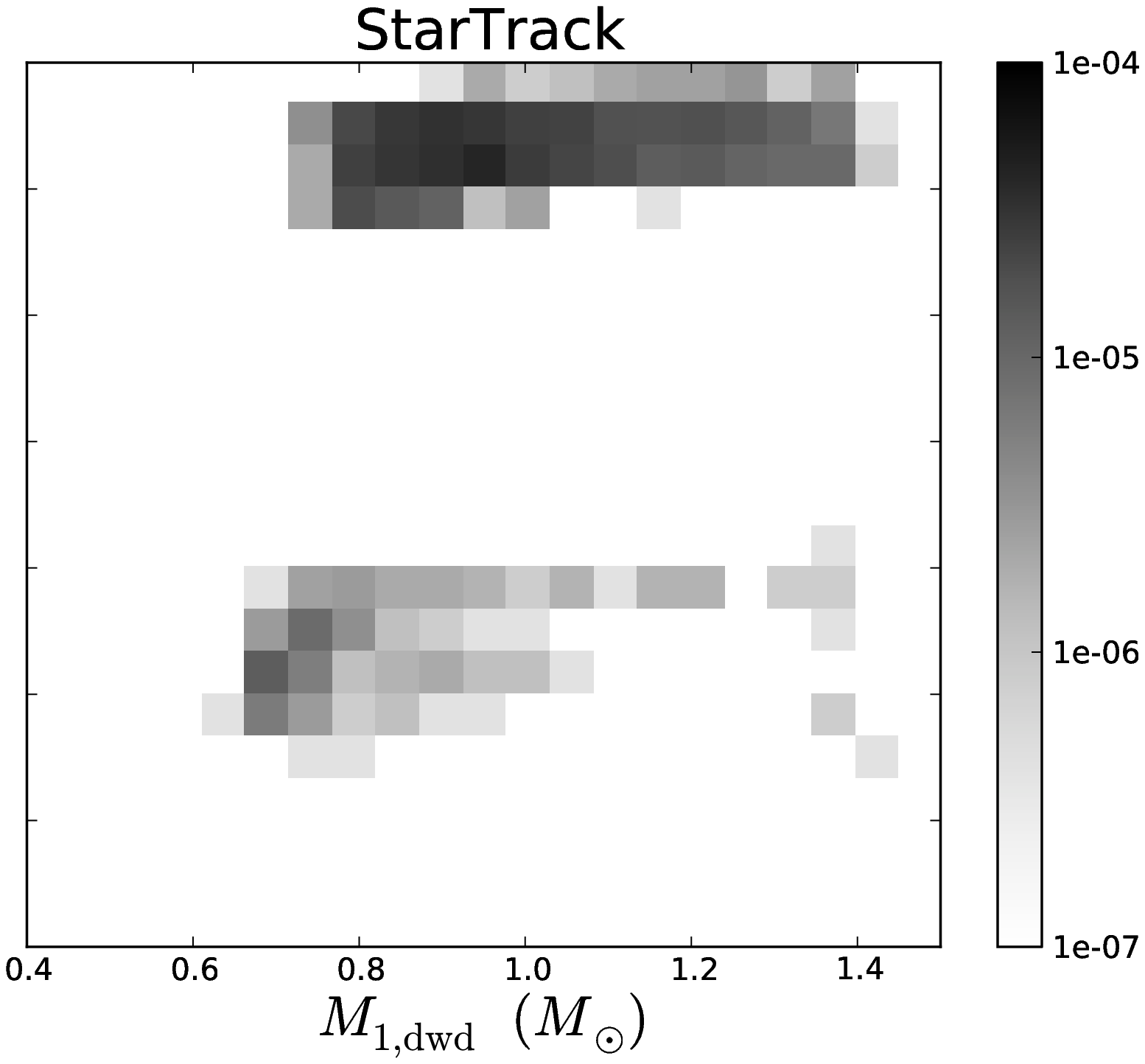} \\	
	\end{tabular}
    \caption{Orbital separation versus primary WD mass for all DWDs in the intermediate mass range at the time of DWD formation.} 
    \label{fig:dwd_a_IM_sin}
    \end{figure*}

    \begin{figure*}[htb]
    \centering
    \setlength\tabcolsep{0pt}
    \begin{tabular}{ccc}
	\includegraphics[height=4.6cm, clip=true, trim =8mm 0mm 48.5mm 5mm]{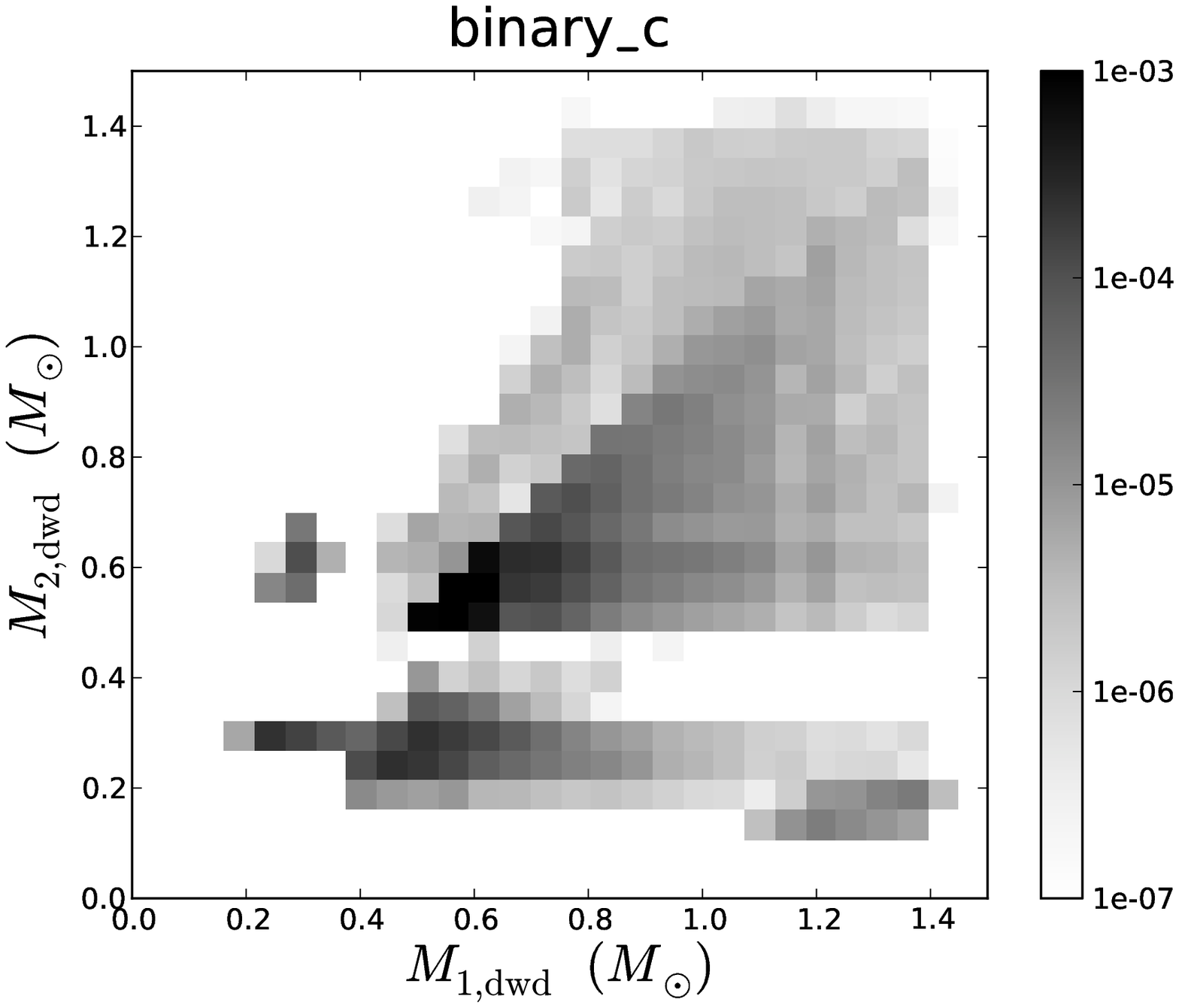} & 
	\includegraphics[height=4.6cm, clip=true, trim =20mm 0mm 48.5mm 5mm]{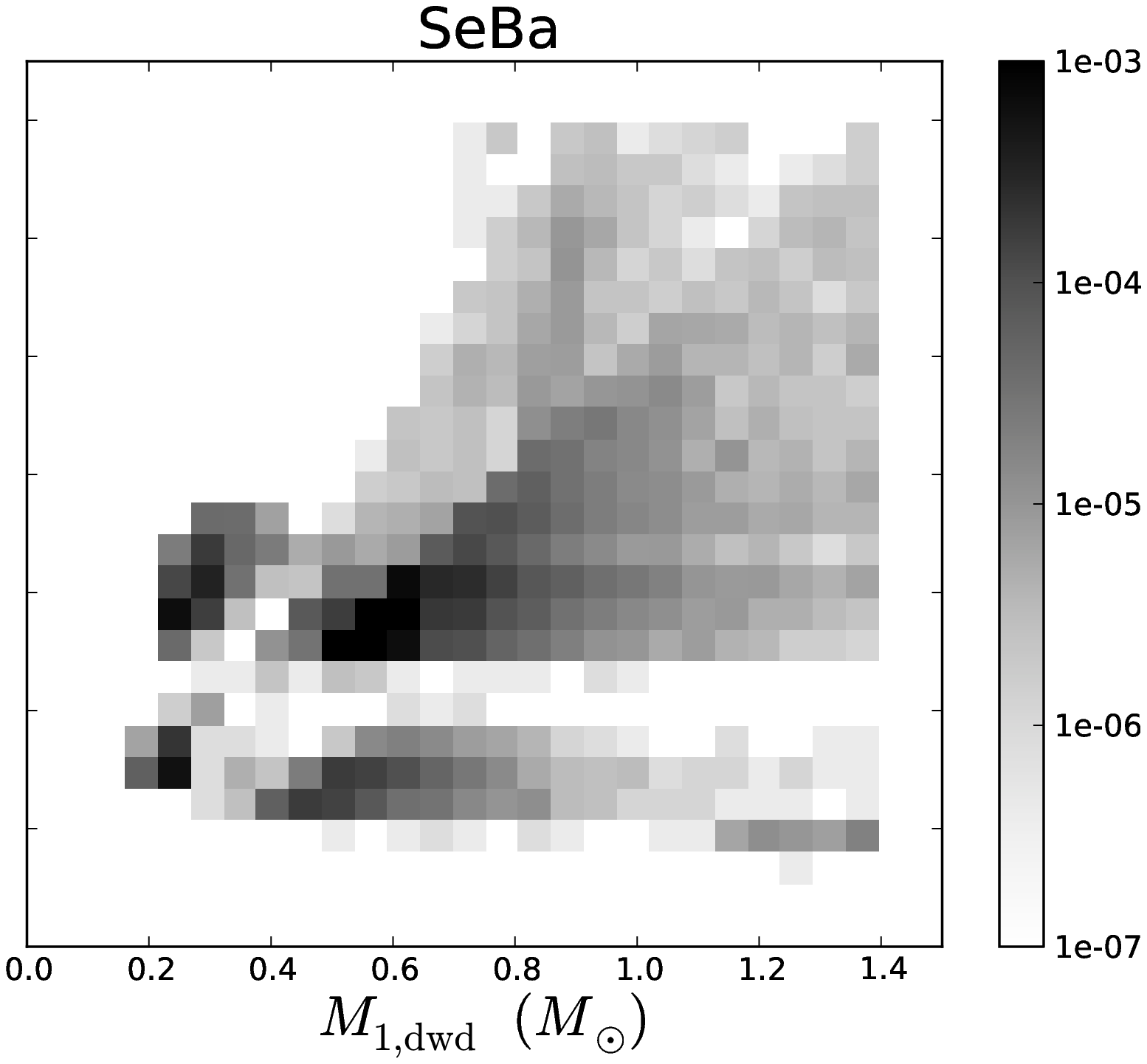} &
	\includegraphics[height=4.6cm, clip=true, trim =20mm 0mm 23mm 5mm]{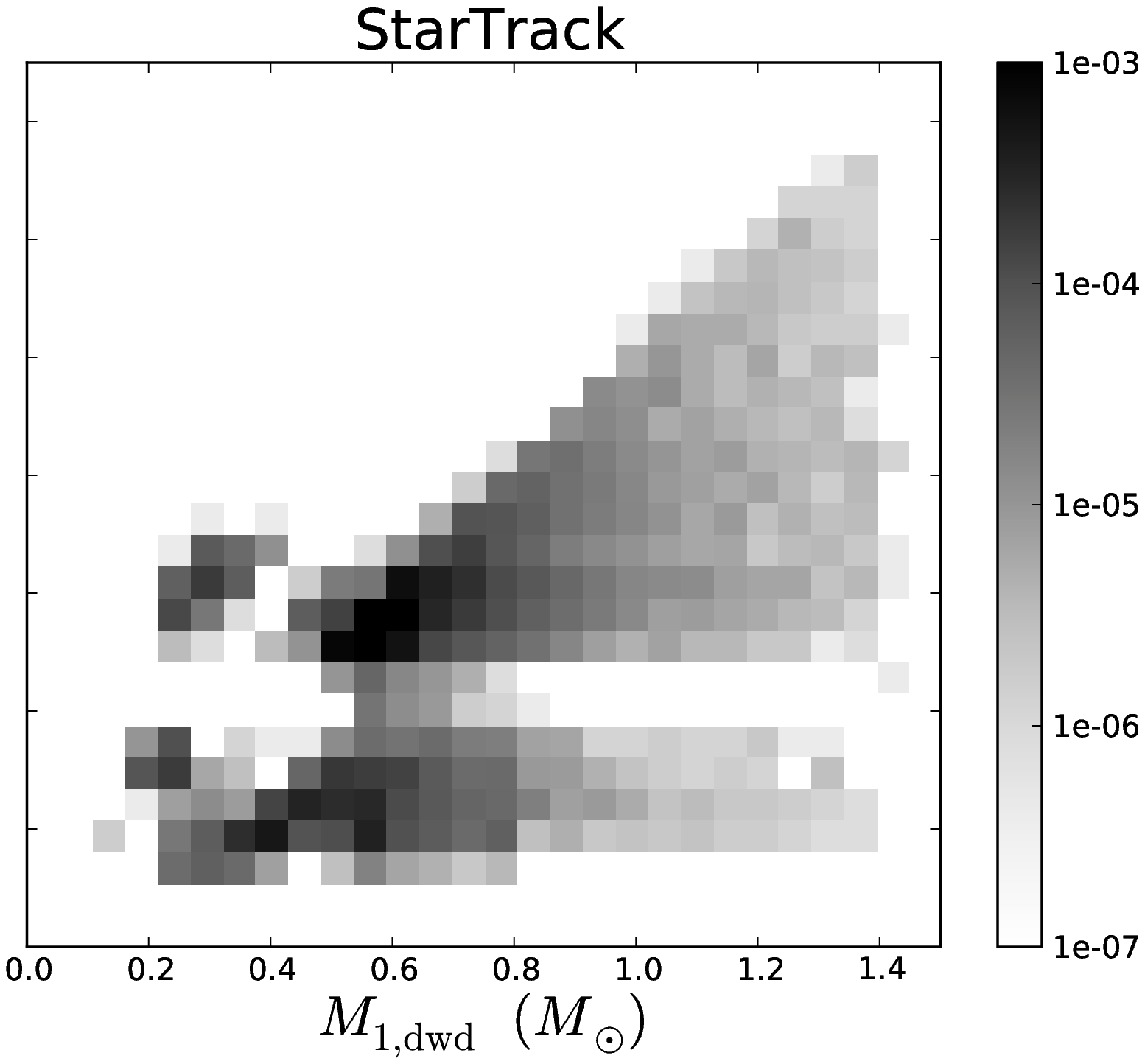} \\	
	\end{tabular}
    \caption{Secondary WD mass versus primary WD mass for all DWDs in the full mass range at the time of DWD formation. } 
    \label{fig:dwd_M2_sin}
    \end{figure*}

    \begin{figure*}[htb]
    \centering
    \setlength\tabcolsep{0pt}
    \begin{tabular}{cccc}
	\includegraphics[height=4.6cm, clip=true, trim =8mm 0mm 48.5mm 5mm]{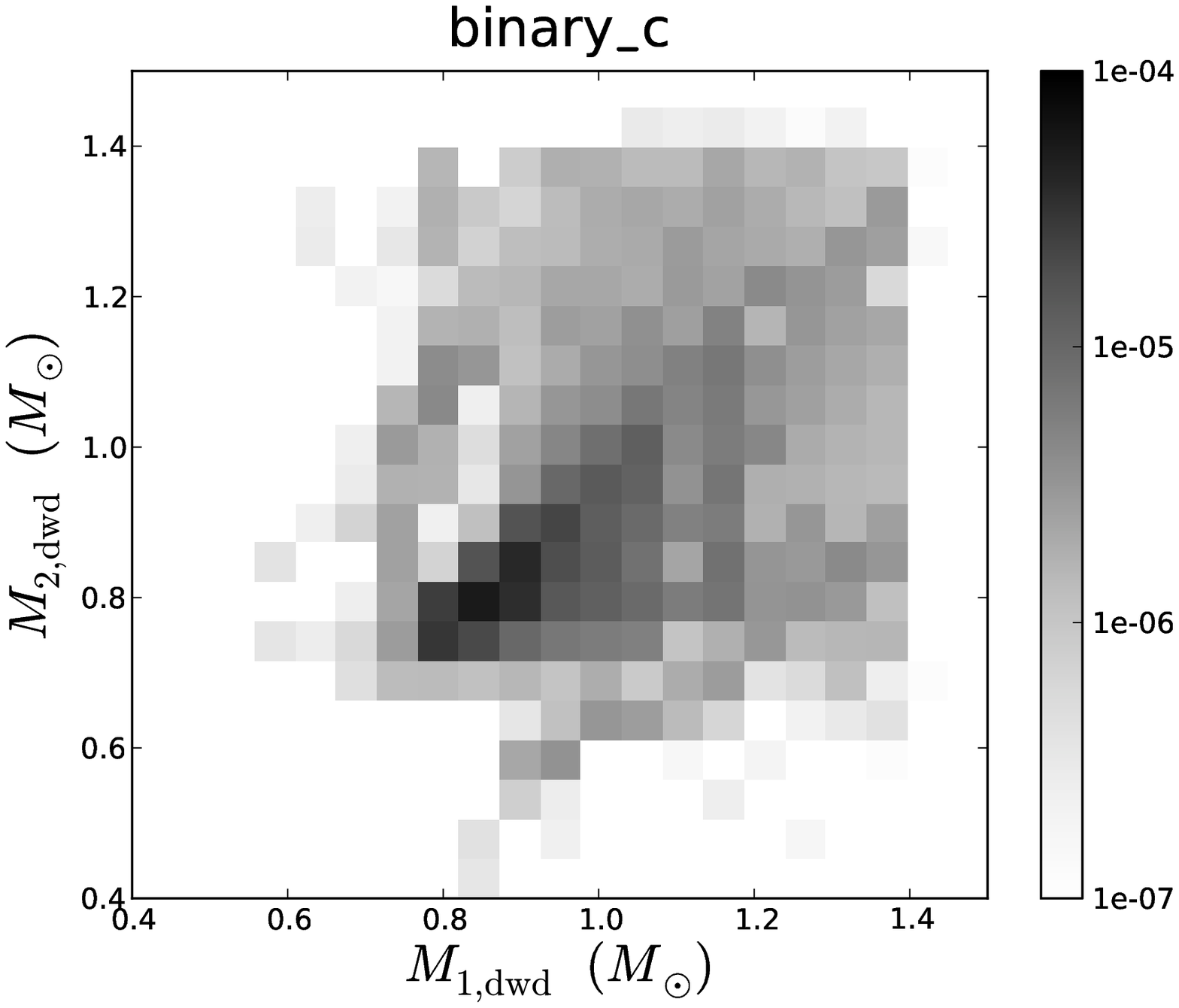} & 
	\includegraphics[height=4.6cm, clip=true, trim =20mm 0mm 48.5mm 5mm]{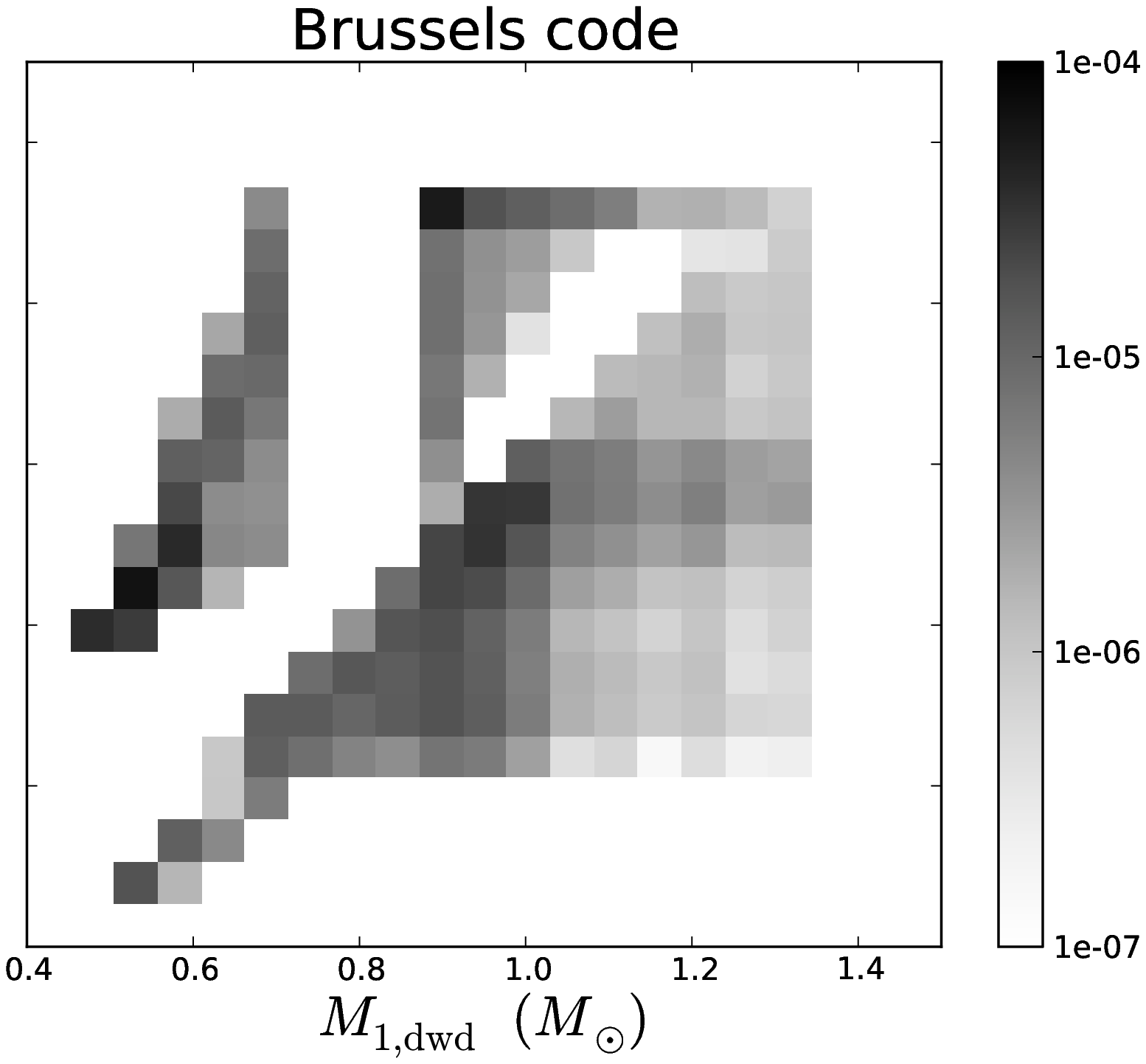} &
	\includegraphics[height=4.6cm, clip=true, trim =20mm 0mm 48.5mm 5mm]{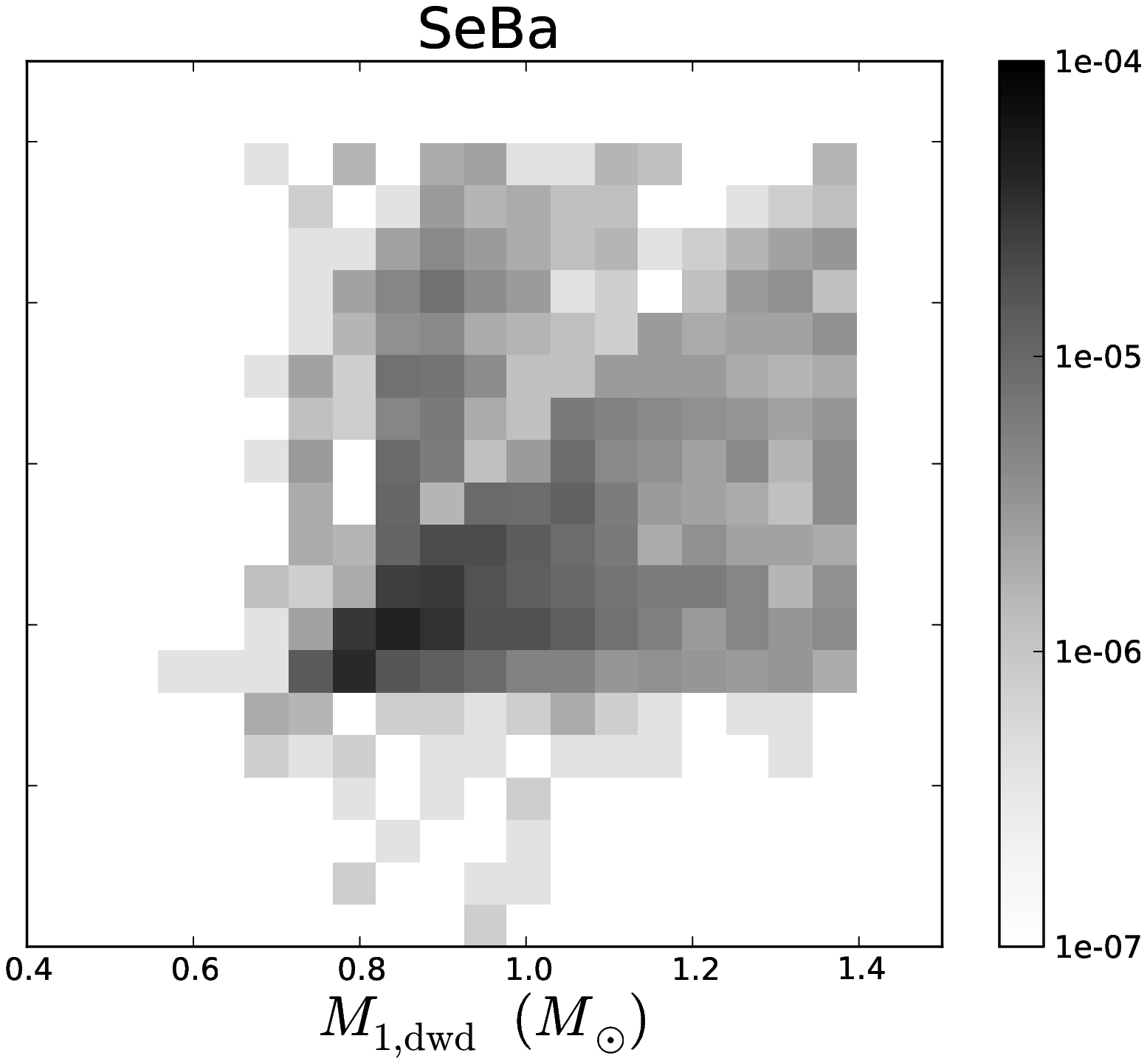} &
	\includegraphics[height=4.6cm, clip=true, trim =20mm 0mm 23mm 5mm]{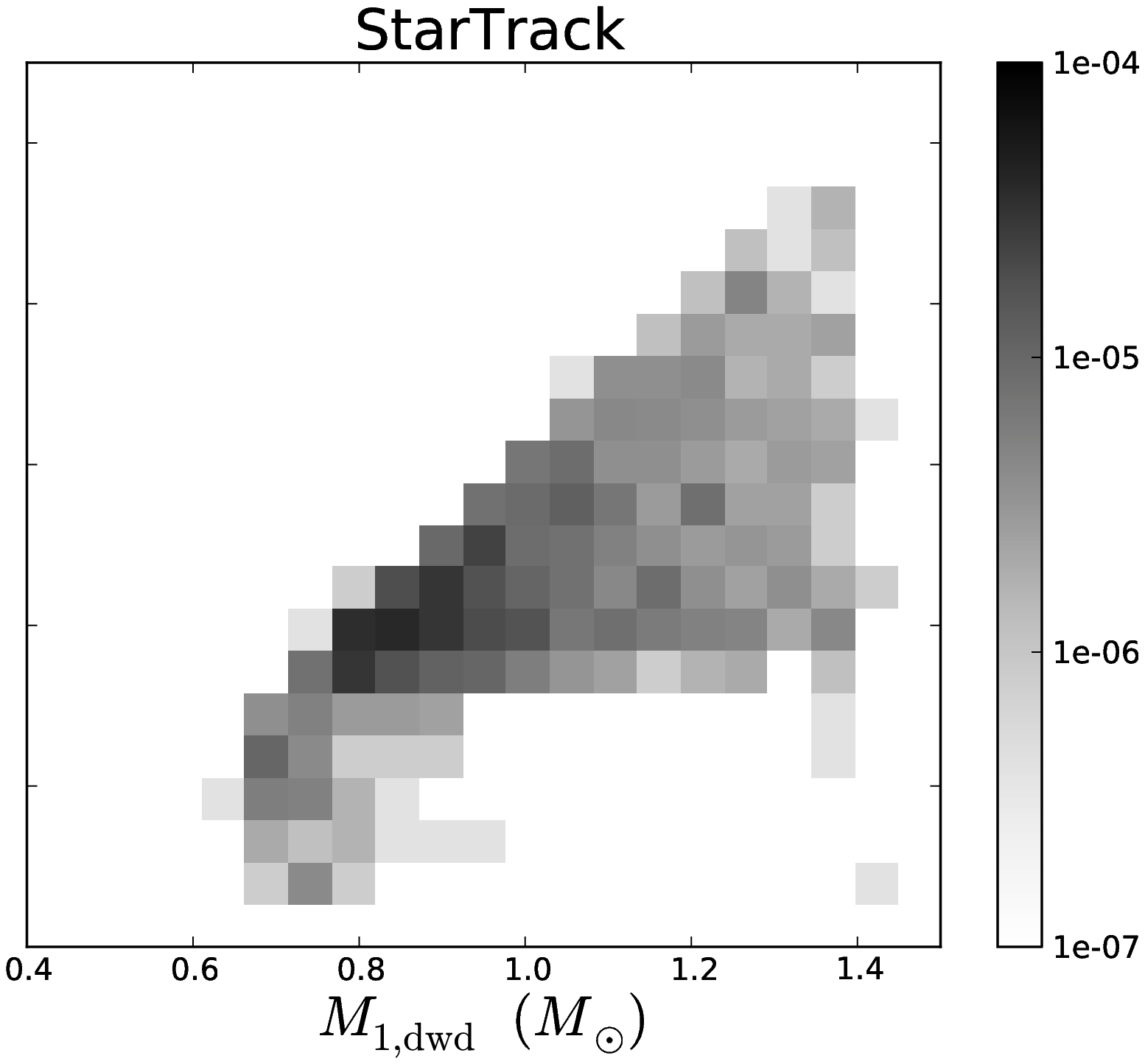} \\	
	\end{tabular}
    \caption{Secondary WD mass versus primary WD mass for all DWDs in the intermediate mass range at the time of DWD formation.} 
    \label{fig:dwd_M2_IM_sin}
    \end{figure*}

    \begin{figure*}[htb]
    \centering
    \setlength\tabcolsep{0pt}
    \begin{tabular}{ccc}
	\includegraphics[height=4.6cm, clip=true, trim =8mm 0mm 48.5mm 5mm]{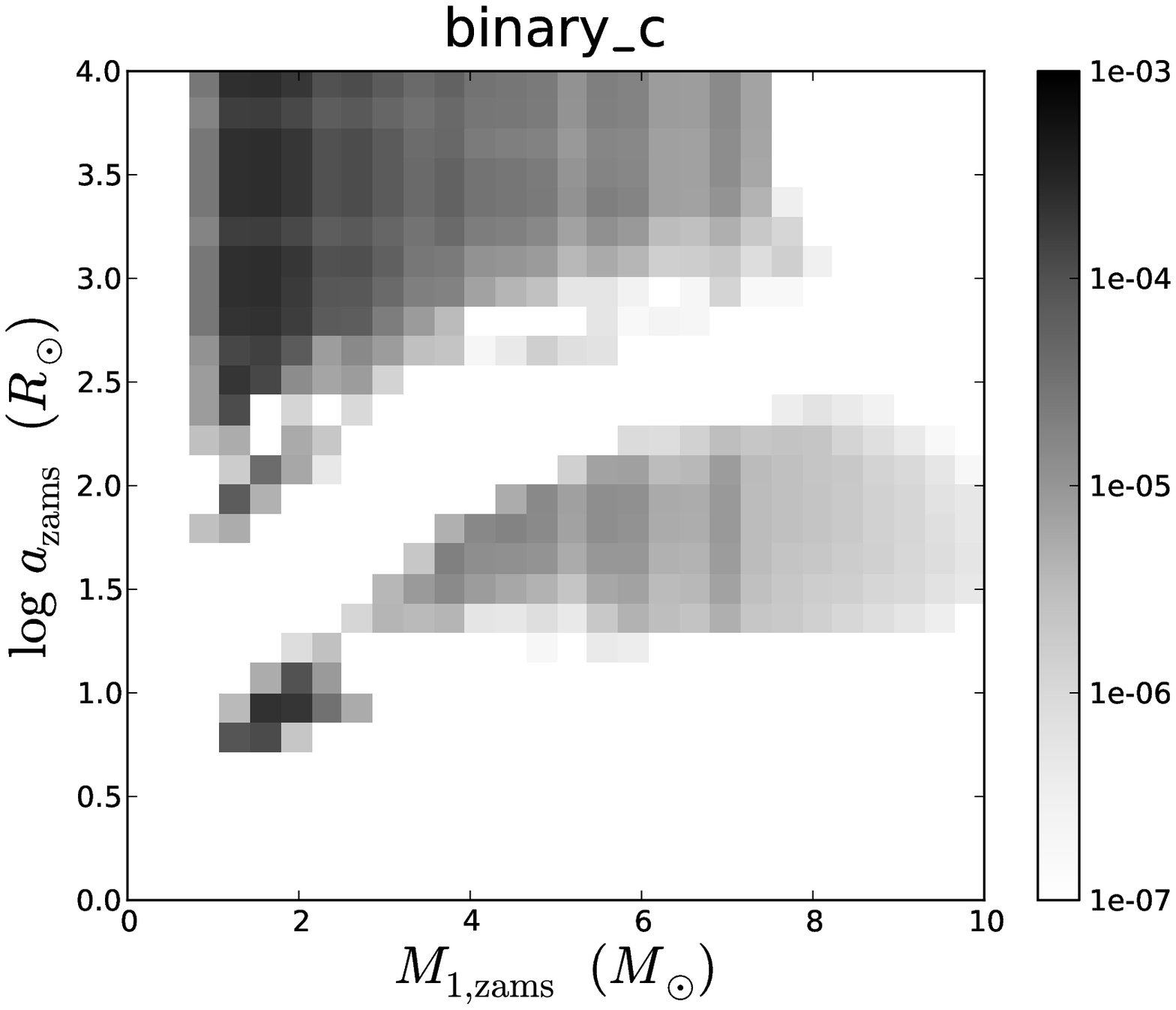} &
	\includegraphics[height=4.6cm, clip=true, trim =20mm 0mm 48.5mm 5mm]{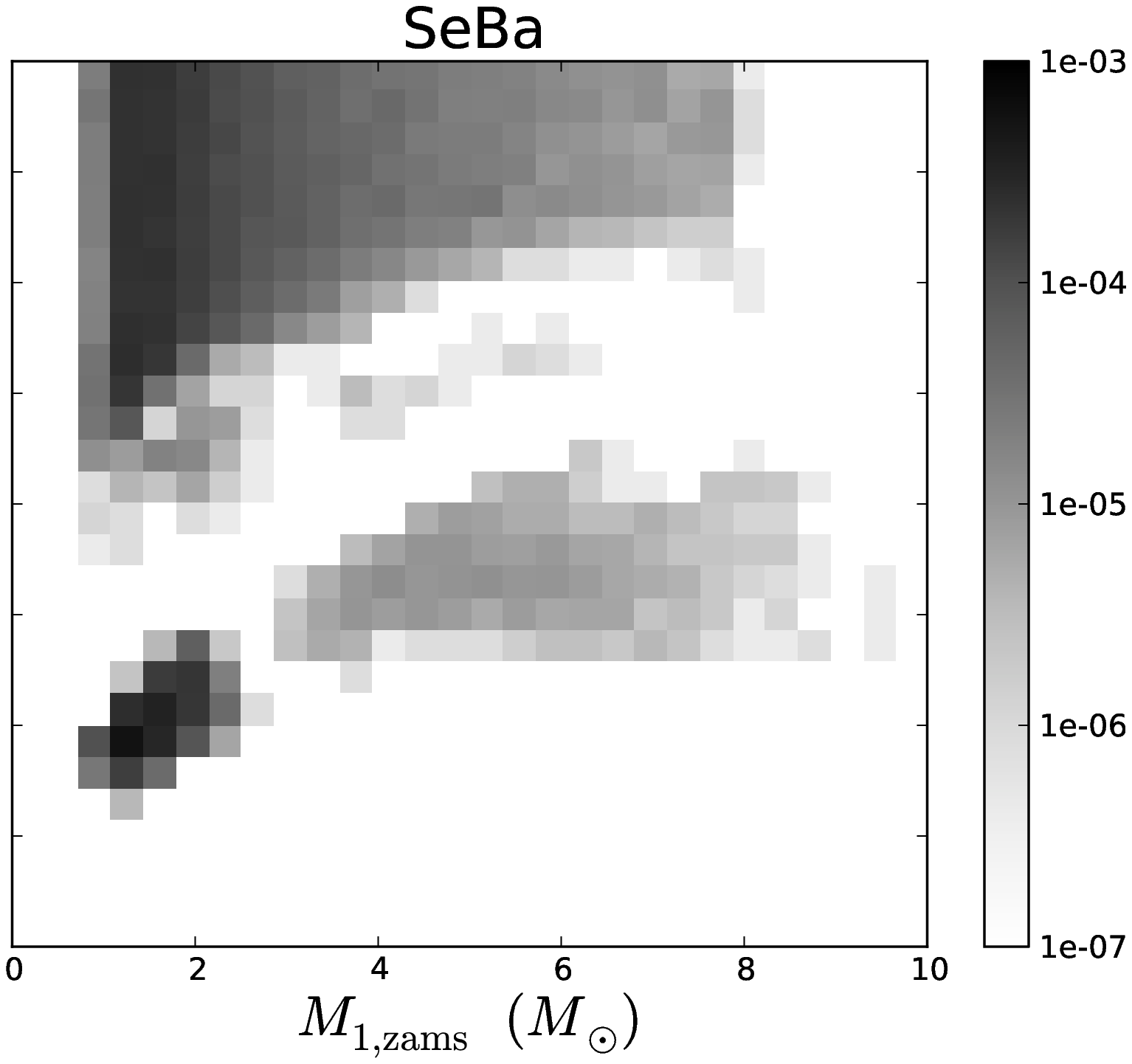} &
	\includegraphics[height=4.6cm, clip=true, trim =20mm 0mm 23mm 5mm]{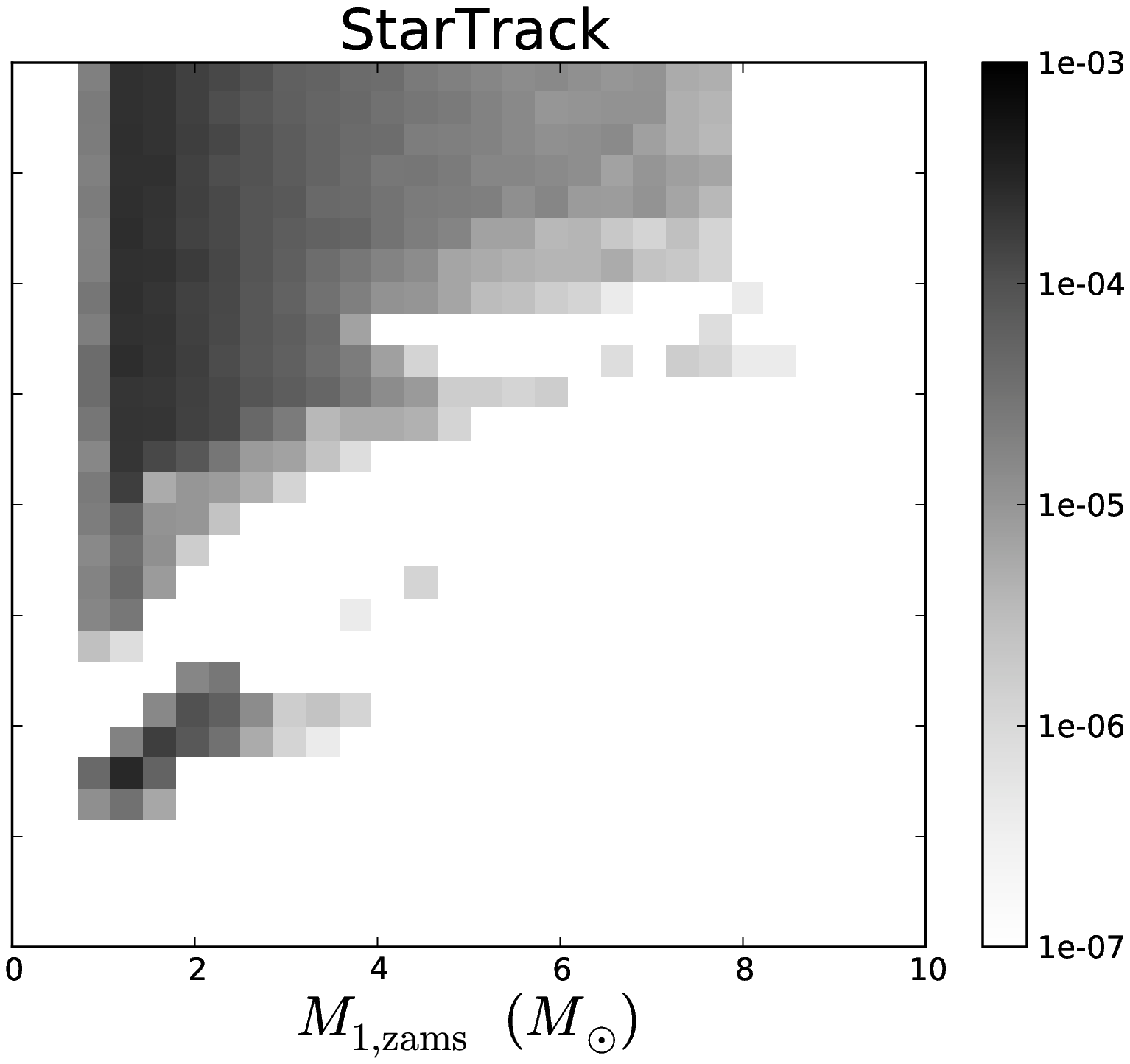} \\	
	\end{tabular}
    \caption{Initial orbital separation versus initial primary mass for all DWDs in the full mass range. } 
    \label{fig:dwd_zams_a_sin}
    \end{figure*}
    
    \begin{figure*}[htb]
    \centering
    \setlength\tabcolsep{0pt}
    \begin{tabular}{cccc}
	\includegraphics[height=4.6cm, clip=true, trim =8mm 0mm 48.5mm 5mm]{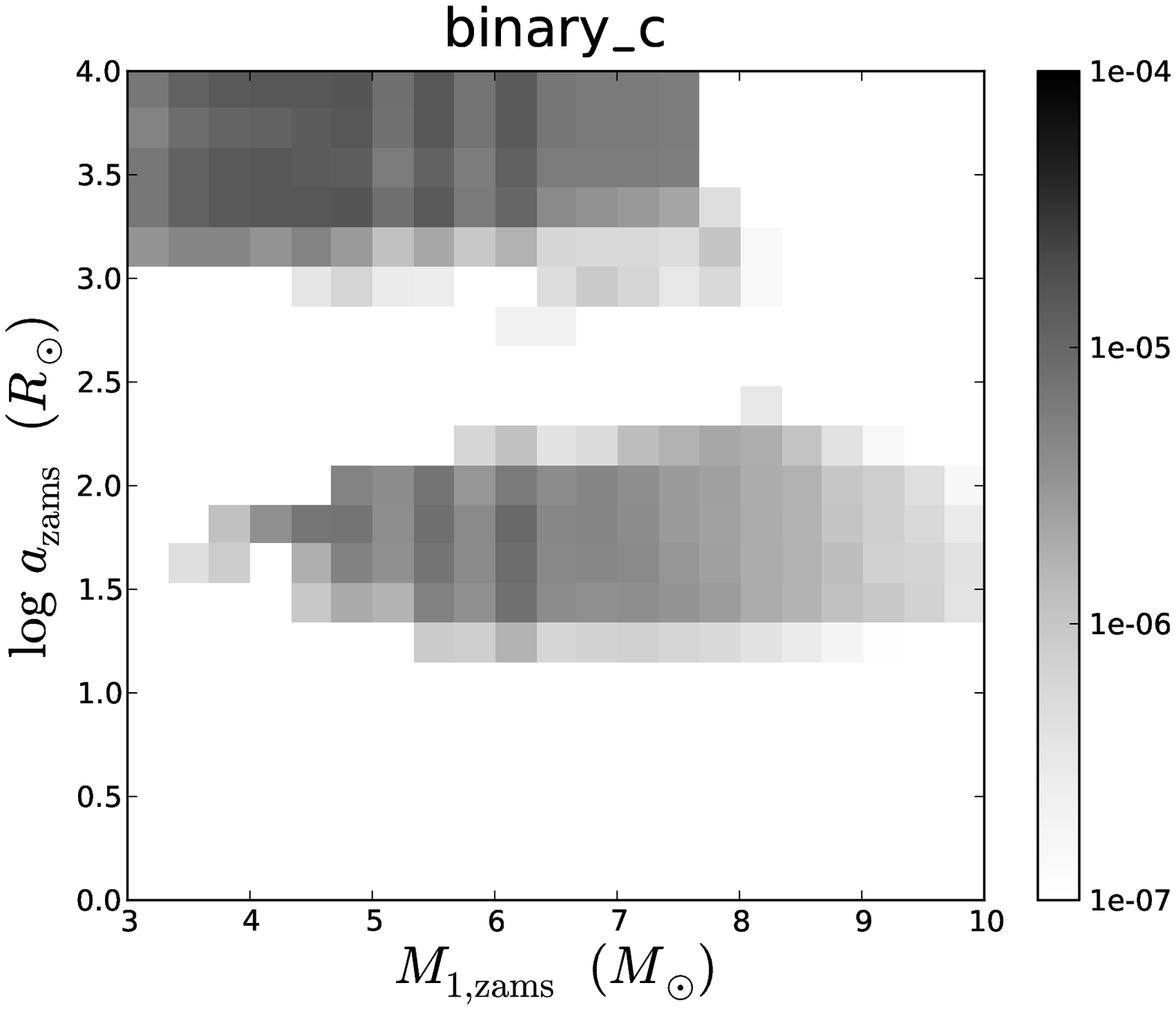} &
	\includegraphics[height=4.6cm, clip=true, trim =20mm 0mm 48.5mm 5mm]{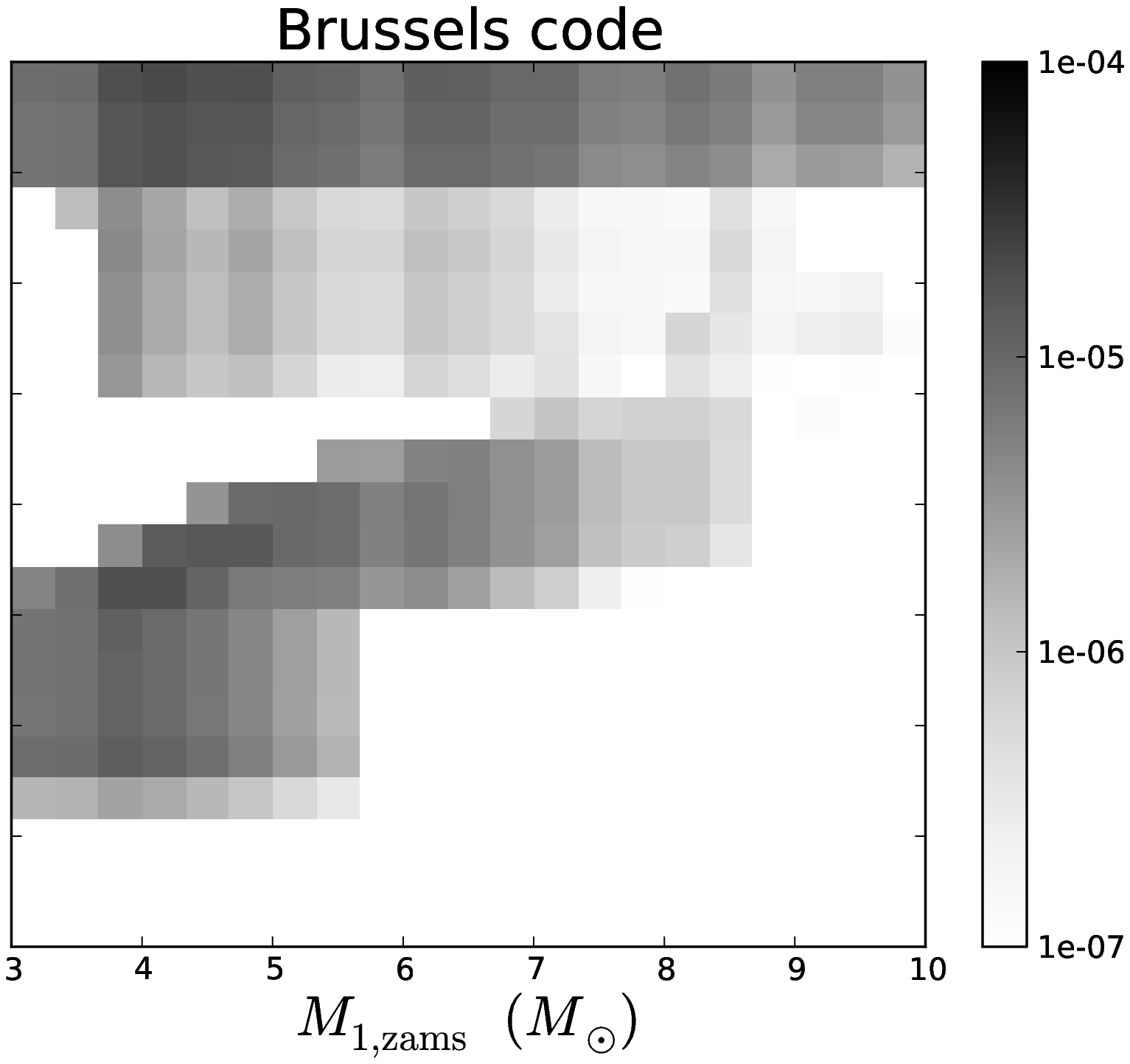} &
	\includegraphics[height=4.6cm, clip=true, trim =20mm 0mm 48.5mm 5mm]{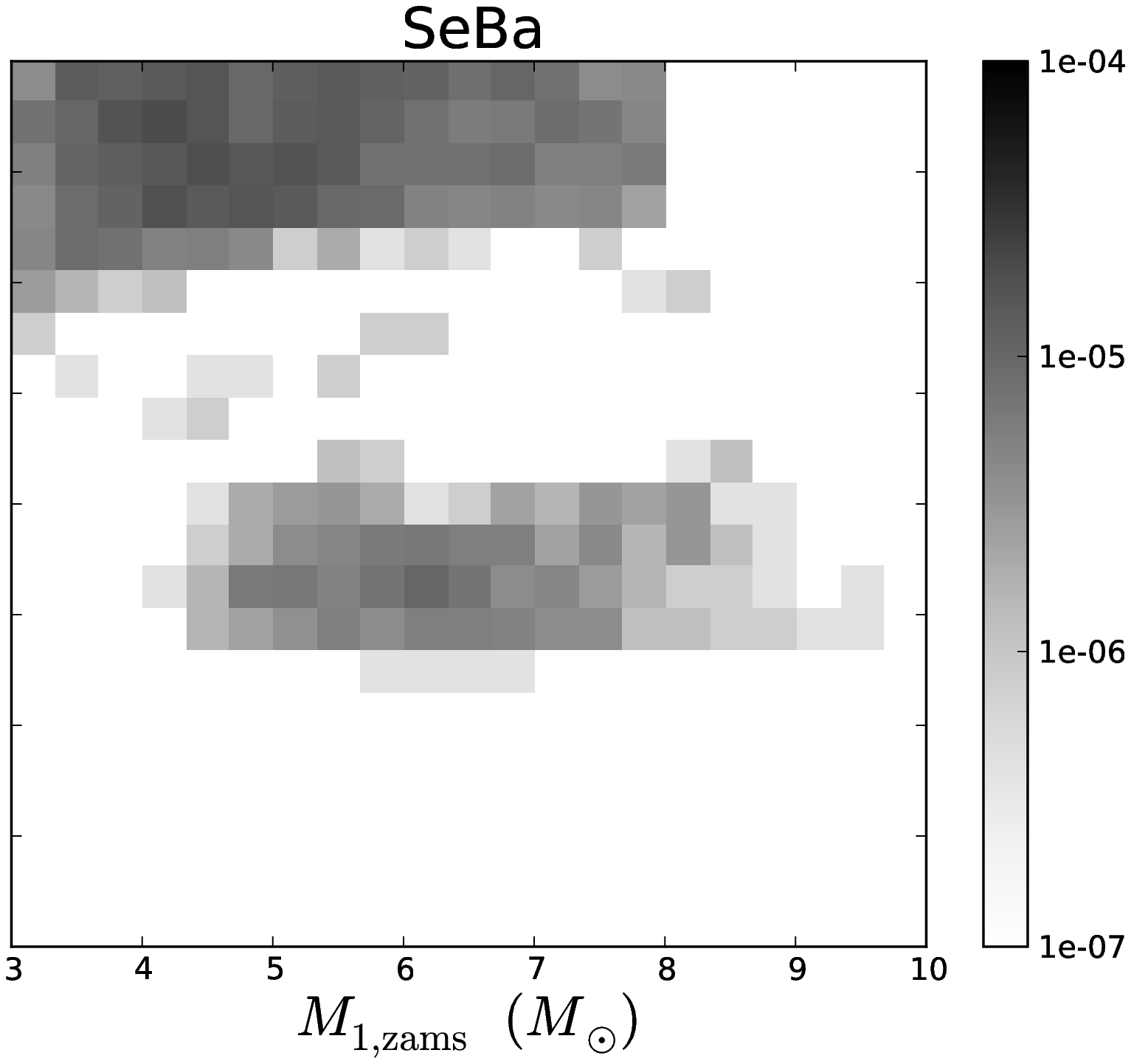} &
	\includegraphics[height=4.6cm, clip=true, trim =20mm 0mm 23mm 5mm]{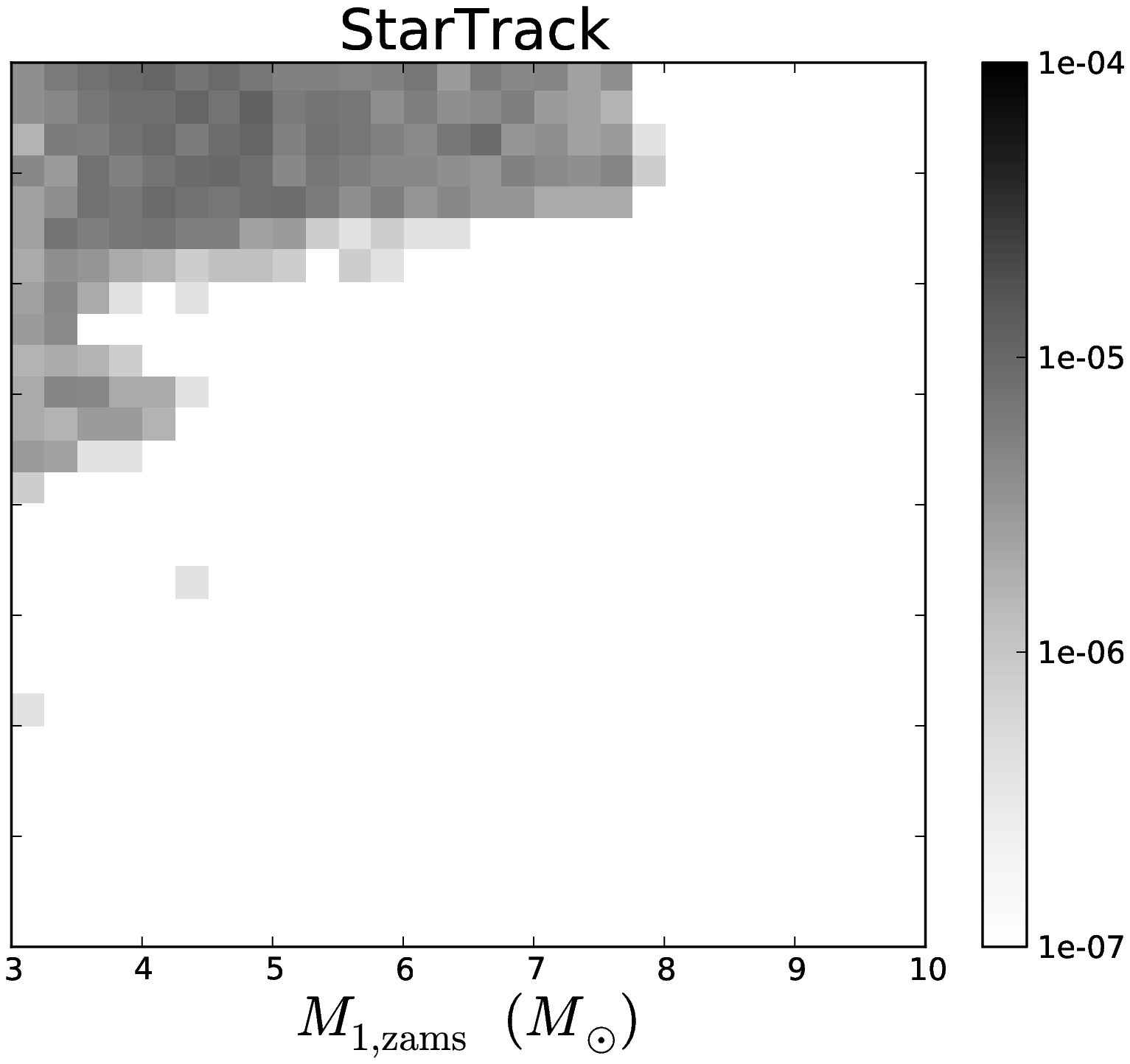} \\	
	\end{tabular}
    \caption{Initial orbital separation versus initial primary mass for all DWDs in the intermediate mass range. } 
    \label{fig:dwd_zams_a_IM_sin}
    \end{figure*}

In this section we compare and discuss the population of DWDs as predicted by binary\_c, the Brussels code, SeBa and StarTrack. 
Prior to the formation of a second degenerate component, DWDs undergo the evolution as described in the previous section. Subsequently, they undergo a second intrusive (series of) event(s) at the time the secondary fills its Roche lobe. 
As a consequence the processes that influence the evolution of SWDs influence the DWD population as well. Here we will point out the evolutionary processes that are specifically important for DWDs.

The population of DWDs at DWD formation is shown in
Fig.\,\ref{fig:dwd_a_sin},\,\ref{fig:dwd_a_IM_sin},\,\ref{fig:dwd_M2_sin},~and~\ref{fig:dwd_M2_IM_sin}
where orbital separation and secondary mass respectively is shown as a function of primary mass for the full and intermediate mass range. The ZAMS progenitors of these systems are shown in Fig.\,\ref{fig:dwd_zams_a_sin}~and~\ref{fig:dwd_zams_a_IM_sin} for the full and intermediate mass range respectively. 

In the full mass range, the population of DWDs is comparable between binary\_c, SeBa and StarTrack with WD masses of $M_{\rm 1, dwd} \approx 0.2-1.4$\Msolar~and~$M_{\rm 2, dwd}\approx 0.1-1.4$\Msolar. At large separations (0.1-5)$\cdot 10^4$\Rsolar~the codes find systems which are formed without any interaction, see Fig.\,\ref{fig:dwd_a_sin}. This figure also shows a population of interacting systems at lower separations, where the majority has separations of $a\approx 0.1-10$\Rsolar.
Furthermore there is a good agreement on which progenitors lead to a DWD system and which do not. Figure\,\ref{fig:dwd_zams_a_sin} shows several subpopulations of DWD progenitors with comparable binary parameters for binary\_c, SeBa and StarTrack; a group of non-interacting systems (at $a_{\rm dwd} \gtrsim 5\cdot 10^2$\Rsolar), a group of systems for which the first phase of mass transfer is stable (at $a_{\rm dwd} \lesssim 25$\Rsolar~for low mass primaries and $a_{\rm dwd} \lesssim 2.5\cdot 10^2$\Rsolar~for the full mass range), and a group of systems at intermediate separations that predominantly undergoes a CE-phase for the first phase of mass transfer.

Effects that play a role when comparing DWDs in the full mass range are the stability of mass transfer and differences in the interpretation of the double CE-phase in which both stars lose their envelopes (eq.\,\ref{eq:Egr_web}). The most pronounced effect of the differences in the stability of mass transfer is the 
decrease of systems that underwent stable mass transfer in the StarTrack data compared to binary\_c and SeBa. This can be seen in Fig.\,\ref{fig:dwd_zams_a_sin} in the lack of systems at $M_{\rm 1, zams}> 3$\Msolar~and $a_{\rm zams}< 2.5\times 10^2$\Rsolar~in the StarTrack data compared to binary\_c and SeBa, and in Fig.\,\ref{fig:dwd_M2_sin} in the lack of systems with $M_{\rm 2, dwd}>M_{\rm 1, dwd}$. Furthermore differences in the stability of mass transfer lead to an increase in systems at $a_{\rm dwd}\approx 10-50$\Rsolar~and $M_{\rm 1, dwd}\approx 0.4-0.47$\Msolar~according to SeBa and StarTrack.
Differences in the modelling of the double CE-phase result in larger separations at DWD formation and less mergers in StarTrack compared to binary\_c and SeBa (Fig.\,\ref{fig:dwd_a_IM_sin} and Appendix\,\ref{sec:channelII}). At the same time, the initial separations of systems evolving through a double CE-phase can be smaller in the StarTrack data compared to binary\_c and SeBa ($a_{\rm dwd}\approx 25-100$\Rsolar, see Fig.\,\ref{fig:dwd_zams_a_sin}).

In the intermediate mass range at DWD formation, two groups of systems can be distinguished in all simulations (Fig.\,\ref{fig:dwd_a_IM_sin}). Similar to the full mass range, there is one group of non-interacting systems at separations higher than $6\cdot 10^3$\Rsolar~and a group of interacting systems with separations $\lesssim 20$\Rsolar. However, the distribution of systems in the latter range varies between the codes. Most DWD systems have primary and secondary WD masses above 0.6\Msolar~in all the codes. The progenitors in the intermediate mass range show the same division in separation in three groups as the progenitors in the full mass range. DWD progenitors with separations $a_{\rm zams}\lesssim 3\cdot 10^2$\Rsolar~undergo a stable first phase of mass transfer. The components of DWD progenitors with $a_{\rm zams} \gtrsim 1.5\cdot 10^3 $\Rsolar~do not interact. At intermediate separations the first phase of mass transfer is predominantly a CE-phase. 

Comparing the Brussels code with binary\_c and SeBa (differences with StarTrack have the same origin as discussed in previous paragraphs), the most important causes for differences in the DWD population in the intermediate mass range are the MiMf-relation, the MiMwd-relation, the modelling of the stable mass transfer phase and the survival of mass transfer. The effect of the first three causes on the DWD population is similar to the effect on the SWD population. 
Firstly, the differences in the MiMf-relation can be seen in the progenitor population of non-interacting binaries in Fig.\,\ref{fig:dwd_zams_a_IM_sin} as an extension to higher primary masses in the Brussels data (8-10\Msolar, see also Appendix\,\ref{sec:channelI}). Secondly, differences in the MiMwd-relation can be seen in Fig.\,\ref{fig:dwd_a_IM_sin} as an extension to lower primary WD masses $M_{\rm 1,dwd}\lesssim 0.64$\Msolar~and the discontinuity in primary WD masses around 0.7-0.9\Msolar~(Appendix\,\ref{sec:channelII}~and~\ref{sec:channelIII}). The MiMwd-relation also effects the orbital separation distribution at DWD formation and results in a higher maximum separation in the Brussels code compared to binary\_c and SeBa. 
Finally, due to the method of modelling of mass transfer there is a disagreement between the codes regarding which systems survive mass transfer, see Fig.\,\ref{fig:dwd_zams_a_IM_sin} at $a_{\rm dwd}\lesssim 20$\Rsolar~(Appendix\,\ref{sec:channelIII}). The survival of mass transfer is more important for the DWD population than for the SWD population, as the average orbital separation of DWDs is lower (Sect.\,\ref{sec:swd} and also Appendix\,\ref{sec:channelII}~and~\ref{sec:channelIII}).
As the formation of DWDs involves more phases of mass transfer than for SWDs, the differences in the SWD population carry through and are larger in the DWD population. The DWD population in the full and intermediate mass range are discussed in more detail in Appendix\,\ref{sec:ev_path_dwd}.

\section{Overview of critical assumptions in BPS studies}
\label{sec:sum}

In the previous section we compared simulations from four different BPS codes and investigated the causes for the differences. The causes that we found are not numerical effects, but are inherent to the codes. In this section we list and discuss the underlying physical principles of the differences described in Sect.\,\ref{sec:Results}. 
The implementations of these principles in each code are described in Appendix\,\ref{sec:TNS_inherent}.

\begin{itemize}
\item Initial-WD mass-relation;\\
For single stars or non-interacting stars, the initial-final mass relation for WDs (Fig.\,\ref{fig:ifm}) is determined by the trade off between the growth of the core and how much mass is lost in stellar winds. The amount of mass a low or intermediate mass star loses in a stellar wind is small on the MS, but significant in later stages of its evolution. The amount of mass that is lost in the wind and in the planetary nebula influences the orbit directly, and indirectly through angular momentum loss (Sect.\,\ref{sec:wind}~and~\ref{sec:AMLwind}). 

The WD mass of primary stars is further affected by the mass transfer event, the moment and the timescale of the removal of the envelope mass. If the primary star becomes a hydrogen-poor helium burning star before turning into WD, the MiMwd-relation is influenced by helium star evolution. Of importance are the core mass growth versus the mass loss from helium stars and a possible second phase of mass transfer. A related issue, of particular importance for supernova Type Ia rates,
concerns the composition of WDs; what is the range of initial
masses for carbon-oxygen WDs or other types of WDs?

\item The stability of mass transfer; \\
For which systems does mass transfer occur in a stable manner and for
which systems is it unstable? As binary population synthesis codes do not solve the stellar structure equations, and cannot model stars that are not in hydrostatic or thermal equilibrium, BPS codes rely on parametrisations or interpolations to determine the stability of mass transfer. Theoretical stability criteria for polytropes exist \citep{HW87}, but are lacking for most real stars \citep[but see][for MS stars]{Mink07, Ge10, Ge13}. 

The critical mass ratio for stable mass transfer with hydrogen shell-burning donors differs between the codes from $q\gtrsim 0.2$ in the Brussels code to $q\gtrsim 0.6$ in StarTrack. A difference for low mass stars between binary\_c, SeBa and StarTrack arises from the uncertainty of the mass transfer stability of donors with shallow convective envelopes. In a recent paper, \citet{Woo12} show that mass transfer between a hydrogen shell-burning donor ($M_{\rm 1,zams}=1-1.3$\Msolar) and a main-sequence star can be stable when non-conservative. The effect on the orbit is a modest widening. 

\item Survival of mass transfer; \\
For which systems does mass transfer lead to a merger and which system survive the mass transfer phase, in particular when mass transfer is unstable? Different assumptions for the properties (e.g. radii) of stripped stars lead to differences in the results of the four BPS codes (e.g. channel~II~and~III in Appendix\,\ref{sec:ev_path_dwd}). 
For donor stars in which the removal of the envelope due to mass transfer leads to an end in nuclear (shell) burning and a WD is formed directly, it is unclear how much the core is bloated just after mass transfer ceases compared to a zero-temperature WD \citep{Hur00}. 
For donor stars that are stripped of their hydrogen envelopes due to mass transfer, but helium burning continues, it is unclear how fast the transition takes place from an exposed core to an (evolved) helium star (channel~2b in Appendix\,\ref{sec:ev_path_swd}).

\item Stable mass transfer;\\
Modelling of the stable mass transfer phase in great detail is not possible in BPS codes, as for the stability of mass transfer. Therefore BPS codes rely on simplified methods to simulate stable mass transfer events. 
The evolution of the mass transfer rate during the mass transfer phase can have a strong effect on the resulting binary. However, in the current set-up of this project that assumes conservative mass transfer, the importance is greatly reduced. The mass transfer rates are only important when the timescale of other effects (e.g. wind mass loss or nuclear evolution) become comparable to the mass transfer timescale (channel~3b in Appendix\,\ref{sec:ev_path_swd}).

A result of the approach is
that mergers are less likely to happen in the Brussels code compared
to the other codes (e.g. channel~5 in Appendix\,\ref{sec:ev_path_swd}). The approach of binary\_c, SeBa and StarTrack is
to follow the mass transfer phase in time, with approximations of the
mass transfer rate. In the Brussels code, the mass transfer phases are
not followed in detail. Instead it only considers the initial and final
situation from interpolations of a grid of detailed
calculations. 
Furthermore, it is important to better understand which contact systems lead to a merger and which do not. From observations, many Algol systems are found which have undergone and survived a phase of shallow contact.

\item The evolution of helium stars;\\ 
A large fraction of interacting systems go through a phase in which one of the stars is a helium star, for SWDs roughly 15\% in the full mass range and roughly 50\% in the intermediate mass range. 
These objects are not well studied and there remain several uncertainties, e.g. mass transfer stability.
Also the mass transfer rate is important, in particular for evolved helium stars whose evolutionary and wind loss timescales can become comparable to the mass transfer timescales. Therefore small differences in the mass transfer rate can lead to large differences in the resulting WD. This is especially important for massive WDs, e.g. SNIa rates. 
\end{itemize}

The influence of the parameters that were equalised in this project has not been studied here, neither qualitatively nor quantitatively. These parameters will lead to a larger diversity in the simulated populations as different groups make different assumptions in their codes (Appendix\,\ref{sec:TNS_equalized}) and these should be taken into account when interpreting BPS results. These assumptions are:

\begin{itemize}
\item the CE-prescription and efficiency;
\item accretion efficiency;
\item angular momentum loss during RLOF;
\item tidal effects;
\item magnetic braking; 
\item the initial distributions of primary mass, mass ratio and orbital separation.
\end{itemize}
Despite the significance of these phenomena to binary evolution and the efforts of the community to understand and quantify them, there remain questions about these phenomena. Several prescriptions exist for these phenomena and the effect on a binary population can be severe. 
For example regarding the CE-phase, it is unclear how efficient orbital energy can be used to expel the envelope and if other sources of energy can be used, 
such as recombination, rotational, tidal or magnetic energy \citep{Ibe93, Han95, Pol07, Web08, Zor10, DeM11, Zor11, Dav12, Iva13}. 
Also, predictions for the efficiency of mass accretion onto WDs vary strongly and the SNIa rate is severely affected by this uncertainty \citep{Bou13}. Furthermore, the adopted mode of angular momentum loss has a strong effect on the evolution of the orbit (Fig.\,\ref{fig:AML} and Appendix\,\ref{sec:AMLRLOF}). 
It is also not clear how the different prescriptions for tidal evolution affect the populations. However, in Appendix\,\ref{sec:channel3} we find that spin-orbit coupling (assuming orbits are continuously synchronised), only has a small effect on the final separation of the SWD systems.
The effect of different initial distributions \citep[see e.g.][]{Duq91, Kou07} of binary parameters can be severe with respect to the birthrate of a stellar population \citep[see e.g.][]{Egg89, deK93, Dav10, Cla13}.
Furthermore, the importance of a certain channel is affected by the boundaries of the distribution through the normalisation of the simulation.

\section{Conclusion}
\label{sec:concl}
In this paper we studied and compared four binary population synthesis
codes. The codes involved are the binary\_c code \citep[][]{Izzard04,
  Izzard06, Izzard09, Cla13}, the Brussels code \citep{DeD04, Men10,
  Men13}, SeBa \citep{Por96, Nel01, Too12, Too13} and StarTrack \citep{Bel02a,
  Bel08a, Rui09, belczynski2010c}. We focused on low and intermediate mass binaries that evolve into single white dwarf systems (containing a WD and a non-degenerate companion) and double white dwarf systems. These populations are interesting for e.g. post-CE binaries, cataclysmic variables, single degenerate as well as double degenerate supernova Type Ia progenitors. For this project input assumptions in the BPS codes were equalised as far as the codes permit. This was done to simplify the complex problem of comparing BPS codes that are based on many (often different) assumptions. In this manner inherent differences between and numerical effects within the codes were investigated. 

Regarding the SWD population, there is a general agreement on what initial parameters of $M_{\rm 1, zams}$, $M_{\rm 2, zams}$ and $a_{\rm zams}$ lead to SWD binaries and which parameters do not lead to SWDs. 
When the SWD system is formed, there is an agreement on the orbital separation range for those systems having undergone stable or unstable mass transfer. Furthermore there is a general agreement on the stellar masses after a phase of stable or unstable mass transfer and between the populations of the most common evolutionary channels.

Regarding the DWD population, there is an agreement on which primordial binaries lead to DWD systems through stable and unstable mass transfer respectively, and a rough agreement on the characteristics ($M_{\rm 1, dwd}$, $M_{\rm 2, dwd}$ and $a_{\rm dwd}$) of the DWD population itself. DWD systems go through more phases of evolution than SWD systems and therefore the uncertainty in their evolution builds up after each mass transfer phase. The WDs are formed with comparable masses, but at different separations.
The most important evolutionary paths leading to DWDs are similar between the BPS codes.

We found that differences between the simulated populations are not due to numerical differences, but due to different inherent assumptions. The most important ones that lead to differences are the MiMf-relations (of single stars), the MiMwd-relation (of binary stars), the stability of mass transfer, the modelling of the mass transfer rate and the modelling of helium star evolution. Different assumptions between the codes are made for these topics as theory is poorly understood and sometimes poorly studied. Further research into these topics is necessary to eliminate the differences between BPS codes e.g. with a detailed (binary) stellar evolution code.

In addition some assumptions that affect the results of the codes were equalised for the comparison. These are the initial binary distributions, the CE-prescription and efficiency, the accretion efficiency, angular momentum loss during RLOF, tidal effects, magnetic braking and wind accretion. We leave the study of their effects on stellar populations for another paper.

In Sect.\,\ref{sec:BPS} a short description is given of each code. In Appendix\,\ref{sec:TNS_inherent}~and~\ref{sec:TNS_equalized}, a more detailed overview is given of the typical assumptions of each code outside the current project. These should be taken into account when interpreting results from the BPS codes. 
Furthermore, we recommend using these sections as a guideline when deciding which code or results to use for which project.  
Finally we would like to encourage other groups involved in BPS simulations, to do the same test as described in this paper and compare the results with the figures given in this paper. More detailed figures and information are available on request and on the website http://www.astro.ru.nl/$\sim$silviato/popcorn. 

Concluding, we found that when the input assumptions are equalised as far as possible within the codes, we find very similar populations and birthrates. 
Differences are caused by different assumptions for the physics of binary evolution, not by numerical effects. So although the four BPS codes use very different ways to simulate the evolution of these systems, the codes give similar and consistent results and are adequate for studying populations of low- and intermediate mass stars.

\begin{acknowledgements}      
The authors would like to thank G. Nelemans, M. van der Sluys,
O. Pols, D. Vanbeveren, W. van Rensbergen, \& R. Voss for very helpful
discussions on this project. 
ST thanks S. Portegies Zwart, JSWC thanks R.G. Izzard and AJR thanks I. Seitenzahl and
K. Belczynski for thoughtful comments and general discussion. 
The authors are very grateful to those involved in the Lorentz
Center workshops ``Stellar mergers'' (2009) and 
``Observational signatures of Type Ia supernova
progenitors'' (2010), who provided the inspiration to carry
out this long-awaited work. 
Furthermore the authors are grateful for the contributions to an early version of this project in 2002-2003 from K. Belczynski, C. Fryer, J. Hurley, V. Kalogera, G. Nelemans, S. Portegies Zwart and C. Tout, and in 2009-2010 from A. Bogomazov, P. Kiel, B. Wang and L. Yungelson. 
The work of ST and JSWC was 
supported by the Netherlands Research School for Astronomy (NOVA). 
\end{acknowledgements}

\begin{appendix} 
\section{Detailed comparison}
\label{sec:ev_path}

\begin{table*}
\caption{Birthrates in $\peryr$ for different evolutionary channels
  (described in Sect.\,\ref{sec:Results}) of single and double white
  dwarf systems for the three BPS codes for the full mass range and
  the intermediate mass range. }
\begin{tabular}{|l||c|c|c||c|c|c|c|}\hline
Evolutionary channels & \multicolumn{3}{|c||}{Full mass range}  & \multicolumn{4}{|c|}{Intermediate mass range} \\
\hline
 & binary\_c & SeBa & StarTrack & binary\_c & Brussels code & SeBa & StarTrack\\
 \hline  \hline 
SWD systems & 0.048 & 0.052 & 0.048 &   5.1e-3 &  7.8e-3 & 5.2e-3 & 4.4e-3\\
Channel 1 & 0.026 & 0.026 & 0.026 &   2.2e-3 &  1.9e-3 &  2.5e-3 & 2.3e-3\\
Channel 2a &   6.9e-3 &    6.5e-3 & 6.8e-3 &   1.1e-3 &  2.6e-3 &   1.1e-3 & 1.1e-3\\
Channel 2b &  5.7e-4   &  5.8e-4  & 5.0e-4 &   5.7e-4   & - &  5.8e-4  &4.8e-4\\
Channel 3a &   1.4e-3 &    4.2e-3 & 9.8e-4 &   4.0e-4   & 1.0e-3  & 2.9e-4  &8.7e-5\\
Channel 3b &   5.7e-4   &  4.6e-4  & 1.3e-4 &   5.7e-4   & 8.2e-4  & 4.6e-4  &1.3e-4\\
Channel 4a & 0.012 &   0.012 & 0.012 & 1.8e-6 & 3.6e-6 & 2.4e-6& 1.6e-6\\		
Channel 4b &   1.8e-4   &   8.9e-5 & 1.8e-4 &   1.8e-4   & 1.8e-4  &
8.9e-5 & 1.8e-4\\
Channel 5 &    2.4e-4   &   5.6e-4  & 3.6e-4 & 9.1e-6 &  1.2e-3 &
5.4e-5 & 2.9e-5\\
&&& &&&&\\
DWD systems & 0.012 & 0.014 & 0.015 &   8.4e-4   &  1.1e-3 & 8.7e-4  &  6.6e-4\\
Channel I &   8.4e-3 &  8.8e-3 & 8.4e-3 &   4.9e-4   & 5.5e-4  & 5.5e-4  & 5.1e-4\\
Channel II &   2.0e-3 &  1.3e-3 & 4.5e-3 & 4.5e-5 & 7.6e-5 & 3.5e-5 & 7.8e-5\\
Channel III &   1.3e-3 &  3.0e-3 & 9.9e-4 &    2.5e-4   & 4.9e-4  & 1.8e-4  &8.1e-7\\
Channel IV &   1.6e-4   & 5.5e-5 & $\lesssim$ 4e-7 & $\lesssim $ 4e-7 & - & $\lesssim$ 4e-7  & $\lesssim$ 4e-7 \\
 \hline 
 \end{tabular}
\label{tbl:birthrates_all}
\end{table*}

\subsection{Single white dwarf systems}
\label{sec:ev_path_swd}
    \begin{figure*}
    \centering
    \setlength\tabcolsep{0pt}
    \begin{tabular}{ccc}
	\includegraphics[height=4.6cm, clip=true, trim =8mm 0mm 48.5mm 5mm]{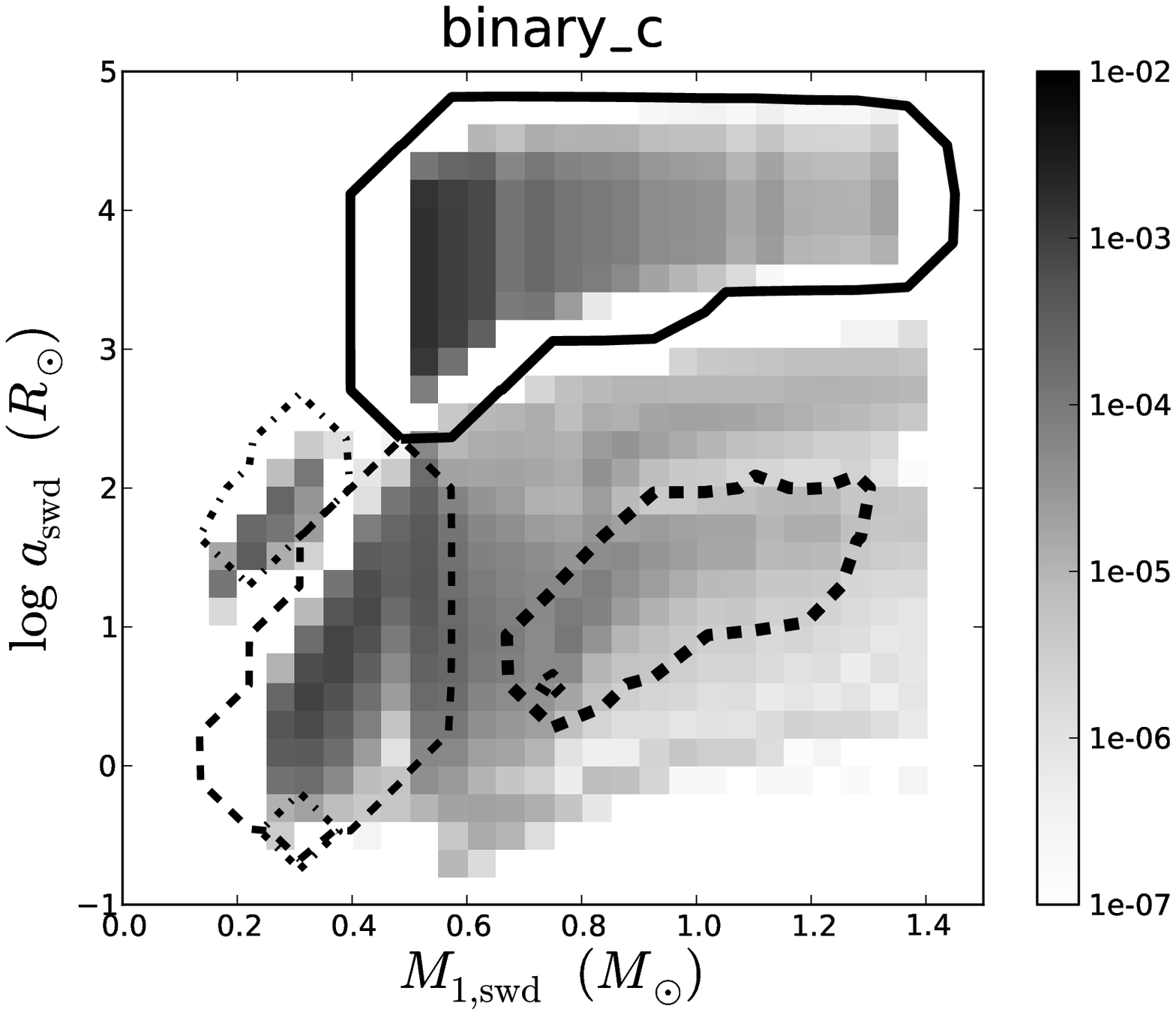} &
	\includegraphics[height=4.6cm, clip=true, trim =20mm 0mm 48.5mm 5mm]{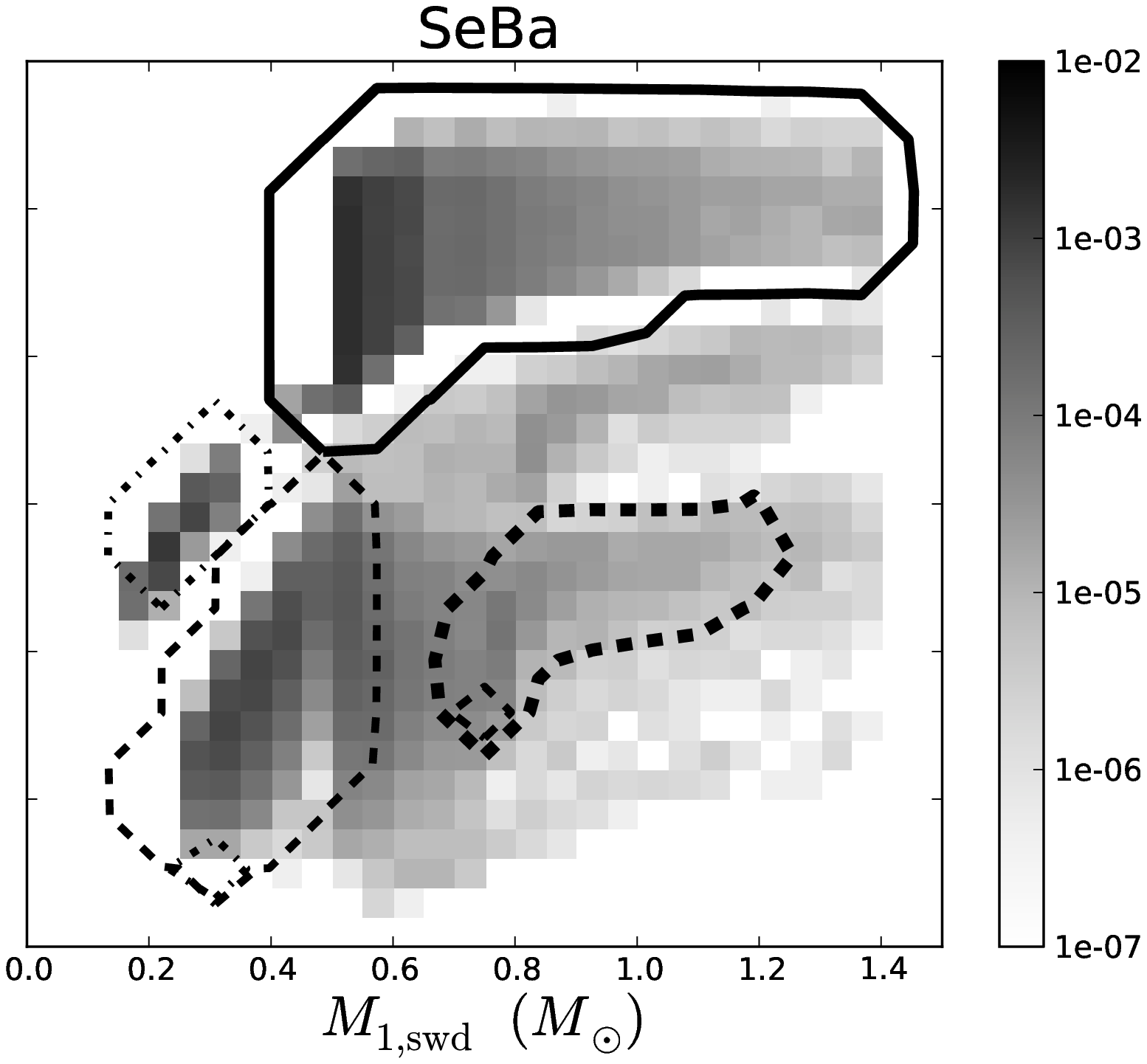} & 
	\includegraphics[height=4.6cm, clip=true, trim =20mm 0mm 23mm 5mm]{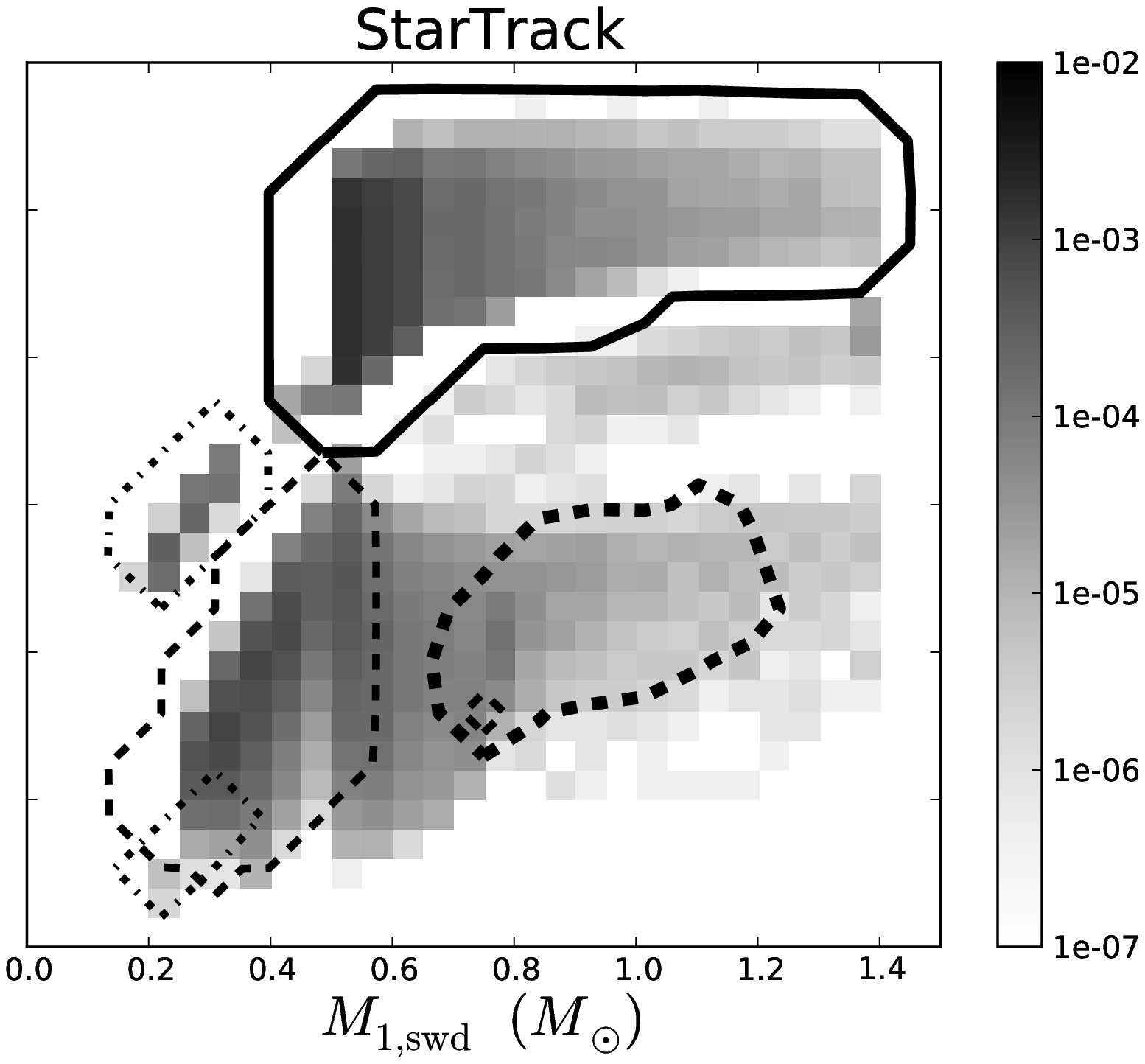} \\
	\end{tabular}
    \caption{Orbital separation versus WD mass for all SWDs in the full mass range at the time of SWD formation. The contours represent the SWD population from a specific channel: channel~1 (solid line), channel~4a (thin dashed line), channel~4b (thick dashed line) and channel~5 (dash-dotted line).} 
    \label{fig:swd_final_a_R1}
    \end{figure*}

    \begin{figure*}
    \centering
    \setlength\tabcolsep{0pt}
    \begin{tabular}{cccc}
	\includegraphics[height=4.6cm, clip=true, trim =8mm 0mm 48.5mm 5mm]{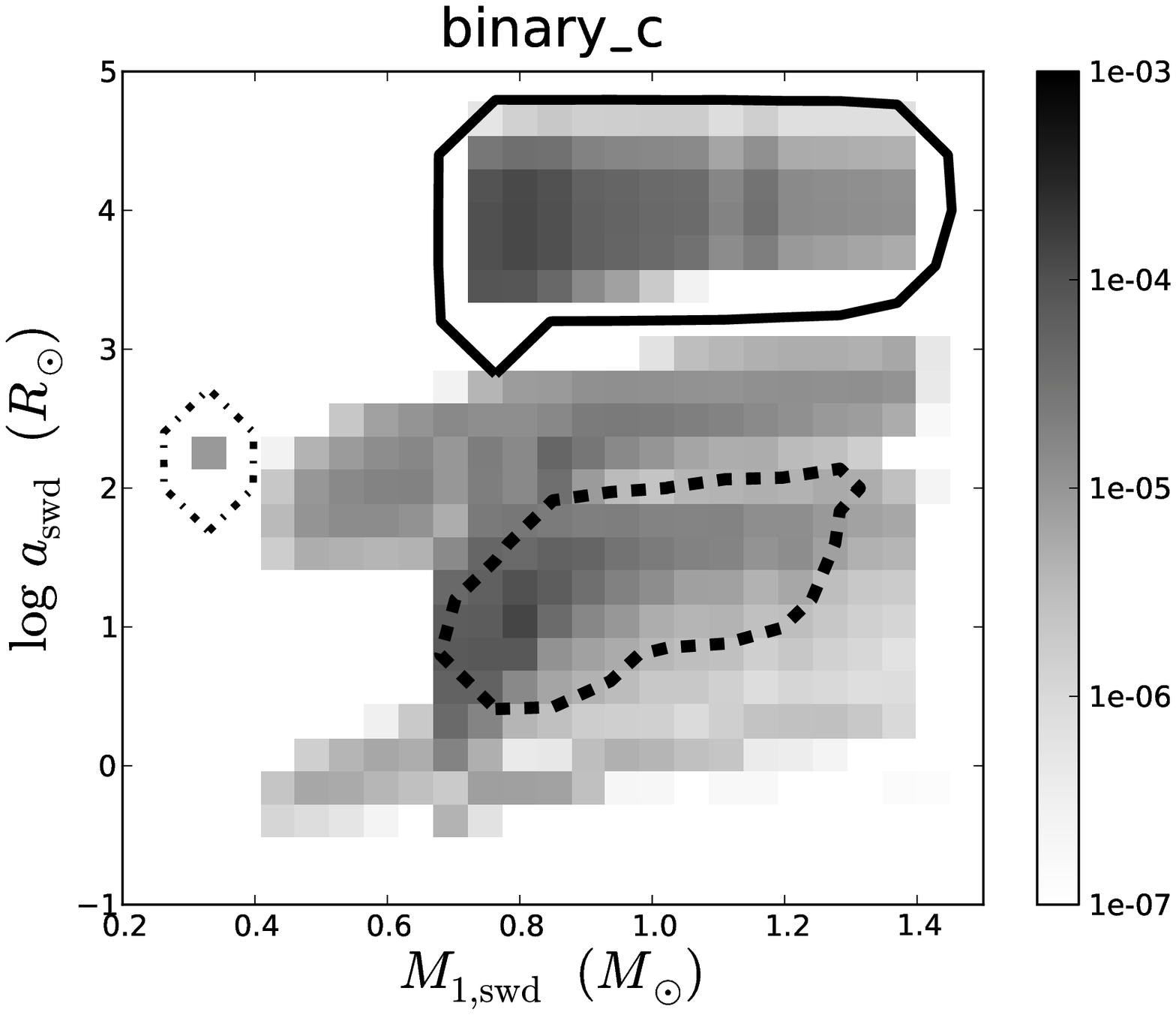} &
	\includegraphics[height=4.6cm, clip=true, trim =20mm 0mm 48.5mm 5mm]{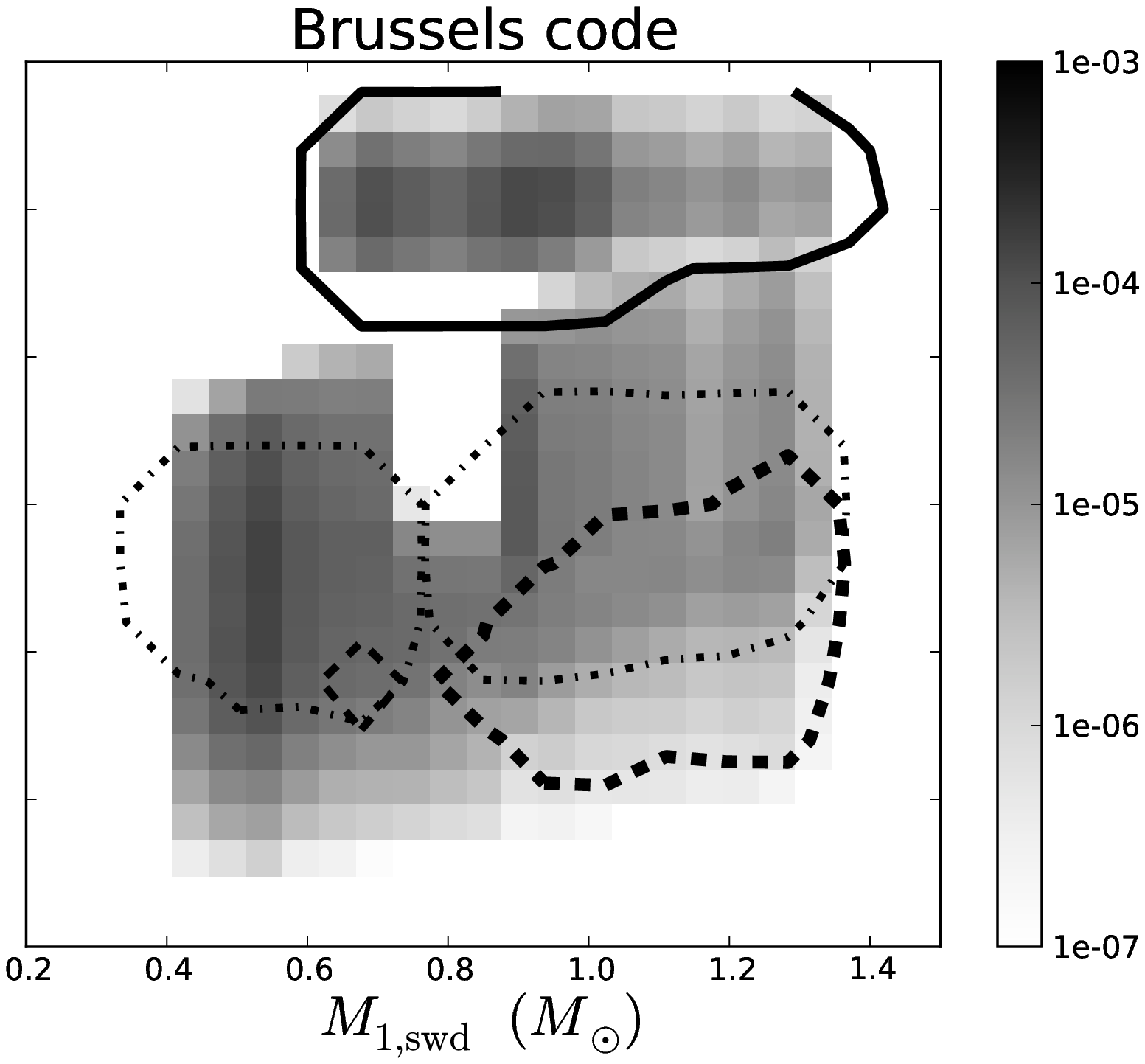} &
	\includegraphics[height=4.6cm, clip=true, trim =20mm 0mm 48.5mm 5mm]{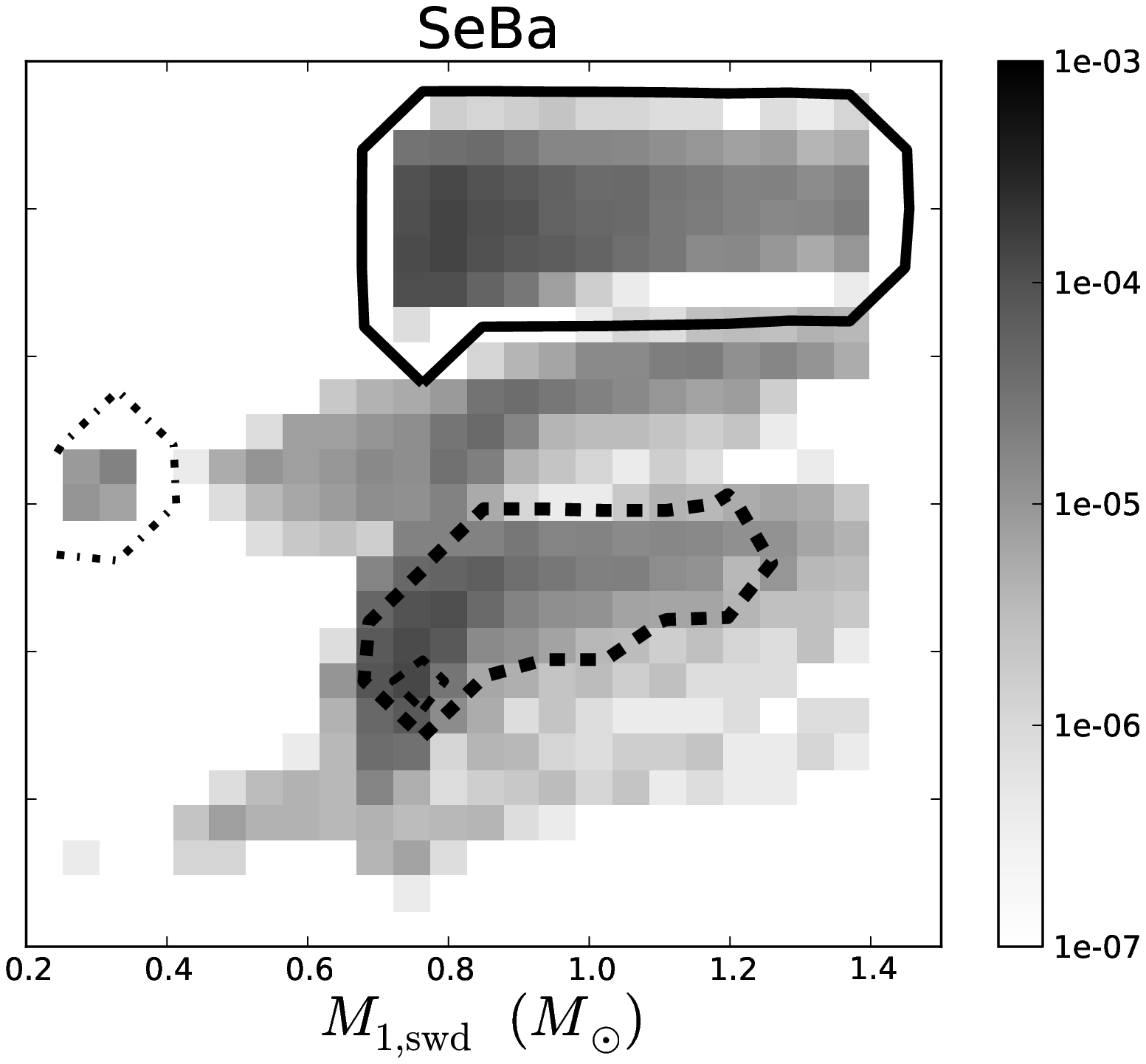} &
	\includegraphics[height=4.6cm, clip=true, trim =20mm 0mm 23mm 5mm]{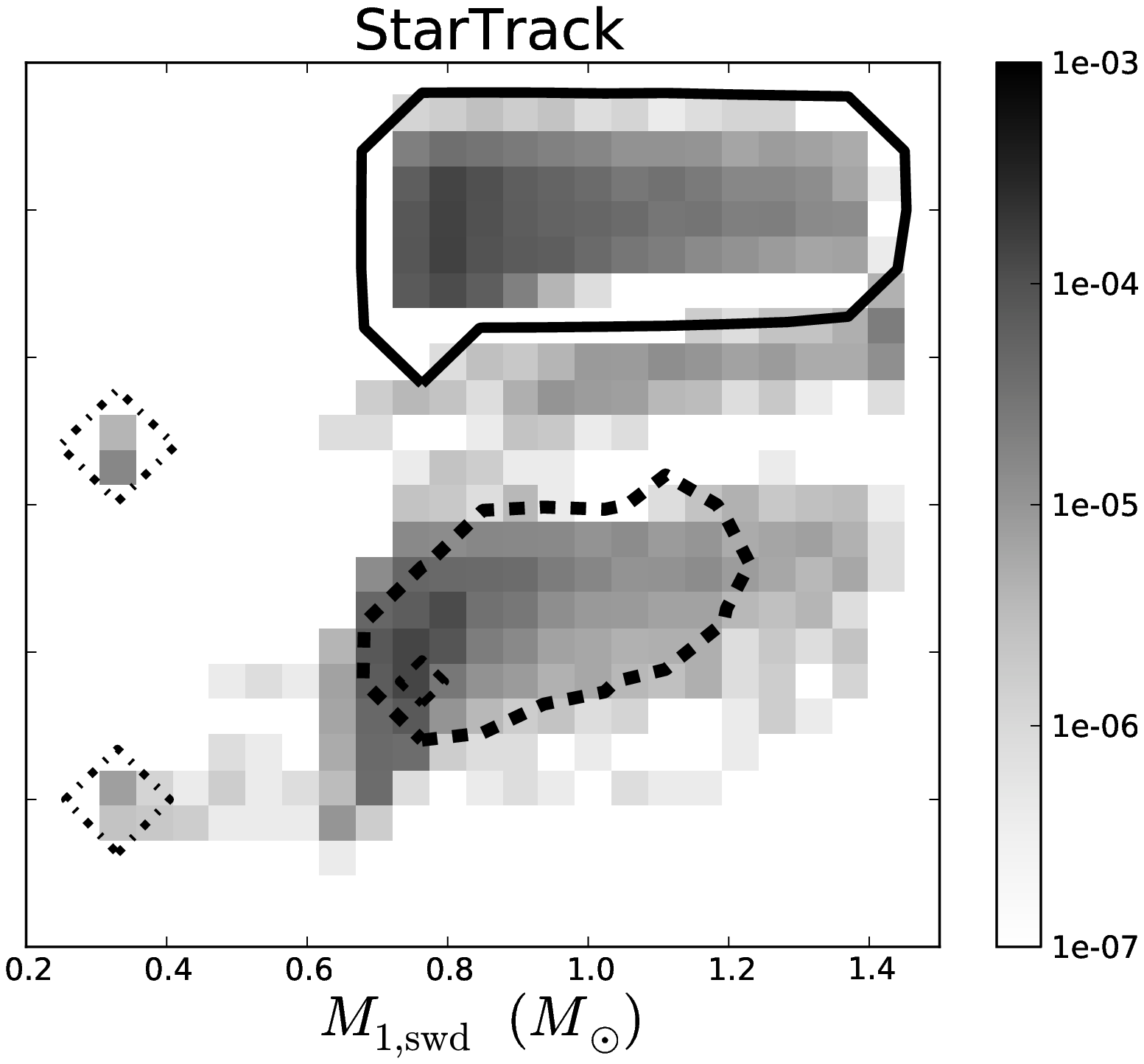} \\
	\end{tabular}
    \caption{Orbital separation versus WD mass for all SWDs in the intermediate mass range at the time of SWD formation. The contours represent the SWD population from a specific channel: channel~1 (solid line), channel~4a (thin dashed line), channel~4b (thick dashed line) and channel~5 (dash-dotted line). }
    \label{fig:swd_final_a_R1_IM}
    \end{figure*}

     \begin{figure*}
    \centering
    \setlength\tabcolsep{0pt}
    \begin{tabular}{ccc}
	\includegraphics[height=4.6cm, clip=true, trim =8mm 0mm 48.5mm 5mm]{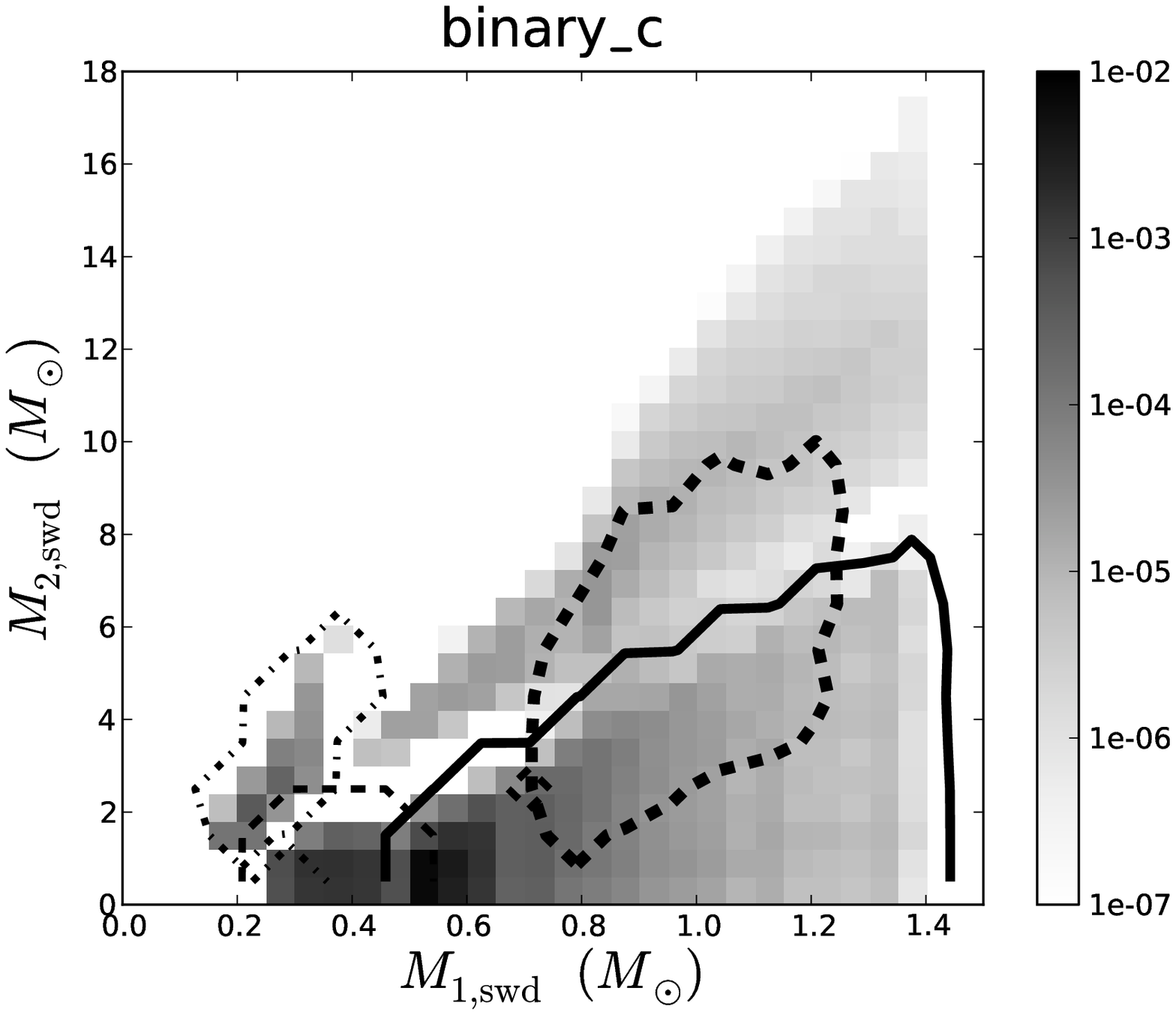} &
	\includegraphics[height=4.6cm, clip=true, trim =20mm 0mm 48.5mm 5mm]{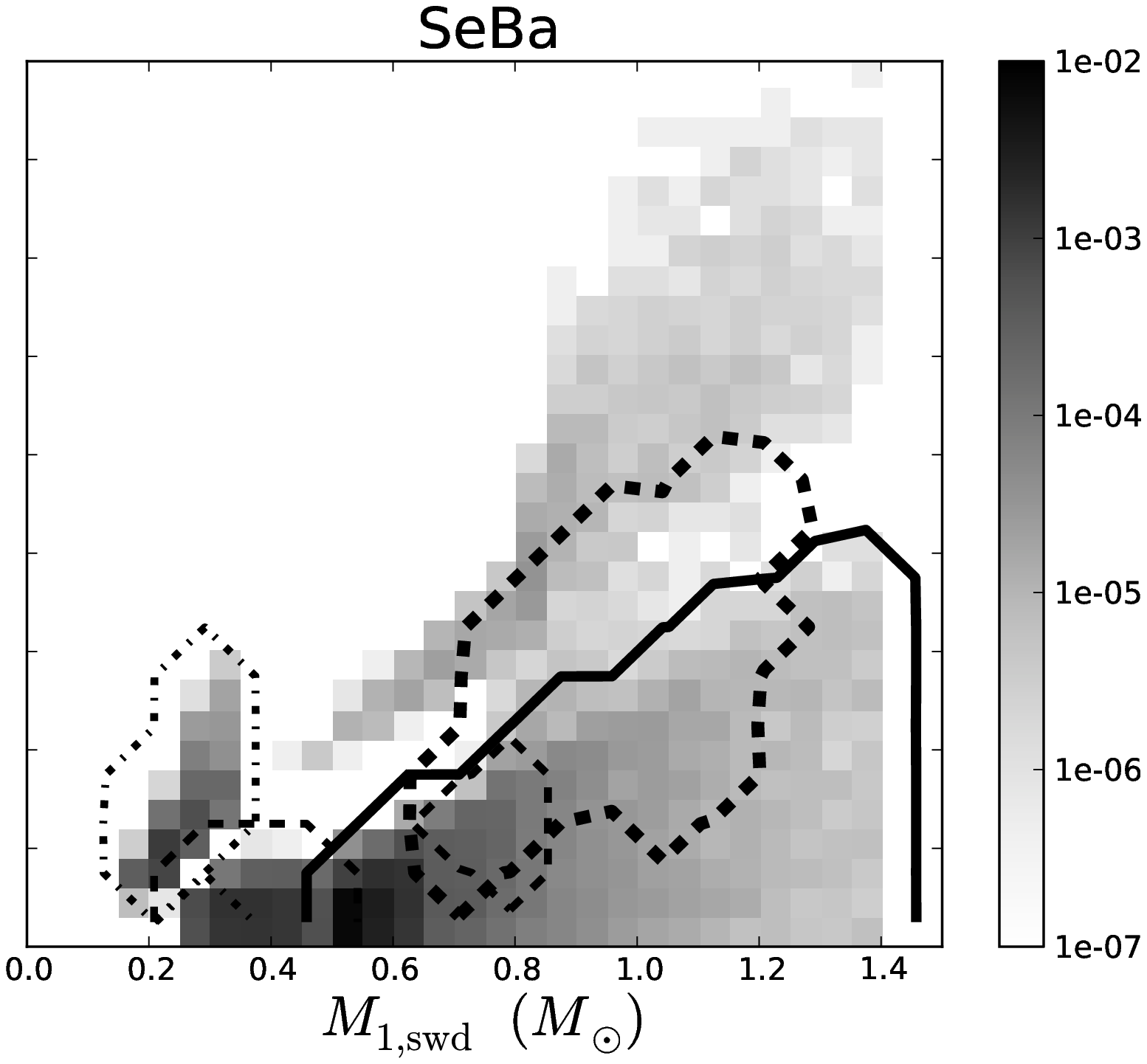} &
	\includegraphics[height=4.6cm, clip=true, trim =20mm 0mm 23mm 5mm]{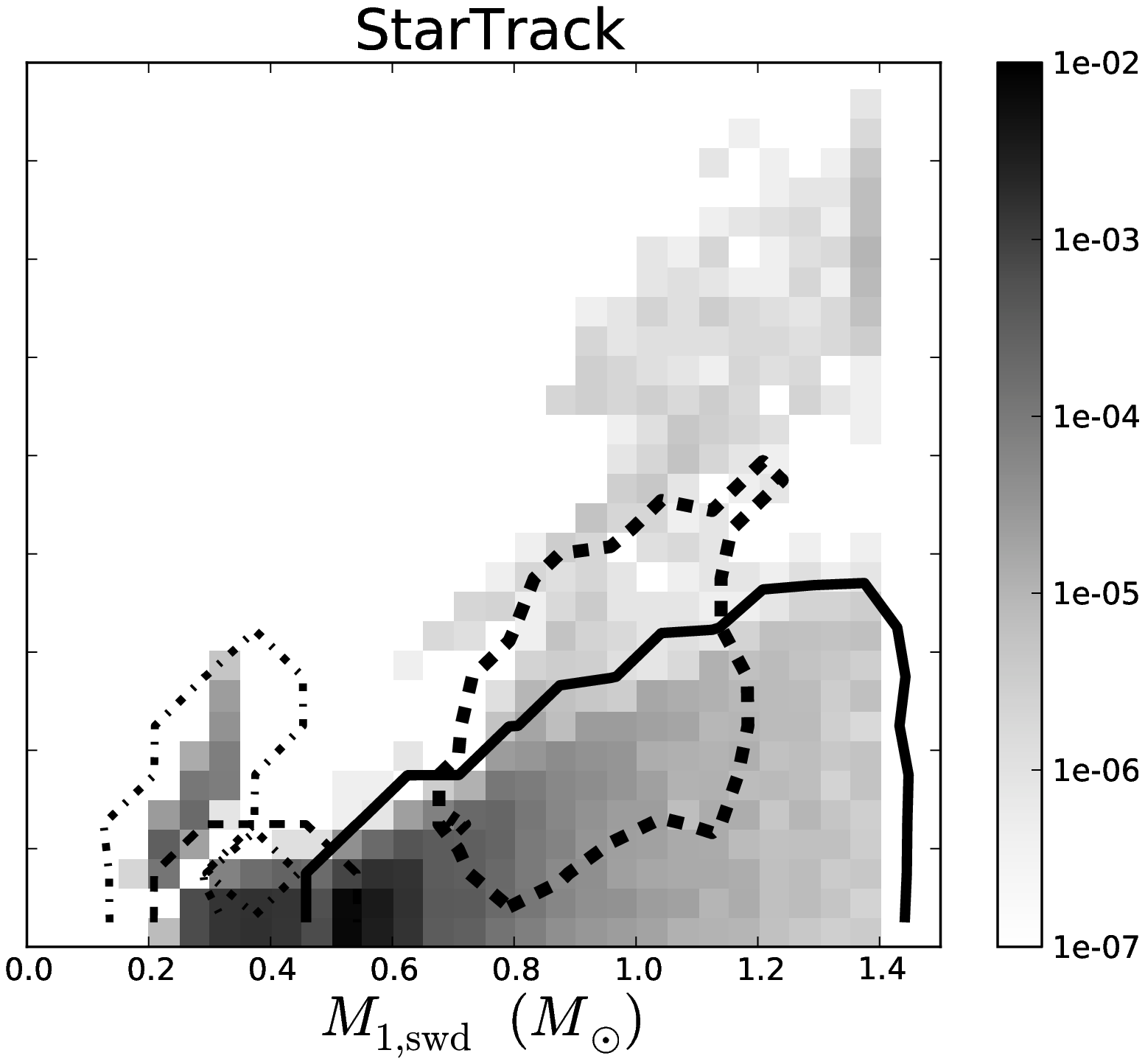} \\
	\end{tabular}
    \caption{Secondary mass versus WD mass for all SWDs in the full mass range at the time of SWD formation. The contours represent the SWD population from a specific channel: channel~1 (solid line), channel~4a (thin dashed line), channel~4b (thick dashed line) and channel~5 (dash-dotted line). }
    \label{fig:swd_final_m2_R1}
    \end{figure*}

     \begin{figure*}
    \centering
    \setlength\tabcolsep{0pt}
    \begin{tabular}{cccc}
	\includegraphics[height=4.6cm, clip=true, trim =8mm 0mm 48.5mm 5mm]{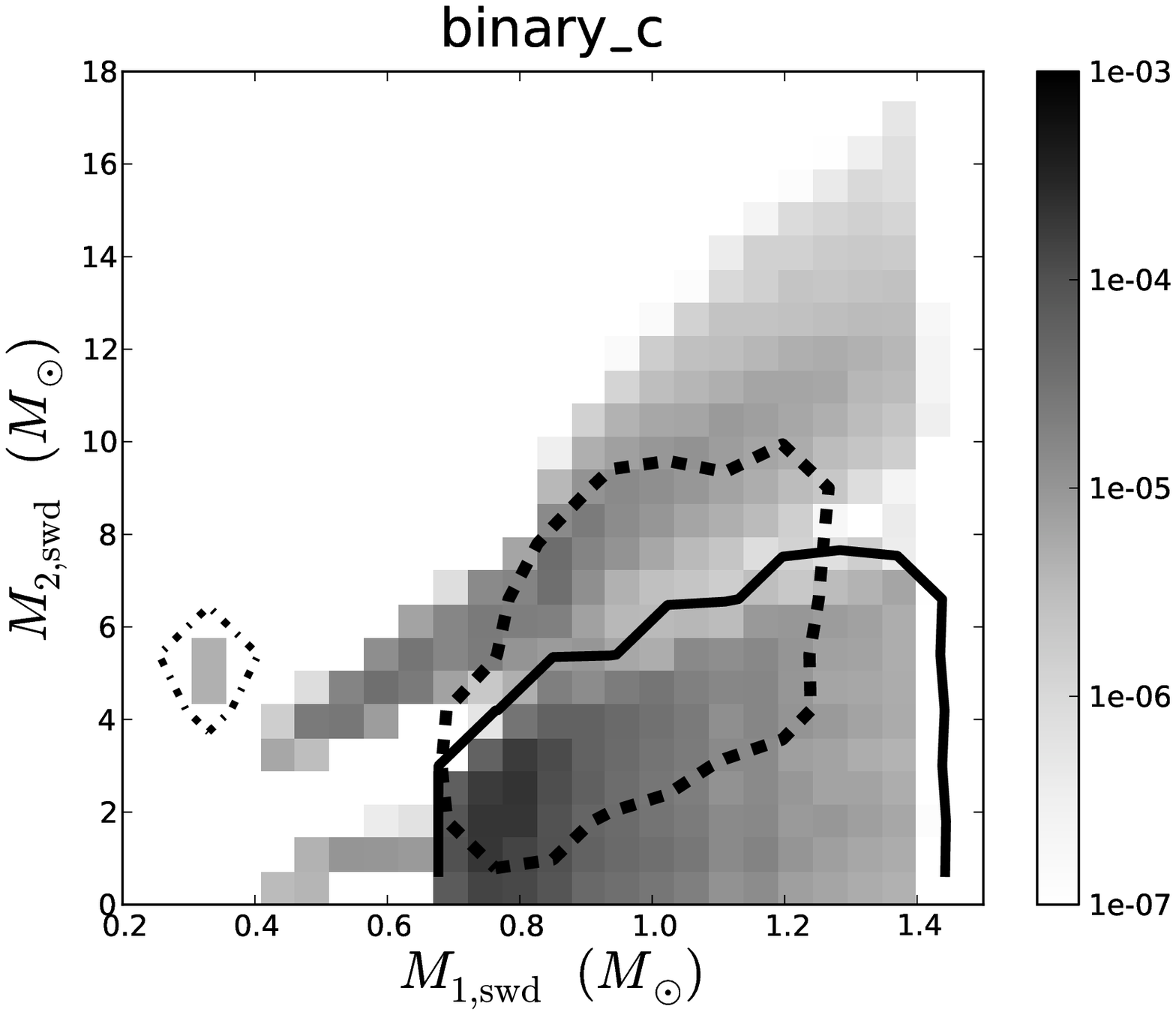} &
	\includegraphics[height=4.6cm, clip=true, trim =20mm 0mm 48.5mm 5mm]{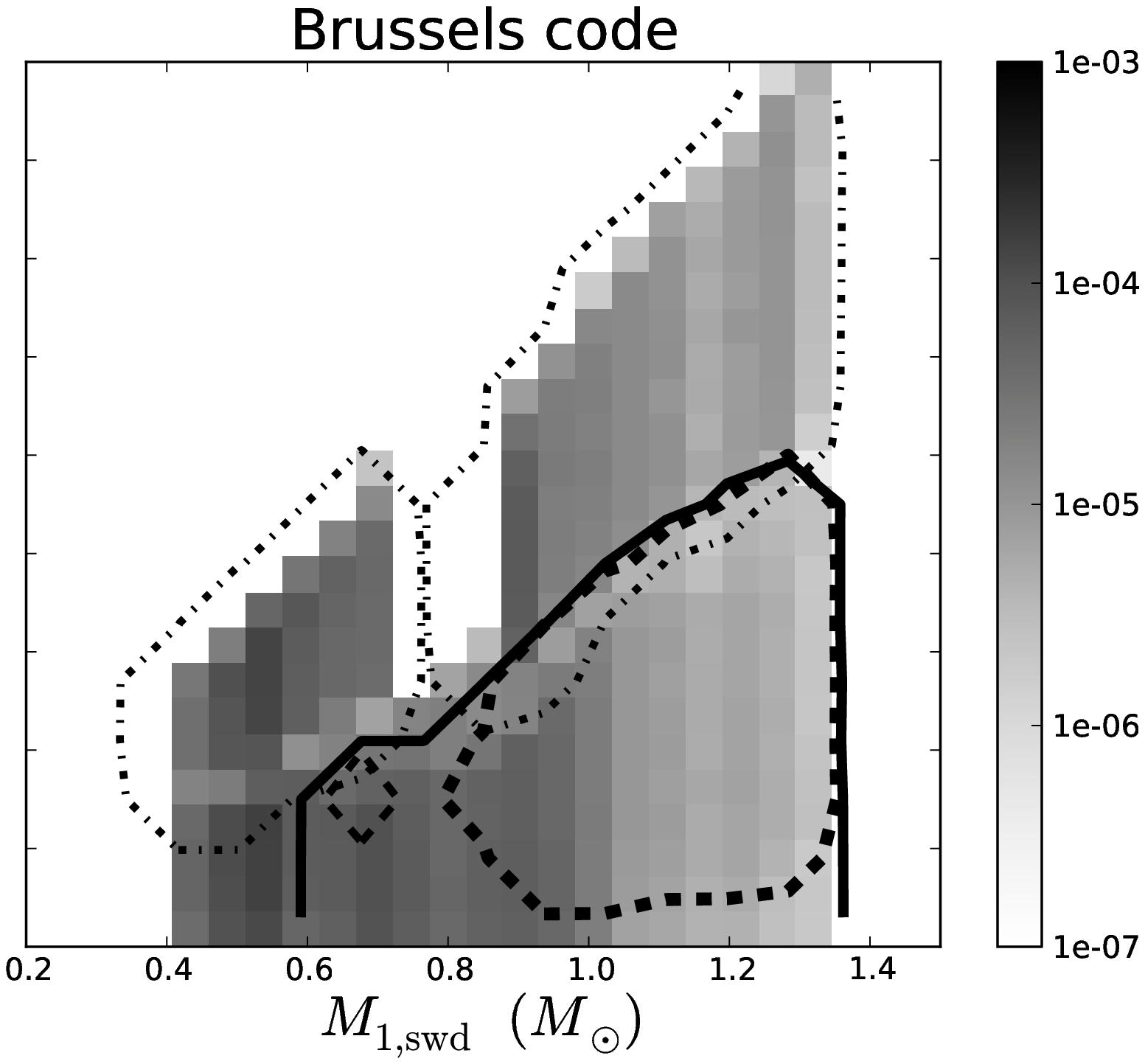} &
	\includegraphics[height=4.6cm, clip=true, trim =20mm 0mm 48.5mm 5mm]{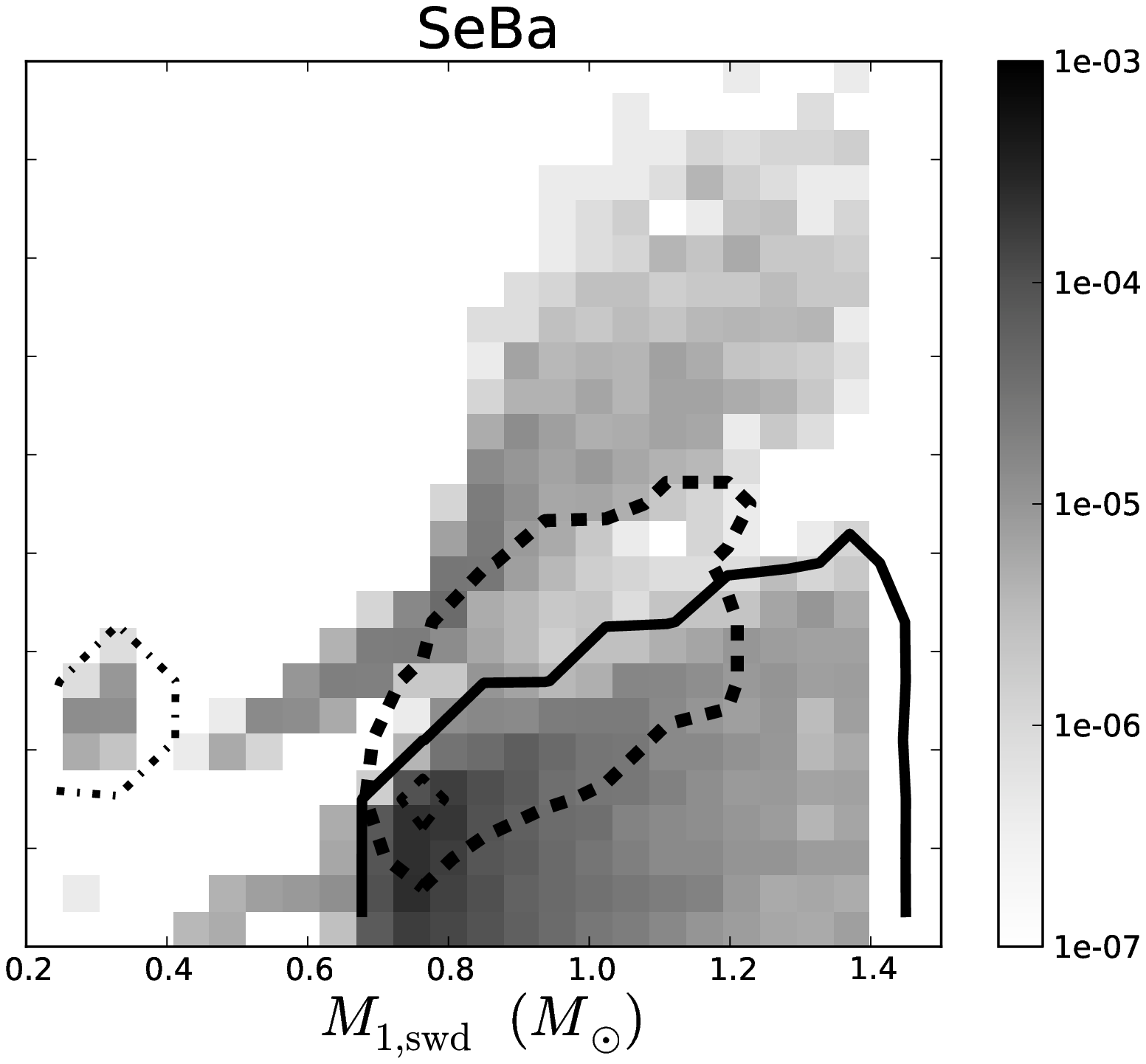} &
	\includegraphics[height=4.6cm, clip=true, trim =20mm 0mm 23mm 5mm]{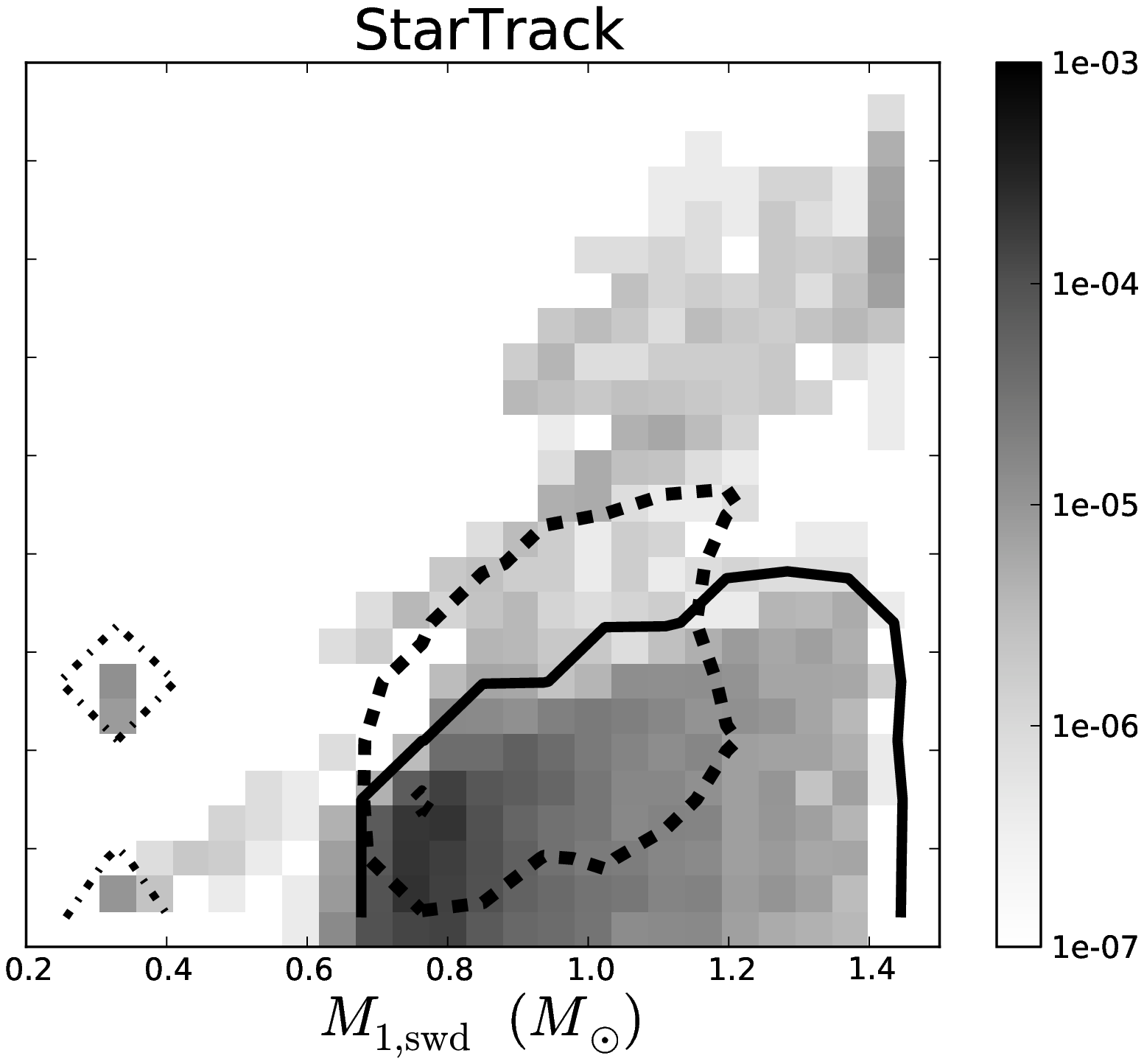} \\	
	\end{tabular}
    \caption{Secondary mass versus WD mass for all SWDs in the intermediate mass range at the time of SWD formation. The contours represent the SWD population from a specific channel: channel~1 (solid line), channel~4a (thin dashed line), channel~4b (thick dashed line) and channel~5 (dash-dotted line). }
    \label{fig:swd_final_m2_R1_IM}
    \end{figure*}

    \begin{figure*}
    \centering
    \setlength\tabcolsep{0pt}
    \begin{tabular}{ccc}
	\includegraphics[height=4.6cm, clip=true, trim =8mm 0mm 48.5mm 5mm]{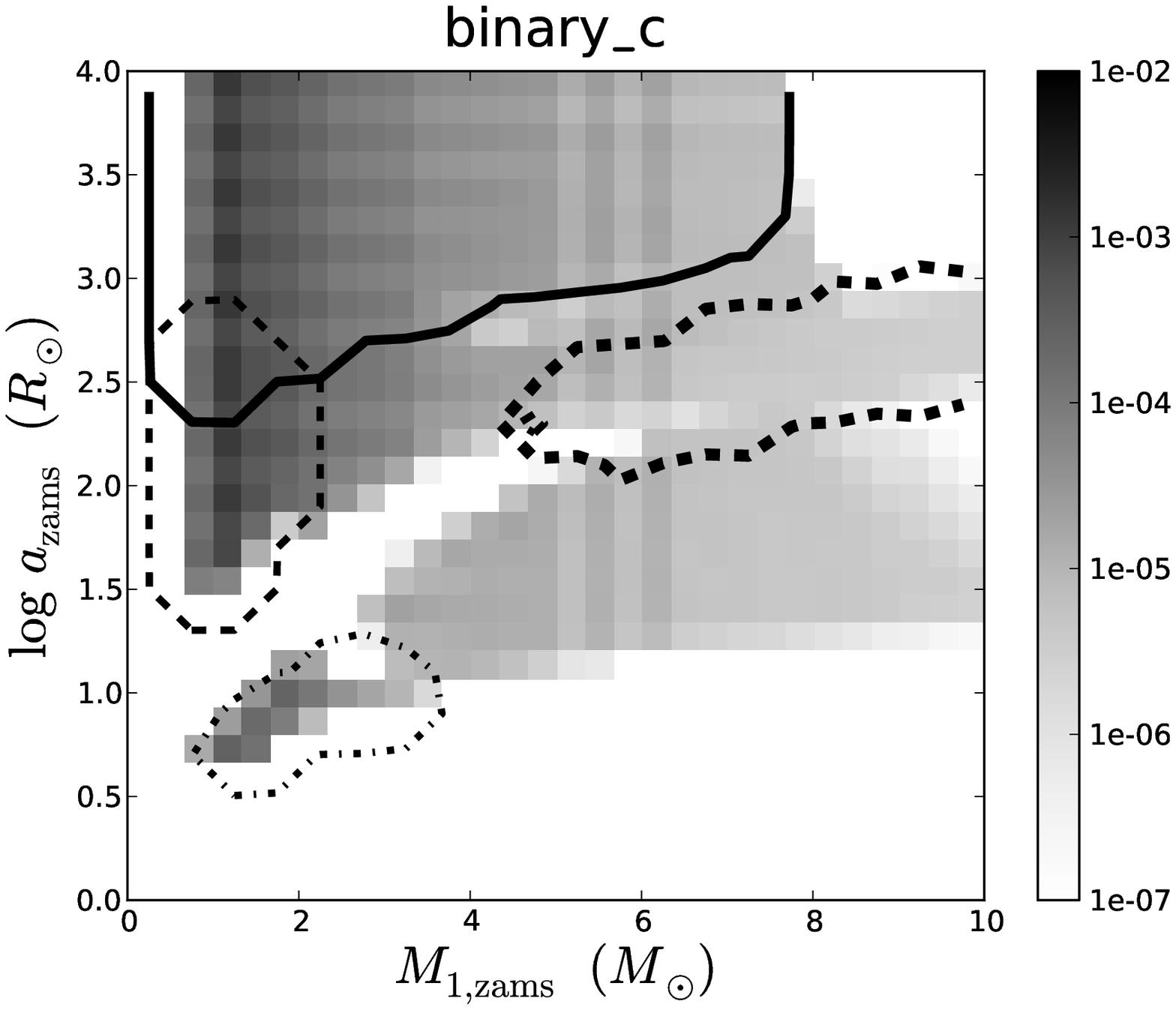} &
	\includegraphics[height=4.6cm, clip=true, trim =20mm 0mm 48.5mm 5mm]{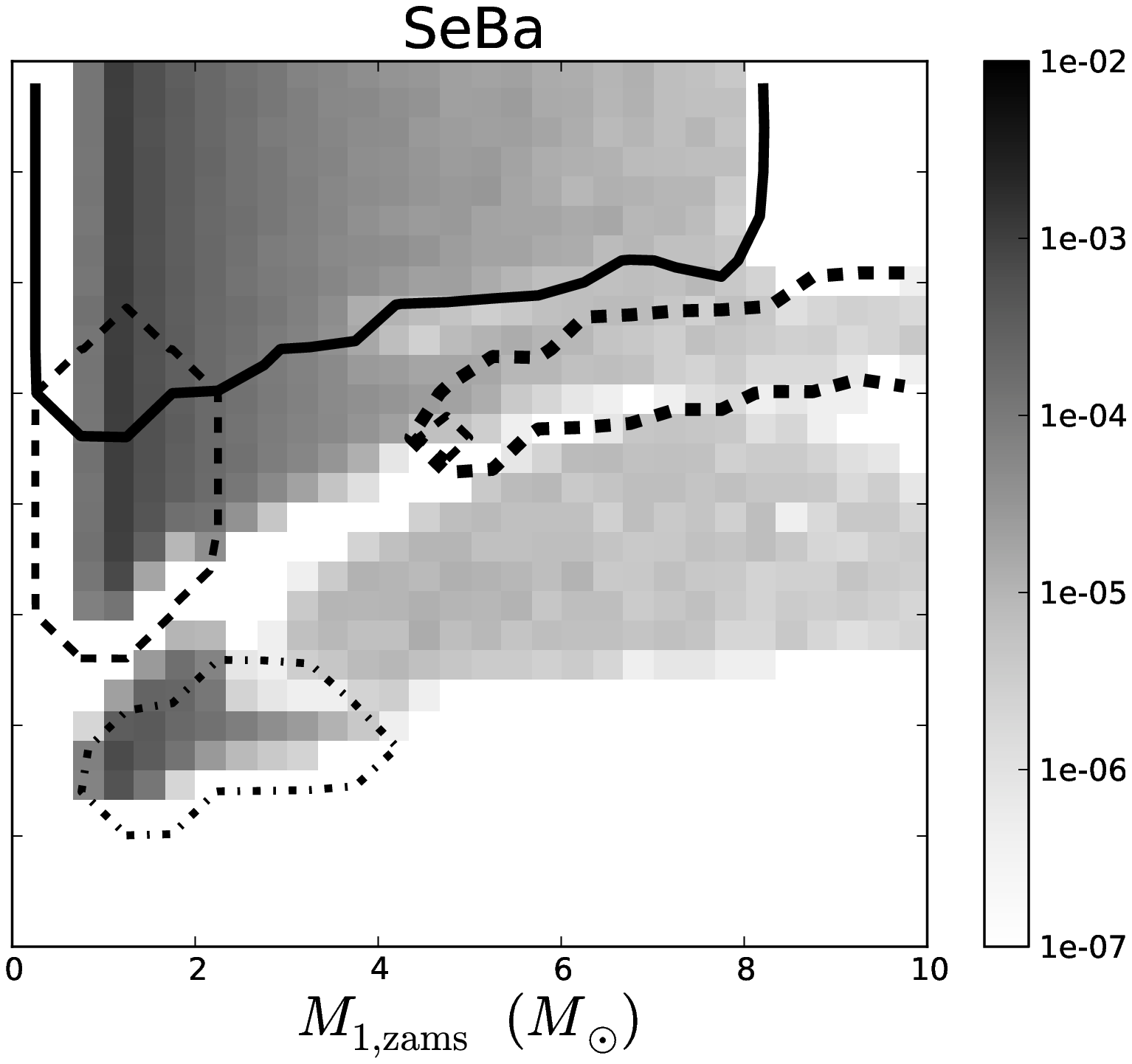} & 
	\includegraphics[height=4.6cm, clip=true, trim =20mm 0mm 23mm 5mm]{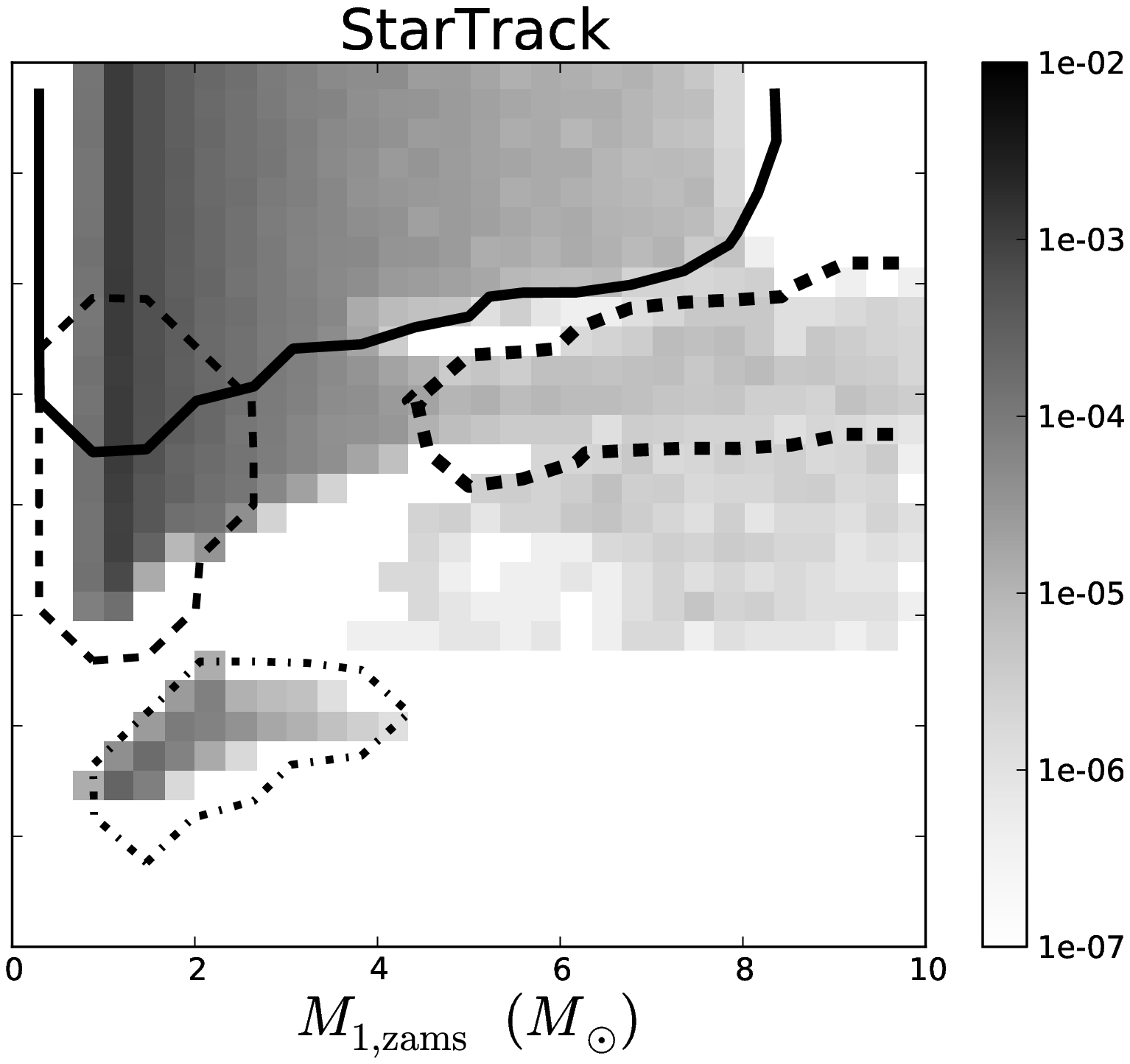} \\
	\end{tabular}
    \caption{Initial orbital separation versus initial primary mass for all SWDs in the full mass range. The contours represent the SWD population from a specific channel: channel~1 (solid line), channel~4a (thin dashed line), channel~4b (thick dashed line) and channel~5 (dash-dotted line).} 
    \label{fig:swd_zams_a_R1}
    \end{figure*}

    \begin{figure*}
    \centering
    \setlength\tabcolsep{0pt}
    \begin{tabular}{cccc}
	\includegraphics[height=4.6cm, clip=true, trim =8mm 0mm 48.5mm 5mm]{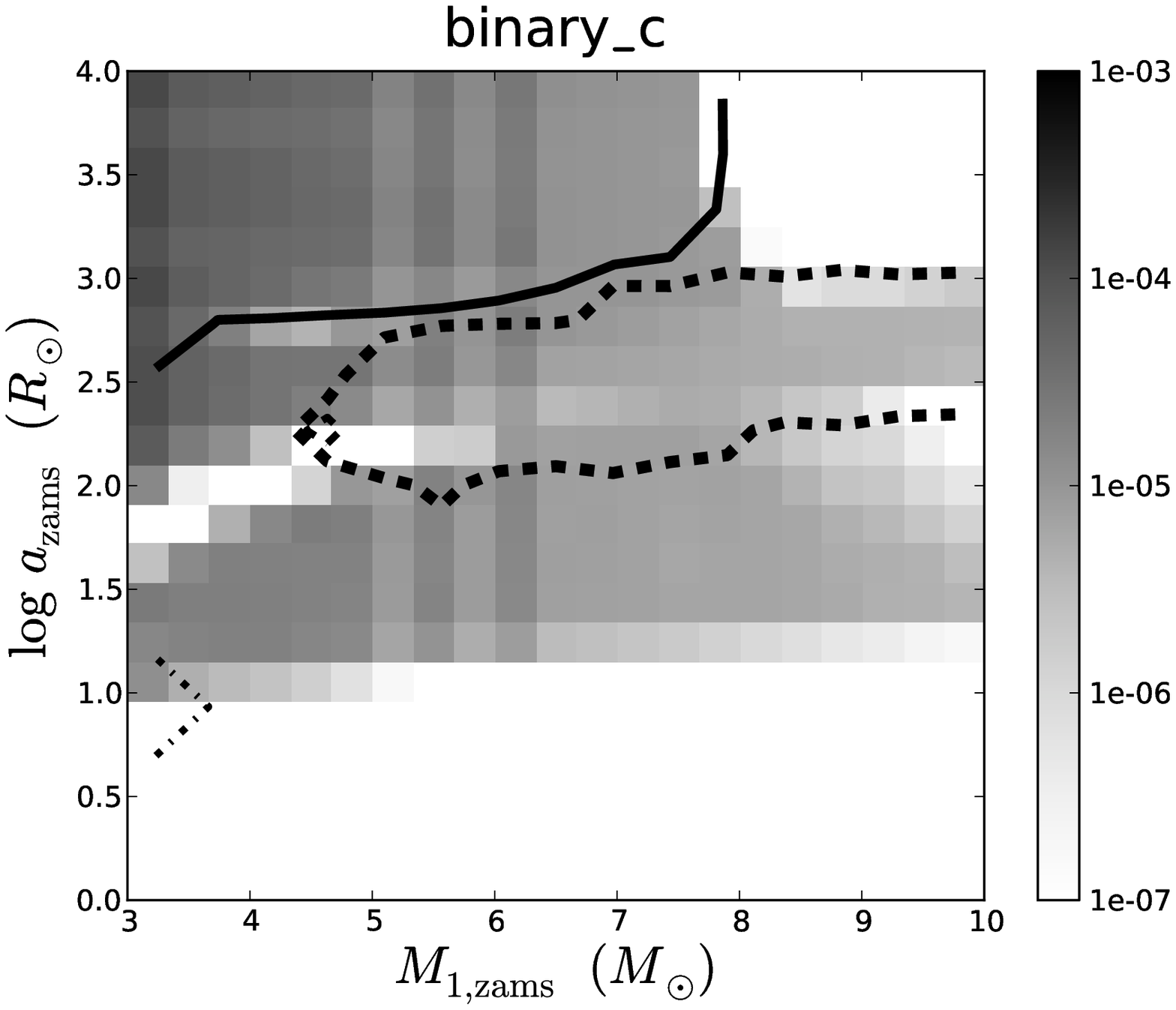} &
	\includegraphics[height=4.6cm, clip=true, trim =20mm 0mm 48.5mm 5mm]{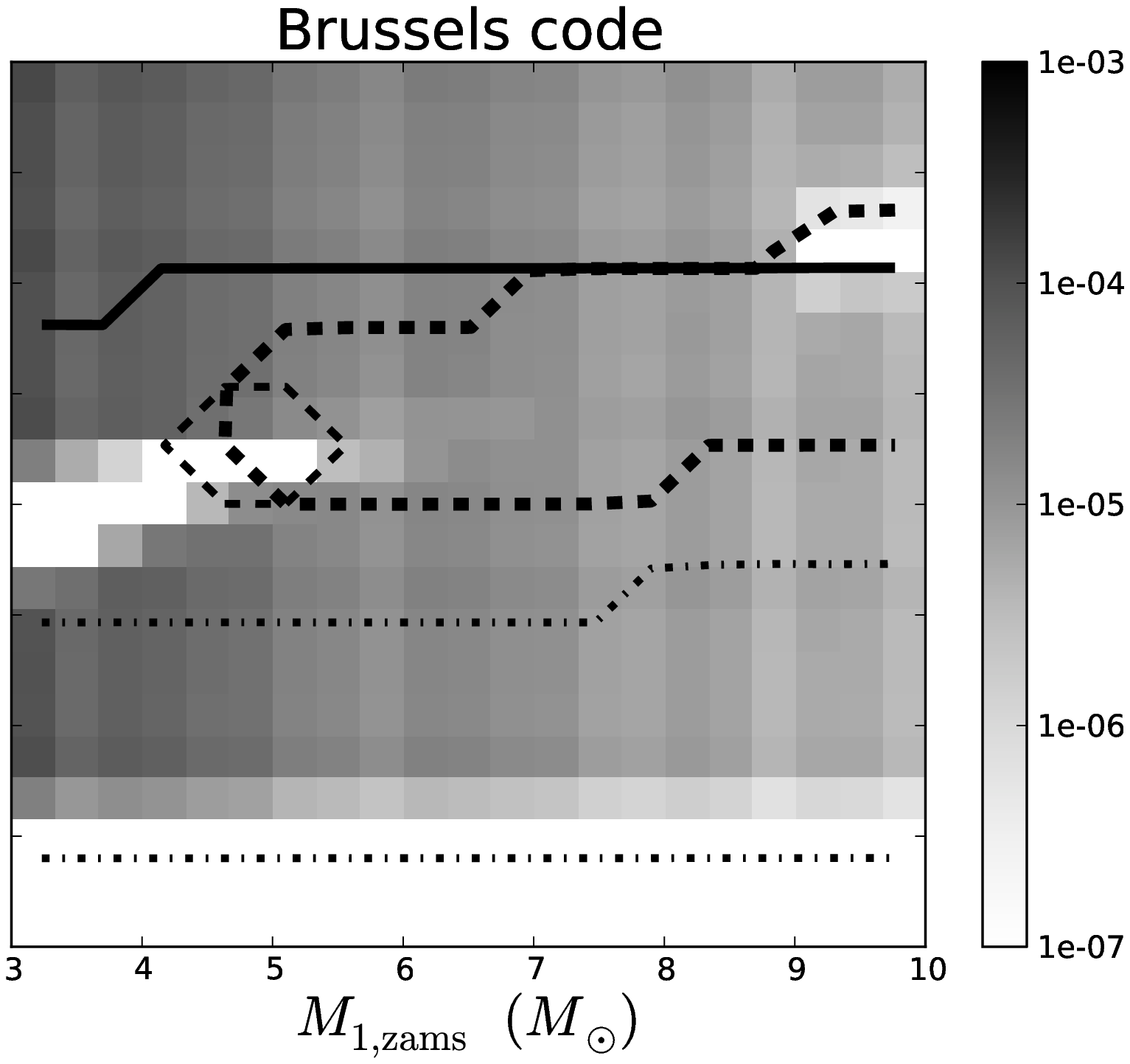} &
	\includegraphics[height=4.6cm, clip=true, trim =20mm 0mm 48.5mm 5mm]{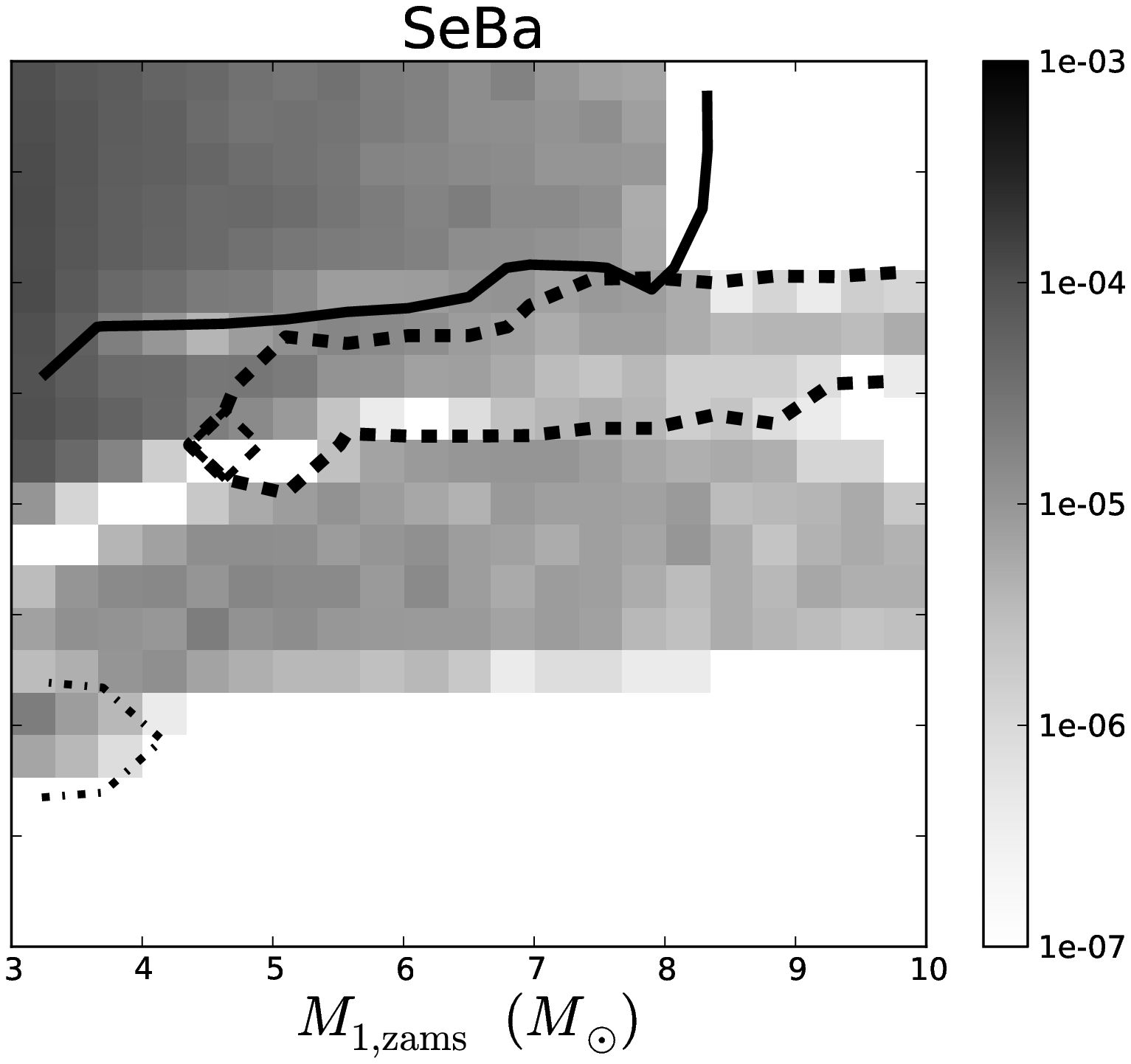} & 
	\includegraphics[height=4.6cm, clip=true, trim =20mm 0mm 23mm 5mm]{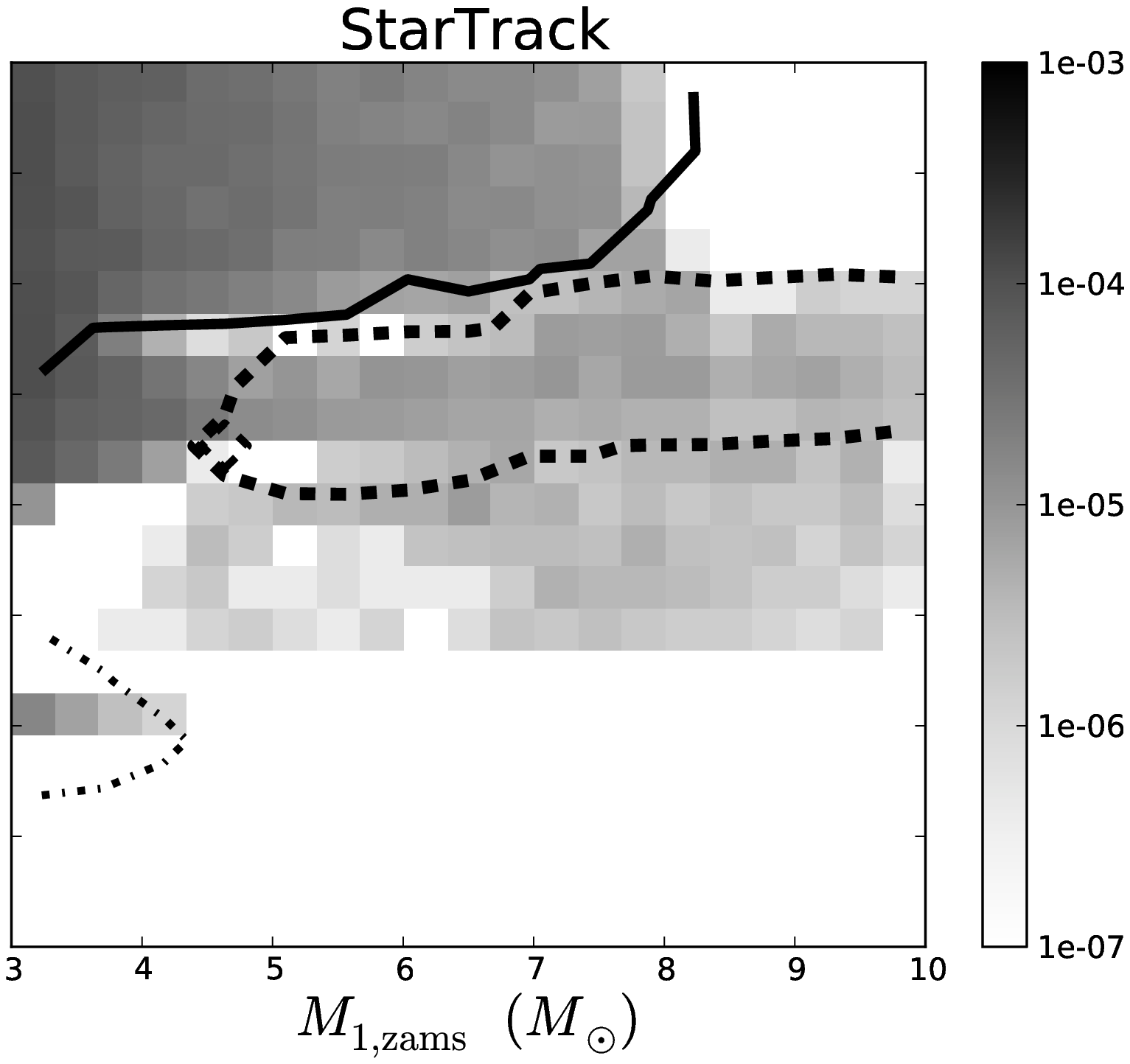} \\
	\end{tabular}
    \caption{Initial orbital separation versus initial primary mass for all SWDs in the intermediate mass range. The contours represent the SWD population from a specific channel: channel~1 (solid line), channel~4a (thin dashed line), channel~4b (thick dashed line) and channel~5 (dash-dotted line).} 
    \label{fig:swd_zams_a_R1_IM}
    \end{figure*}

In the next sections, we make a more detailed comparison between the
simulated populations of SWDs of the four codes. We distinguish
between the most commonly followed evolutionary paths with birthrates
larger than $1.0\cdot 10^{-3}\peryr$ (Table\,\ref{tbl:birthrates_all}).
We describe each evolutionary path, the similarities and differences,
and investigate the origin of these differences. Specific examples
are given and discussed for the most common paths. Abbreviations of stellar types are shown in Table\,\ref{tbl:star_type}. Paragraphs explaining the evolutionary path, an example evolution and the comparison of the simulated populations for each evolutionary channel are indicated with \emph{Evolutionary path}, \emph{Example} and \emph{Population}, respectively. For some channels, causes for differences between the populations are discussed separately in paragraphs that are indicated by \emph{Effects}. 
Masses and orbital separations according to each code are given in
vector form $[c_1, c_2, c_3, c_4]$ where $c_1$ represents the value according to the
binary\_c code, $c_2$ according to the Brussels code, $c_3$ according to SeBa, and $c_4$
according to StarTrack.  
The examples are given to illustrate the evolutionary path and relevant
physical processes. However, note that when comparing different BPS
codes, achieving similar results for specific binary populations is more 
desirable and important than achieving a perfect match between
specific, individual binary systems.

\begin{table*}
\caption{Definitions of abbreviations of stellar types used in the text and figures.}
\begin{tabular}{|l|l|}\hline 
Abbreviation &  Type of star  \\
 \hline  \hline 
MS & Main-sequence star\\
HG & Hertzsprung-gap star\\
GB & Star on the first giant branch (red giants)\\
AGB & Star on the asymptotic giant branch\\
He-MS & Star on the equivalent of the main-sequence for hydrogen-poor helium-burning stars \\
Ev. He-star & Evolved hydrogen-poor helium-burning star \\
WD & White dwarf \\
 \hline 
 \end{tabular}
\label{tbl:star_type}
\end{table*}

\subsubsection{Channel 1: detached evolution}
\label{sec:channel1}
\emph{Evolutionary path} Most SWD binaries are non-interacting binaries where the stars essentially evolve as single stars. Most binary processes that are discussed in Sect.\,\ref{sec:BinEvol} do not play a role in channel~1.

\emph{Example} As an example of a system in channel 1, we discuss the evolution of a system that initially contains a 5\Msolar~and 4\Msolar~star in an orbit of $10^4$\Rsolar~(and $e_{\rm zams}=0$ by assumption). When the primary star becomes a WD its mass is $[1.0, 0.94,
1.0, 1.0]$\Msolar~in an orbit of $[1.8, 1.8, 1.8, 1.8]\cdot10^4$\Rsolar.  
The differences in the resulting SWD system from different BPS codes are small and mainly due to different initial-final mass (MiMf)-relations (Fig.\,\ref{fig:ifm}). The maximum progenitor mass to form a WD from a single star is $[7.6, 10, 7.9, 7.8]$\Msolar~and corresponding maximum WD mass of $[1.38, 1.34, 1.38, 1.4]$\Msolar~according to the four codes. 
The MiMf-relations of the binary\_c code, SeBa and StarTrack are very similar. The similarities are not surprising as these codes are based on the same single stellar tracks and wind prescriptions of \citet{HPT00}. However, small differences arise in the MiMf-relation as the prescriptions for the stellar wind are not exactly equal. 
The Brussels code is based on different models of single stars e.g. different stellar winds and a different overshooting prescription (Appendix\,\ref{sec:TNS_inherent}). The result is that the core mass of a specific single star is larger according to the Hurley tracks. In other words, the progenitor of a specific single WD is more massive in the Brussels code. 

\emph{Population} Despite differences for individual systems, the population of non-interacting binaries at WD formation is very similar. 
The previously mentioned differences in the MiMf-relations are noticeable in the maximum initial primary mass in Fig.\,\ref{fig:swd_zams_a_R1}~and~\ref{fig:swd_zams_a_R1_IM}. 
The distribution of separations at WD formation (Fig.\,\ref{fig:swd_final_a_R1}~and~\ref{fig:swd_final_a_R1_IM}) are very similar between the
codes. For the intermediate mass range, the separations at SWD formation are
$\gtrsim 4.5 \cdot 10^3$\Rsolar~for the Brussels code and extend to slightly lower values of $\gtrsim 2.0
\cdot 10^3$\Rsolar~for binary$\_$c, SeBa, and StarTrack. 
For the full mass range, the latter three codes agree that the separations can be as low as
$5.0\cdot 10^2$\Rsolar. The progenitor
systems of channel~1 have similar separations of $\gtrsim 3.0 \cdot
10^2$ \Rsolar~for low mass primaries. For intermediate mass stars binary\_c, SeBa and StarTrack find that the initial separation is $\gtrsim 0.7 \cdot 10^3$\Rsolar~where the Brussels code finds a slightly higher value of $\gtrsim 1.6 \cdot 10^3$\Rsolar~(Fig.\,\ref{fig:swd_zams_a_R1}~and~\ref{fig:swd_zams_a_R1_IM}). The minimum separation (at ZAMS and WD
formation) for a given primary mass depends on whether or not the
primary fills its Roche lobe, which in turn depends on the maximum
radius for that star according to the particular single star
prescriptions that are used. Even though the progenitor populations
are not 100\% equal, the characteristics of the SWD population and the
birthrates (Table\,\ref{tbl:birthrates_all}) in this channel are
in excellent agreement.  

\subsubsection{Channel 2: unstable case C}
\label{sec:channel2}
\begin{figure}
\centering
\includegraphics[width = 6cm,angle=270]{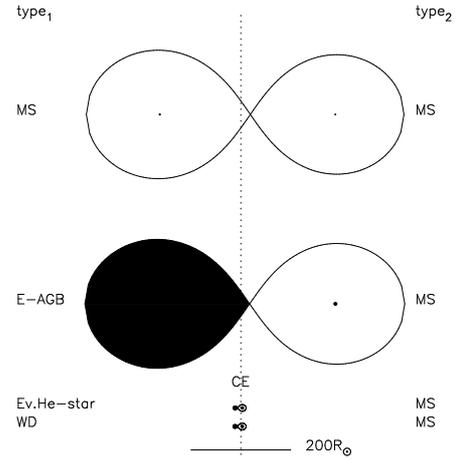} 
\caption{Example of the evolution of a SWD system in channel~2a. Abbreviations are as in Table\,\ref{tbl:star_type}. } 
\label{fig:rl_C}
\end{figure}

\emph{Evolutionary path} One of the most common evolutionary paths of
interacting binaries is channel~2, of which an example is shown in
Fig.\,\ref{fig:rl_C}. In this channel, the primary star fills its
Roche Lobe when helium is exhausted in its core, so-called case C mass
transfer \citep{Lau70}. As the envelope of the donor star is deeply
convective at this stage, generally mass transfer leads to an unstable
situation and a CE-phase develops. While the orbital separation
shrinks severely, the primary loses its hydrogen envelope. By
assumption in this project, the secondary is not affected during the
CE-phase. The primary can either directly become a WD or continue burning helium as an evolved helium star as shown in the example of Fig.\,\ref{fig:rl_C}. If the primary becomes a WD directly, or indirectly but without further interaction, the evolutionary path is called channel~2a. 
Evolution according to channel~2b occurs if the primary fills its Roche lobe for a second time when it is a helium star. The second phase of mass transfer can be either stable or unstable. 

\emph{Example} As an example of channel~2a, we discuss the evolution of the binary system in Fig.\,\ref{fig:rl_C} with initial parameters $M_{\rm 1, zams} = 3.5\Mo$, $M_{\rm 2, zams} = 3\Mo$ and $a_{\rm zams}=350\Ro$ in more detail. The primary star fills the Roche lobe early on the AGB before thermal pulses and superwinds occur. Wind mass loss prior to the CE-phase is small, $[4.4, 0, 4.3, 4.9]\cdot10^{-2}$\Msolar. After the CE-phase the orbital separation is reduced to $[14, 9.1, 14, 14]$\Rsolar. In this example the primary continuous burning helium as an evolved helium star of mass $[0.78, 0.55, 0.78, 0.78]$\Msolar. When the primary exhausts its fuel, it becomes a WD of $[0.76, 0.51, 0.77, 0.76]$\Msolar~in an orbit of $[14, 9.1, 14, 14]$\Rsolar~with a 3\Msolar~MS companion. The most important differences, to be seen between the Brussels code and the other codes, arises from the different single star prescriptions that are used. This affects the resulting mass of a WD from a specific primary, and the resulting orbital separation. 
Note that while the MiMf-relation for single stars depends on the single star prescriptions (i.e. core mass growth and winds), the MiMwd-relation is also affected by the companion mass and separation (which determine when and which kind of mass transfer event takes place), and the single star prescriptions for helium stars. In other words, the MiMwd-relation represents how fast the core grows on one hand, and the envelope is depleted by mass transfer and stellar winds on the other hand. 

\begin{figure*}
    \centering
    \setlength\tabcolsep{0pt}
    \begin{tabular}{ccc}
	\includegraphics[height=4.6cm, clip=true, trim =8mm 0mm 48.5mm 5mm]{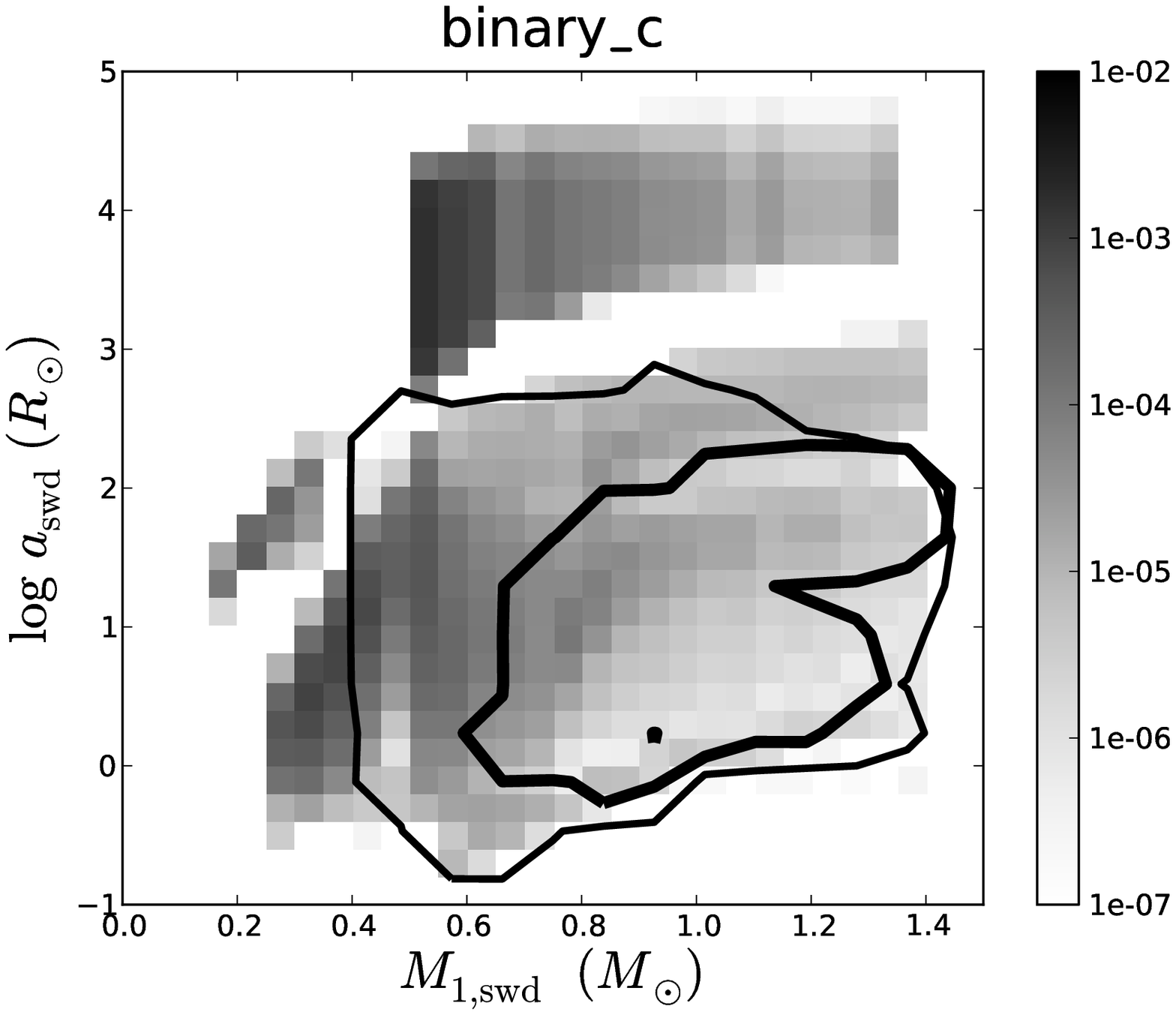} &
	\includegraphics[height=4.6cm, clip=true, trim =20mm 0mm 48.5mm 5mm]{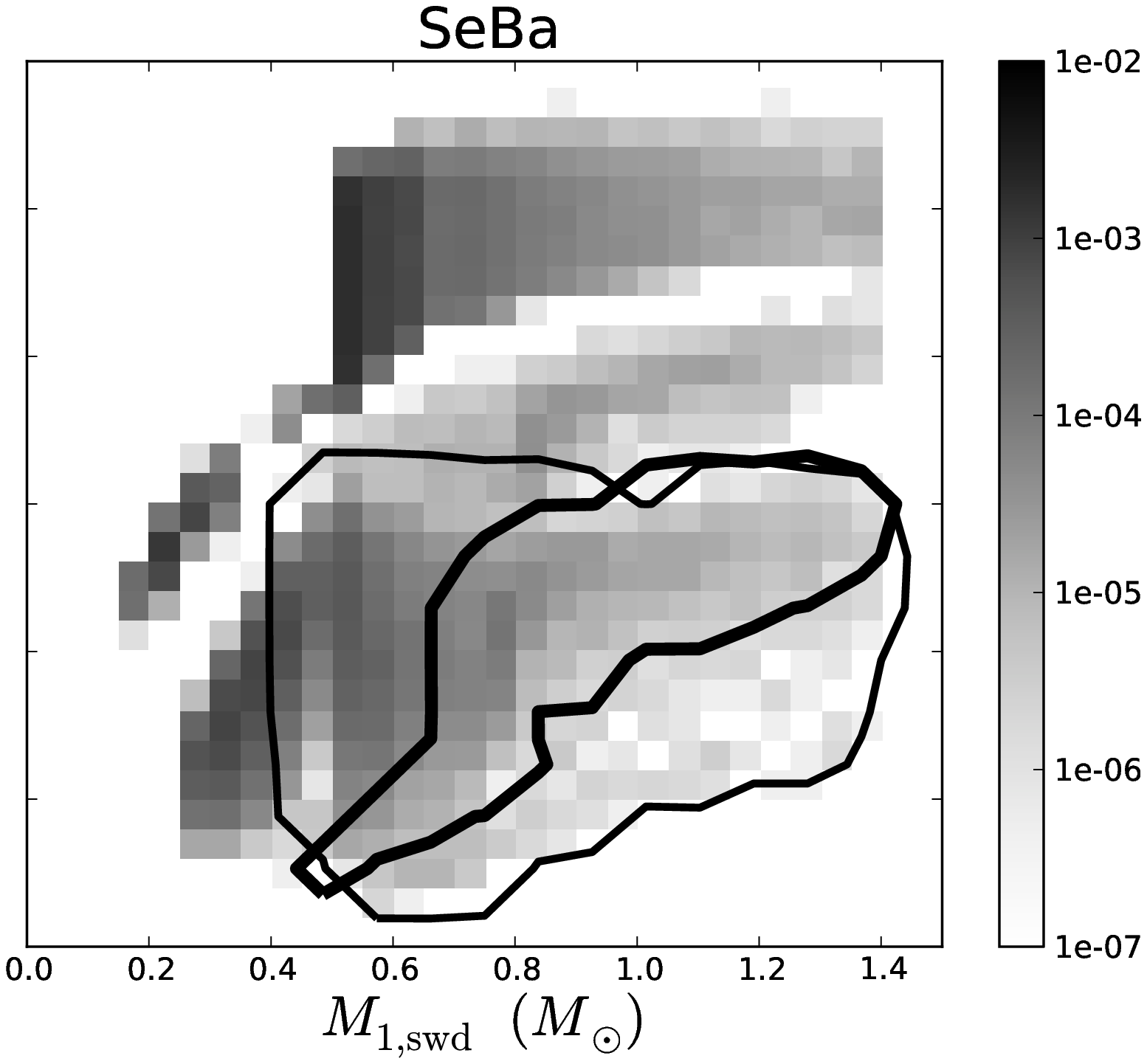} &
	\includegraphics[height=4.6cm, clip=true, trim =20mm 0mm 23mm 5mm]{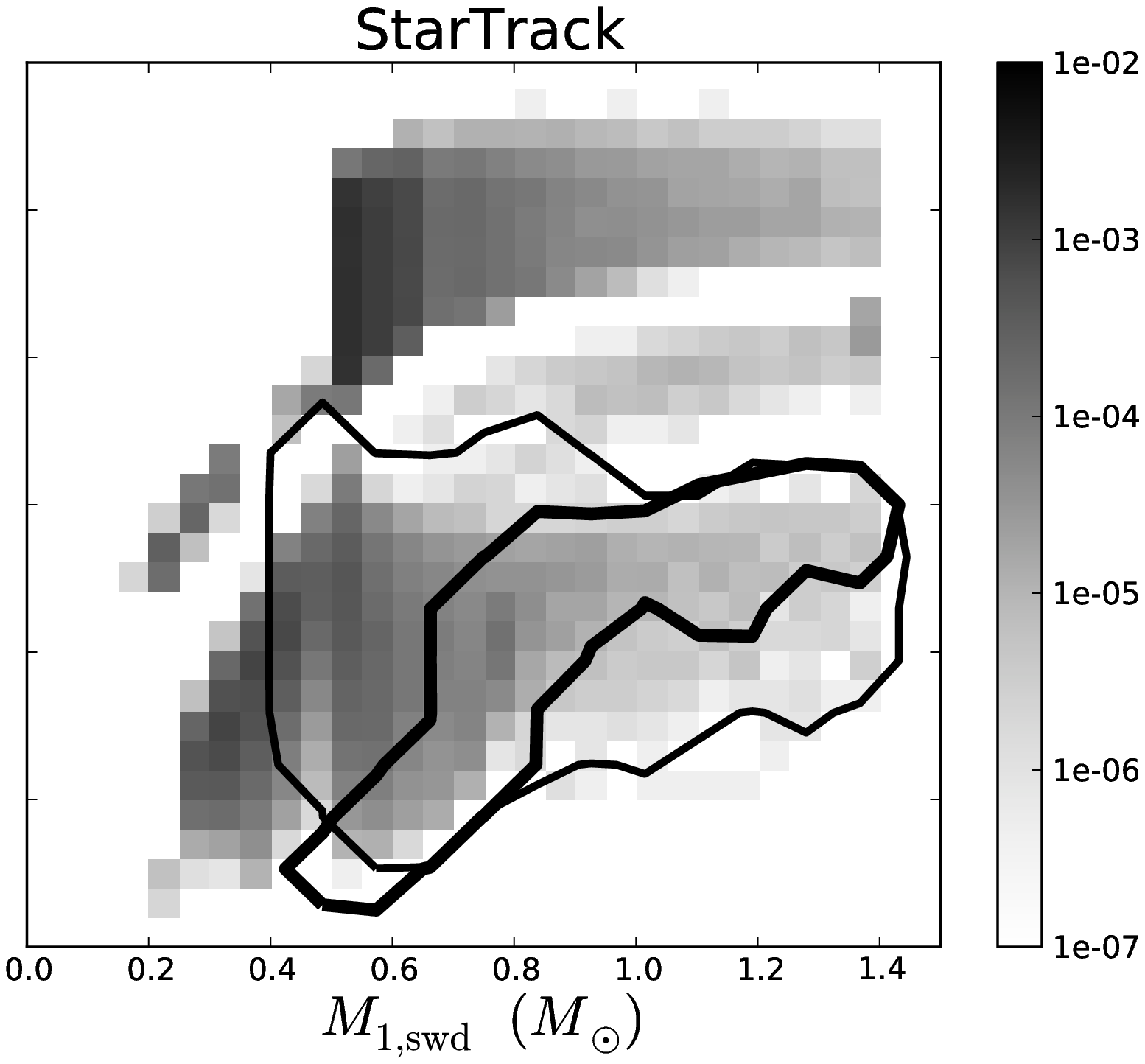} \\
	\end{tabular}
    \caption{Orbital separation versus WD mass for all SWDs in the full mass range at the time of SWD formation. The contours represent the SWD population from a specific channel: channel~2a (thin line) and channel~2b (thick line).} 
    \label{fig:swd_final_a_R2}
    \end{figure*}

\begin{figure*}
    \centering
    \setlength\tabcolsep{0pt}
    \begin{tabular}{cccc}
	\includegraphics[height=4.6cm, clip=true, trim =8mm 0mm 48.5mm 5mm]{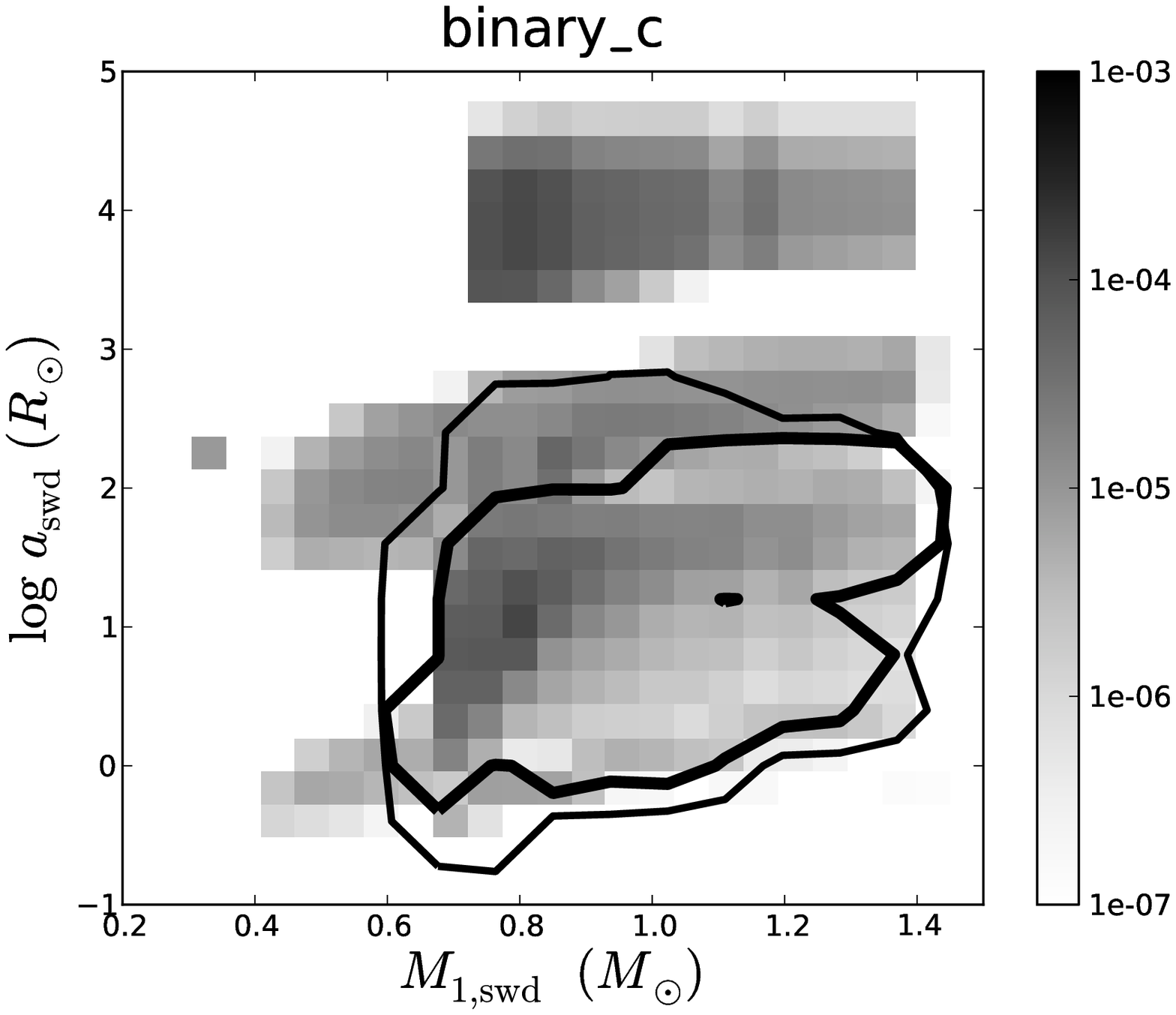} &
	\includegraphics[height=4.6cm, clip=true, trim =20mm 0mm 48.5mm 5mm]{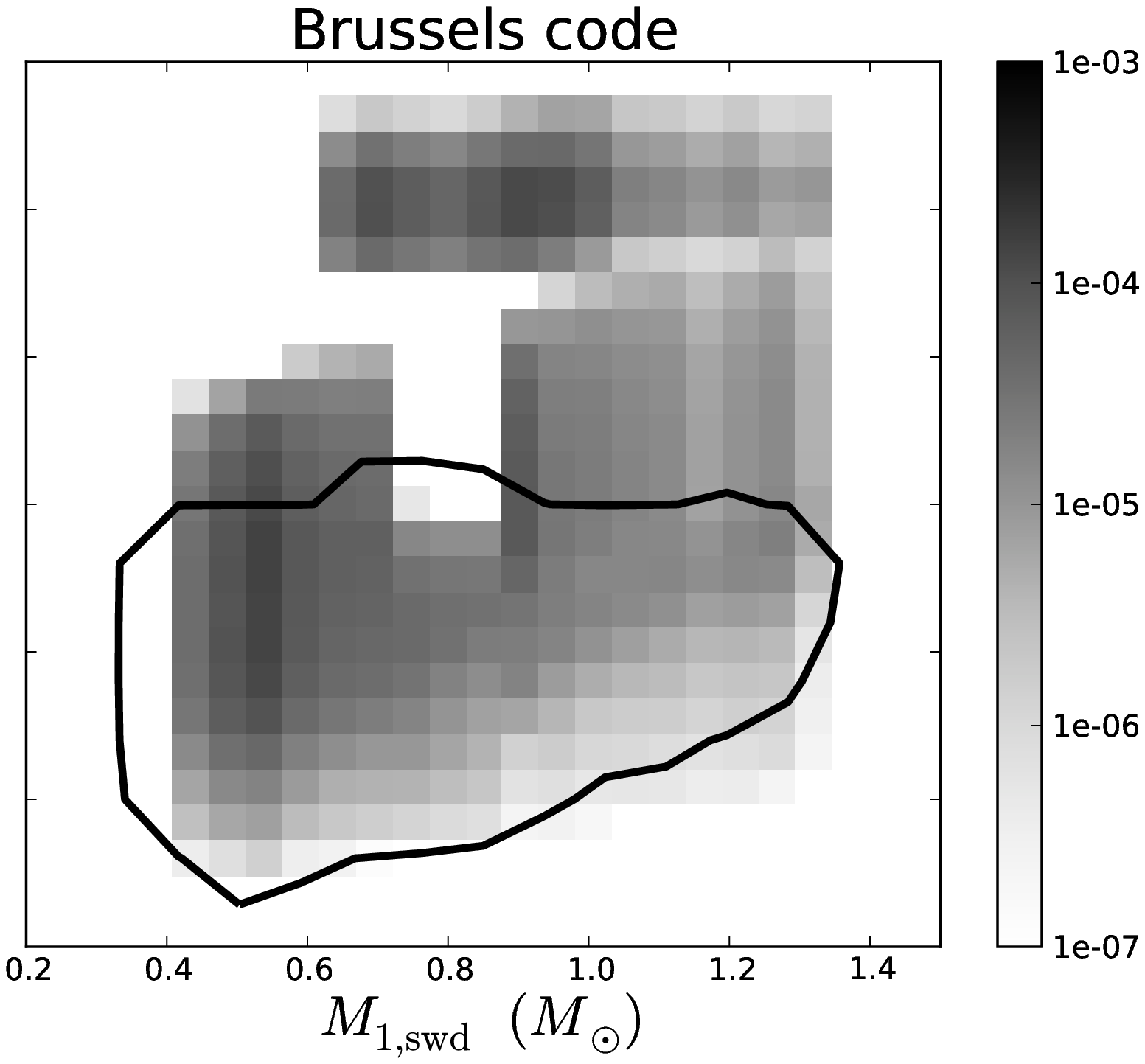} &
	\includegraphics[height=4.6cm, clip=true, trim =20mm 0mm 48.5mm 5mm]{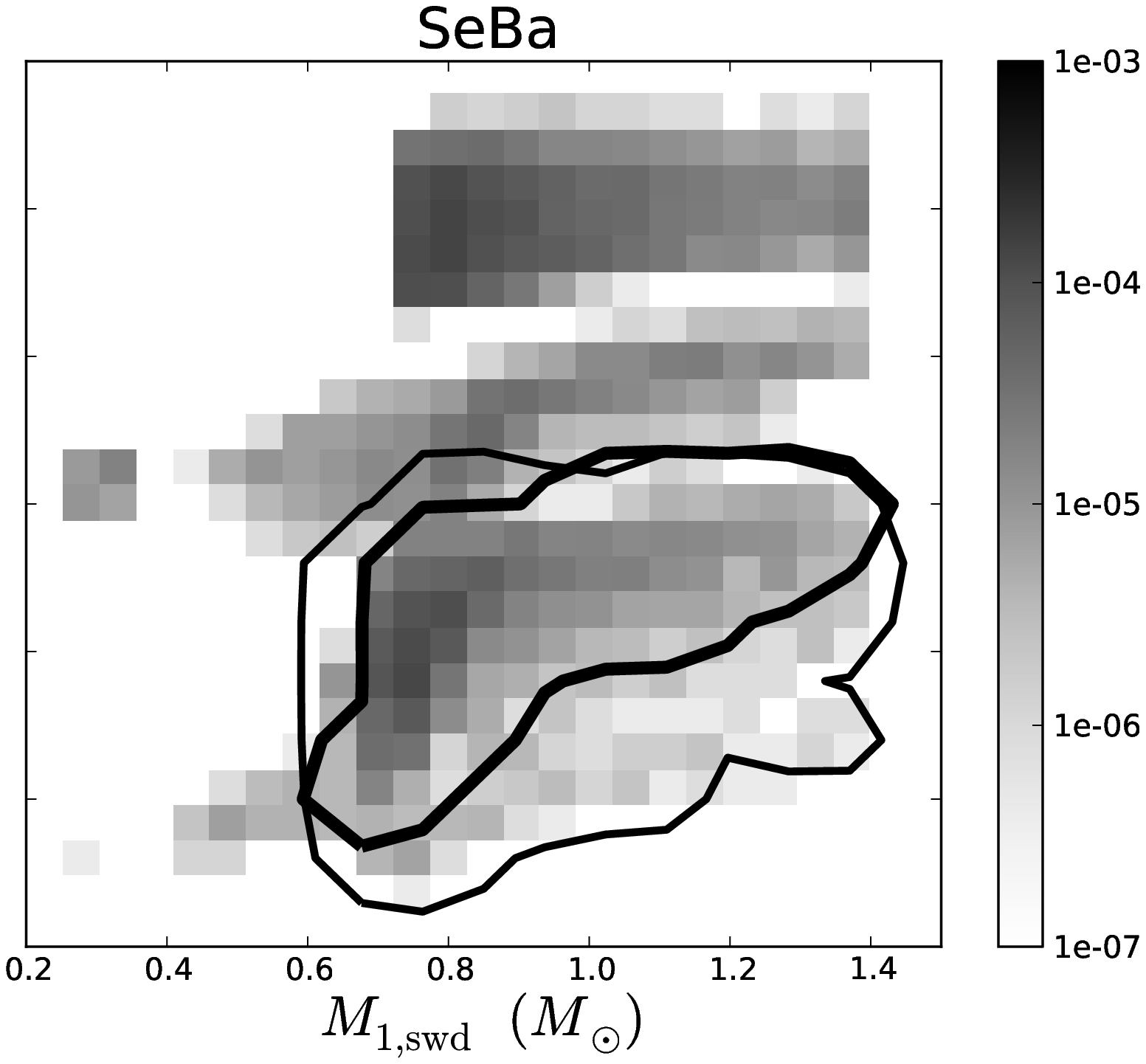} &
	\includegraphics[height=4.6cm, clip=true, trim =20mm 0mm 23mm 5mm]{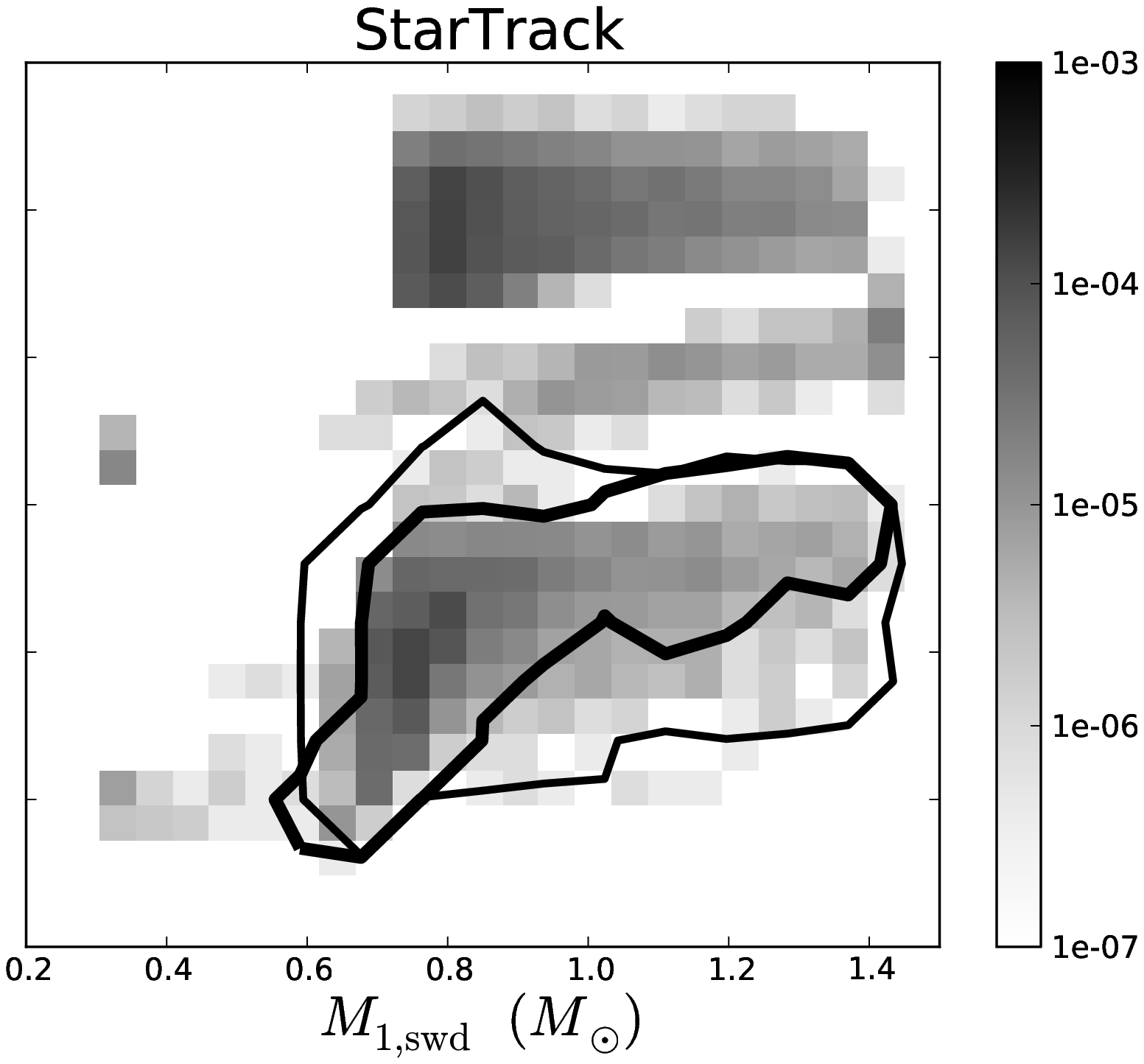} \\
	\end{tabular}
    \caption{Orbital separation versus WD mass for all SWDs in the intermediate mass range at the time of SWD formation. The contours represent the SWD population from a specific channel: channel~2a (thin line) and channel~2b (thick line).} 
    \label{fig:swd_final_a_R2_IM}
    \end{figure*}

\begin{figure*}
    \centering
    \setlength\tabcolsep{0pt}
    \begin{tabular}{ccc}
	\includegraphics[height=4.6cm, clip=true, trim =8mm 0mm 48.5mm 5mm]{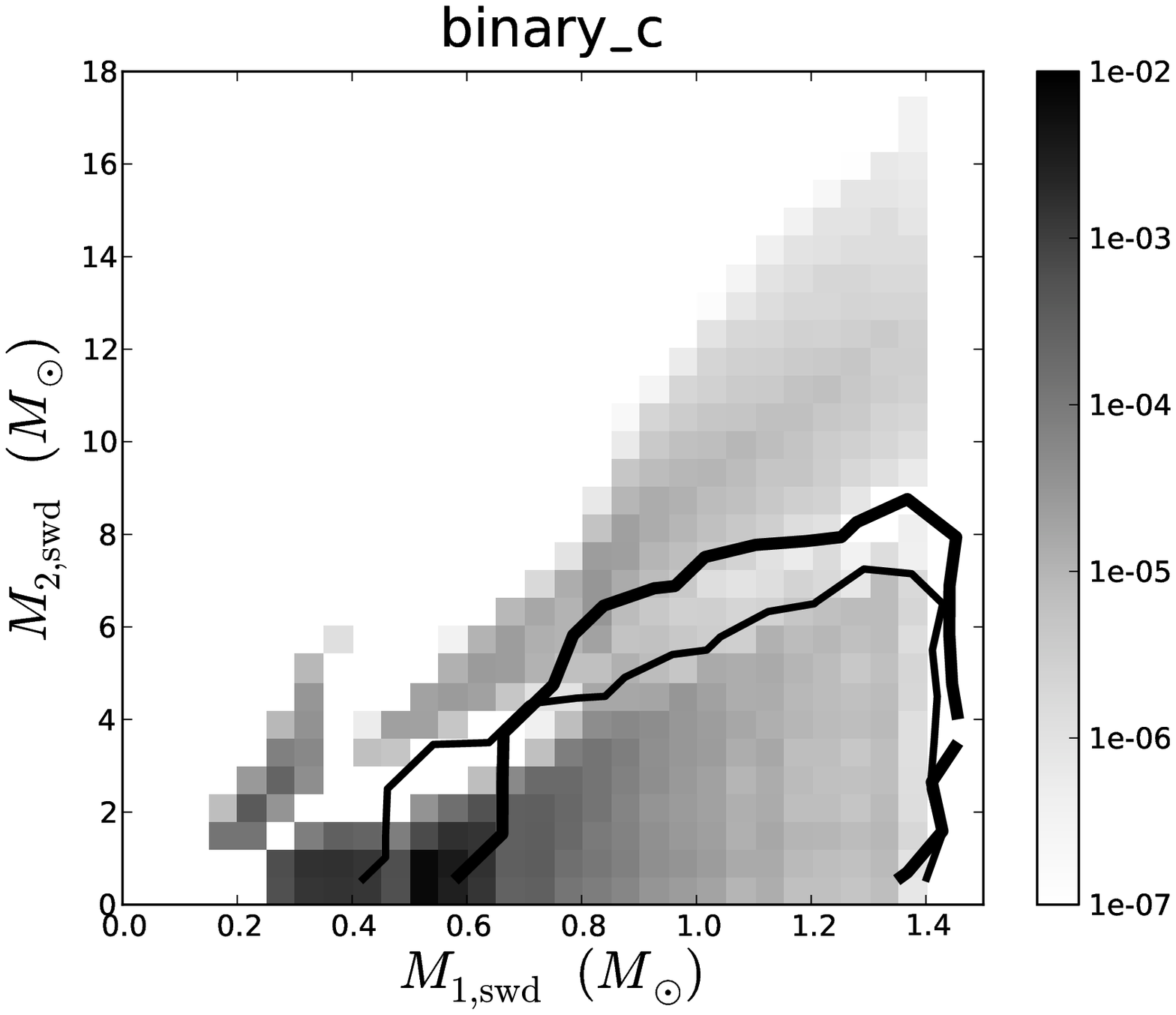} &
	\includegraphics[height=4.6cm, clip=true, trim =20mm 0mm 48.5mm 5mm]{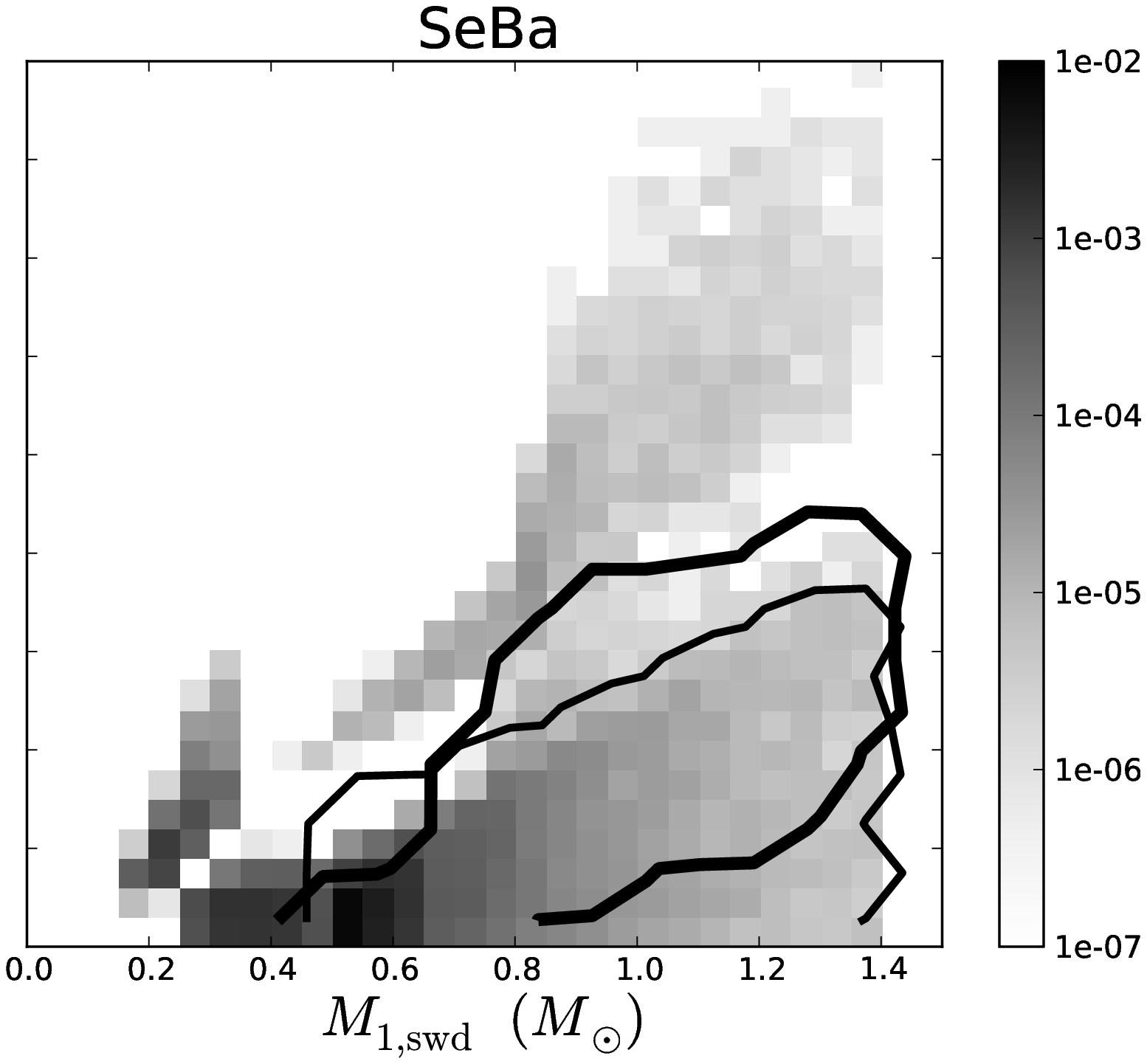} &
	\includegraphics[height=4.6cm, clip=true, trim =20mm 0mm 23mm 5mm]{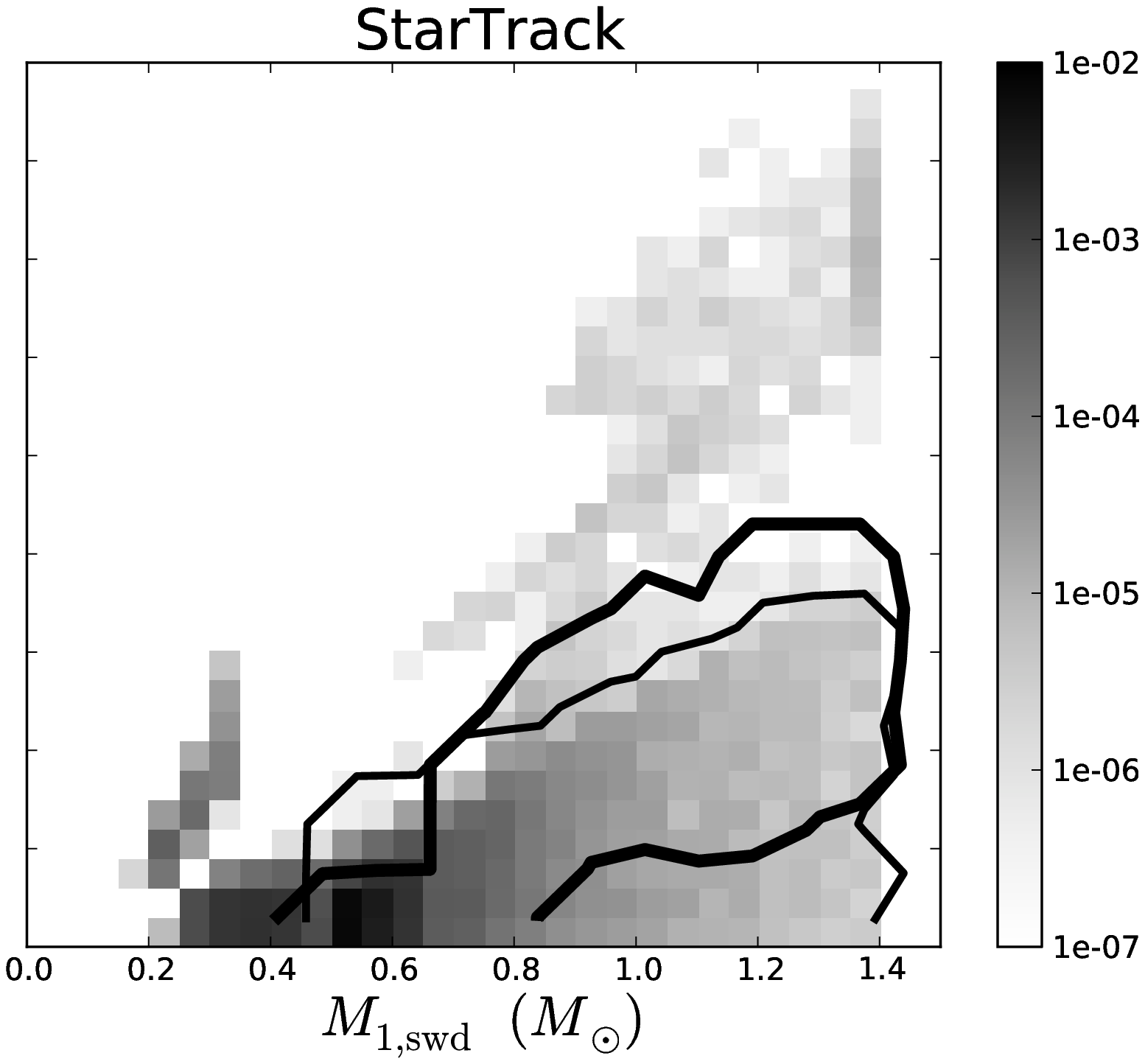} \\
	\end{tabular}
    \caption{Secondary mass versus WD mass for all SWDs in the full mass range at the time of SWD formation. The contours represent the SWD population from a specific channel: channel~2a (thin line) and channel~2b (thick line).} 
    \label{fig:swd_final_m2_R2}
    \end{figure*}

\begin{figure*}
    \centering
    \setlength\tabcolsep{0pt}
    \begin{tabular}{cccc}
	\includegraphics[height=4.6cm, clip=true, trim =8mm 0mm 48.5mm 5mm]{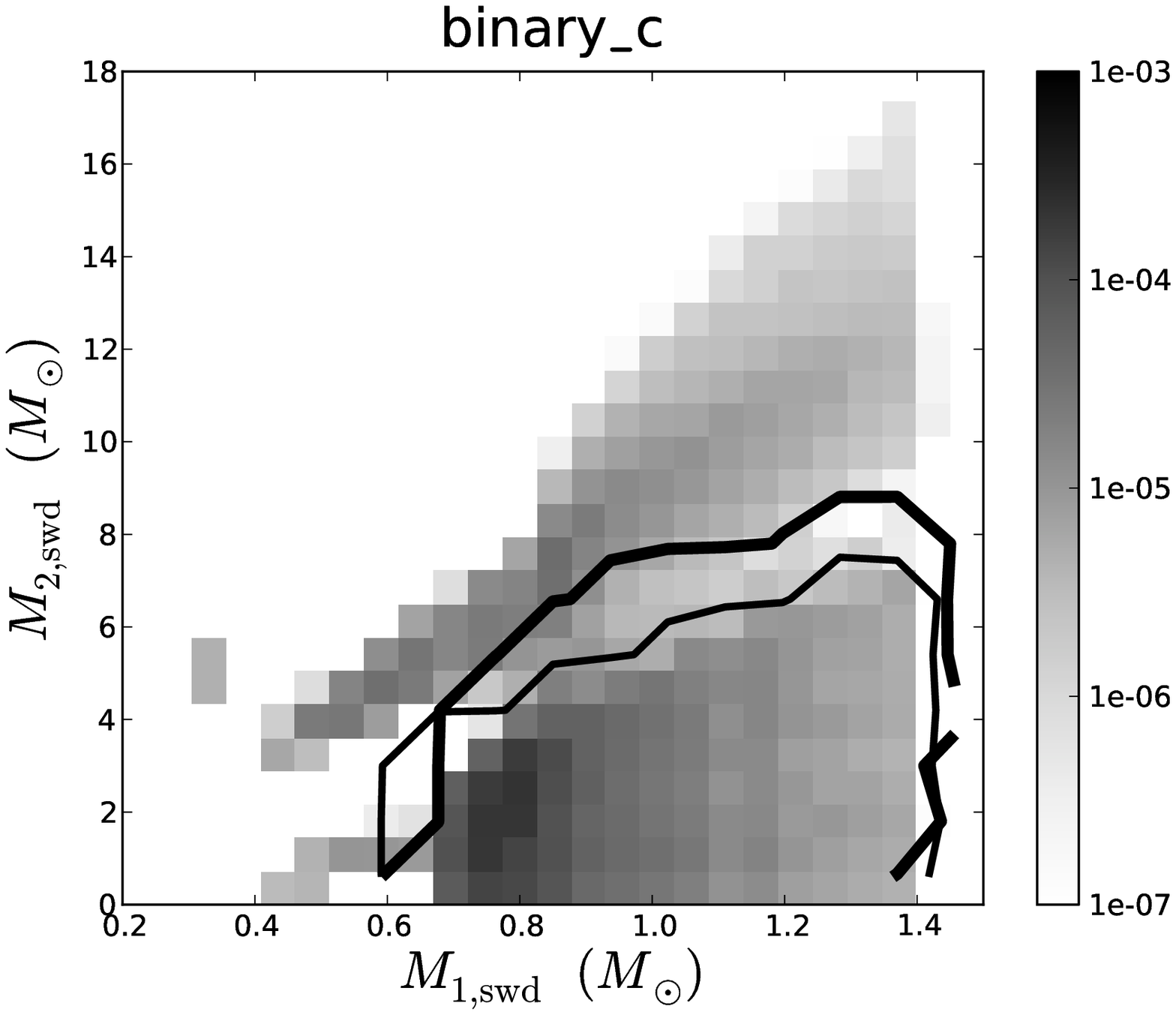} &
	\includegraphics[height=4.6cm, clip=true, trim =20mm 0mm 48.5mm 5mm]{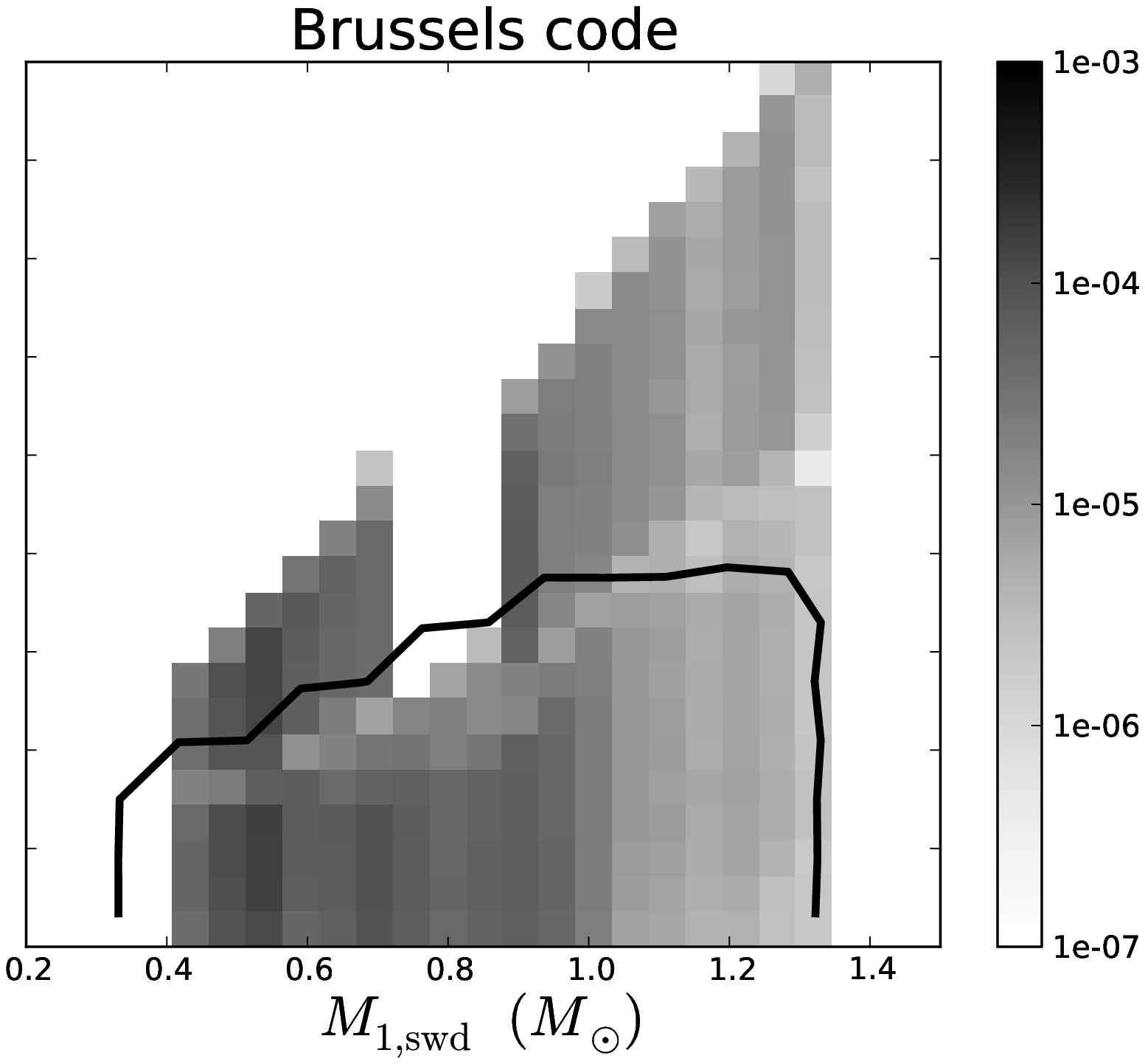} &
	\includegraphics[height=4.6cm, clip=true, trim =20mm 0mm 48.5mm 5mm]{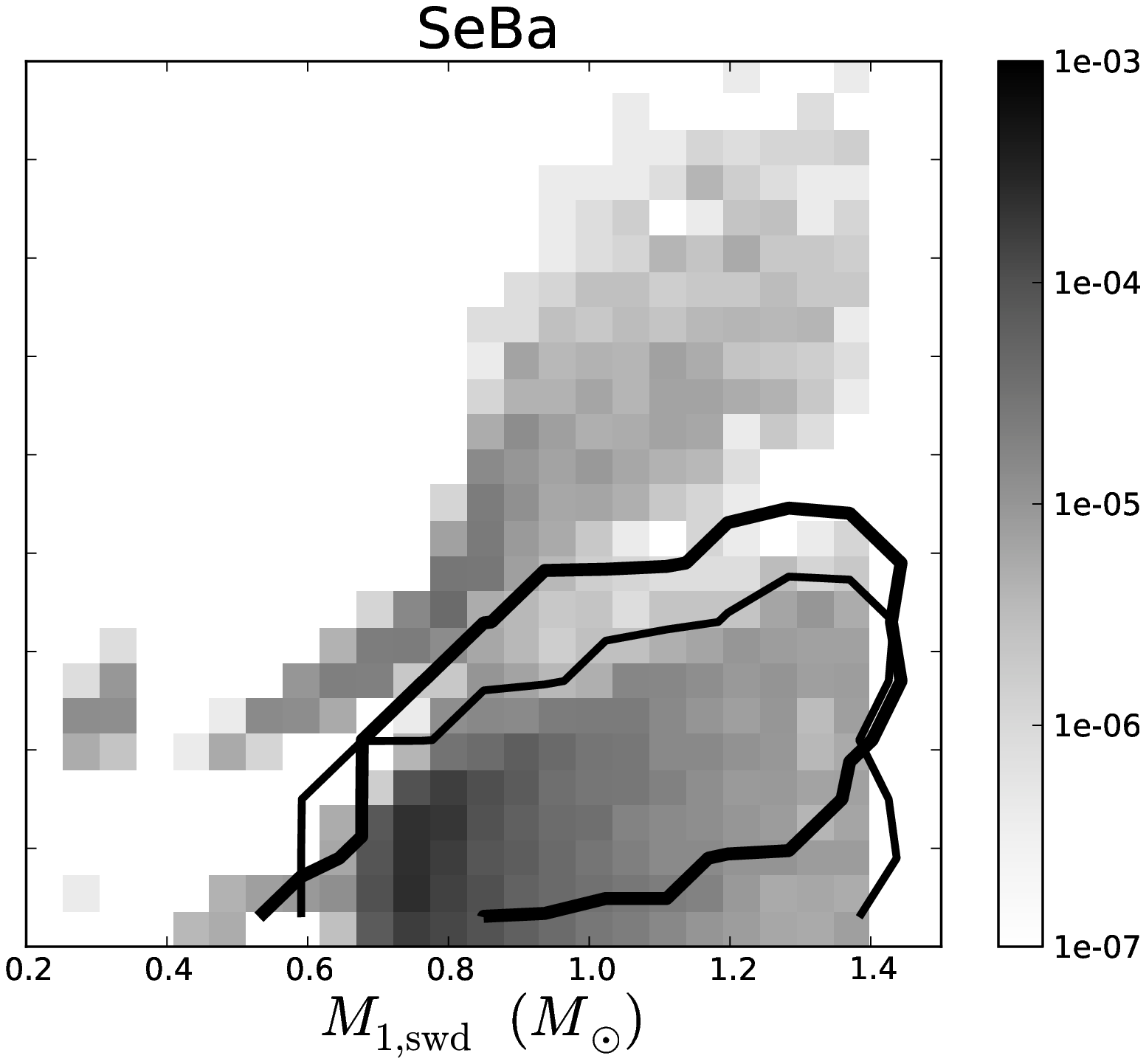} &
	\includegraphics[height=4.6cm, clip=true, trim =20mm 0mm 23mm 5mm]{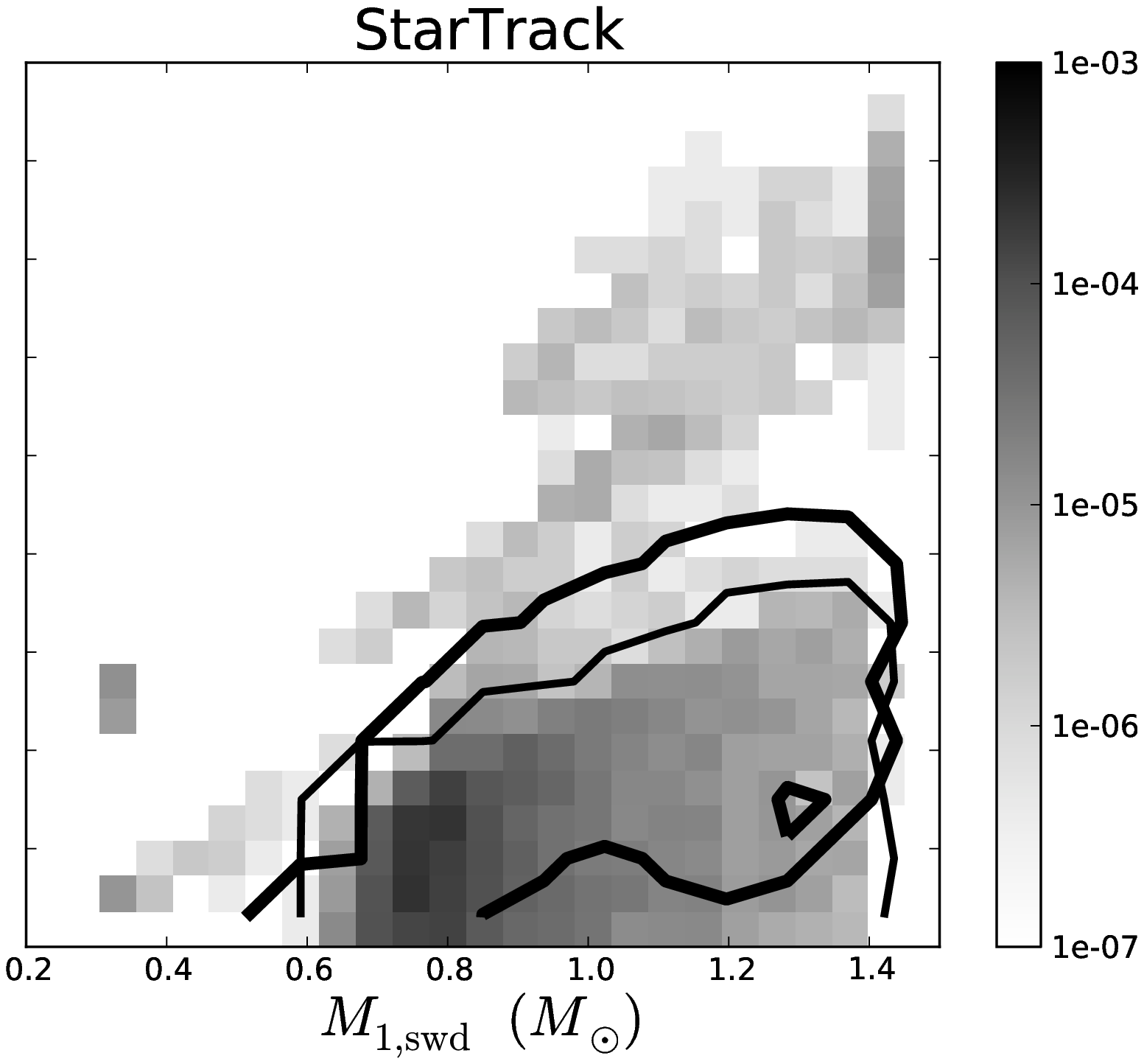} \\
	\end{tabular}
    \caption{Secondary mass versus WD mass for all SWDs in the intermediate mass range at the time of SWD formation. The contours represent the SWD population from a specific channel: channel~2a (thin line) and channel~2b (thick line).} 
    \label{fig:swd_final_m2_R2_IM}
    \end{figure*}

\begin{figure*}
    \centering
    \setlength\tabcolsep{0pt}
    \begin{tabular}{ccc}
	\includegraphics[height=4.6cm, clip=true, trim =8mm 0mm 48.5mm 5mm]{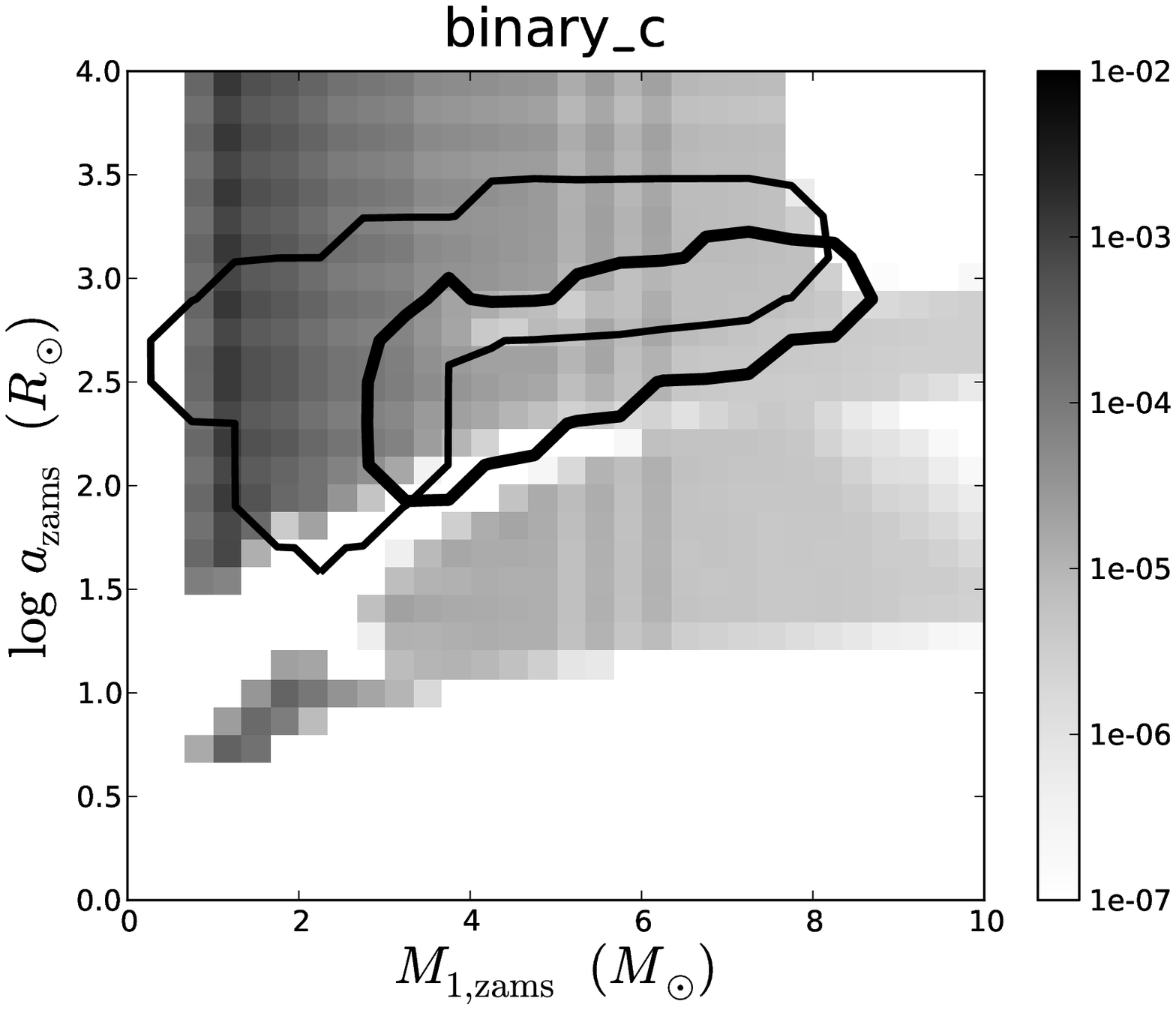} &
	\includegraphics[height=4.6cm, clip=true, trim =20mm 0mm 48.5mm 5mm]{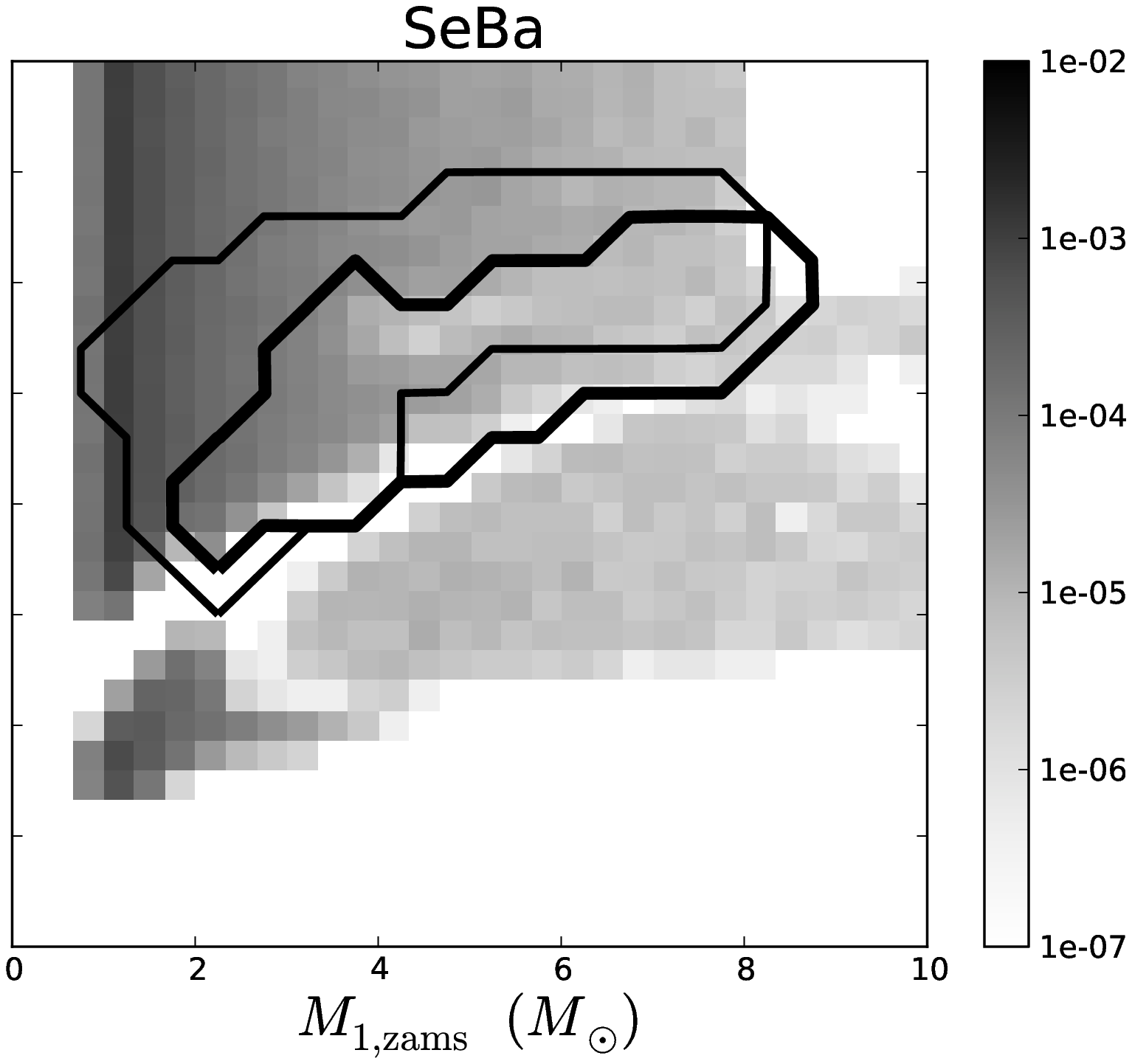} & 
	\includegraphics[height=4.6cm, clip=true, trim =20mm 0mm 23mm 5mm]{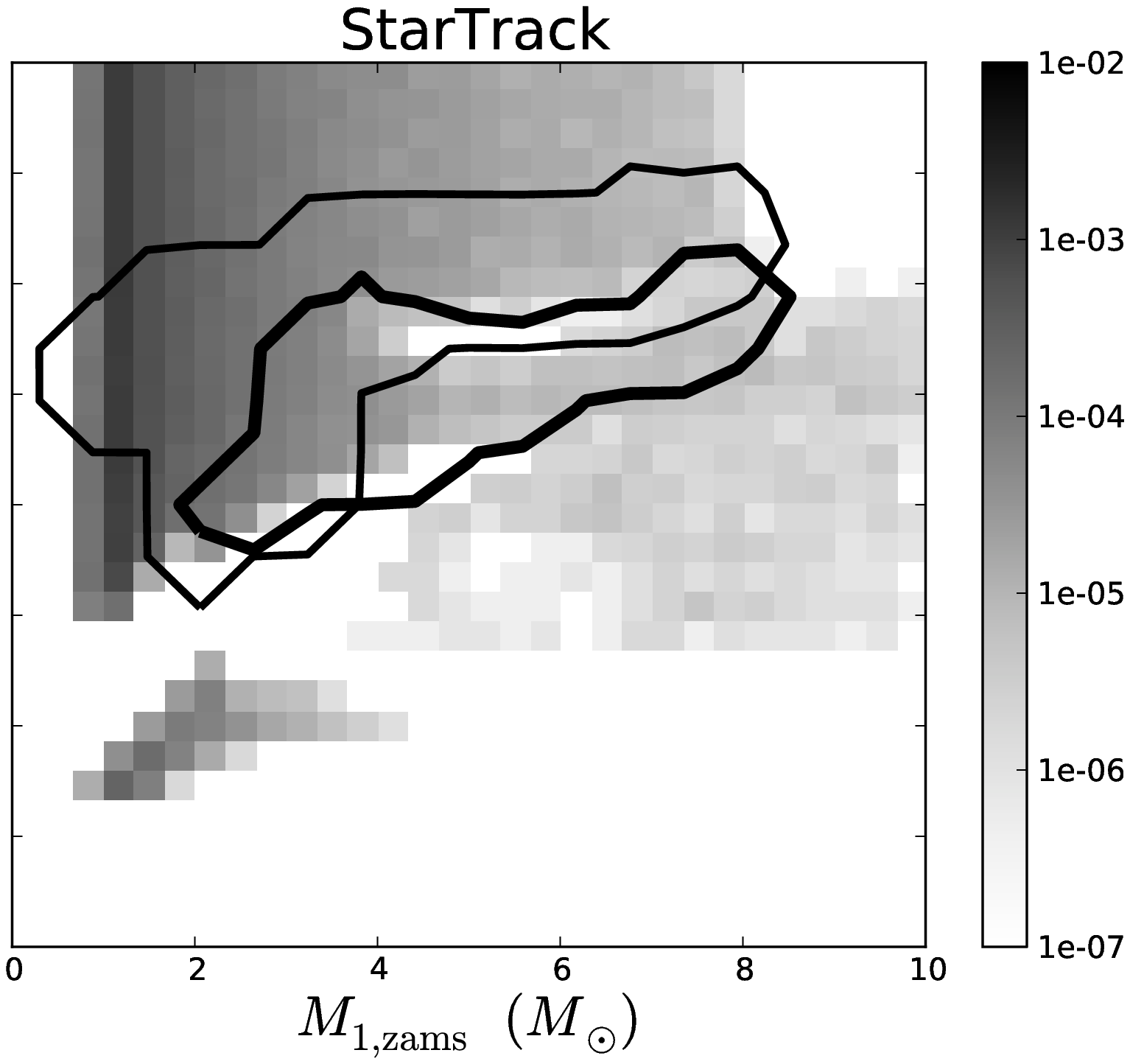} \\
	\end{tabular}
    \caption{Initial orbital separation versus initial primary mass for all SWDs in the full mass range. The contours represent the SWD population from a specific channel: channel~2a (thin line) and channel~2b (thick line).} 
    \label{fig:swd_zams_a_R2}
    \end{figure*}

\begin{figure*}
    \centering
    \setlength\tabcolsep{0pt}
    \begin{tabular}{cccc}
	\includegraphics[height=4.6cm, clip=true, trim =8mm 0mm 48.5mm 5mm]{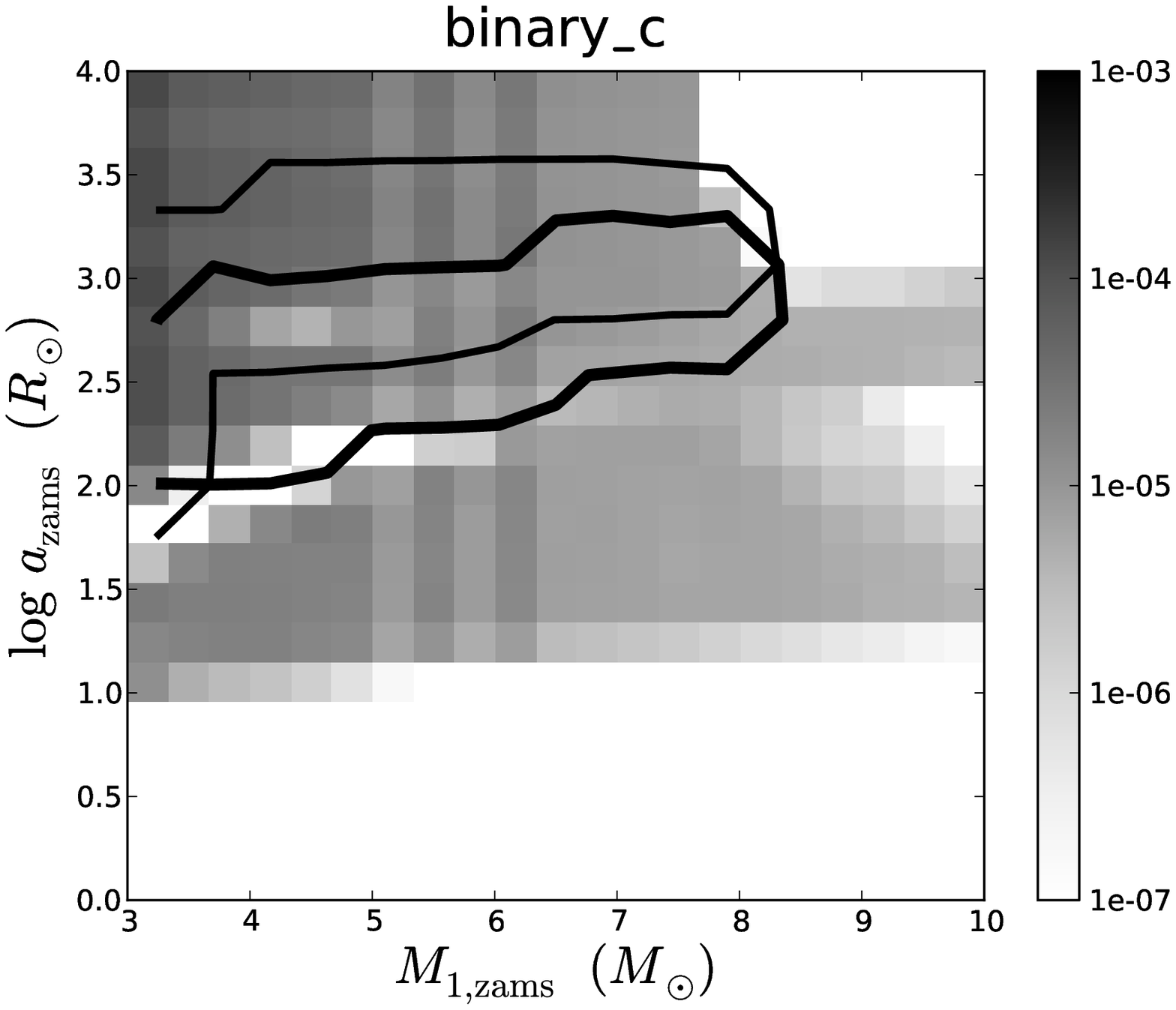} &
	\includegraphics[height=4.6cm, clip=true, trim =20mm 0mm 48.5mm 5mm]{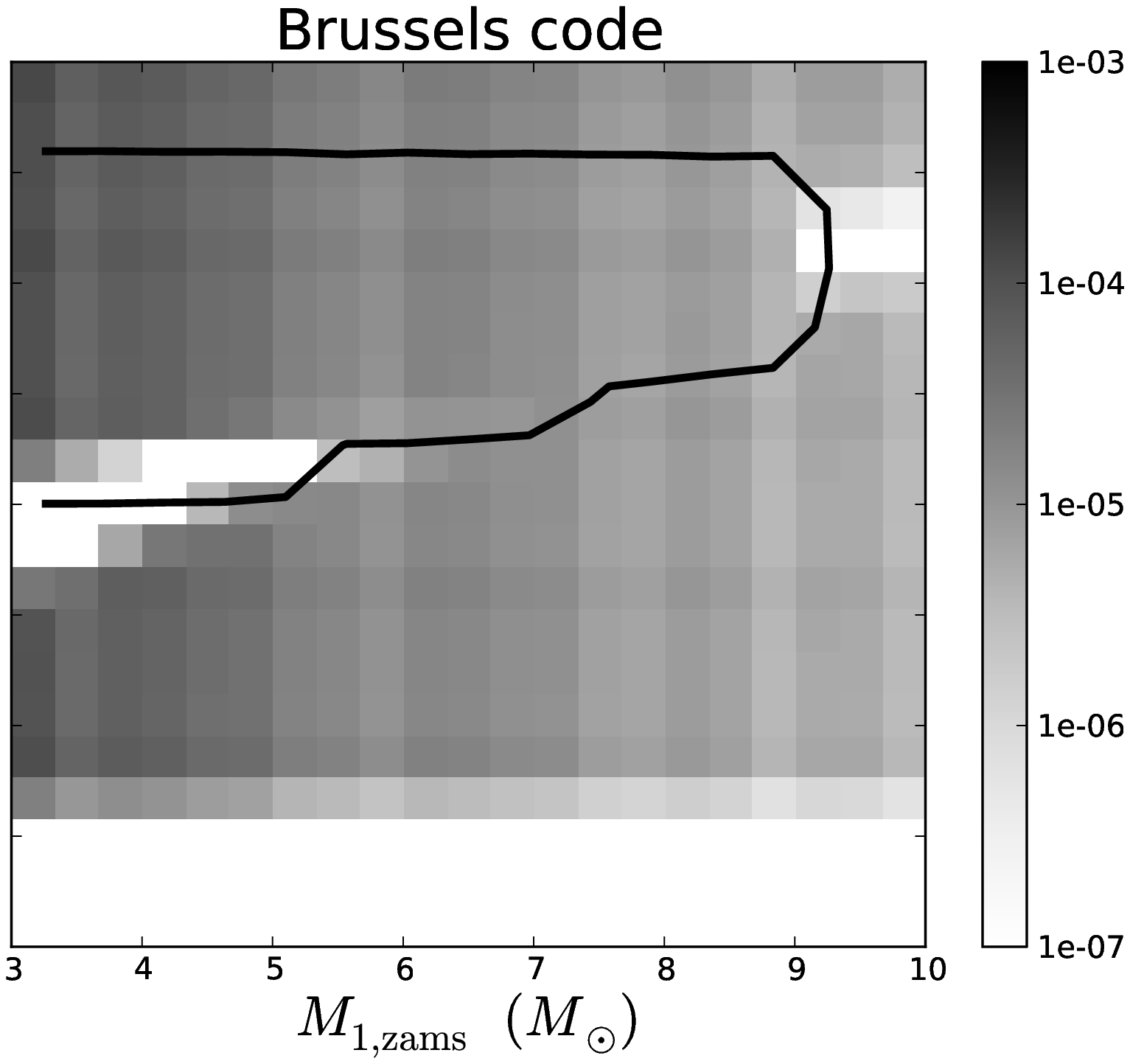} &
	\includegraphics[height=4.6cm, clip=true, trim =20mm 0mm 48.5mm 5mm]{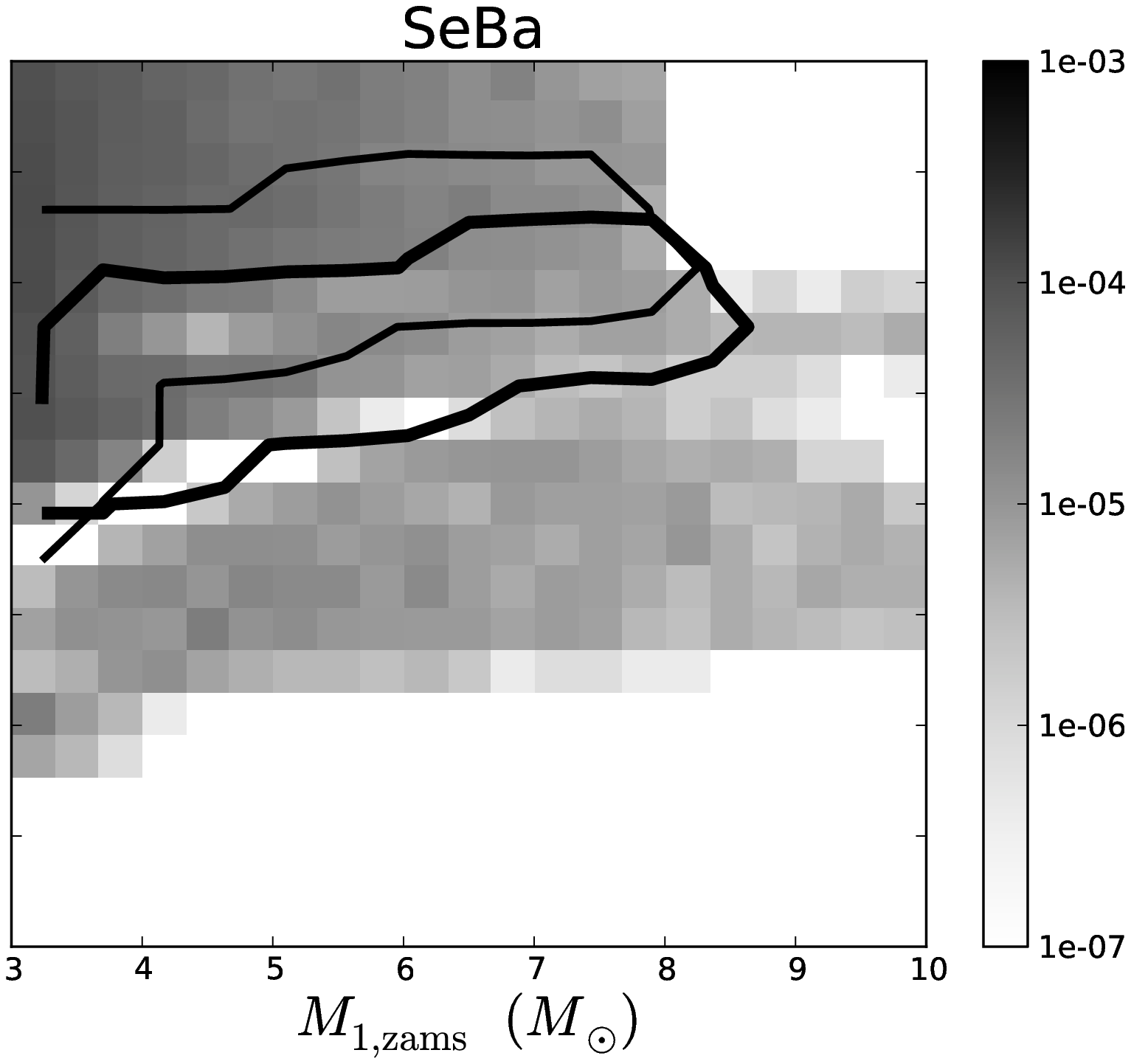} & 
	\includegraphics[height=4.6cm, clip=true, trim =20mm 0mm 23mm 5mm]{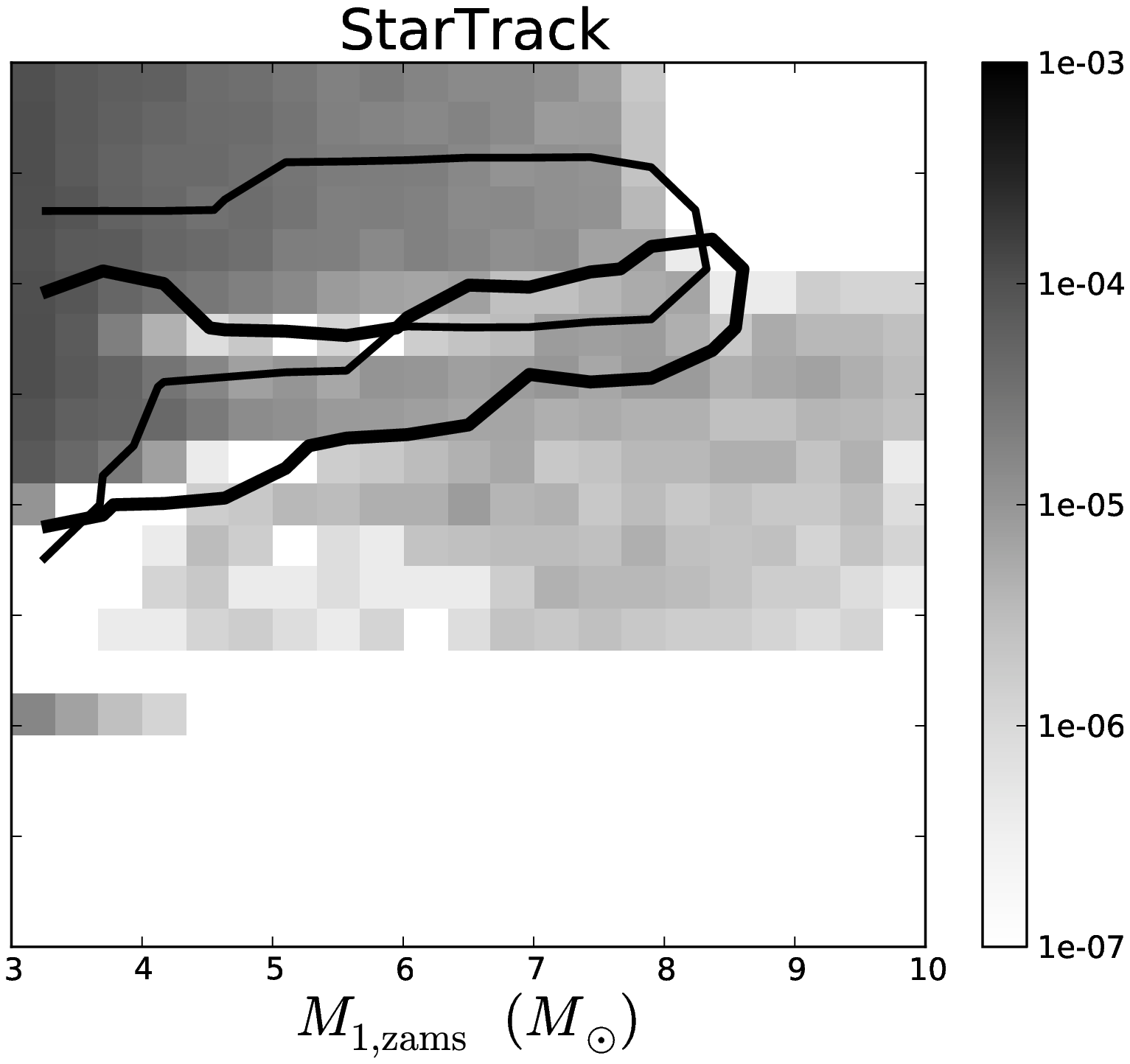} \\	
	\end{tabular}
    \caption{Initial orbital separation versus initial primary mass for all SWDs in the intermediate mass range. The contours represent the SWD population from a specific channel: channel~2a (thin line) and channel~2b (thick line).} 
    \label{fig:swd_zams_a_R2_IM}
    \end{figure*}

\emph{Population} Despite the differences between individual systems, the different BPS codes agree in which regions of phase space ($M_{\rm 1, swd}$,
$M_{\rm 2, swd}$, $a_{\rm swd}$) in Fig.\,\ref{fig:swd_final_a_R2},\,\ref{fig:swd_final_a_R2_IM},\,\ref{fig:swd_final_m2_R2}~and~\ref{fig:swd_final_m2_R2_IM} the systems from channel~2 lie. The systems of channel~2 evolve towards small separations, with the majority in the range
$0.2-150$\Rsolar~at WD formation. In addition, the codes agree on the masses of both stars at formation of the single WD system. In the low mass range binary\_c, SeBa and StarTrack find and agree that $M_ {1,swd} \approx 0.5-0.7$\Msolar~and $M_ {2,swd} \approx 0.1-2.7$\Msolar. 
In the intermediate mass range the different codes find that $M_{1,swd} \gtrsim 0.64$\Msolar, however, the Brussels code finds primary WD masses down to 0.5\Msolar~due to differences in MiMwd-relation. For secondary masses the codes find $M_ {2,swd}\approx 0.1-7.0$\Msolar. The binary\_c, SeBa, and StarTrack codes agree on the initial separation for low mass binaries, which is between $(0.6-12) \cdot 10^2$\Rsolar~(Fig.\,\ref{fig:swd_zams_a_R2}), $M_ {1,zams} \approx 1.0-3.0$\Msolar~and $M_{2,zams} \approx 0.1-3.0$\Msolar. For intermediate mass binaries in channel~2, there is an agreement between all codes that the initial primary masses lie between 
$M_{1,zams} \approx 3-8.5$\Msolar~and $M_ {2,zams} \approx 0.1-7.7$\Msolar. Due to the MiMwd-relation, the maximum initial primary mass extends to slightly higher values for the Brussels code in comparison with the other codes (Fig.\,\ref{fig:swd_zams_a_R2_IM}). 
However, for 
massive primary progenitors e.g. $M_{\rm 1, zams}>9$\Msolar~in the
Brussels code, the envelope mass of the donor is large and therefore a
merger is more likely to happen in the simulations of the Brussels
code compared to those of the other three codes.
The initial orbital separation lies between $(0.1-2.4)\cdot 10^3$\Rsolar~(Fig.\,\ref{fig:swd_zams_a_R2_IM}) according to binary\_c, SeBa and StarTrack, however, the range is extended to $3.2\cdot 10^3$\Rsolar~in the Brussels code due to the single  star prescriptions of stellar radii.

\emph{Effects} Comparing channel~2a~and~2b separately, the birthrates of SWDs (Table\,\ref{tbl:birthrates_all}) in the full mass range are close
between the codes binary\_c, SeBa, and StarTrack. 
In the intermediate mass range for channel~2a, the birthrates of
binary\_c, SeBa, and StarTrack are essentially identical, and within a
factor of 2.5 lower compared to that of the Brussels code. The larger difference with the
Brussels code are caused because this code assumes a priori that a
WD is formed without a second interaction, thus there is
  no entry for the Brussels code in Table\,\ref{tbl:birthrates_all}
  for channel~2b. 
The birthrates for channel~2b are very similar within a factor of about 1.2 between binary\_c, SeBa and StarTrack. 
Comparing the total birthrate in channel~2 between all codes, the rate of binary\_c, SeBa and StarTrack is only lower by about a factor 1.5 compared to the Brussels code, as some systems merge in the
second interaction in the simulations of the former codes. Other
differences in the simulated populations from this channel are due to
the MiMwd-relation as seen in the example, but also due to differences
in the criteria for the stability of mass transfer and the
prescriptions for the wind mass loss (see below).

The effect of the stellar wind in the example above is negligible, but the effect of wind mass loss becomes more important for systems with more evolved donors. Mass loss from the primary either in the CE-phase or in foregoing wind mass loss episodes affects the maximum orbital separation of the SWD systems directly and through angular momentum loss. 
In the simulations of the Brussels code, the maximum orbital separations at WD formation are lower ($a_{\rm swd}\lesssim80$ \Rsolar~compared to$\lesssim 150$\Rsolar~for the main group of systems in binary\_c, SeBa and StarTrack), as winds are not taken into account and more mass is removed during the CE-phase in this code. 
More mass loss during a CE-phase leads to a greater shrinkage of the orbit, where as more wind mass loss with the assumption of specific angular momentum loss from the donor (Jeans-mode, see eq.\,\ref{eq:Jloss_donor}), leads to an orbital increase. 

Another effect arises from the stellar wind in combination with the stability criterion of mass transfer. For systems with high wind mass losses in which the mass ratio has
reversed, the first phase of mass transfer can become stable according to binary$\_$c, SeBa, and StarTrack. Systems
in which this happen are not included in channel~2, however, the
birthrates are low ($[1.3, -, 6.5, 4.7]\cdot 10^{-4} \peryr$ in the full
mass range and $[5.4, -, 10, 9.1]\cdot 10^{-5} \peryr$ in the
intermediate mass range).
In general, when a AGB star initiates mass transfer, stable mass transfer is more readily realised in SeBa and StarTrack than in binary\_c. Therefore the maximum separation of SWDs in channel~2 is highest in the binary\_c data (up to 650\Rsolar). However, only about 1\% of systems in channel~2 in the binary\_c code lie in the region with a separation larger than 70\Rsolar~and a WD mass higher than 0.6\Msolar. 

The stability of mass transfer is another important effect for the population of
systems in channel~2b during the second phase of mass transfer. We
only compare the binary\_c code, SeBa, and StarTrack, as the Brussels
code does not consider this evolutionary path. Whether or not the second phase of mass transfer is
stable affects the resulting distribution of orbital separations. This
effect is shown in Fig.\,\ref{fig:swd_final_a_R2} as an extension to lower separations $a_{\rm swd} \lesssim 10$\Rsolar~for $M_{\rm 1,swd}\gtrsim 0.8$\Msolar~in the binary\_c data due to unstable mass transfer. 

There is a difference between StarTrack on one hand, and binary\_c and SeBa on the other hand regarding the survival of systems in channel~2b during the first phase of mass transfer. Due to a lack of understanding of the CE-phase, generally BPS codes assume for simplicity that when the stars fit in their consecutive Roche lobes after the CE is removed, the system survives the CE-phase. However, this depends crucially on the evolutionary state of the stars after the CE. For channel~2b in which the primary continues helium burning in a shell as a non-degenerate helium star, the response of the primary to a sudden mass loss in the CE-phase is not well known.
The StarTrack code assumes the stripped star immediately becomes an evolved helium star and corresponding radius, while binary\_c and SeBa assume the stripped star is in transition from an exposed core to an evolved helium star with a radius that can be a factor of about 1-15 smaller. The uncertainty in the radii of the stripped star mostly affect systems with $M_{1,zams} \gtrsim 5$\Msolar~at separations $\gtrsim 450$\Rsolar~that merge according to StarTrack, and survive according to binary\_c and SeBa.

Included in channel~2 are systems that evolve through a double CE-phase\footnote{Note that systems in which the double CE-phase results directly in a DWD system are not taken into account for the comparison of SWD systems.} in which both stars lose their envelope described in Sect.\,\ref{sec:BinEvol} and in eq.\,\ref{eq:ce_dspi}. The double CE-mechanism is taken into account by the binary\_c, SeBa and StarTrack code. However, there is a difference between StarTrack on one hand, and binary\_c and SeBa on the other hand regarding the binding energy of the envelope of the secondary star. In StarTrack the binding energy is calculated according to eq.\,\ref{eq:Egr_web}) with $R_2=R_{\rm RL,2}$, where as in binary\_c and SeBa the instantaneous radius at the start of the double CE-phase is taken for the secondary radius. This can have a significant effect on the orbit of the post-double CE-system, leading to an increase of systems at low separations (approximately 1\Rsolar) in the binary\_c and SeBa data compared to the StarTrack data.

\subsubsection{Channel 3: stable case B}
\label{sec:channel3}
\emph{Evolutionary path} For channel~3, mass transfer starts when a hydrogen shell burning star fills its Roche lobe in a stable way before core helium-burning starts \citep[][case Br]{Kip67}. This can occur when the envelope is radiative or when the convective zone in the upper layers of the envelope is shallow.
In this project we assume that stable mass transfer proceeds conservatively and so the secondary significantly grows in mass. Because mass transfer is conservative, the orbit first shrinks and when the mass ratio has reversed the orbit widens. Mass transfer continues until the primary has lost (most of) its hydrogen envelope. 
At this stage the primary can become a helium WD or, if it is massive enough, ignite helium in its core. In the latter scenario the 
primary is a He-MS star. Like for channel~2, there are two sub-channels depending on whether the primary star fills the Roche lobe for a second time as a helium star. If the primary does not go through a helium-star phase or does not fill its Roche lobe as a helium star, the system evolves according to channel~3a. In channel~3b there is a second phase of mass transfer. 

\begin{figure*}
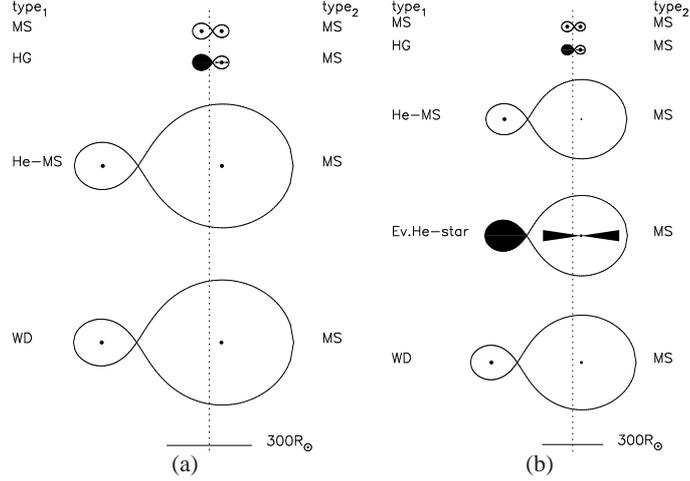

\centering
\begin{tabular}{c c}
 	\includegraphics[width = 6cm,angle=270]{CaseBr.ps} &
 	\includegraphics[width = 6cm,angle=270]{CaseBrB.ps} \\
	(a) & (b) \\
\end{tabular}
\caption{Example of the evolution of a SWD system in channel~3a (left) and channel~3b (right). 
Abbreviations are as in Table\,\ref{tbl:star_type}.}
\label{fig:rl_B}
\end{figure*}

\emph{Example of channel~3a} Figure\,\ref{fig:rl_B}a shows an example of the evolution of a binary system of channel~3a with initial parameters $M_{\rm 1, zams}=4.8\Mo$, $M_{\rm 1, zams}=3\Mo$ and $a_{\rm zams}=70\Ro$. The masses of He-MS and secondary star are very similar in the BPS codes $[0.82, 0.83, 0.82, 0.82]$\Msolar~and $[6.9, 7.0, 7.0, 7.0]$\Msolar~respectively. The separations at the moment the helium star forms are $[4.2, 4.3, 4.3,4.7]\cdot 10^2$\Rsolar~and are similar as well. 
In the subsequent evolution, the primary star effectively evolves as a single helium star before becoming a carbon-oxygen WD and loses 
$[0.038, 0.14, 0.043, 0.038]$\Msolar~during that time and the orbit does not change significantly.  changes by [2.1,, 2.5,4.2]\Rsolar.

\begin{figure*}
    \centering
    \setlength\tabcolsep{0pt}
    \begin{tabular}{ccc}
	\includegraphics[height=4.6cm, clip=true, trim =8mm 0mm 48.5mm 5mm]{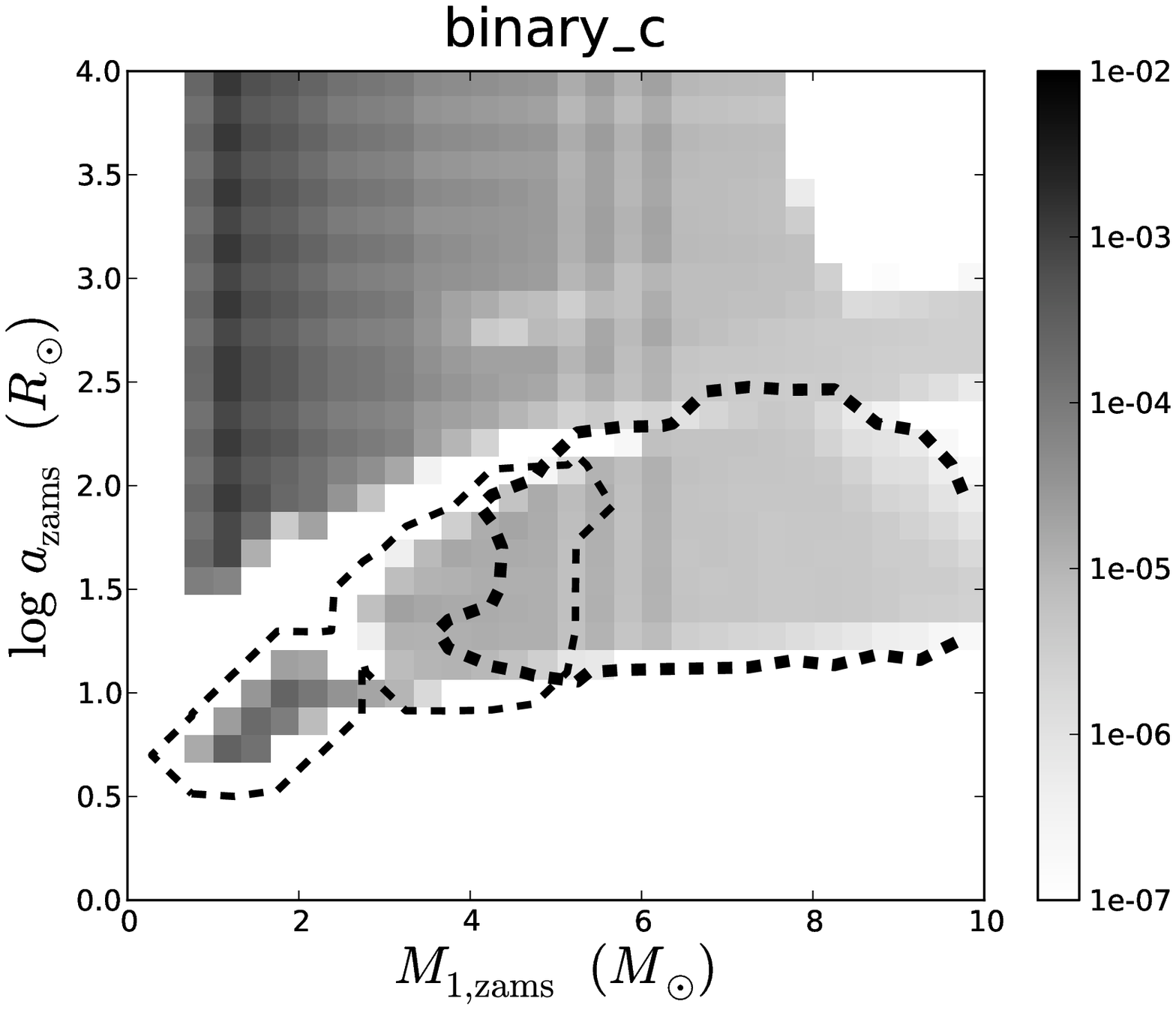} &
	\includegraphics[height=4.6cm, clip=true, trim =20mm 0mm 48.5mm 5mm]{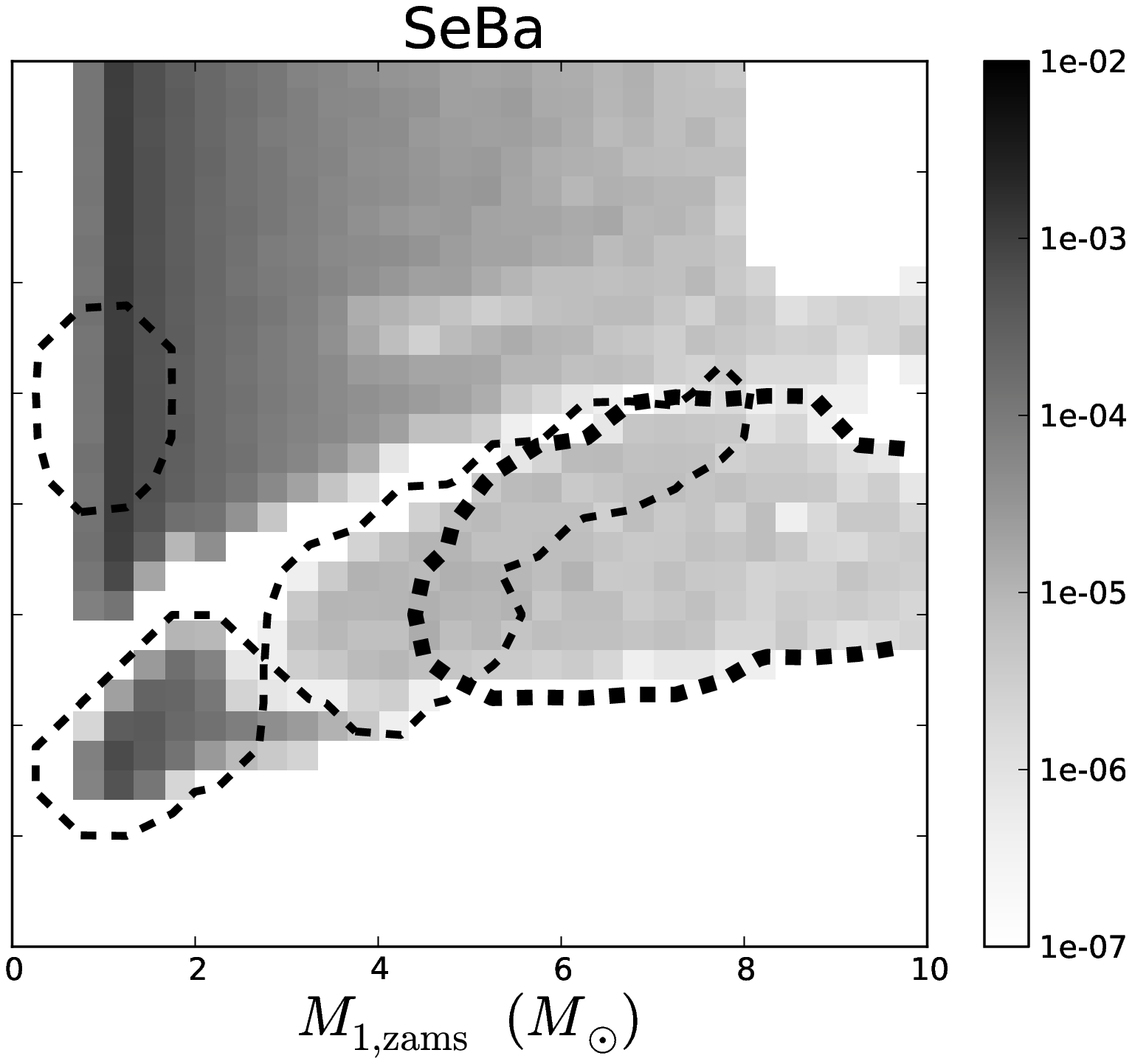} & 
	\includegraphics[height=4.6cm, clip=true, trim =20mm 0mm 23mm 5mm]{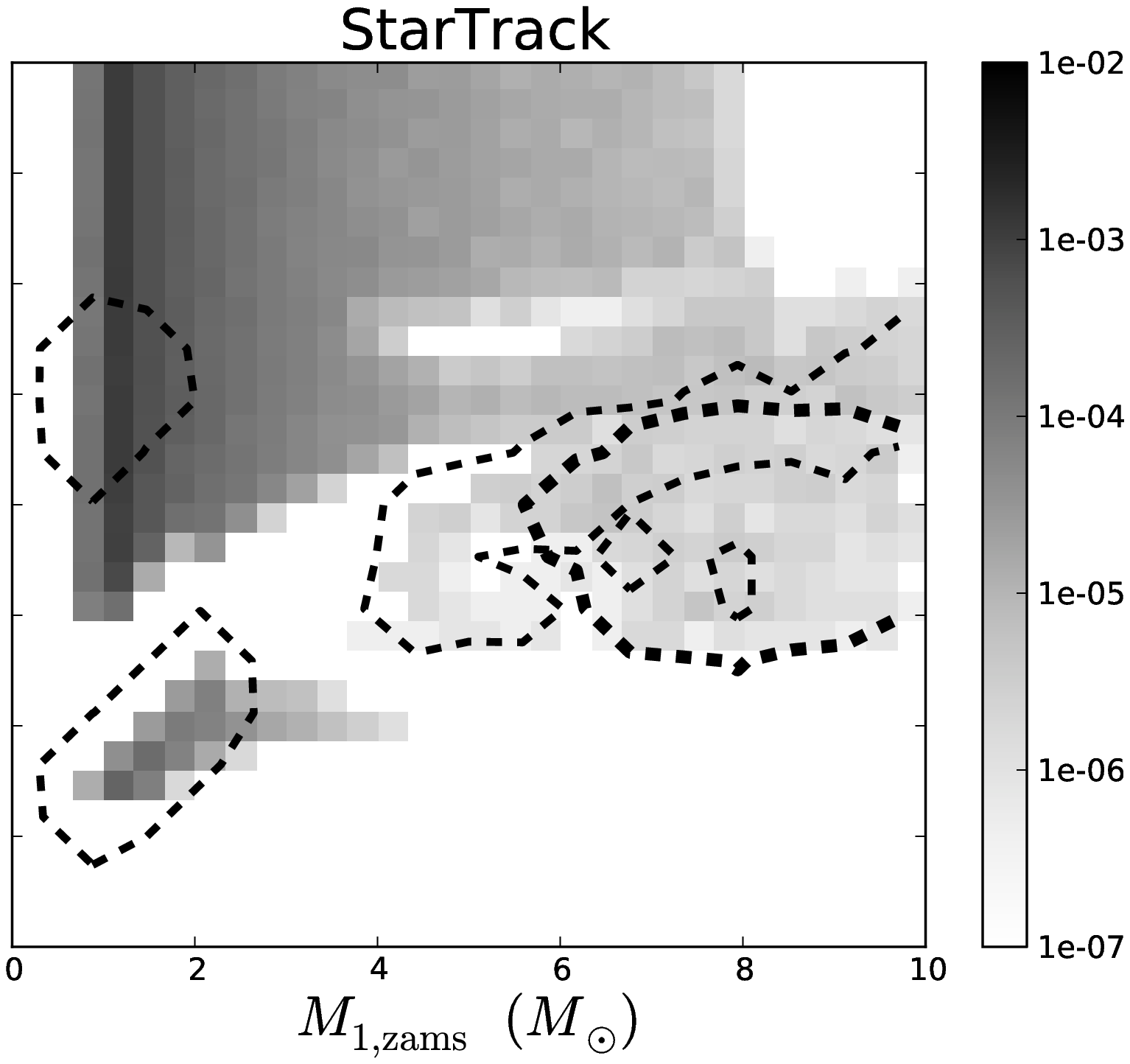} \\
	\end{tabular}
    \caption{Initial orbital separation versus initial primary mass for all SWDs in the full mass range. The contours represent the SWD population from a specific channel: channel~3a (thin line) and channel~3b (thick line).} 
    \label{fig:swd_zams_a_R3}
    \end{figure*}

\begin{figure*}
    \centering
    \setlength\tabcolsep{0pt}
    \begin{tabular}{cccc}
	\includegraphics[height=4.6cm, clip=true, trim =8mm 0mm 48.5mm 5mm]{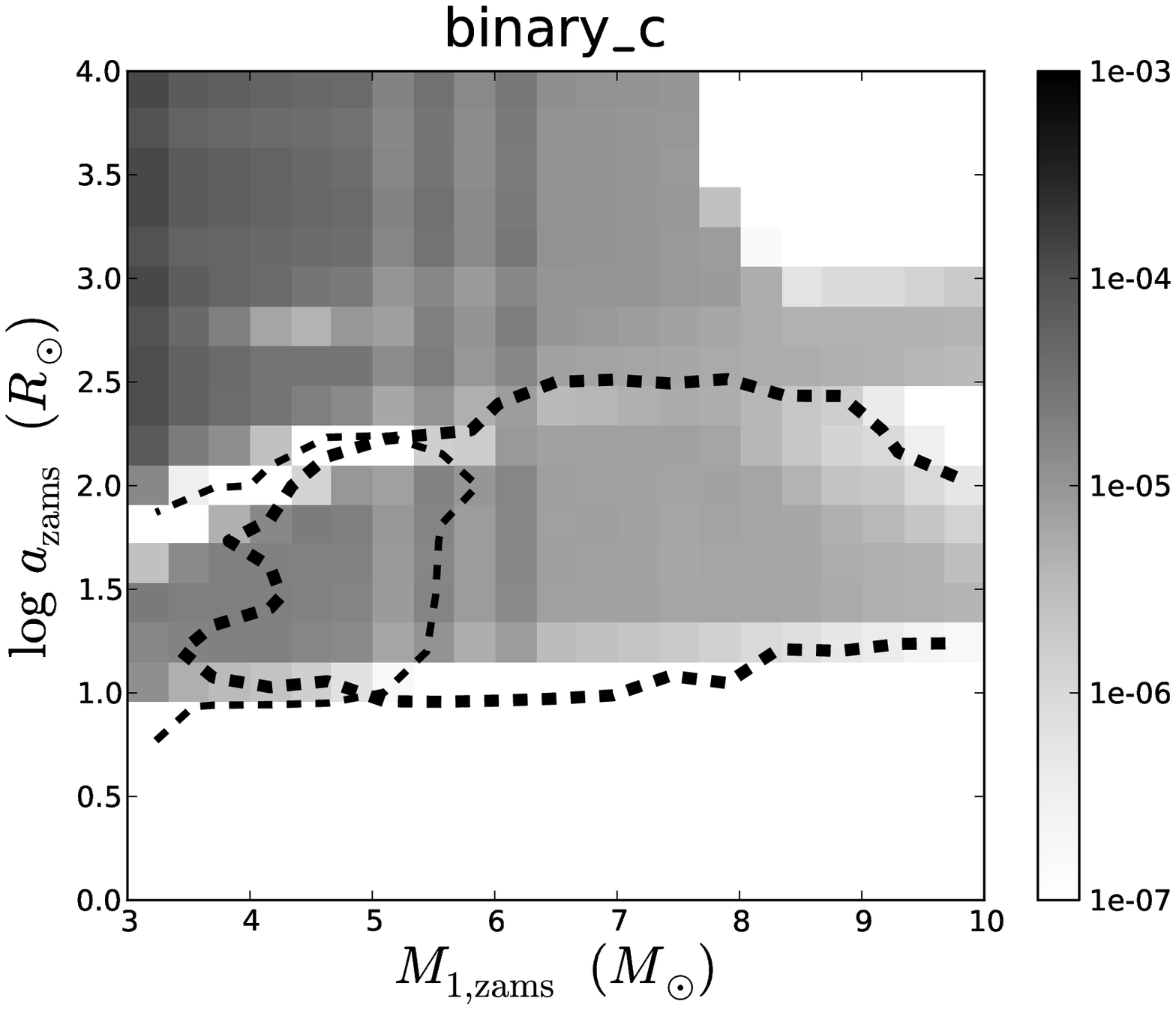} &
	\includegraphics[height=4.6cm, clip=true, trim =20mm 0mm 48.5mm 5mm]{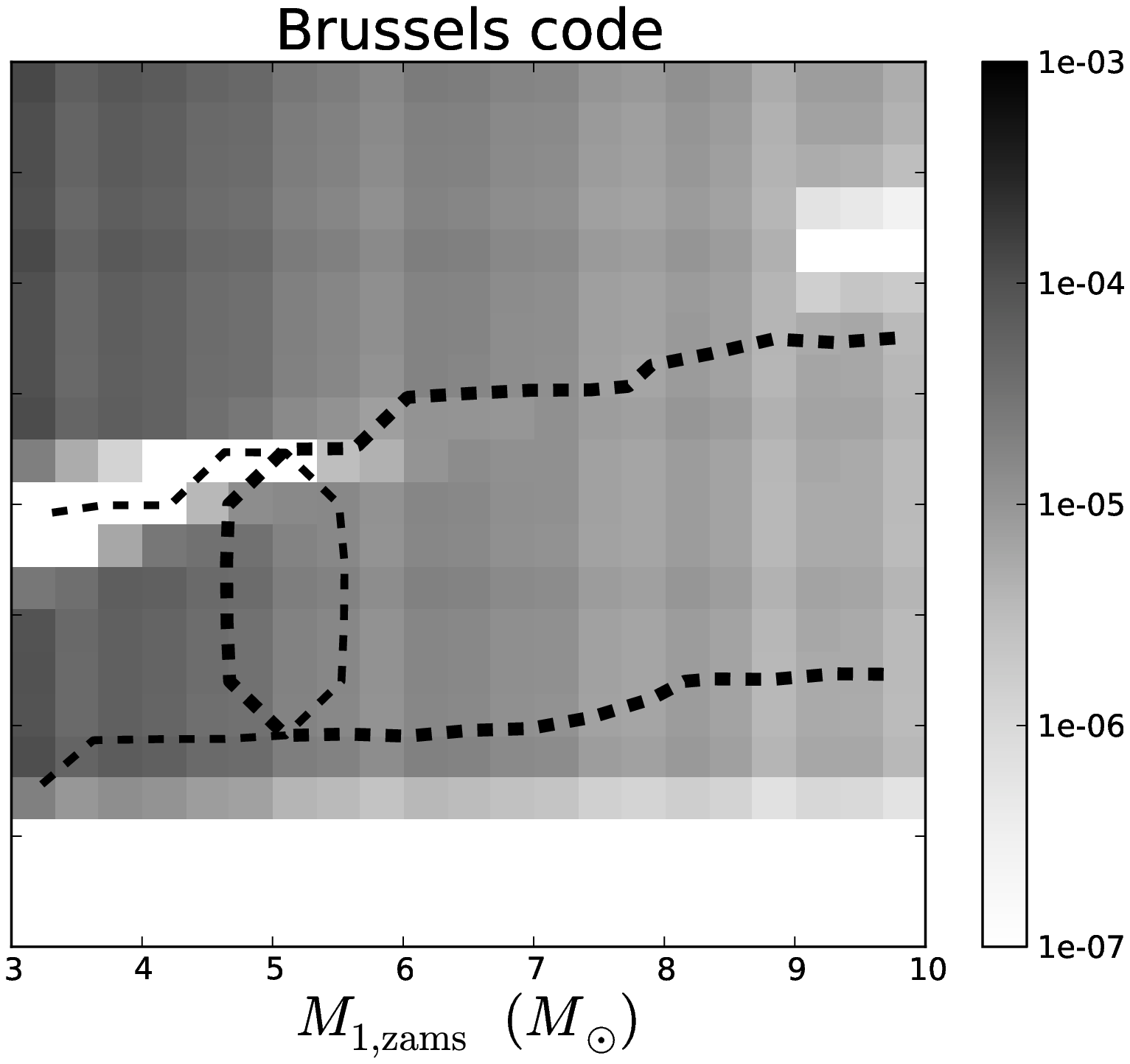} &
	\includegraphics[height=4.6cm, clip=true, trim =20mm 0mm 48.5mm 5mm]{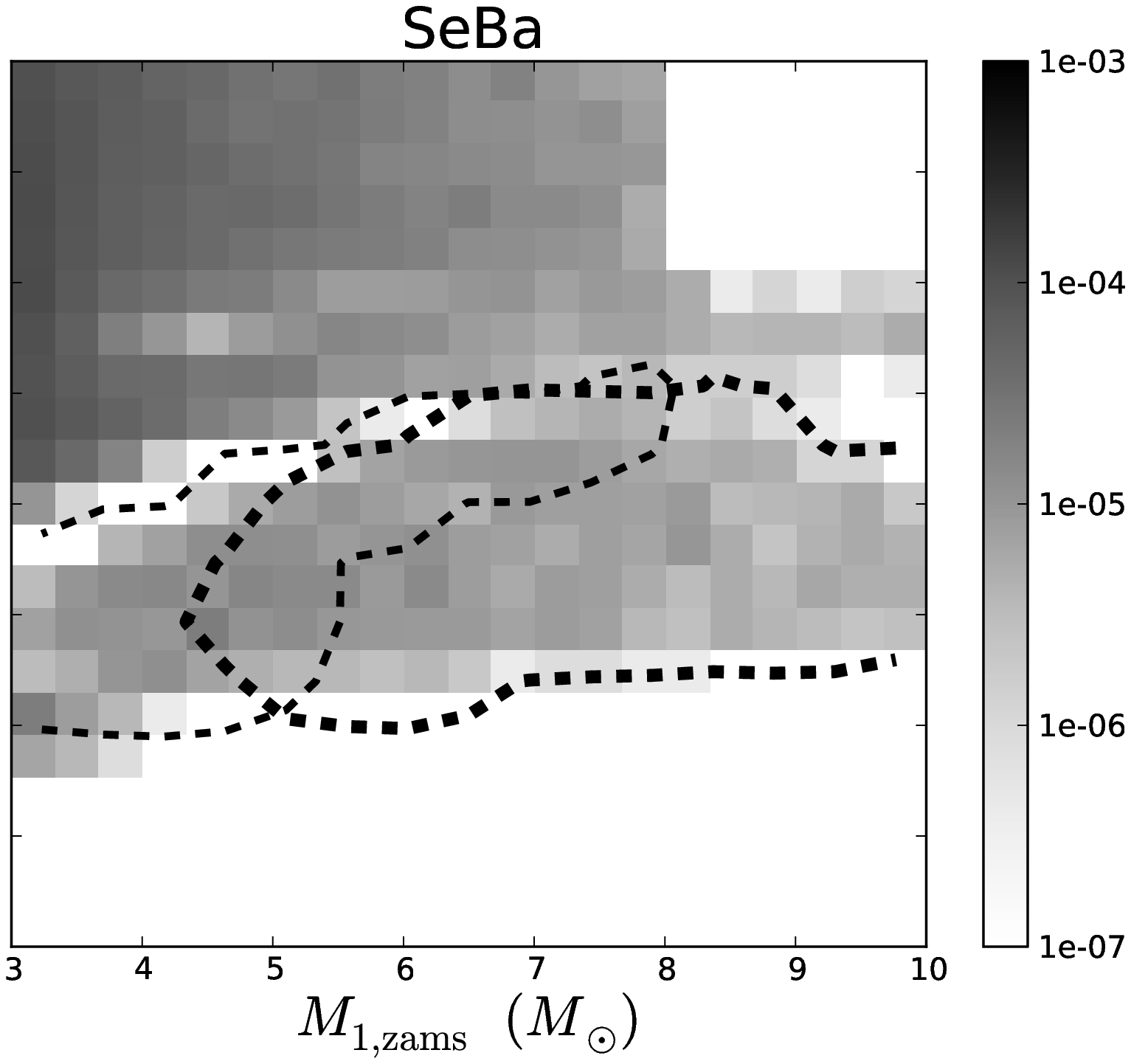} & 
	\includegraphics[height=4.6cm, clip=true, trim =20mm 0mm 23mm 5mm]{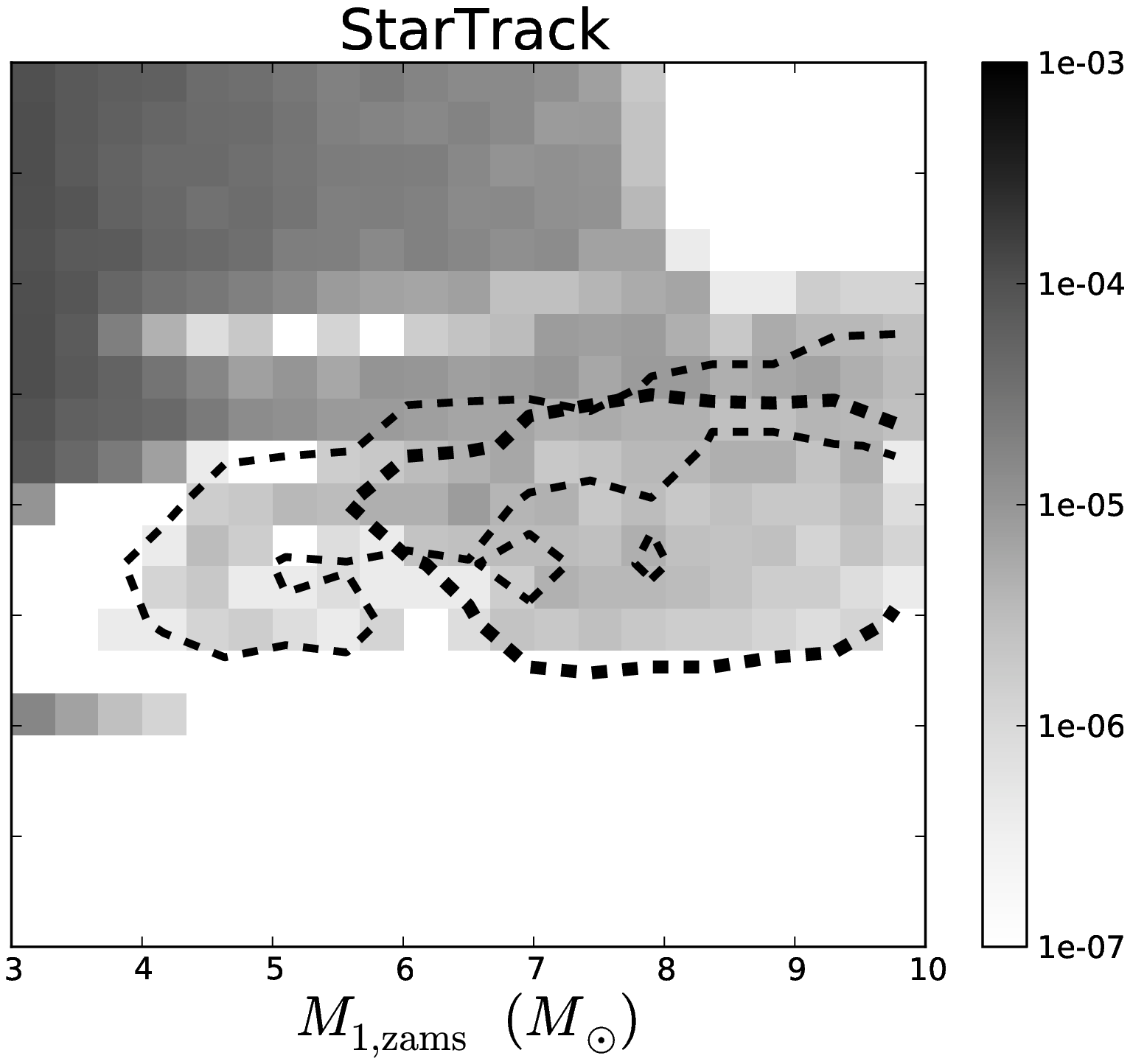} \\
	\end{tabular}
    \caption{Initial orbital separation versus initial primary mass for all SWDs in the intermediate mass range. The contours represent the SWD population from a specific channel: channel~3a (thin line) and channel~3b (thick line).} 
    \label{fig:swd_zams_a_R3_IM}
    \end{figure*}

\emph{Population from channel~3a} Regarding channel~3a, not all codes agree on the ranges of separation and masses (Fig.\,\ref{fig:swd_zams_a_R3}~and~\ref{fig:swd_zams_a_R3_IM}). However, there is an agreement between binary\_c, the Brussels code and SeBa that majority of intermediate mass systems originate from systems with $M_{\rm 1, zams}$ between 3 and 5\Msolar~and $a_{\rm zams}$ between 10 and 100\Rsolar. The SWD population at WD formation is centred around systems with $M_{\rm 1, swd}\approx 0.6$\Msolar~for the 
binary\_c, Brussels and SeBa codes, and with the majority of separations between about $20-1000$\Rsolar. 
The SWD systems and their progenitors that are just described are not SWD progenitors according to StarTrack. According to this code, mass transfer is unstable and the system merges. The birthrates of binary\_c, the Brussels code and SeBa differ within a factor of about 4
(Table\,\ref{tbl:birthrates_all}). In addition binary\_c, SeBa
and StarTrack show a good agreement on the different sub-populations
for the full mass range. At WD formation these codes show a subpopulation between 15
to about 200\Rsolar~with WD masses of between 0.17 and 0.35\Msolar. There is a
second subpopulation at about 1\Rsolar~with most systems having a WD
between 0.4 and 0.8\Msolar. A third population shows mainly WD masses of more than 0.8\Msolar~at separations of more than 300\Rsolar, where the population is extended to higher separations and WD masses in the results of SeBa and StarTrack. 
The third population is also visible in the progenitor population in Fig.\,\ref{fig:swd_zams_a_R3} with primary masses of more than 5\Msolar~and separations of more than about 70\Rsolar. Again this population is more extended to high masses and separations according to SeBa and StarTrack. The low mass range of the progenitor population shows predominantly systems in orbits of 5-15\Rsolar. SeBa and StarTrack agree that there is an extra group at high orbital separations $a_{\rm zams} \approx (1.3-4.6)\cdot 10^2$\Rsolar.

\emph{Example of channel~3b} An example of the evolution in channel~3b
is shown in Fig.\,\ref{fig:rl_B}b. 
Initially the system has
$M_1=7\Mo$, $M_2=5\Mo$ and $a=65\Ro$. After the first phase of mass
transfer the primary masses $M_1=[1.4,1.5,1.4,1.4]$\Msolar, the secondary masses $M_2=[11,11,11,11]$\Msolar~and separations $a=[3.8,3.3,3.8,4.1]\cdot 10^2$\Rsolar. When the primary fills its Roche lobe again, it has lost $[5.8,-,6.8,7.3]\cdot10^{-2}$\Msolar~in the wind.
The mass transfer phase is stable and the secondary increases in mass to $[11,11,11,11]$\Msolar. The primary becomes a WD of $[1.1,1.0,0.99,1.0]$\Msolar~in an orbit of $[4.5,6.5,5.9,6.2]\cdot 10^2$\Rsolar. 

\emph{Population from channel~3b} The binary\_c, Brussels and SeBa codes agree well on the initial systems leading to SWDs through channel~3b. This holds for both the initial mass, namely between about 5 and 10\Msolar~and the initial separation between $0.1-3.0\cdot 10^2$\Rsolar. The population of progenitors of channel~3b according to the StarTrack code lies inside the previously mentioned ranges, however, the parameter space is smaller. In addition the four codes agree that at WD formation the majority of companions that are formed through channel~3b are massive, about 6 to 18\Msolar~ (for StarTrack 8-18\Msolar.) The orbits of these systems are wide around $10^3$\Rsolar, however, the ranges in separation and WD mass differs between the codes and will be discussed in the next paragraphs. 
Binary\_c, SeBa and StarTrack also show a group of lower mass companions. For binary\_c and SeBa these lie in the range 0.8-4.5\Msolar~with separations of 0.5-30\Rsolar~and $M_{\rm 1,swd}$ mainly between 0.6 and 1.0\Msolar. The population of StarTrack agrees with these ranges, however, the parameter space for this population is smaller.

\begin{figure*}
    \centering
    \setlength\tabcolsep{0pt}
    \begin{tabular}{cccc}
	\includegraphics[height=4.6cm, clip=true, trim =8mm 0mm 48.5mm 5mm]{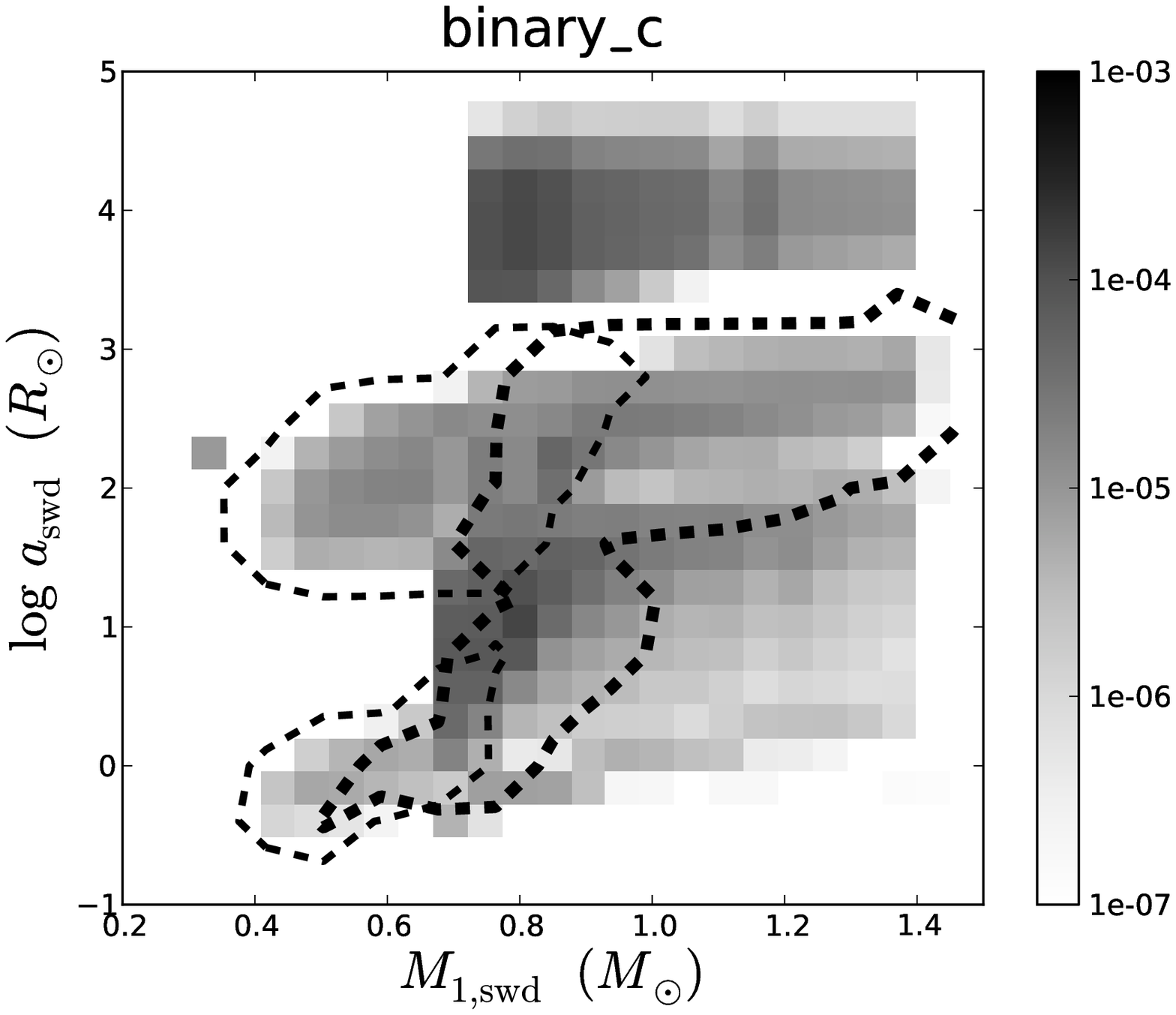} &
	\includegraphics[height=4.6cm, clip=true, trim =20mm 0mm 48.5mm 5mm]{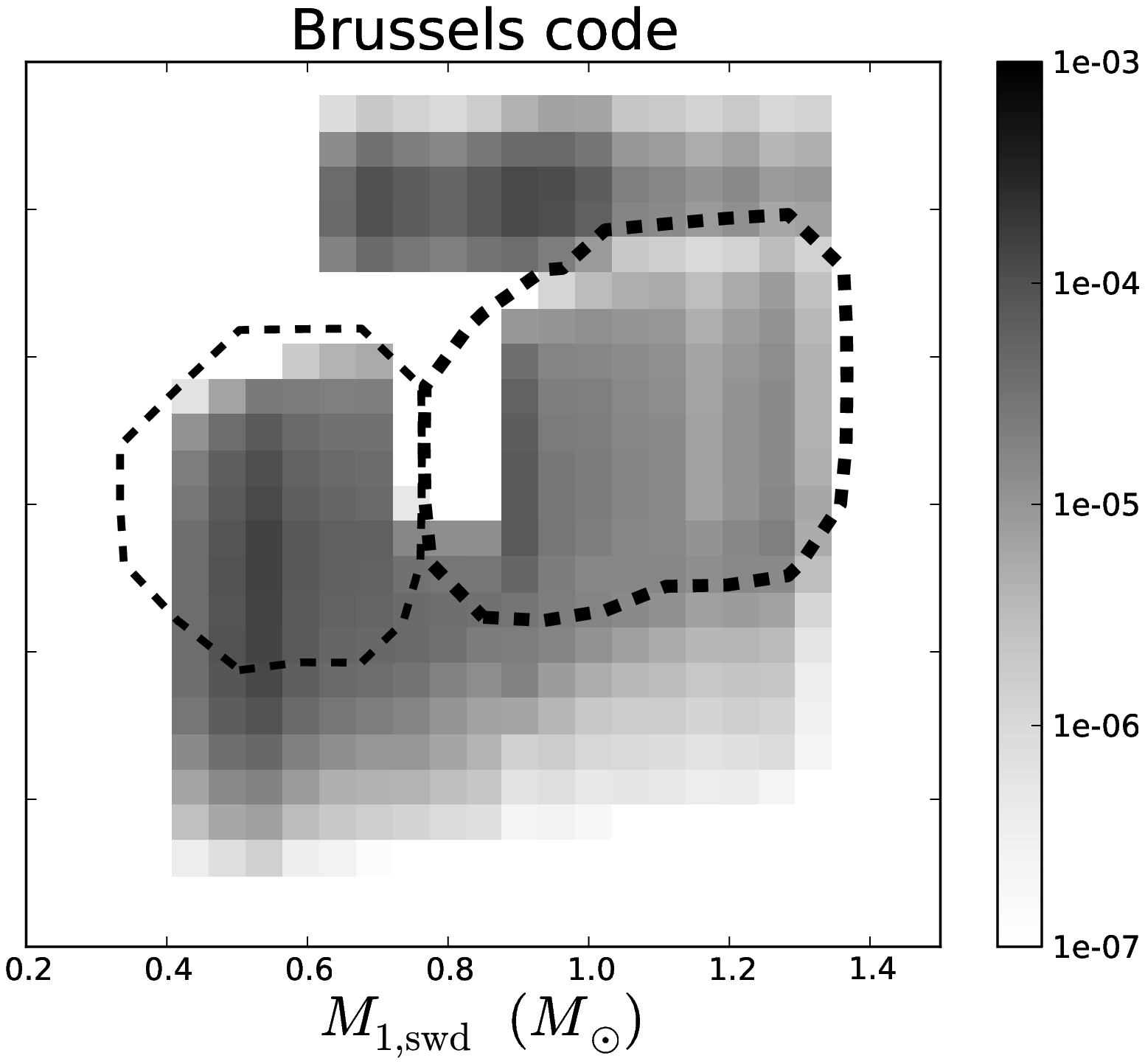} &
	\includegraphics[height=4.6cm, clip=true, trim =20mm 0mm 48.5mm 5mm]{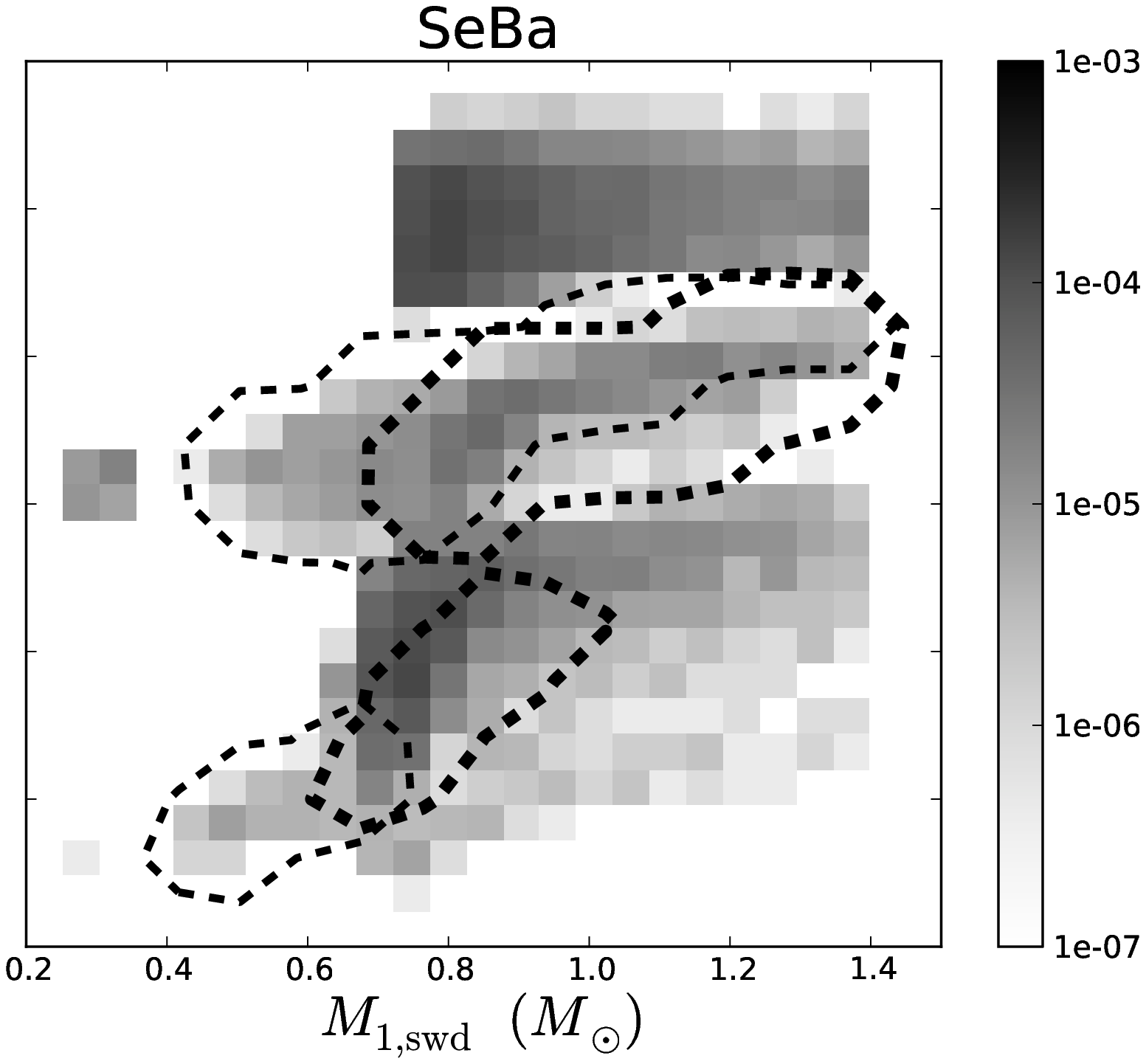} &
	\includegraphics[height=4.6cm, clip=true, trim =20mm 0mm 23mm 5mm]{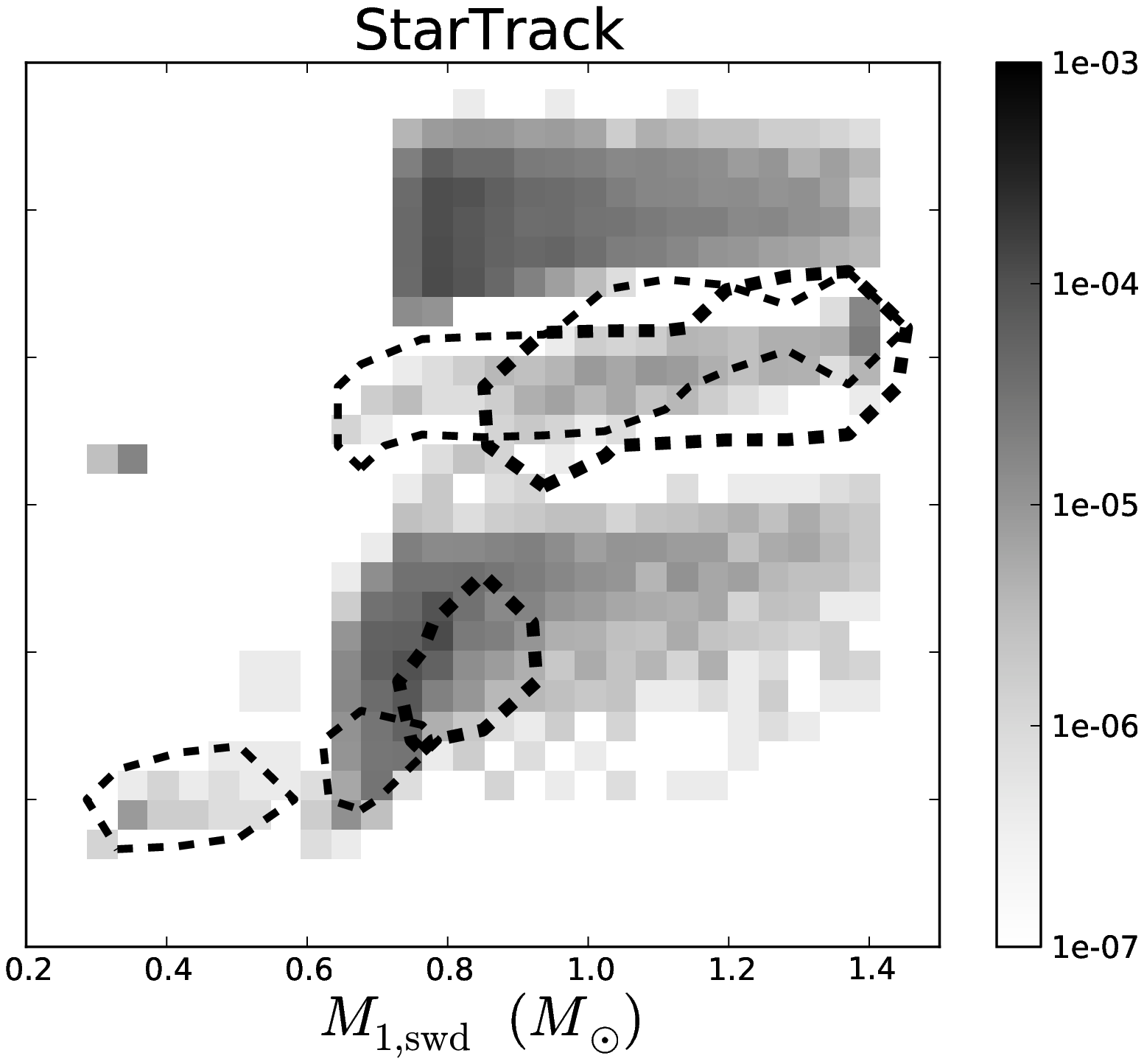} \\
	\end{tabular}
    \caption{Orbital separation versus WD mass for all SWDs in the intermediate mass range at the time of SWD formation. The contours represent the SWD population from a specific channel: channel~3a (thin line) and channel~3b (thick line).} 
    \label{fig:swd_final_a_R3_IM}
    \end{figure*}

\emph{Effects}
The population of SWDs from channel~3a~and~3b are influenced by the
MiMwd-relation. An important contribution to the MiMwd-relation comes
from the assumed mass losses for helium stars and its mechanism,
i.e. in a fast spherically symmetric wind or in planetary nebula (Appendix\,\ref{sec:TNS_inherent}). There is not much known about the
mass loss from helium stars either observationally or theoretically. The differences in the MiMwd-relation affect for example the distribution of separations in Fig.\,\ref{fig:swd_final_a_R3_IM}. For channel~3b the separation is $\lesssim$ 1400\Rsolar~for binary\_c, SeBa, and StarTrack, but is extended to 6600\Rsolar~in the Brussels code. Binaries become wider in the Brussels code, as the WD masses in channel~3 are in general smaller compared to the other three codes. 

There is also a difference in the MiMwd-relation between StarTrack on
one hand, and binary\_c and SeBa on the other hand regarding primaries that after losing their hydrogen envelopes become helium
stars. For massive helium stars, binary\_c and SeBa find that these stars will collapse to neutron stars, where as in StarTrack these stars form WDs. For channel~3a the difference occurs for the range of helium star masses of 1.6-2.25\Msolar.  
As a result, systems containing massive helium stars are not considered to become SWD systems in binary\_c and SeBa. 
These systems are present in the SWD data of StarTrack at $M_{\rm 1, swd} \gtrsim 1.38$\Msolar~in Fig.\,\ref{fig:swd_final_m2_R1} for channel~3a~and~3b. The progenitors lie at $M_{\rm 1,zams} \gtrsim 8$\Msolar~with mostly $a_{\rm zams} \approx 65-220$\Rsolar~for channel~3a~and~3b.

Another effect on the SWD population is the modelling of the mass transfer phases which is inherent to the BPS codes. The value of the mass transfer rate or the length of the mass transfer phase, however, do not have a large effect on the population or the evolution of individual systems from channel~3b in the set-up of the current study. This is because a priori conservative mass transfer is assumed, and therefore the accretion efficiency is not affected by the mass transfer rate. The mass transfer timescale only affects the binary evolution when other evolutionary timescales (such as the wind mass loss timescale or nuclear evolution timescale) are comparable. 
For example, while for $M_{\rm 1} \ll M_{\rm 2}$ the orbit increases strongly during RLOF, the orbit increases only moderately during wind mass loss assuming Jeans mode angular momentum loss. The range of separations in Fig.\,\ref{fig:swd_final_a_R3_IM} is therefore, besides the MiMwd-relation, also affected by the amount of wind mass and wind angular momentum leaving the system during RLOF.
The binary\_c, SeBa, and StarTrack codes assume wind mass takes with it the specific angular momentum of the donor star (Jeans mode), where as the Brussels code does not take wind mass loss into account during stable mass transfer. 

Generally, no significant evolution of the donor star takes place during the mass transfer phase. Therefore with the current set-up, the post-mass transfer masses are determined by their initial mass and for binary\_c, SeBa and StarTrack also the evolutionary moment the donor star fills its Roche lobe. However, an exception to this occurs for channel~3b during the second phase of mass transfer. Here the length of the mass transfer phase is important, as the evolutionary time scale of an evolved helium star is very short (of the order of few Myr) and the core grows significantly during this period. As a result the duration of the mass transfer phase becomes important for the resulting WD mass and separation in binary\_c, SeBa and StarTrack (e.g. the example of channel~3b). 

\begin{figure*}
    \centering
    \setlength\tabcolsep{0pt}
    \begin{tabular}{ccc}
	\includegraphics[height=4.6cm, clip=true, trim =8mm 0mm 48.5mm 5mm]{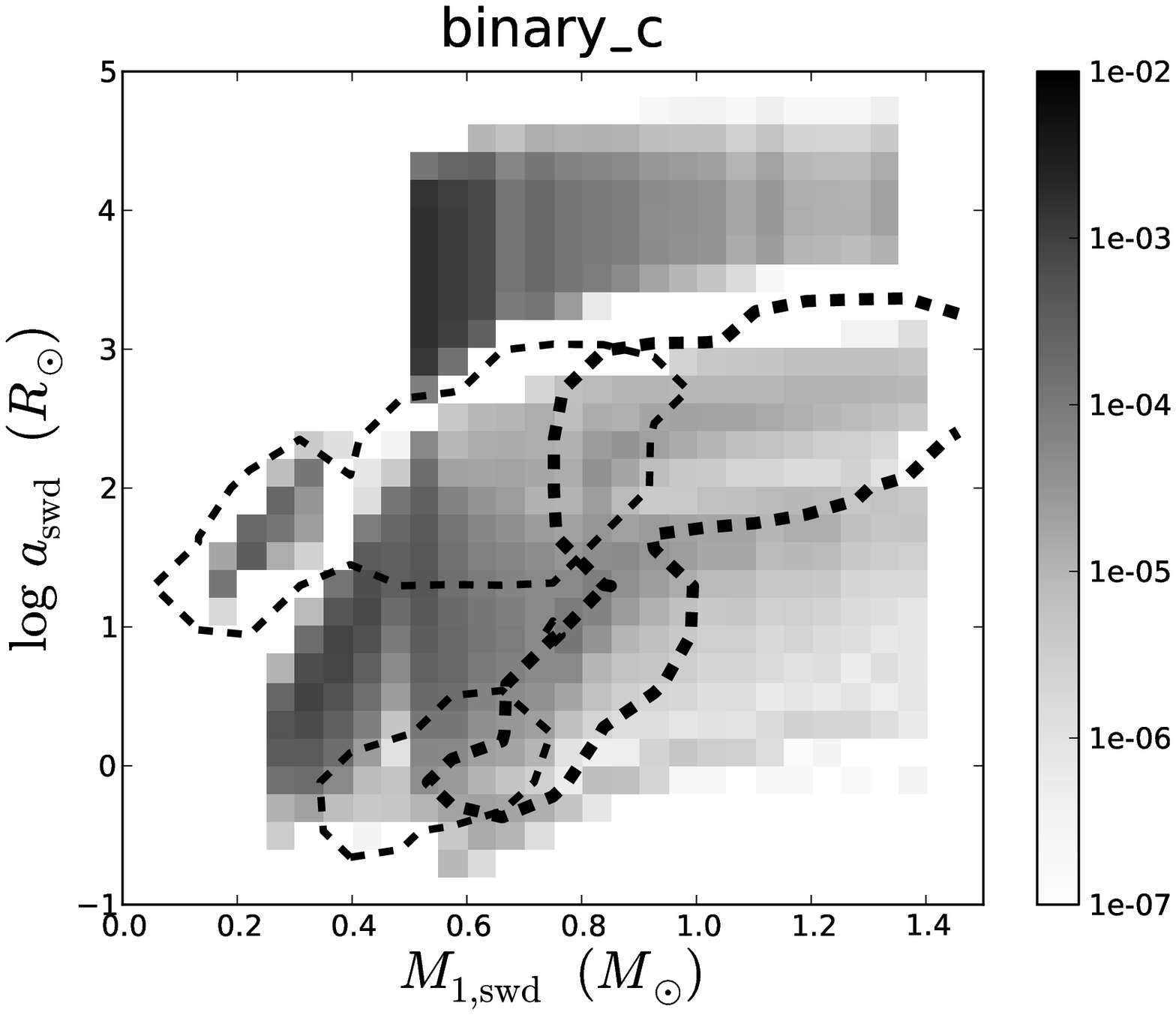} &
	\includegraphics[height=4.6cm, clip=true, trim =20mm 0mm 48.5mm 5mm]{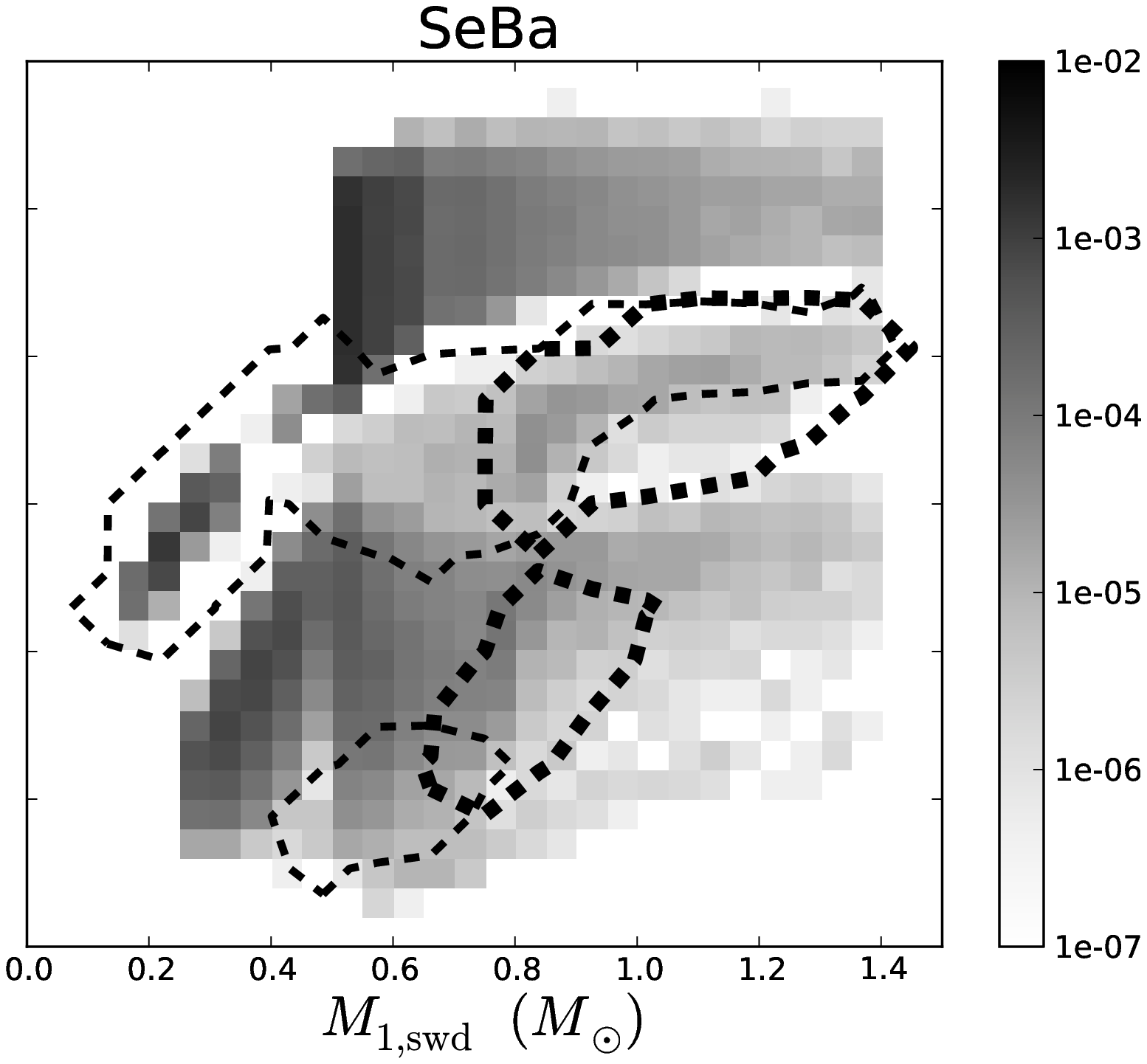} &
	\includegraphics[height=4.6cm, clip=true, trim =20mm 0mm 23mm 5mm]{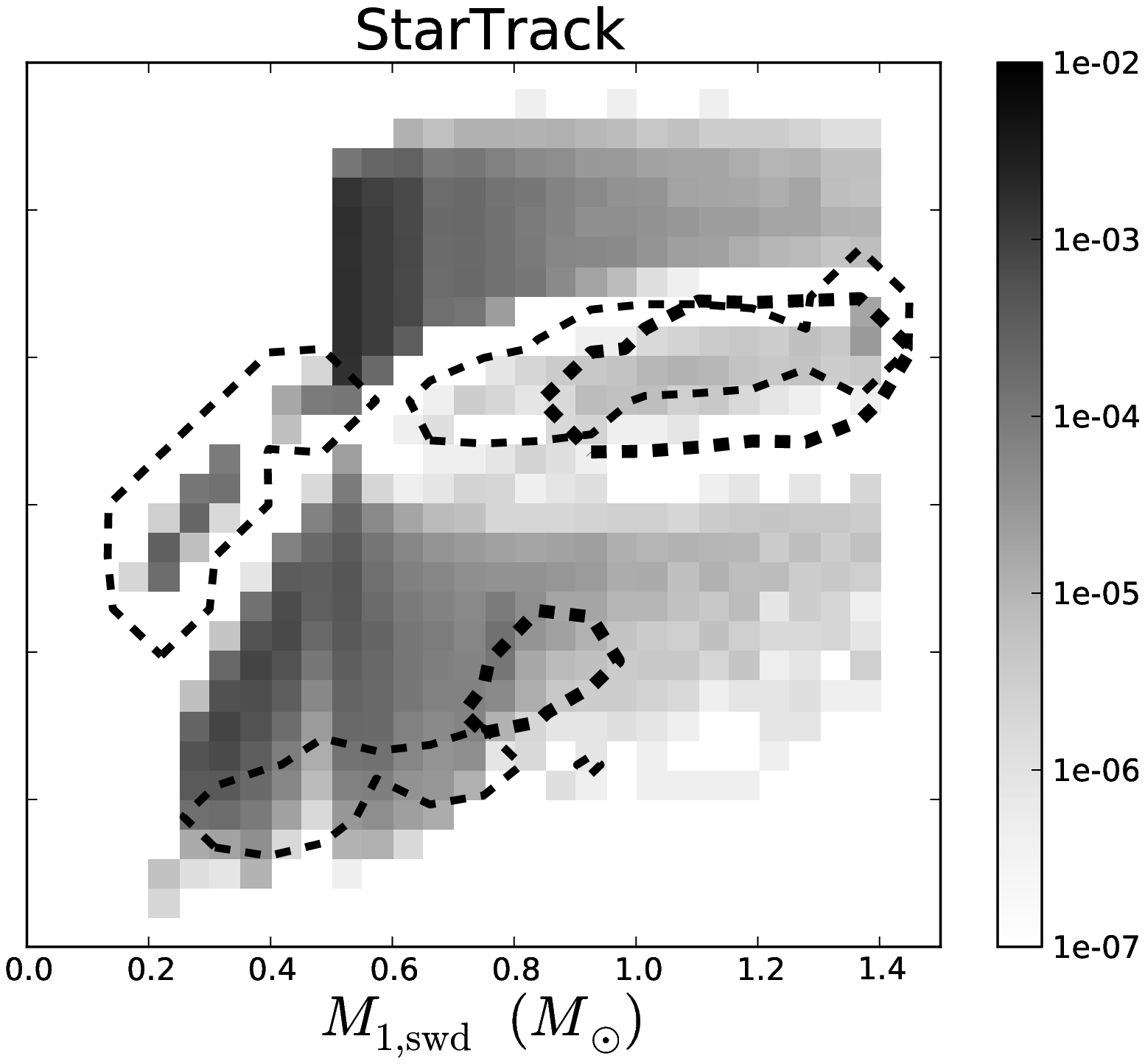} \\
	\end{tabular}
    \caption{Orbital separation versus WD mass for all SWDs in the full mass range at the time of SWD formation. The contours represent the SWD population from a specific channel: channel~3a (thin line) and channel~3b (thick line).} 
    \label{fig:swd_final_a_R3}
    \end{figure*}

\begin{figure*}
    \centering
    \setlength\tabcolsep{0pt}
    \begin{tabular}{ccc}
	\includegraphics[height=4.6cm, clip=true, trim =8mm 0mm 48.5mm 5mm]{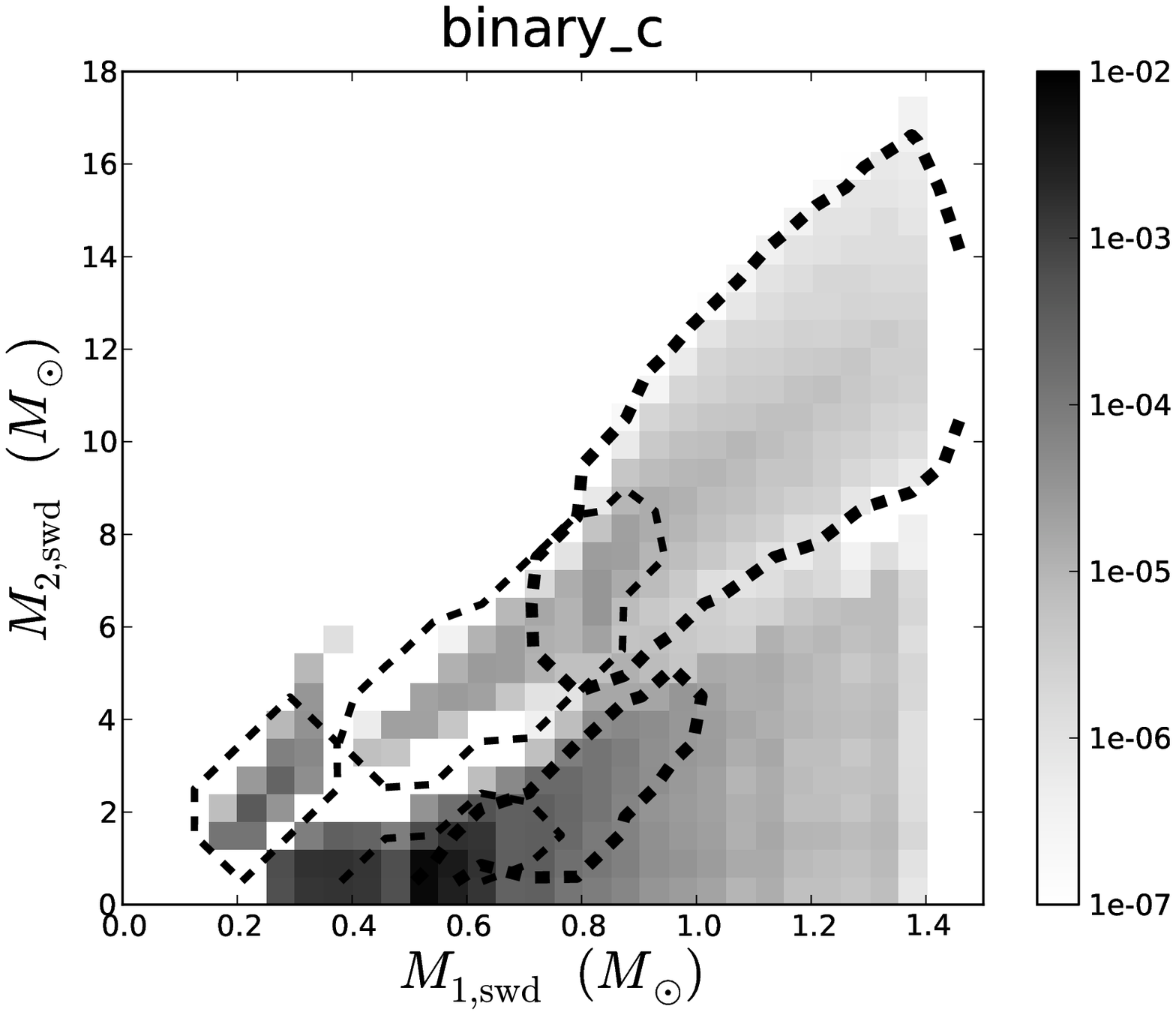} &
	\includegraphics[height=4.6cm, clip=true, trim =20mm 0mm 48.5mm 5mm]{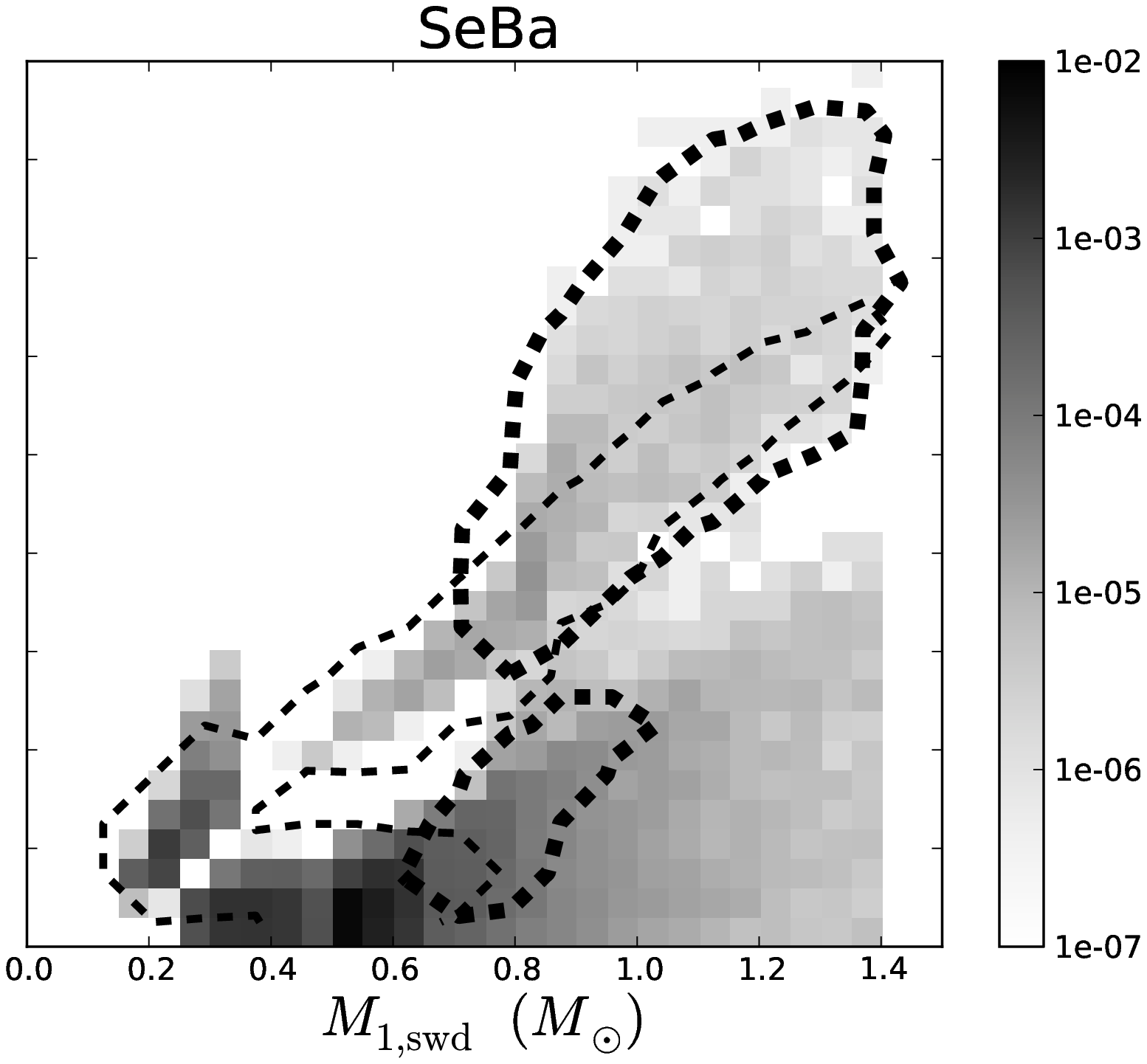} &
	\includegraphics[height=4.6cm, clip=true, trim =20mm 0mm 23mm 5mm]{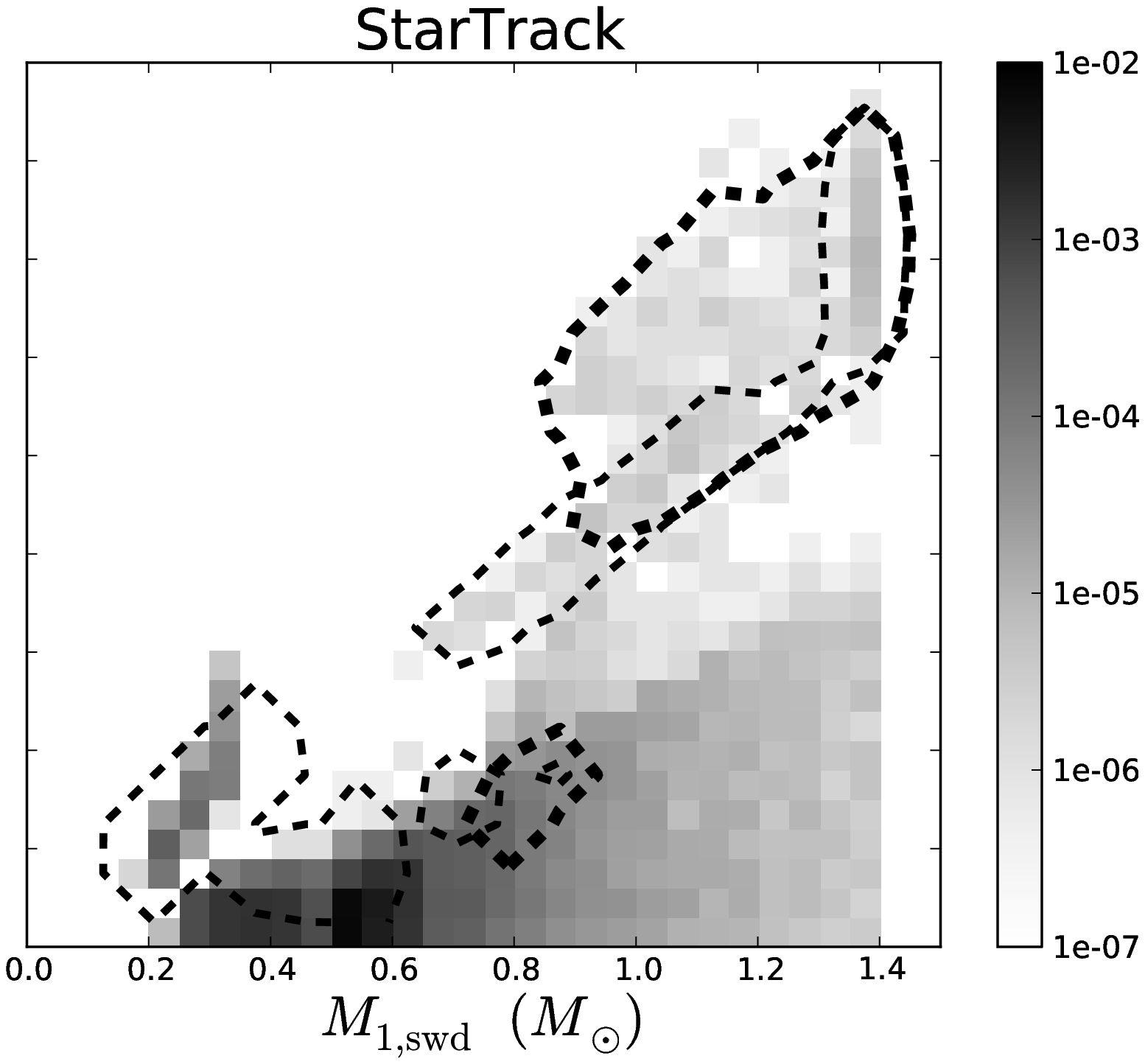} \\	
	\end{tabular}
    \caption{Secondary mass versus WD mass for all SWDs in the full mass range at the time of SWD formation. The contours represent the SWD population from a specific channel: channel~3a (thin line) and channel~3b (thick line).} 
    \label{fig:swd_final_m2_R3}
    \end{figure*}

\begin{figure*}
    \centering
    \setlength\tabcolsep{0pt}
    \begin{tabular}{cccc}
	\includegraphics[height=4.6cm, clip=true, trim =8mm 0mm 48.5mm 5mm]{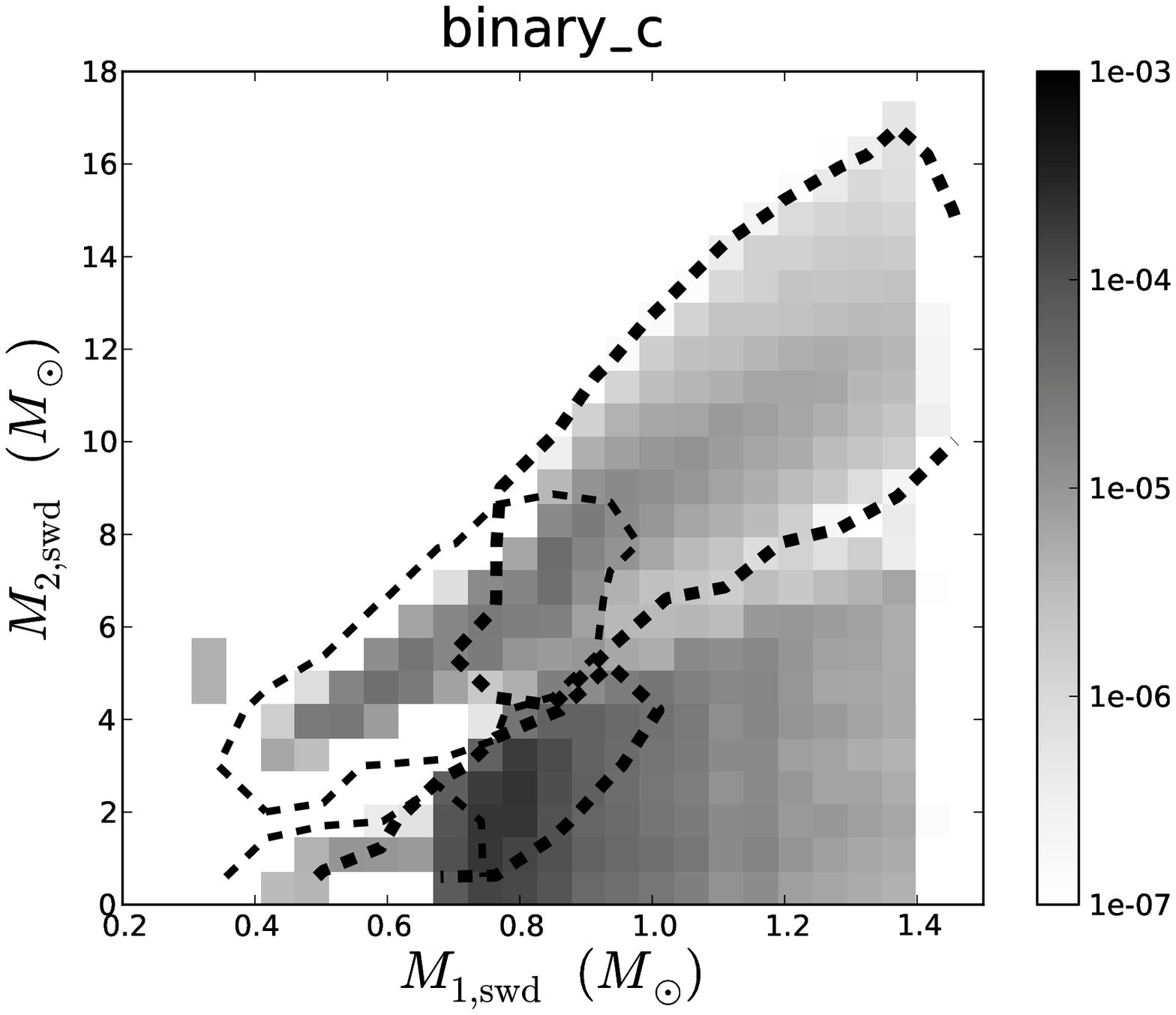} &
	\includegraphics[height=4.6cm, clip=true, trim =20mm 0mm 48.5mm 5mm]{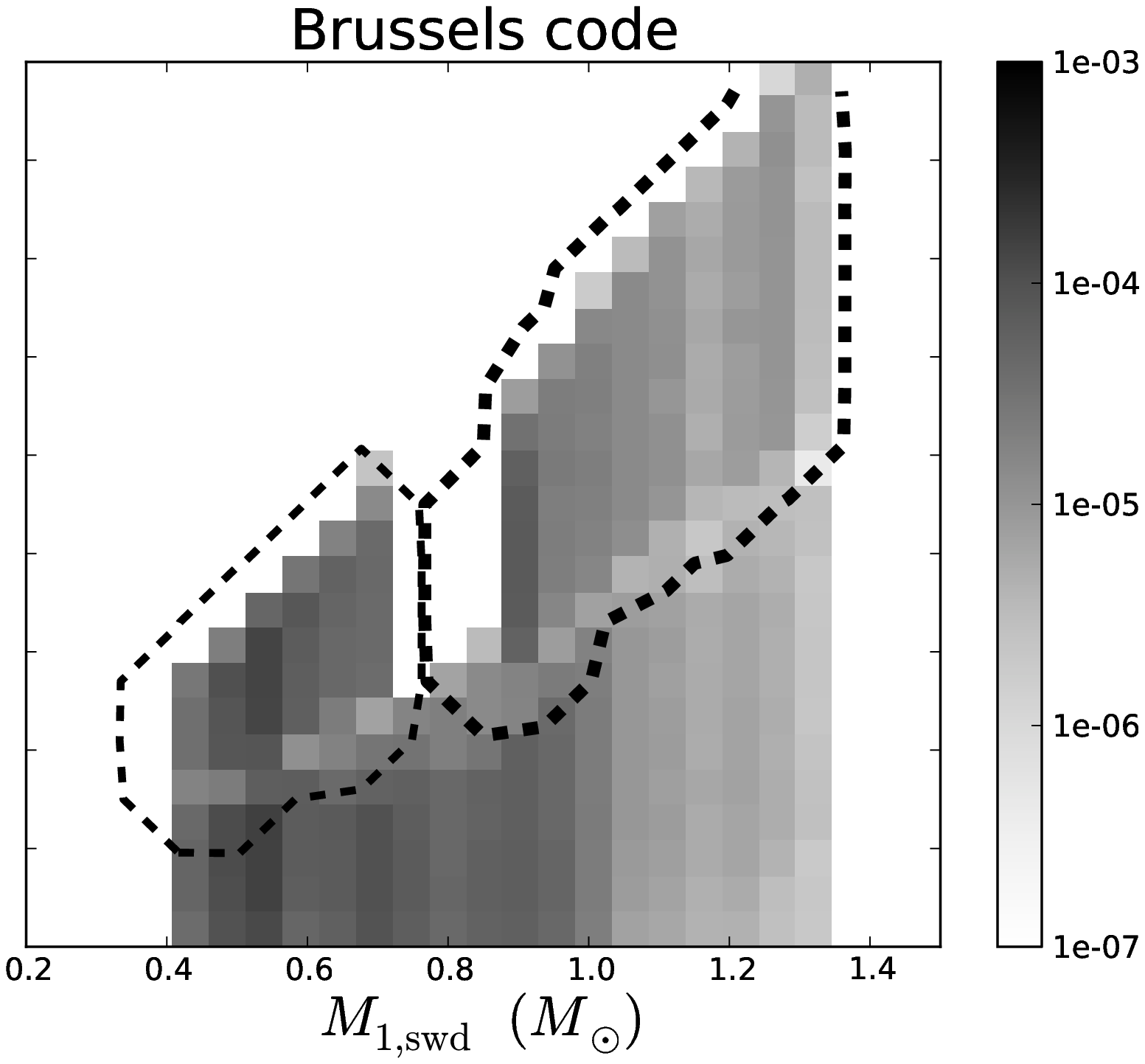} &
	\includegraphics[height=4.6cm, clip=true, trim =20mm 0mm 48.5mm 5mm]{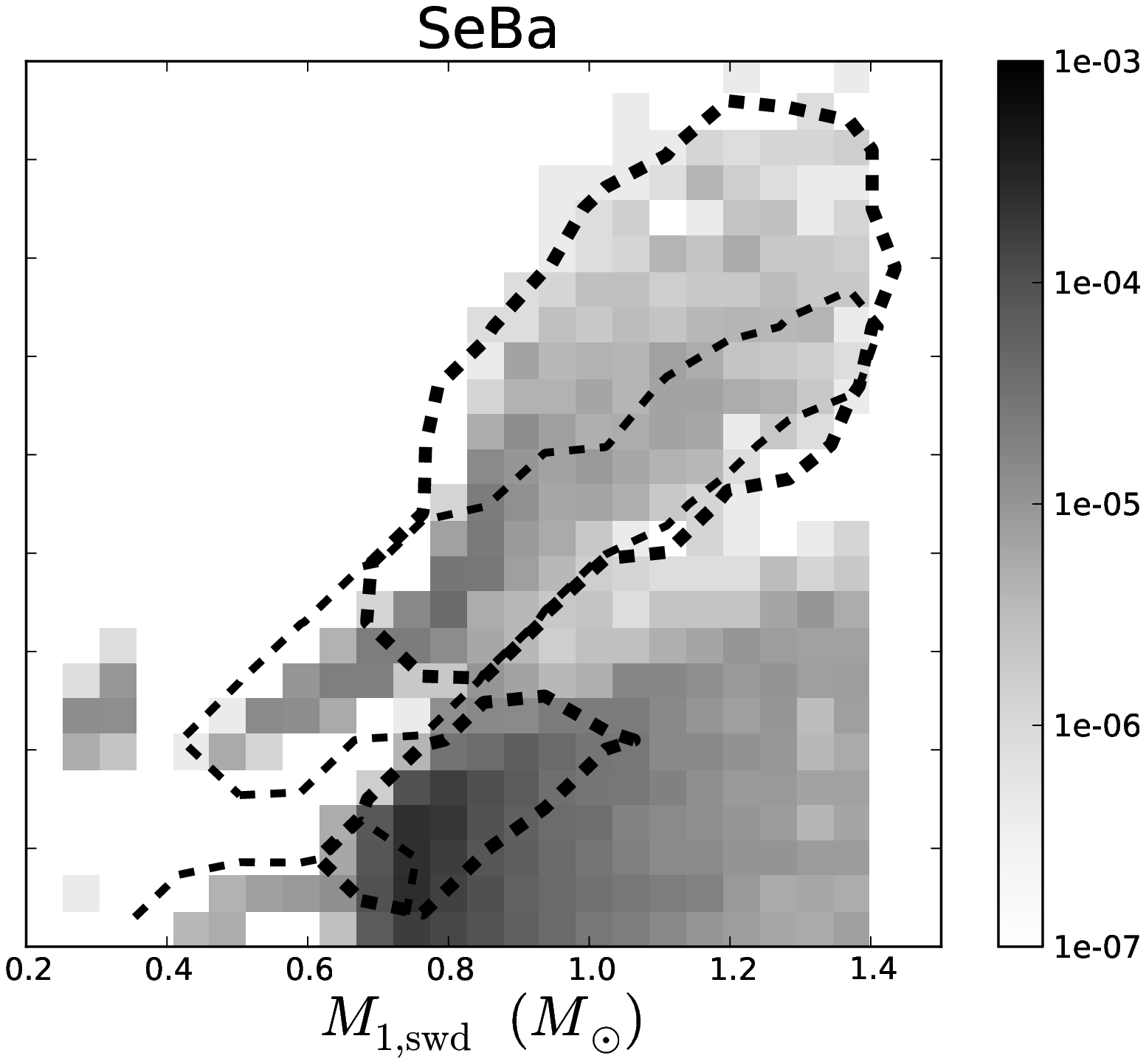} &
	\includegraphics[height=4.6cm, clip=true, trim =20mm 0mm 23mm 5mm]{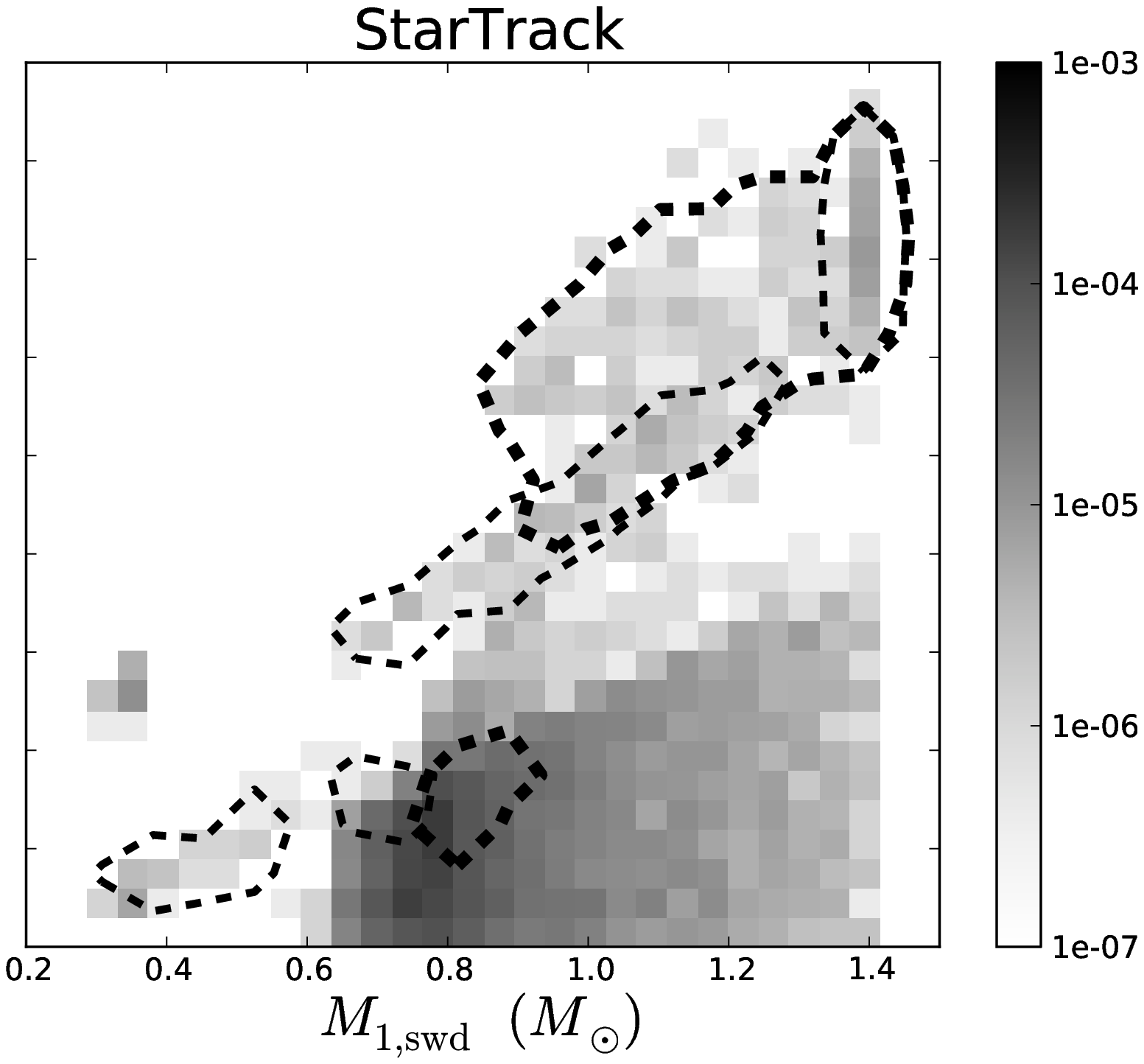} \\
	\end{tabular}
    \caption{Secondary mass versus WD mass for all SWDs in the intermediate mass range at the time of SWD formation. The contours represent the SWD population from a specific channel: channel~3a (thin line) and channel~3b (thick line).} 
    \label{fig:swd_final_m2_R3_IM}
    \end{figure*}

A crucially important assumption for the evolutionary outcome of channel~3 are the adopted stability criteria. Despite the importance of the stability criteria, the various implementations have not been compared until this study. 
We find a clear disagreement between the codes; stable mass transfer is possible in systems with mass ratios $q_{\rm zams}\gtrsim 0.6$ according to StarTrack, in SeBa $q_{\rm zams}>0.35$, in binary\_c $q_{\rm zams}>0.25$ and $q_{\rm zams}>0.2$ in the Brussels code. The effect of the relative large critical mass ratio for StarTrack results in a low birthrate in particular in the intermediate mass range (Table\,\ref{tbl:birthrates_all} and Fig.\,\ref{fig:swd_zams_a_R3_IM}), which results in a lack of SWD systems lower than 300\Rsolar~(Fig.\,\ref{fig:swd_final_a_R3}) ~and fewer SWD systems with $M_{\rm 2, swd}\lesssim 1.0$\Msolar~(Fig.\,\ref{fig:swd_final_m2_R3}). The effect of the relative low critical mass ratio for the Brussels code can be seen in Fig.\,\ref{fig:swd_final_m2_R3_IM} as an extension in the Brussels code to lower separations $a_{\rm swd} \lesssim 50$. Systems with lower mass companions initially, go through mass ratio reversal and subsequent expansion of the orbit later in the mass transfer phase.

In the low mass range we find that the stability criteria vary most strongly for donor stars that are early on the first giant branch when they have shallow convective zones in the upper layers of the envelope. In general, stable mass transfer from this type of donor for the same conditions is more readily realised in StarTrack than in binary\_c, and it is even more readily realised in SeBa. Systems with this kind of donor show in Fig.\,\ref{fig:swd_zams_a_R3} at $M_{\rm 1, zams}<3$\Msolar~a larger range in initial separations for SeBa (5-25\Rsolar) than for binary\_c and StarTrack (5-18\Rsolar).
There is also an extra population of SWD systems in the SeBa and StarTrack data with high initial separation $a_{\rm zams}\approx (1.3-4.6)\cdot 10^{2}$\Rsolar~and high initial mass ratio $q_{\rm zams} \approx M_{\rm 2, zams}/ M_{\rm 1, zams}>0.8$. In these systems the primary fills its Roche lobe stably on the giant branch after the mass ratio has reversed due to wind mass losses.
When donors with shallow convective zones are excluded, the birthrate in the full mass range in channel~3a decreases to $1.4\cdot 10^{-3}\peryr$ for SeBa and $7.3\cdot 10^{-4}\peryr$ for StarTrack, which is comparable to the birthrate predicted by binary\_c (Table\,\ref{tbl:birthrates_all}).

The long-term behaviour of the orbit can be effected by tides. If energy is dissipated, angular momentum can be exchanged between the orbit and the spin of the stars. For this project the binary\_c, Brussels and SeBa code assume that the spin angular momentum of the stars can be neglected compared to the orbital angular momentum\footnote{In binary\_c it is possible to take into account spin angular momentum into the total angular momentum of the system.}. As such in their simulations orbital angular momentum is conserved. In the StarTrack code, the orbital and spin angular momentum combined are conserved, under the assumption that the stars are and remain in a synchronized orbit. As a consequence after the first phase of mass transfer in channel~3, the orbits are slightly larger in StarTrack compared to those of the other codes (see the example system of channel~3a~and~3b).

Whether or not a primary fills its Roche lobe for a second time is modelled different in the Brussels code than in the other three codes. 
In the Brussels code stars with an initial mass less than 5\Msolar~are assumed to evolve through channel~3a, and stars with a higher mass evolve through channel~3b. The binary\_c and SeBa simulations roughly agree with this, see Fig.\,\ref{fig:swd_zams_a_R3}. However, the boundary between
channel~3a~and~3b is determined at run time in binary\_c, SeBa, and StarTrack. 
It is dependent on the evolution of the radii and wind mass loss of helium stars, the stability criterion and the separation after the first phase of mass transfer. Therefore differences exist between these codes in the upper limits for ZAMS masses and separations in channel~3a in Fig.\,\ref{fig:swd_zams_a_R3} and at WD formation in Fig.\,\ref{fig:swd_final_m2_R3}.
Binary\_c, SeBa, and StarTrack also find that systems
that evolve through channel~3a~or~3b overlap in WD mass at WD
formation (Fig.\,\ref{fig:swd_final_a_R3_IM}~and~\ref{fig:swd_final_m2_R3_IM}). 
In the data from the Brussels code, the boundary at WD formation is discontinuous in primary mass causing a gap between 0.7 and 0.9 \Msolar~(Fig.\,\ref{fig:swd_final_a_R3_IM}~and~\ref{fig:swd_final_m2_R3_IM}). The gap in WD
mass in the data from the Brussels code originates as a considerable amount of mass
is lost during the planetary nebula phase of a star that does not
initiate a second mass transfer phase. In the other three codes, the mass loss in winds from helium stars is less strong compared to the mass loss in the planetary nebula phase of helium stars in the Brussels code.

The evolution of helium stars (their radii, core masses, wind mass
losses, and if they fill their Roche lobes also the stability and mass
transfer rates) are important in channel~3. A difference arises
between the Brussels code and the others, because of the way helium
stars are simulated. In binary\_c, SeBa, and StarTrack it is possible
that after the first phase of mass transfer, the secondary fills its
Roche lobe before the primary moves off the He-MS and becomes a white
dwarf. Subsequently the primary becomes a WD before the secondary
evolves significantly\footnote{Note that it is also possible that the secondary becomes a WD before the primary does \citep{Too12, Cla13}. Because of the evolutionary reversal, these systems are not shown in Fig.\,\ref{fig:swd_final_a_R3}~to~\ref{fig:swd_zams_a_R3_IM} nor included in channel~3. The birthrates, however, are low ($[1.4, -, 5.6, 0.7]\cdot 10^{-4}\peryr$ in the full mass range). }. 
This reversal can occur because the
evolutionary timescale of a low-mass helium -star is very long (about
$10^8$yr), while that of the secondary that gained much mass is
reduced. As a result, when the first WD is formed, the mass of the
secondary and the orbital separation has decreased substantially. 
These systems lie according to binary\_c, SeBa and StarTrack at separations $\lesssim 20$\Rsolar~, primary WD masses of $\lesssim 1.0$\Msolar~and secondary masses of $\lesssim 4.5$\Msolar~in Fig.\,\ref{fig:swd_final_a_R3}~and~\ref{fig:swd_final_m2_R3}.
The birthrates of the systems in binary\_c, SeBa and StarTrack, are low ($[1.1, -, 8.6, 0.4]\cdot 10^{-4}\peryr$ in the full mass range). 
In the Brussels code, the evolution of the stars is not followed in time,
and this evolutionary track is not considered. As a result in the range of 0.45-0.7\Msolar~for the WD mass, the range in secondary masses is broader in the Brussels code.  

\subsubsection{Channel 4: unstable case B}
\label{sec:channel4}
\begin{figure}
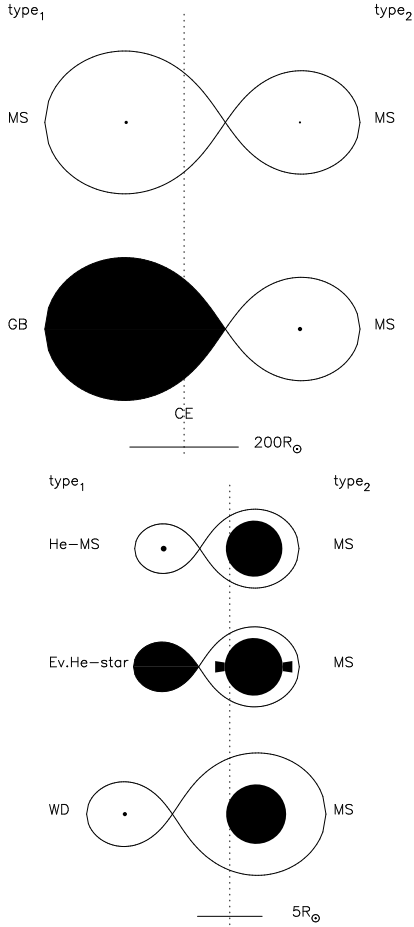

\centering
\begin{tabular}{c}
\includegraphics[width = 6cm,angle=270]{CaseBcB_A.ps} \\
\includegraphics[width = 6cm,angle=270]{CaseBcB_B.ps} \\
\end{tabular}
\caption{Example of the evolution of a SWD system in channel~4b. 
The primary fills its Roche lobe a second time. The top and bottom parts of the figure have different scales due to a CE-phase, denoted as CE in the figure. Abbreviations are as in Table\,\ref{tbl:star_type}.}
\label{fig:rl_BcB}
\end{figure}

\emph{Evolutionary path} In this path, a hydrogen shell burning star fills its Roche lobe \citep[][case Bc]{Kip67}, but the mass transfer is unstable. After the CE-phase the primary becomes a helium WD or a He-MS. Again, we differentiate two evolutionary paths within a channel. In channel~4a, the primary becomes a WD directly or the primary becomes a helium star that will evolve into a WD without any further interaction with the secondary. If the primary star fills its Roche lobe for a second time, the system evolves through subchannel~4b. An example of the evolutionary path of channel~4b in shown in Fig.\,\ref{fig:rl_BcB}. 

\emph{Example} Figure\,\ref{fig:rl_BcB} shows the evolution of a system
of channel~4b, that starts its evolution with $M_{\rm 1, zams}=6\Mo$, $M_{\rm 1, zams}=3\Mo$
and $a_{\rm zams}=320\Ro$. The primary fills its Roche lobe as it ascends the
first giant branch. After mass transfer ceases the primary has become
a He-MS of mass $[1.1, 1.1, 1.1, 1.1]$\Msolar~in an orbit with a separation of $[7.0, 7.1, 7.0, 7.0]$\Rsolar. As the helium star evolves and increases in radius, it initiates the second phase of mass transfer. 
Soon after mass transfer ceases, the primary becomes a WD with 
$M_{\rm 1, swd} = [0.81, 0.91, 0.77, 0.79]$\Msolar. The secondary is still on the MS with $M_{\rm 2, swd} = [3.2,
3.2, 3.3, 3.3]$\Msolar~and the orbital separation is $a_{\rm swd} = [10, 9.4, 11,
11]$\Rsolar. The differences in this example are caused by effects
discussed before; the MiMwd-relation including the mass transfer rates from helium rich donors. 

\emph{Population} The codes agree well on the location of the
SWDs at WD formation from channel~4a~and~4b in
Fig.\,\ref{fig:swd_final_a_R1}, \,\ref{fig:swd_final_a_R1_IM},\,\ref{fig:swd_final_m2_R1}~and~\ref{fig:swd_final_m2_R1_IM}, their
progenitor systems in Fig.\,\ref{fig:swd_zams_a_R1}~and~\ref{fig:swd_zams_a_R1_IM} and the birthrates of the channels (Table\ref{tbl:birthrates_all}). For channel~4a,
which predominantly contains low mass binaries, there is an excellent
agreement between binary\_c, SeBa and StarTrack in the previously mentioned figures as well as in the birthrates (Table\,\ref{tbl:birthrates_all}). The low mass SWDs at WD formation have WDs of $0.25-0.48$\Msolar, companions of $<1.8$\Msolar, in an orbit of $0.5-100$\Rsolar, and progenitor systems with $a_{zams}\approx(0.3-4.0)\cdot 10^2$\Rsolar~for $M_{\rm 1, zams}\approx 1-2$\Msolar. 

The population of systems that evolve through channel~4b are primarily intermediate mass binaries of mass $M_{\rm 1,zams} \approx 4.5-10$\Msolar~that become WDs of $M_{\rm 1, swd}\approx 0.7-1.3$\Msolar. The majority of systems have initial separations of $a_{\rm zams}\approx(0.2-1.0)\cdot 10^3$\Rsolar. At WD formation the range of separations according to binary\_c, SeBa and StarTrack is $4-1.0\cdot 10^2$\Rsolar, however, for the Brussels code it is extended to $0.9-1.4\cdot 10^2$\Rsolar. 

\emph{Effects}
There is a difference between the Brussels code on one hand and the other three codes on the other hand, regarding the survival of systems with low initial secondary masses $M_{\rm 2, zams}<3$\Msolar~in channel~4b. This is predominantly due to the difference in the single star prescriptions for the radii of stars. The radius of low-mass secondary-stars are in general larger in binary\_c, SeBa and StarTrack than in the Brussels code. Therefore in the former three codes, the stars are more likely to fill their Roche lobe at the end of the CE-phase resulting in a merger. In the Brussels data, these systems survive at small separations ($\lesssim 10$\Rsolar~at 1\Msolar, see Fig.\,\ref{fig:swd_final_a_R1_IM}). Note that the Brussels code was written for intermediate mass stars (Sect.\,\ref{sec:BPS}), and in principal the code does not allow for the detailed evolution of stars with initial masses below 3\Msolar.
 
In addition, the stability of the second phase of mass transfer affects the SWD population of channel~4. If this phase is unstable, the system will evolve to a merger. In the Brussels code, it is assumed that the second phase of mass transfer is always stable, however, this is not the case in the three other codes. Differences in the stability criteria affect the orbital separation of SWD formation for all codes.

\subsubsection{Channel 5: case A} 
\label{sec:channel5}
\emph{Evolutionary path} In channel~5 mass transfer starts during the core hydrogen burning phase of the donor \citep[Case A,][]{Kip67}. 

\emph{Population} 
The birthrates in the full mass range differ within a factor 2.5 between binary\_c, SeBa and StarTrack (Table\,\ref{tbl:birthrates_all}). According to these codes, the progenitors of the primaries in channel~5 are stars of low mass (1-4\Msolar) in small orbits (5-13\Rsolar), see Fig.\,\ref{fig:swd_zams_a_R1}. There is a good agreement that the majority of SWDs from channel~5 at WD formation consists of a primary of mass 0.2-0.35\Msolar, a secondary of mass 1.8-5.5\Msolar~in an orbit with a separation of 30-240\Rsolar~(Fig.\,\ref{fig:swd_final_a_R1}~and~\ref{fig:swd_final_m2_R1}).  
Binary\_c, SeBa and StarTrack further agree on a subchannel ($a_{\rm swd} \approx 0.4$\Rsolar~and $M_{\rm 1, swd}\approx 0.3$\Msolar~in Fig.\,\ref{fig:swd_final_a_R1}) in which the secondary is a hydrogen-poor helium-star at WD formation (see also channel~3). The birthrates of this subchannel are low ($[4.5,-,4.8,18]\cdot 10^{-6}\peryr$).

The birthrate of channel~5 in the Brussels code is higher by over a factor 20 compared to binary\_c, SeBa and StarTrack (Table\,\ref{tbl:birthrates_all}). For the Brussels code the intermediate mass primaries have an initial mass $M_{\rm 1, zams} \approx 3-10$\Msolar~and WD mass $M_{\rm 1, swd} \approx 0.45-1.3$\Msolar, while the other codes show smaller ranges: 
for the main group of progenitors $M_{\rm 1, zams}\approx 3-4$\Msolar~and WD mass $M_{\rm 1, swd} \lesssim 0.35$\Msolar~(Fig.\,\ref{fig:swd_final_a_R1}~and~\ref{fig:swd_zams_a_R1}). The initial separation in the Brussels code $a_{\rm zams} \approx 5-22$\Rsolar, while in binary$\_$c, SeBa and StarTrack $a_{\rm zams}\approx  8-13$\Rsolar. The separation at SWD formation $a_{\rm swd}$ in the Brussels code is between 20-350\Rsolar, while in the other codes the separation is mainly between 100-250\Rsolar. 
The range of secondary masses is $M_{\rm 2, swd}\approx 3-18$\Msolar~in the Brussels code, but only $M_{\rm 2, swd}\approx 4-6$\Msolar~in the other codes. Note that the region indicated by the dash-dotted contours in Fig.\,\ref{fig:swd_final_a_R1}, contains systems from channel~5 as well as from channel~3, however, this does not change our conclusion regarding the extended range and birthrates in the Brussels code compared to the other codes.

\emph{Effects} The differences between the Brussels code and the other codes is caused by the fact that the Brussels population code does not follow the mass transfer event and its mass transfer rate in detail. It considers only the initial and final moment of the mass transfer phase, therefore any intermediate steps in which the system can be closer are disregarded. 
For example, during conservative mass transfer to an initially less massive companion, the orbital separation first decreases and then increases again after mass ratio reversal. 
As the orbital separation decreases, the secondary can fill its Roche lobe leading to a contact system, especially as it grows in mass and radius due to the accretion.  
In the binary\_c, SeBa and StarTrack code, it is assumed that the contact phase will lead to a merger or CE-phase for evolved secondaries. The Brussels code assumes that for shallow contact, the merger can be avoided. In other words, the codes have different assumptions for the stability of mass transfer.

\subsection{Double white dwarfs}
\label{sec:ev_path_dwd}
\begin{figure*}
    \centering
    \setlength\tabcolsep{0pt}
    \begin{tabular}{ccc}
	\includegraphics[height=4.6cm, clip=true, trim =8mm 0mm 48.5mm 5mm]{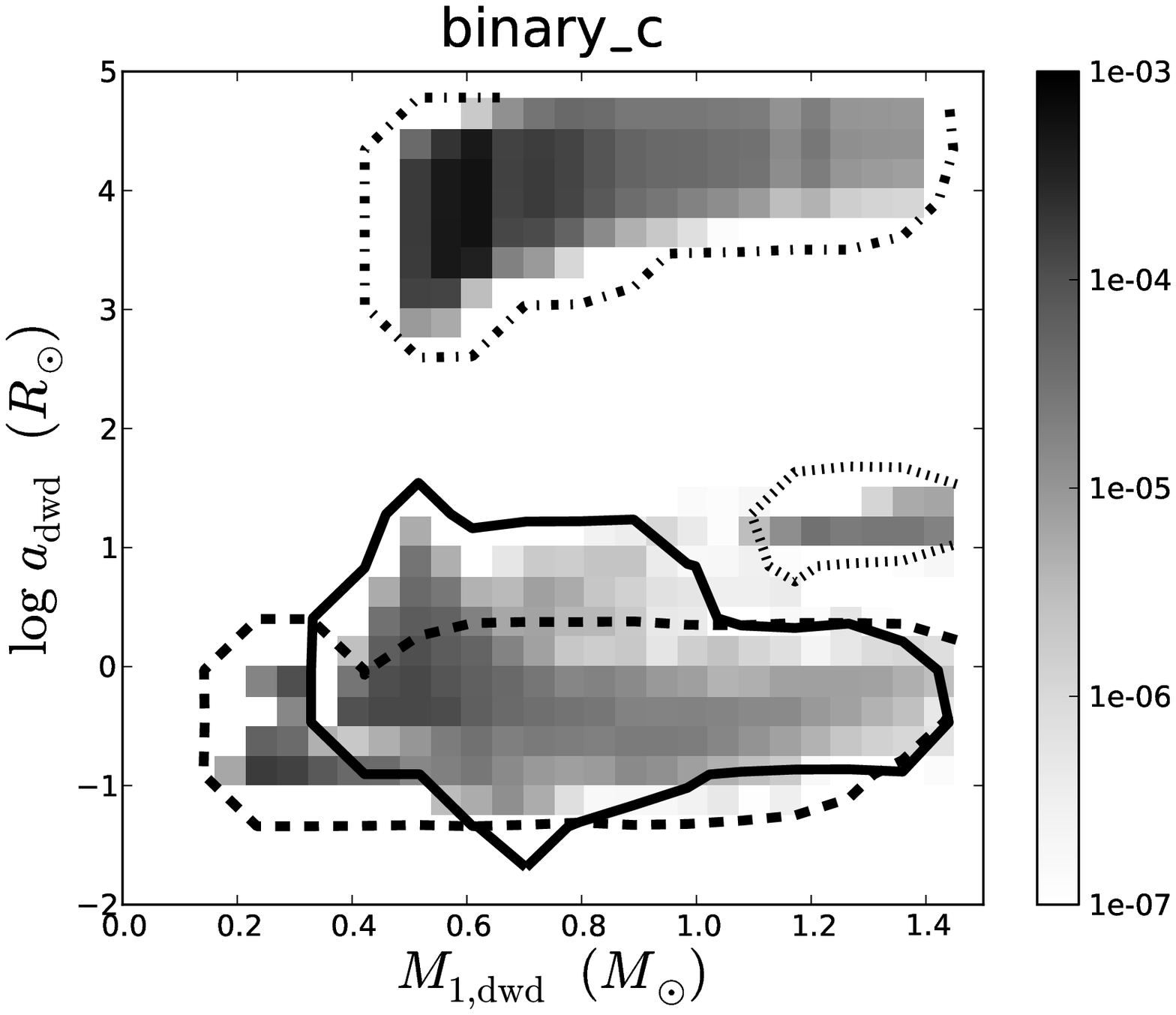} &
	\includegraphics[height=4.6cm, clip=true, trim =20mm 0mm 48.5mm 5mm]{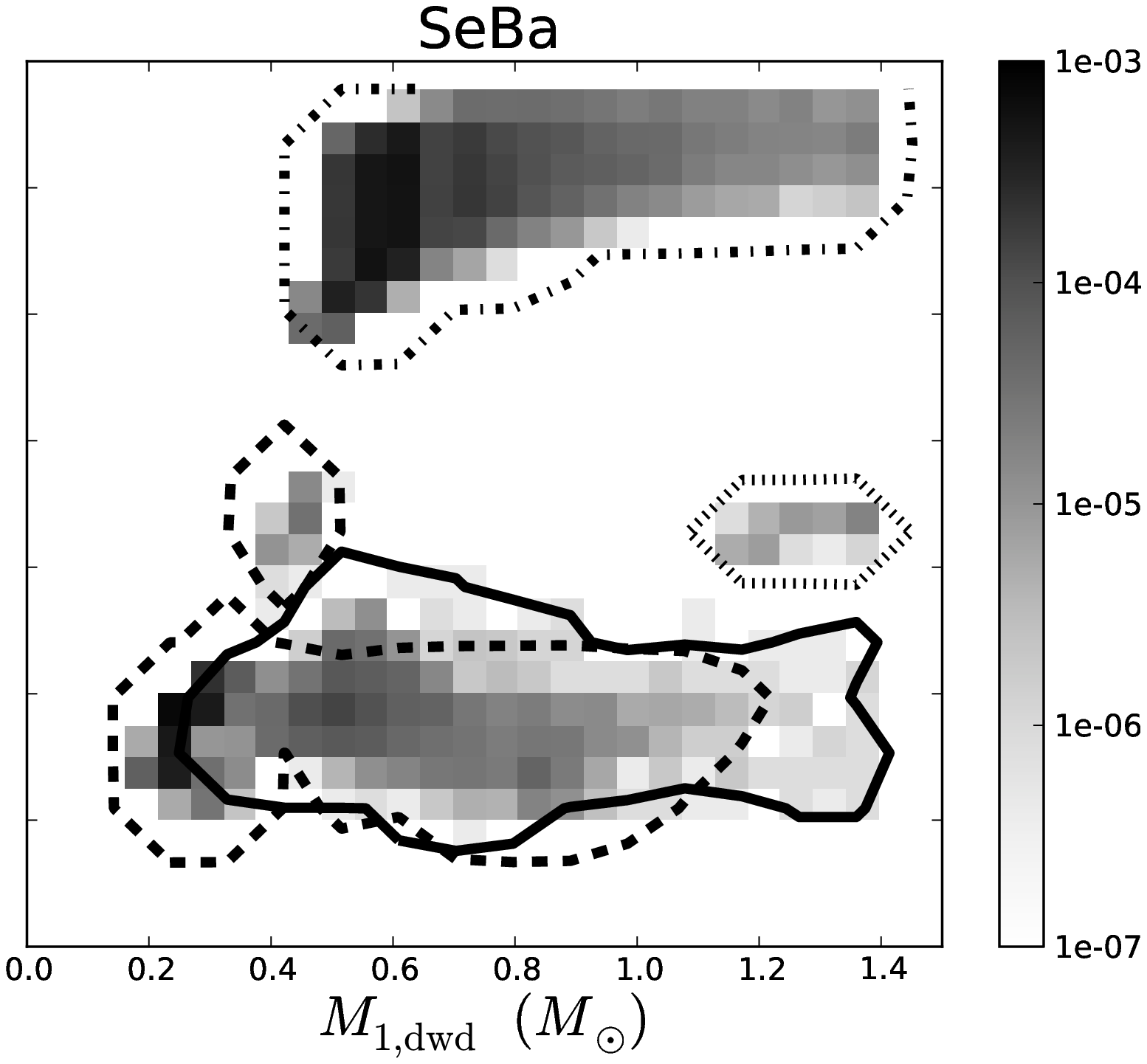}&
	\includegraphics[height=4.6cm, clip=true, trim =20mm 0mm 23mm 5mm]{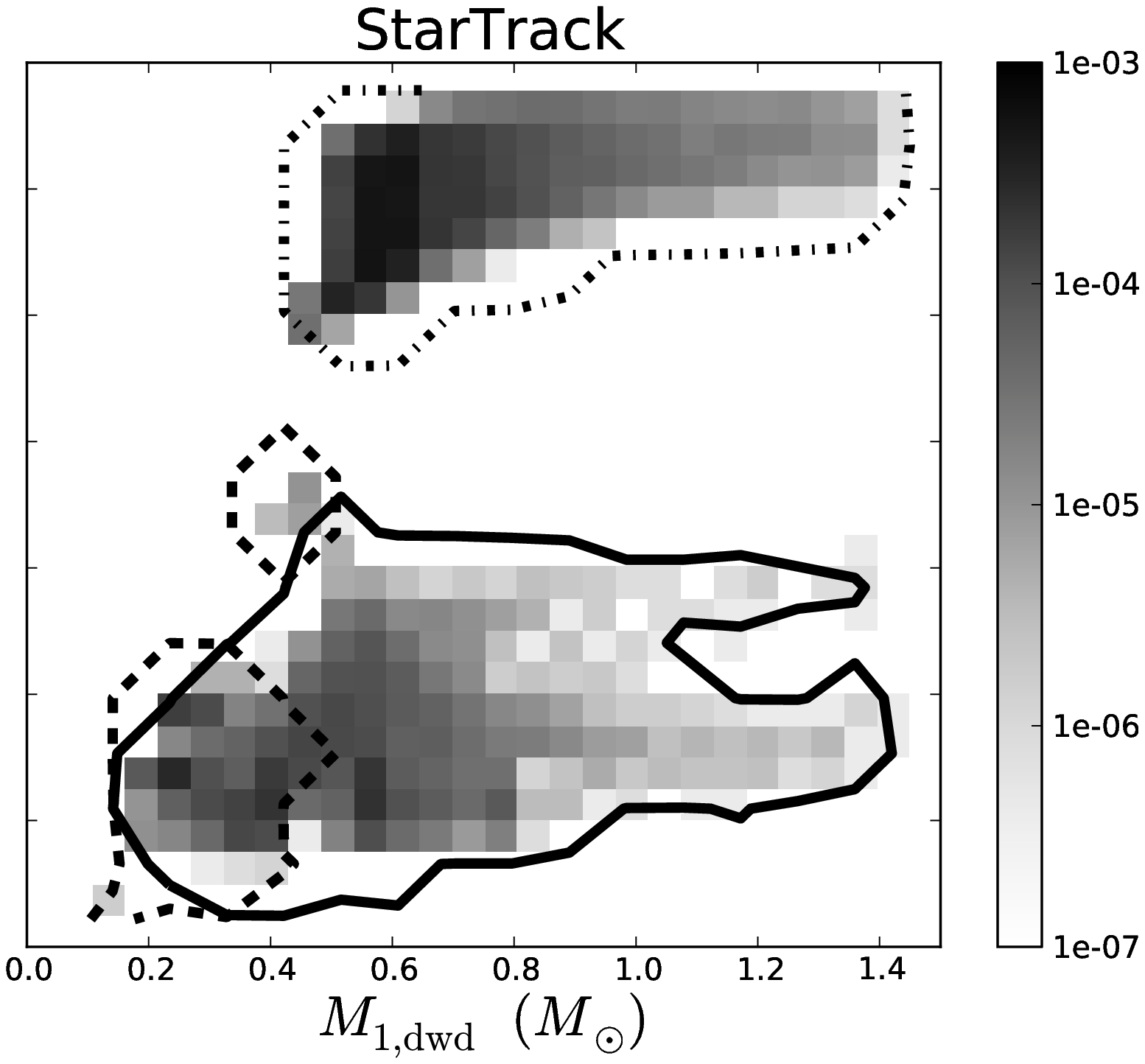} \\	
	\end{tabular}
    \caption{Orbital separation versus primary WD mass for all DWDs in the full mass range at the time of DWD formation. The contours represent the DWD population from a specific channel: channel~I (dash-dotted line), channel~II (solid line), channel~III (dashed line) and channel~IV (dotted line).}
    \label{fig:dwd_a}
    \end{figure*}

\begin{figure*}
    \centering
    \setlength\tabcolsep{0pt}
    \begin{tabular}{ccc}
	\includegraphics[height=4.6cm, clip=true, trim =8mm 0mm 48.5mm 5mm]{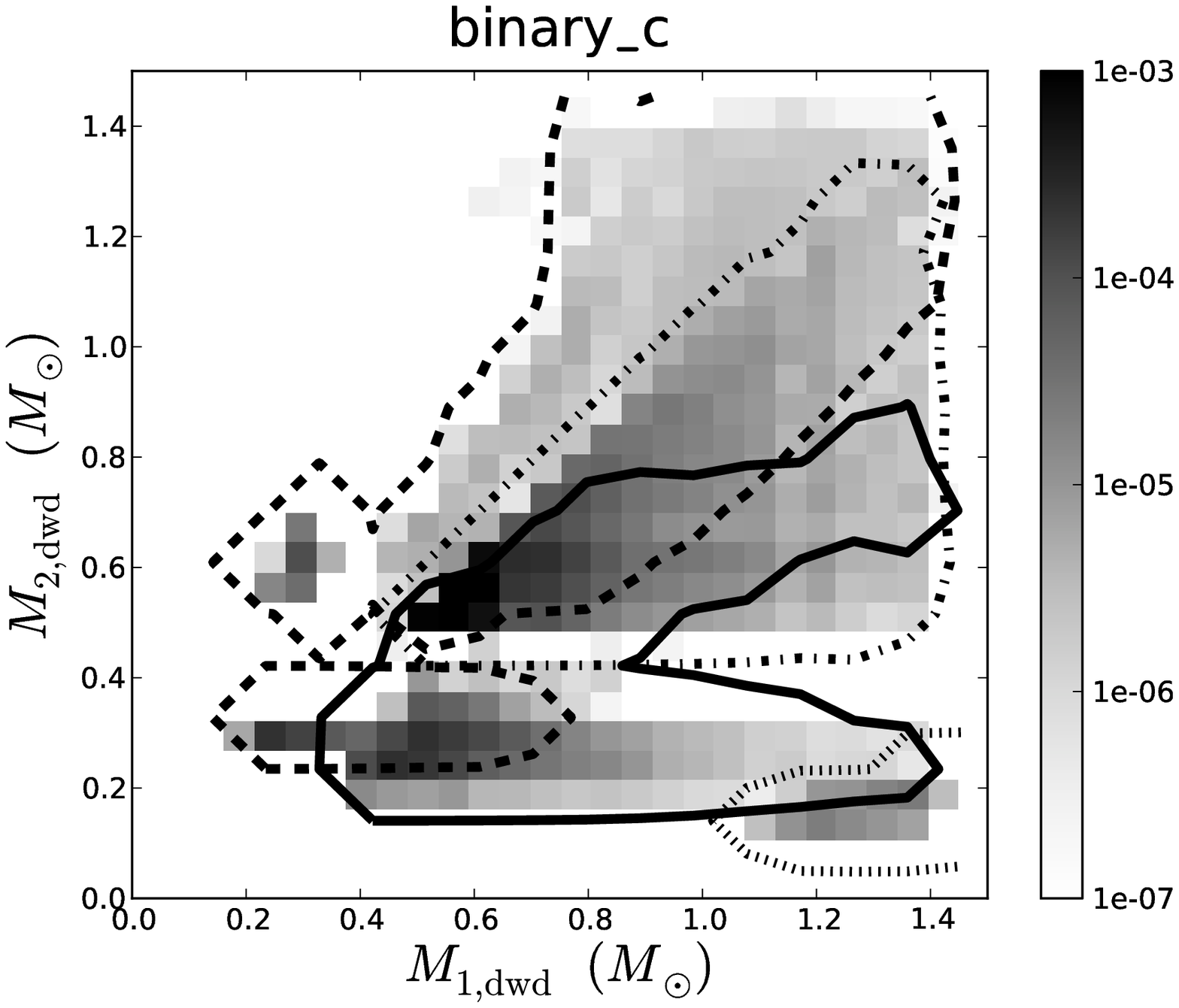} & 
	\includegraphics[height=4.6cm, clip=true, trim =20mm 0mm 48.5mm 5mm]{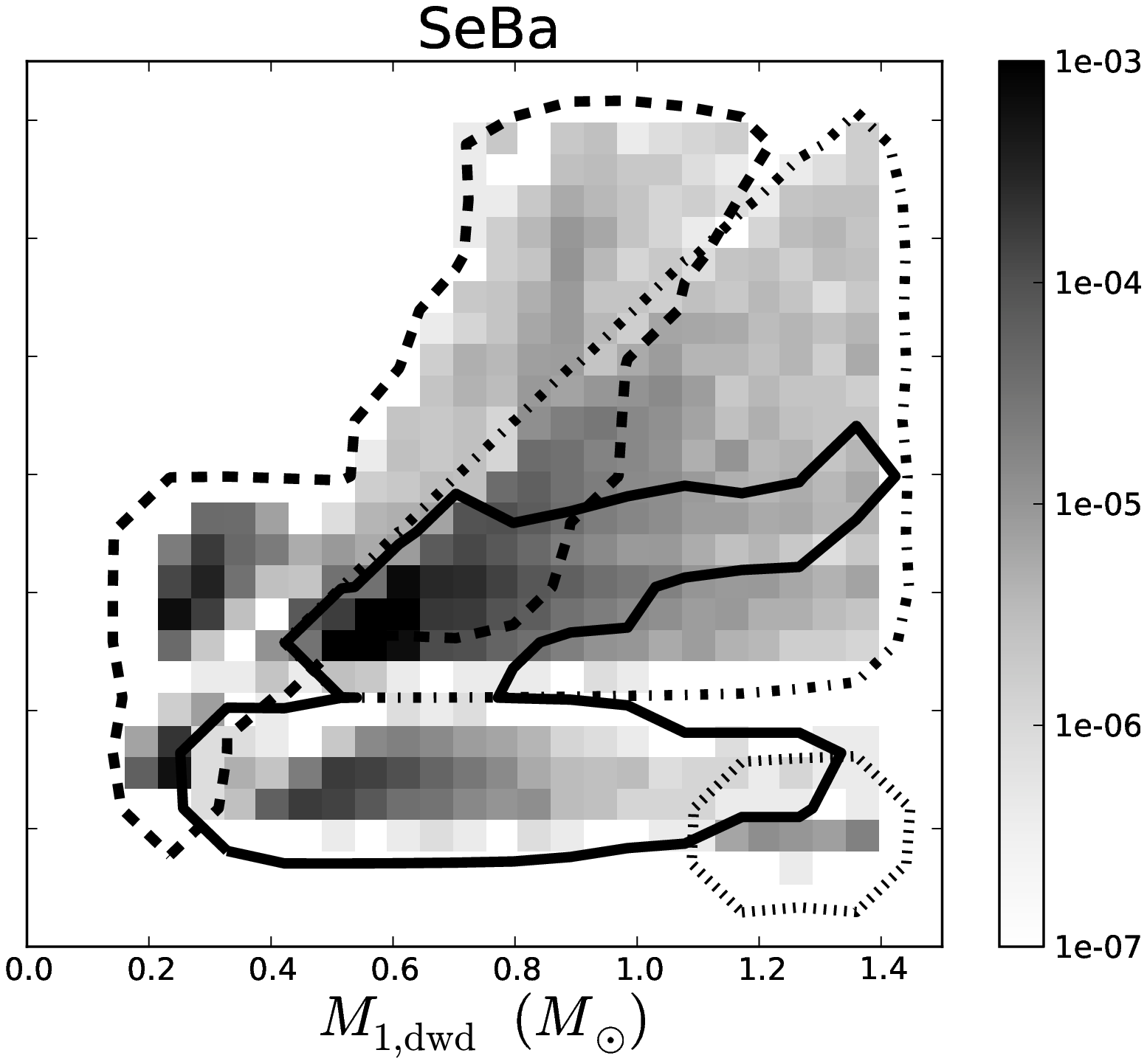} &
	\includegraphics[height=4.6cm, clip=true, trim =20mm 0mm 23mm 5mm]{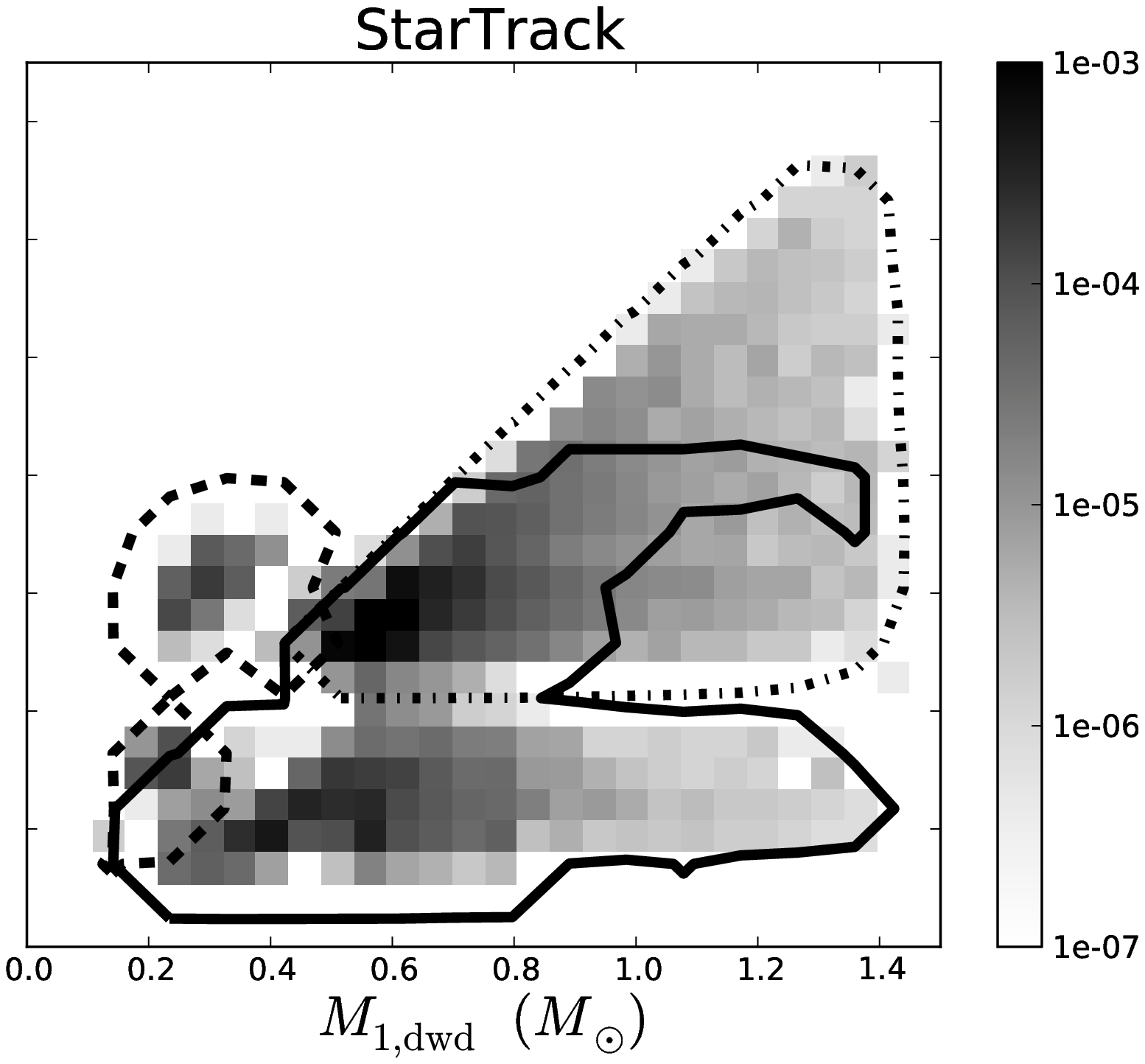} \\	
	\end{tabular}
    \caption{Secondary WD mass versus primary WD mass for all DWDs in the full mass range at the time of DWD formation. The contours represent the DWD population from a specific channel: channel~I (dash-dotted solid line), channel~II (solid line), channel~III (dashed line) and channel~IV (dotted line).} 
    \label{fig:dwd_M2}
    \end{figure*}

In the next sections, we differentiate four different evolutionary paths of DWDs. This is based on whether or not mass transfer occurs and if so, if the mass transfer initiated by the primary and secondary is stable or unstable. 
For clarity we do not distinguish the evolutionary path further e.g. by separating channel~3~and~5, nor channel~3a~and~3b. Channel~I,~II~and~III represent the most commonly followed evolutionary paths with birthrates larger than $1.0\cdot 10^{-3}\peryr$. Channel~IV is included because it stands out in Fig.\,\ref{fig:dwd_a}~and~\ref{fig:dwd_M2}, even though the birthrates in this channel are low (Table\,\ref{tbl:birthrates_all}). 
In each section we describe a specific evolutionary path (marked as \emph{Evolutionary path}), we compare the simulated populations from each code (marked as \emph{Population}) and investigate where differences between the populations come from (marked as \emph{Population} and \emph{Effects}).

    \begin{figure*}
    \centering
    \setlength\tabcolsep{0pt}
    \begin{tabular}{cccc}
	\includegraphics[height=4.6cm, clip=true, trim =8mm 0mm 48.5mm 5mm]{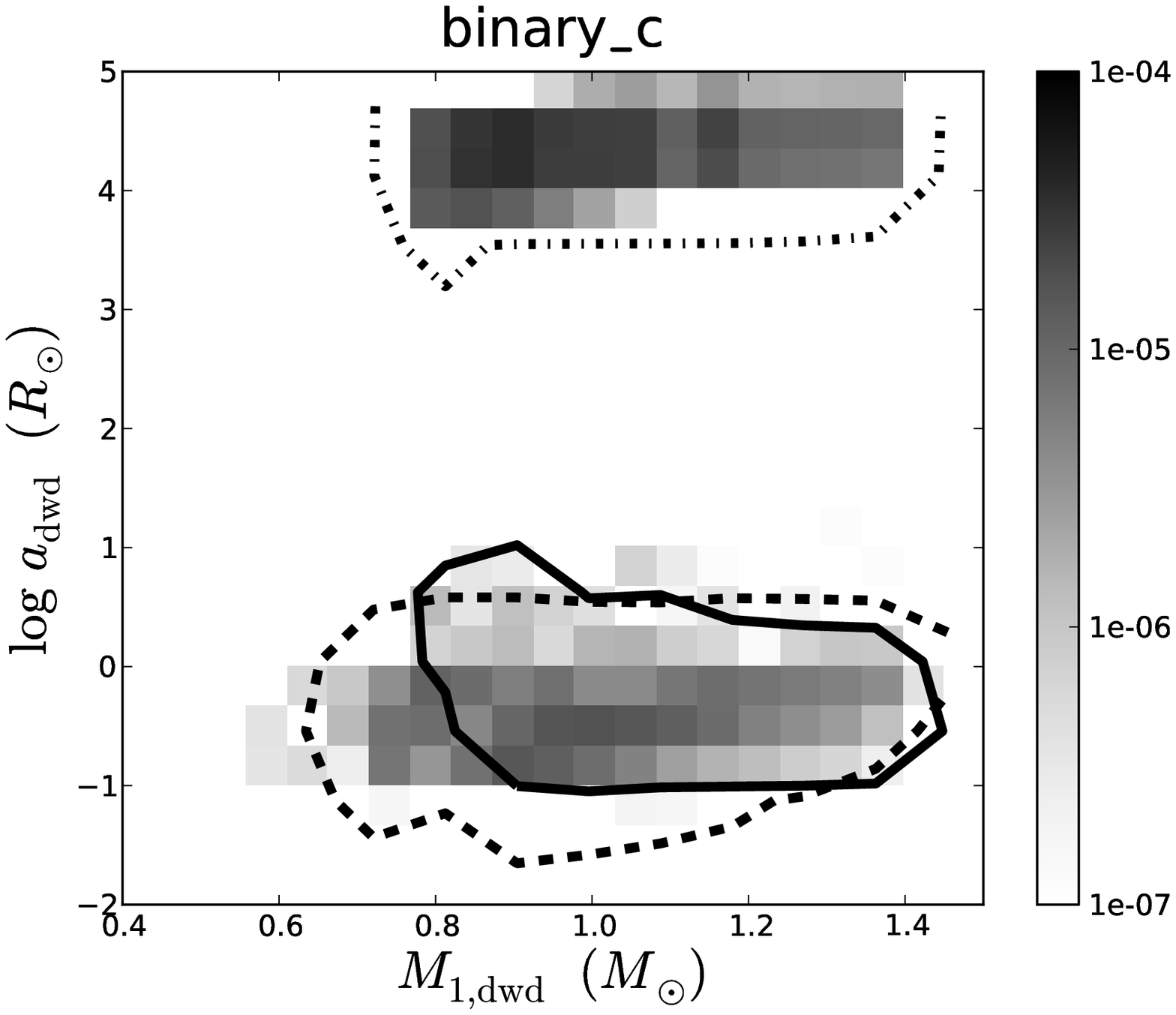} &
	\includegraphics[height=4.6cm, clip=true, trim =20mm 0mm 48.5mm 5mm]{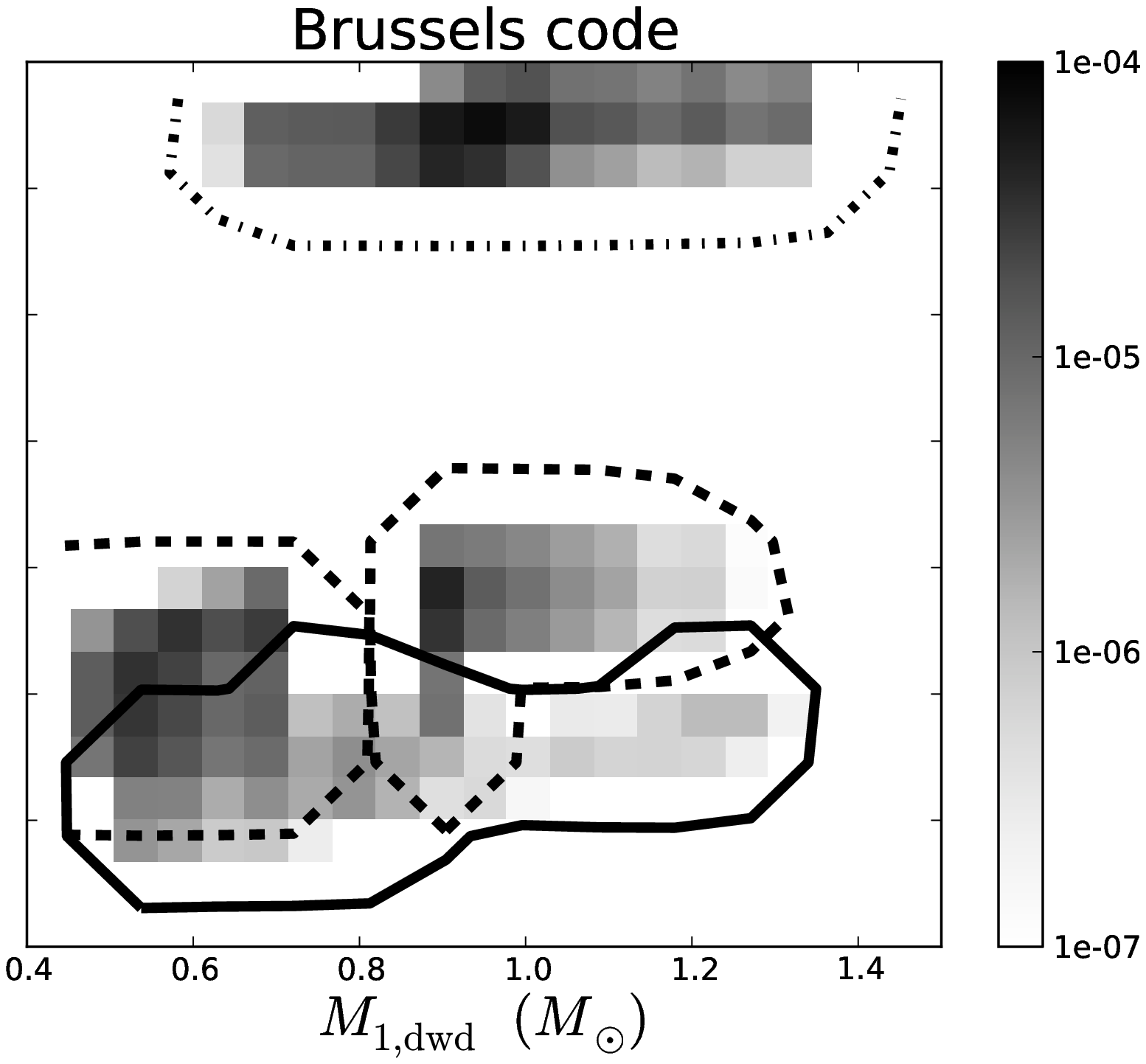} & 
	\includegraphics[height=4.6cm, clip=true, trim =20mm 0mm 48.5mm 5mm]{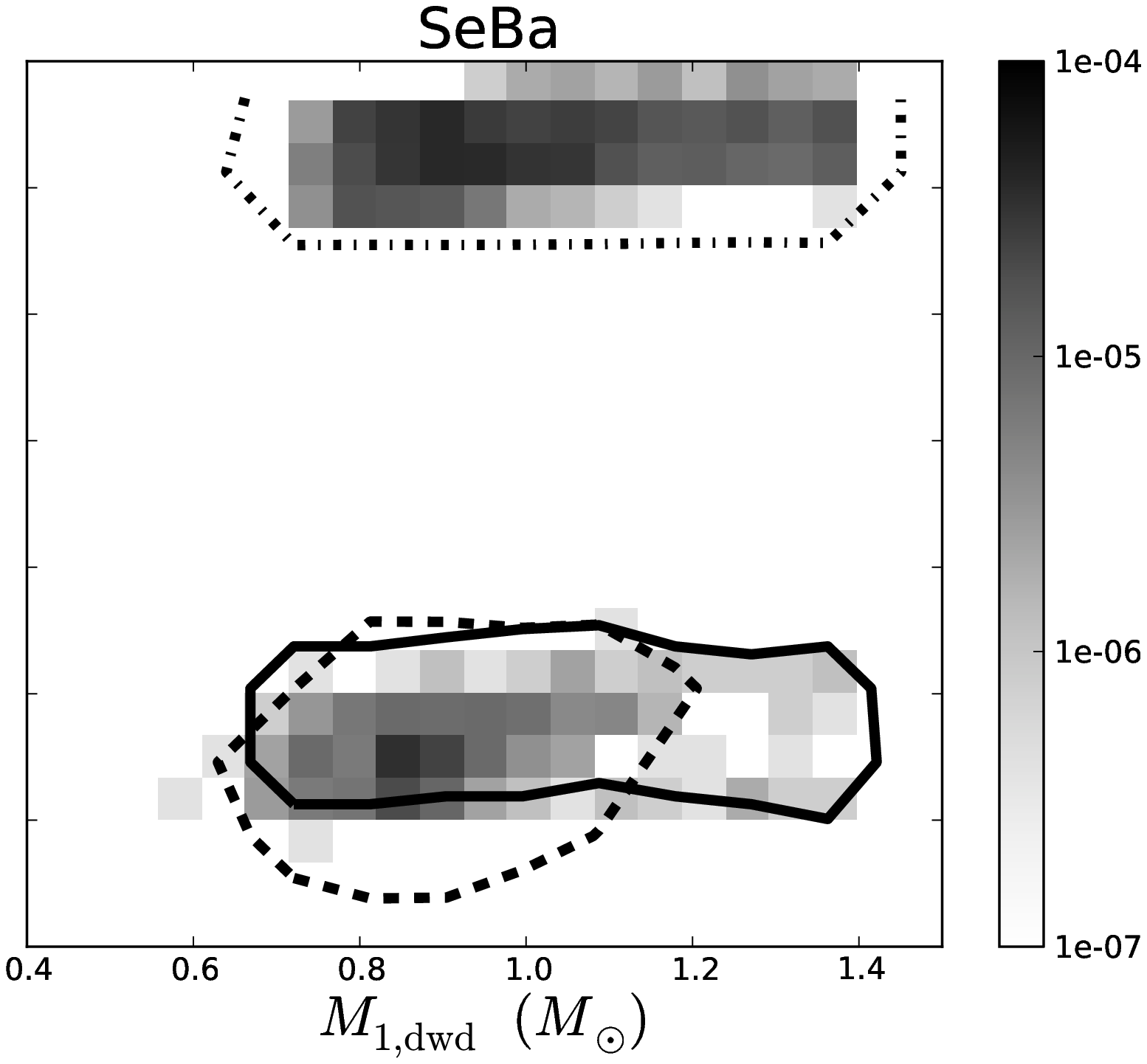} &
	\includegraphics[height=4.6cm, clip=true, trim =20mm 0mm 23mm 5mm]{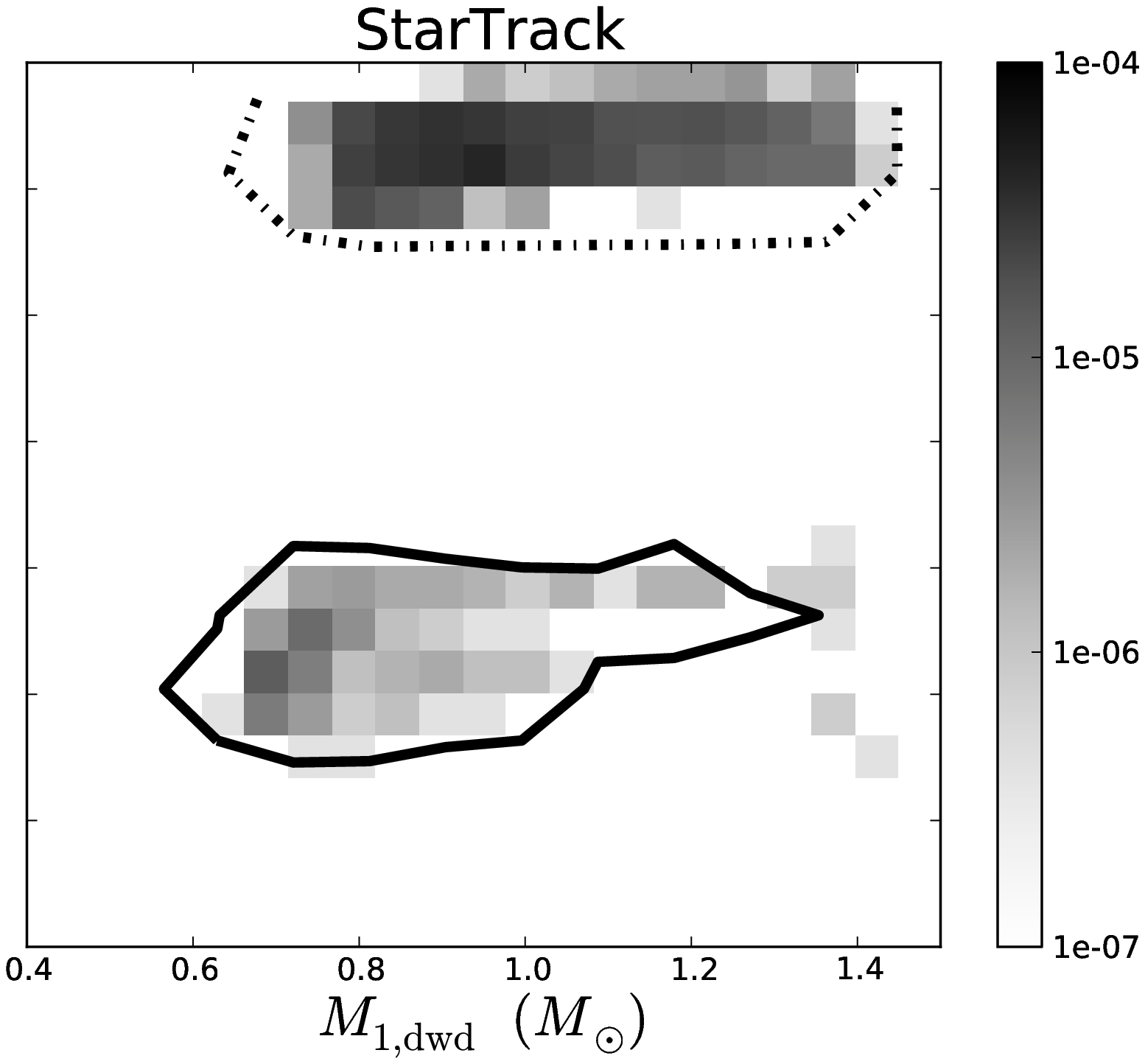} \\	
	\end{tabular}
    \caption{Orbital separation versus primary WD mass for all DWDs in the intermediate mass range at the time of DWD formation. The contours represent the DWD population from a specific channel: channel~I (dash-dotte solid line), channel~II (solid line) and channel~III (dashed line). The contours of the DWD population from channel~III according to StarTrack and channel~IV according to all codes are not shown, as the birthrate from this channel is too low. } 
    \label{fig:dwd_a_IM}
    \end{figure*}

    \begin{figure*}
    \centering
    \setlength\tabcolsep{0pt}
    \begin{tabular}{cccc}
	\includegraphics[height=4.6cm, clip=true, trim =8mm 0mm 48.5mm 5mm]{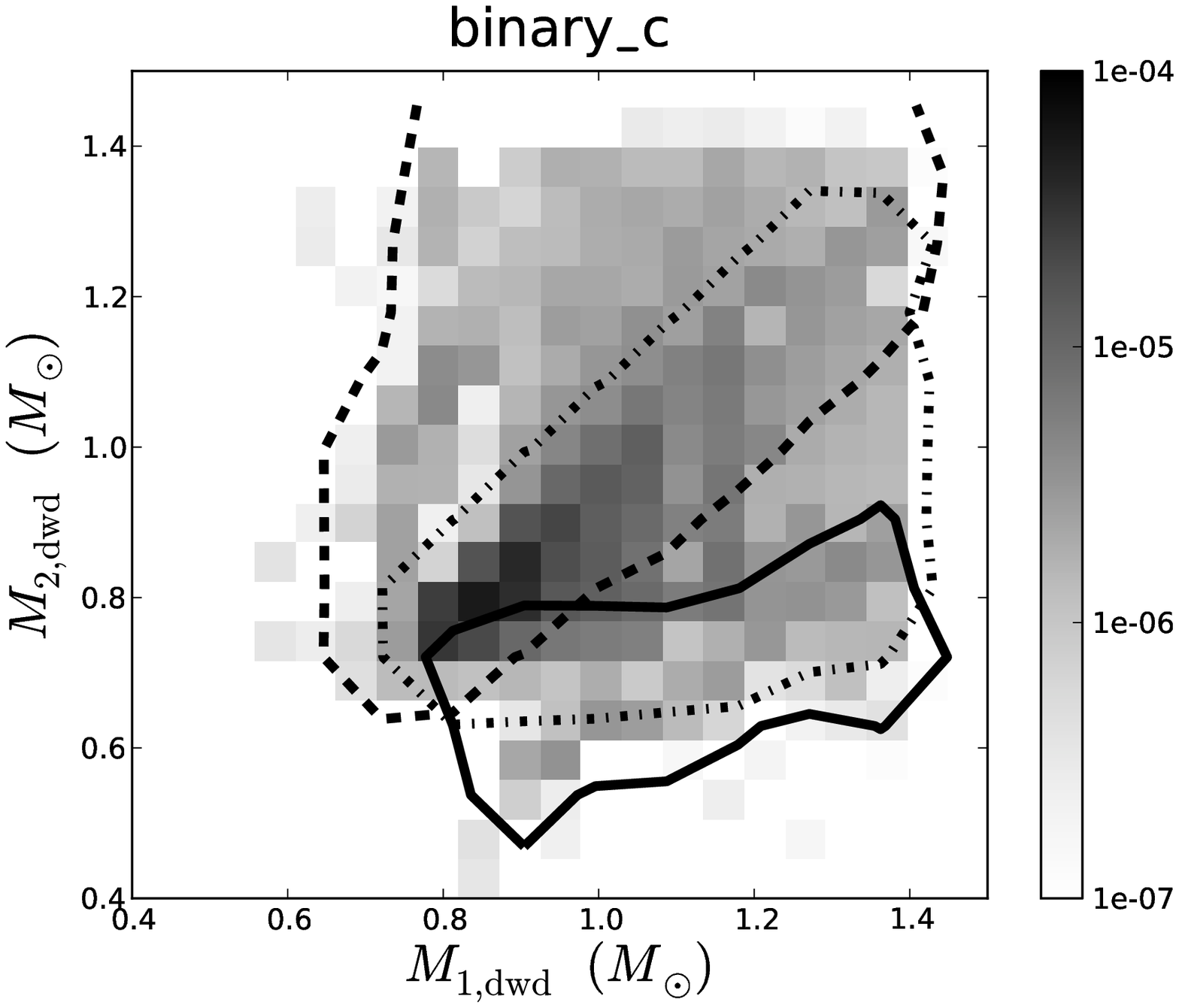} & 
	\includegraphics[height=4.6cm, clip=true, trim =20mm 0mm 48.5mm 5mm]{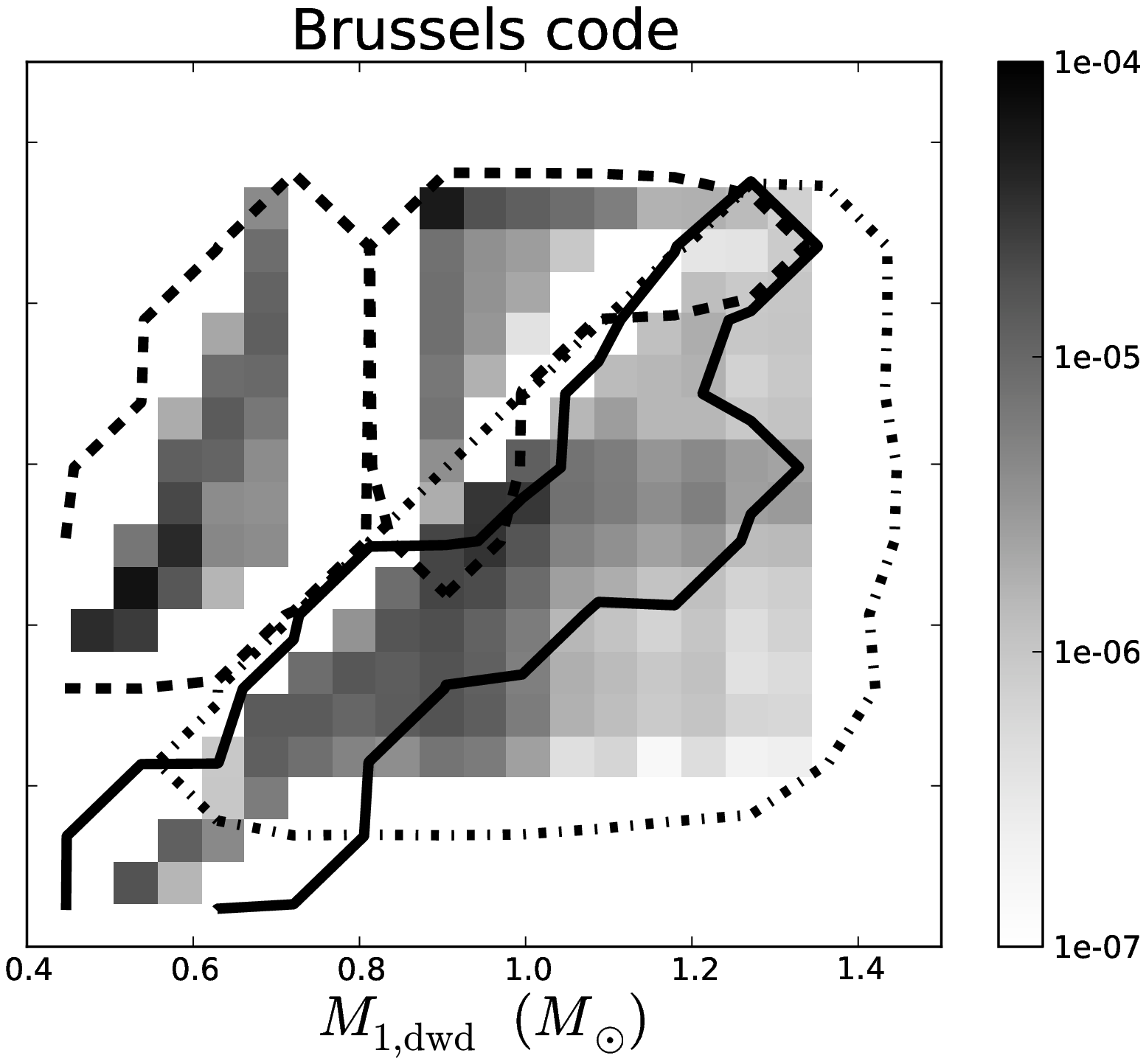} &
	\includegraphics[height=4.6cm, clip=true, trim =20mm 0mm 48.5mm 5mm]{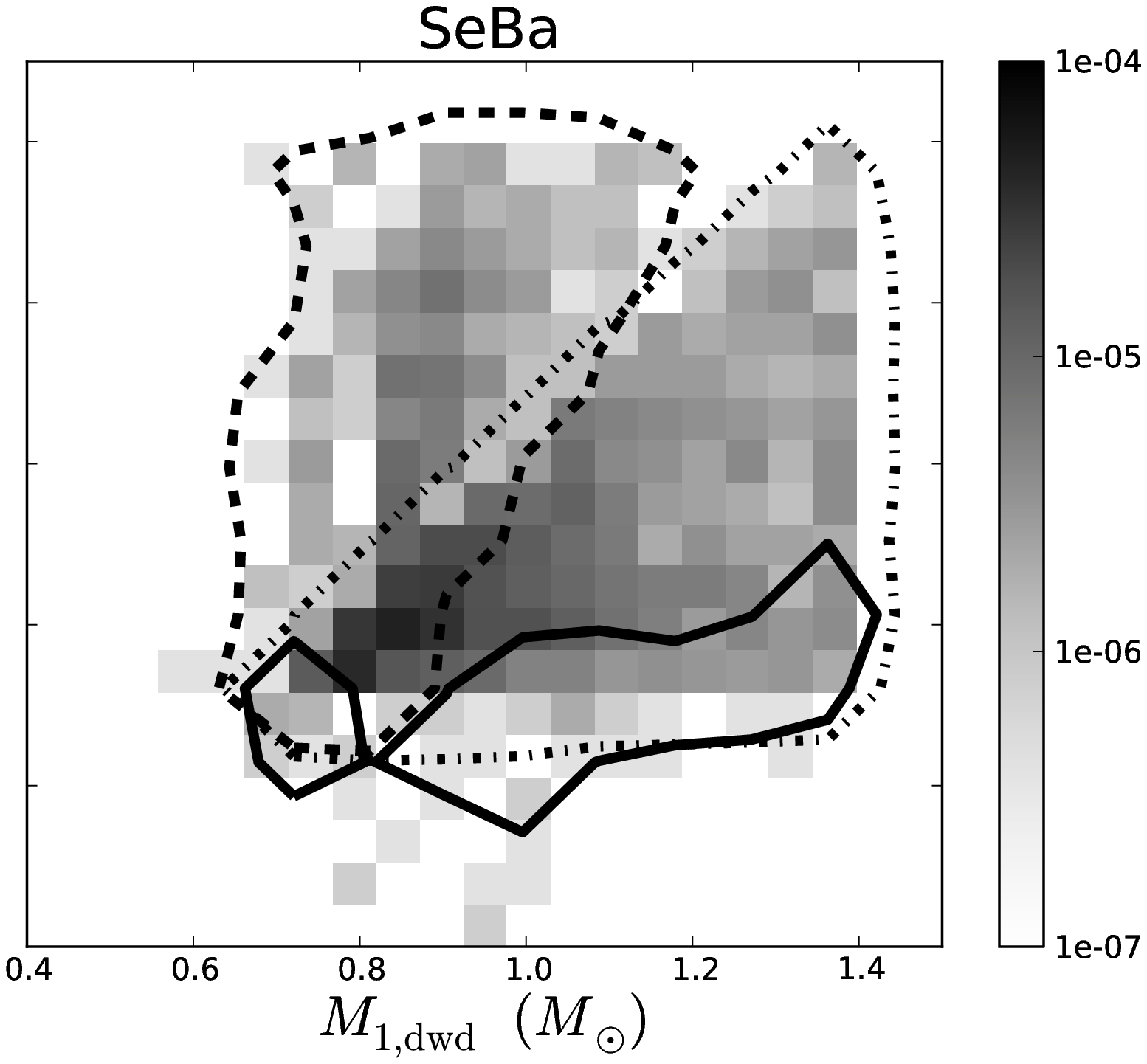} &
	\includegraphics[height=4.6cm, clip=true, trim =20mm 0mm 23mm 5mm]{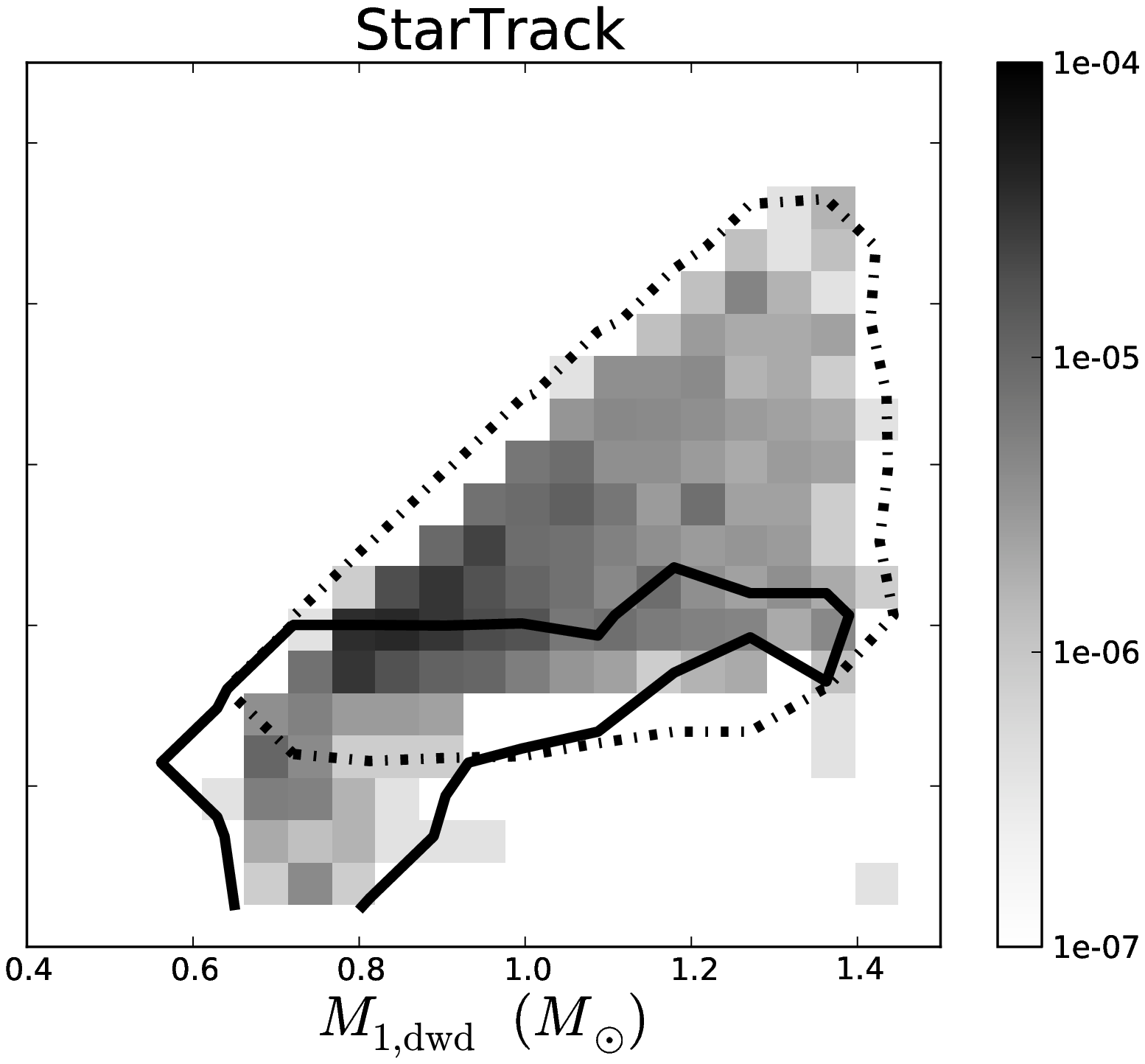} \\	
	\end{tabular}
    \caption{Secondary WD mass versus primary WD mass for all DWDs in the intermediate mass range at the time of DWD formation. The contours represent the DWD population from a specific channel: channel~I (dash-dotted solid line), channel~II (solid line) and channel~III (dashed line). The contours of the DWD population from channel~III according to StarTrack and channel~IV according to all codes are not shown, as the birthrate from this channel is too low.} 
    \label{fig:dwd_M2_IM}
    \end{figure*}

    \begin{figure*}
    \centering
    \setlength\tabcolsep{0pt}
    \begin{tabular}{ccc}
	\includegraphics[height=4.6cm, clip=true, trim =8mm 0mm 48.5mm 5mm]{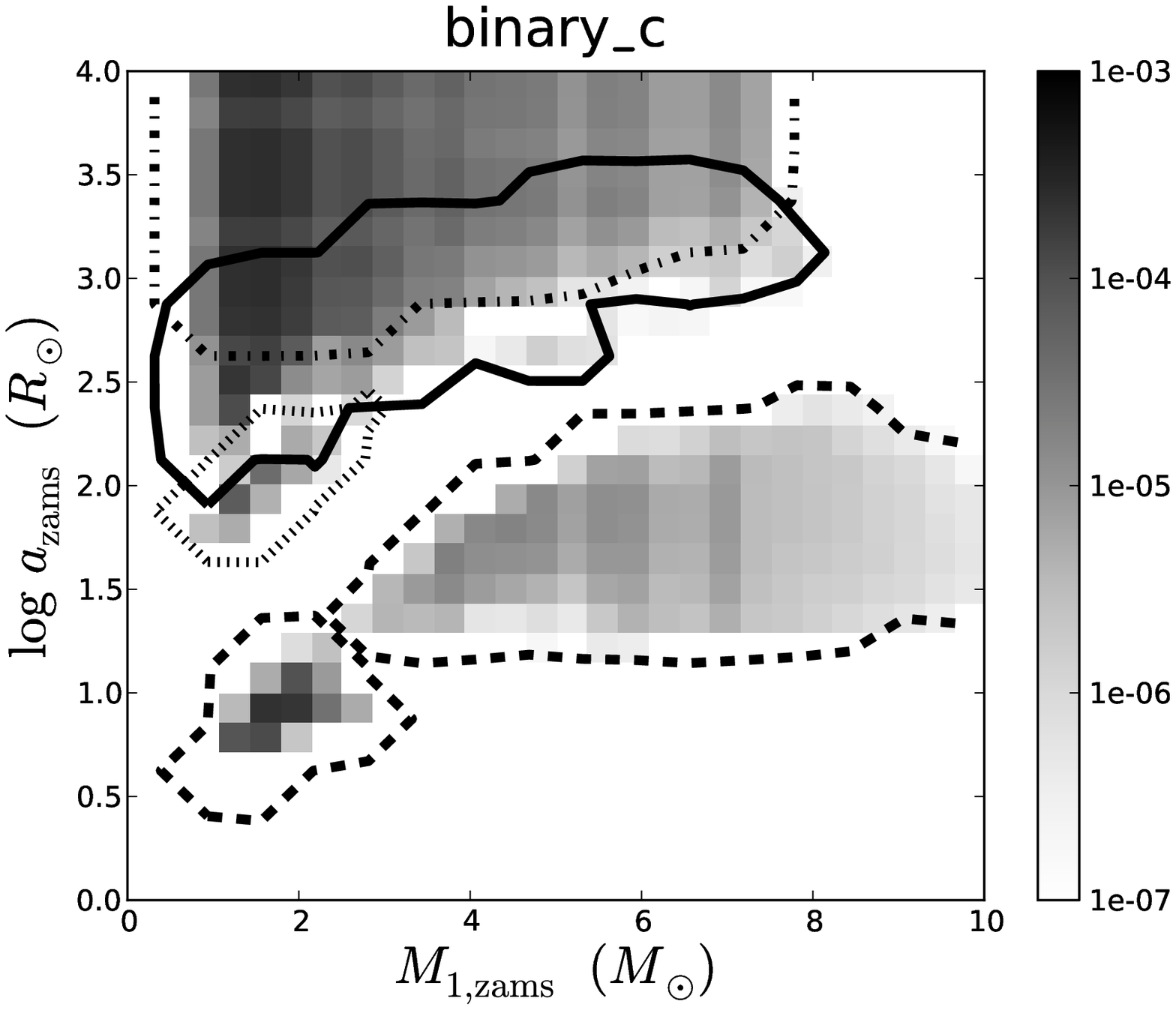} &
	\includegraphics[height=4.6cm, clip=true, trim =20mm 0mm 48.5mm 5mm]{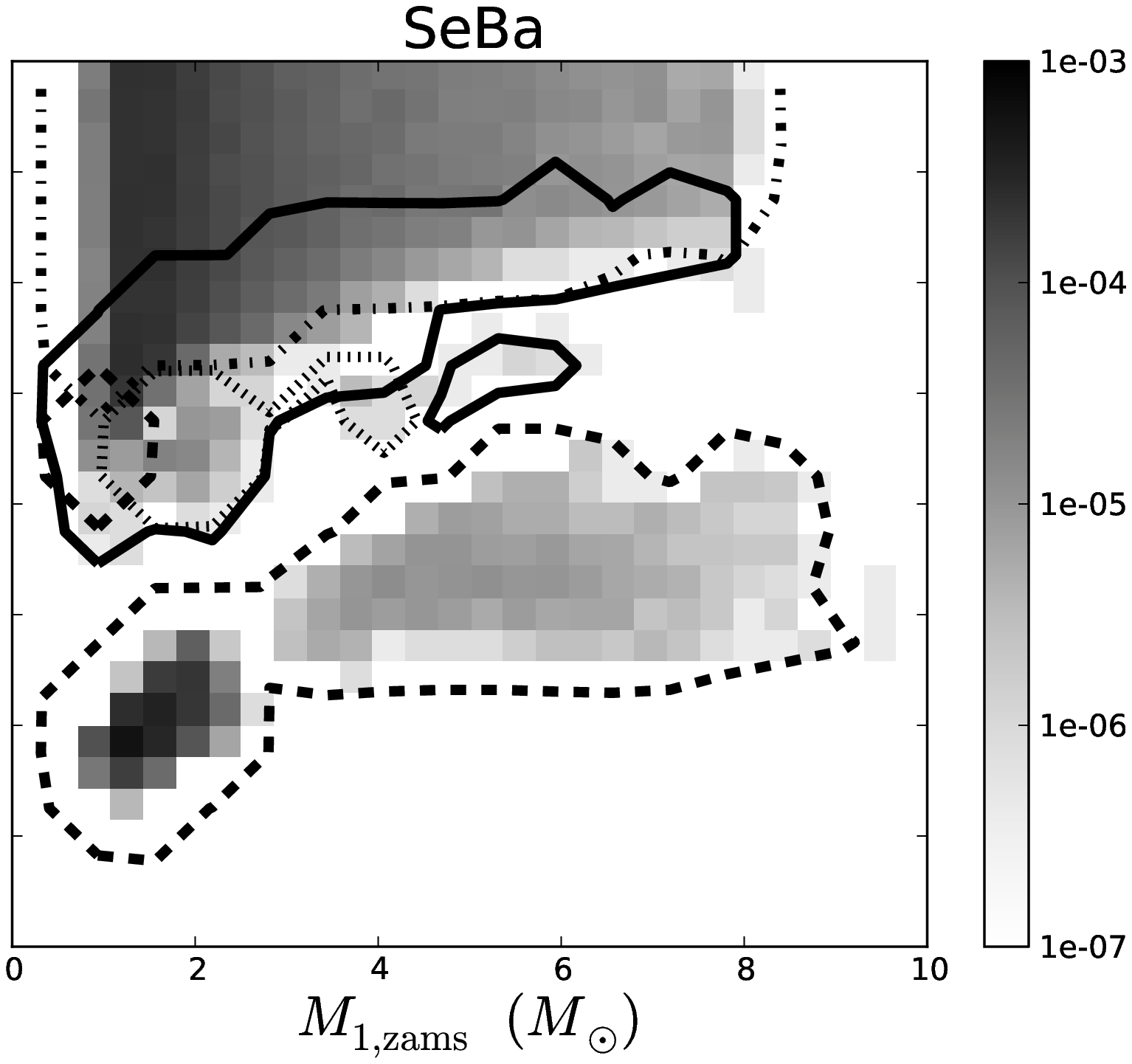} &
	\includegraphics[height=4.6cm, clip=true, trim =20mm 0mm 23mm 5mm]{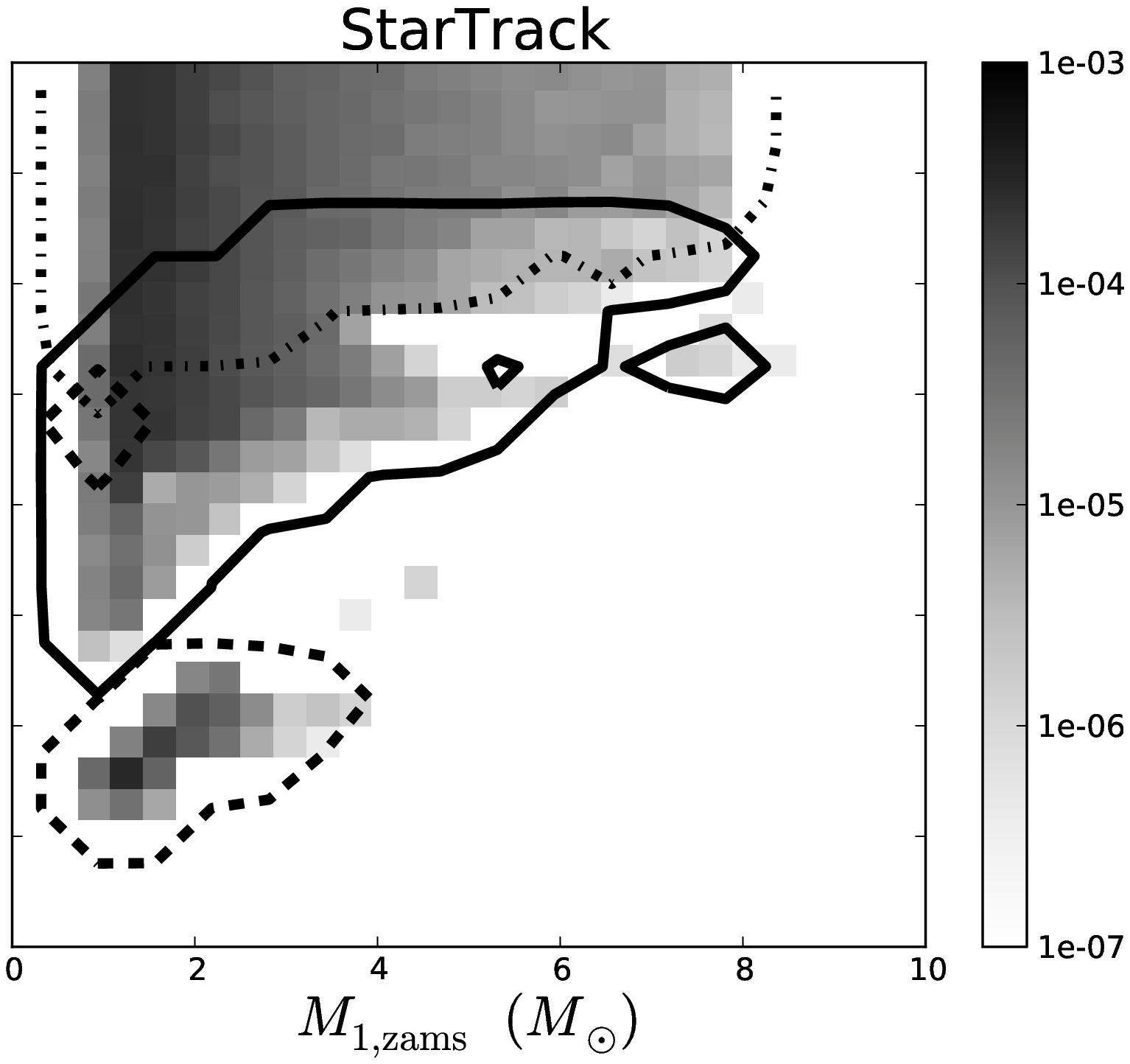} \\	
	\end{tabular}
    \caption{Initial orbital separation versus initial primary mass for all DWDs in the full mass range. The contours represent the DWD population from a specific channel: channel~I (dash-dotted solid line), channel~II (solid line), channel~III (dashed line) and channel~IV (dotted line).} 
    \label{fig:dwd_zams_a}
    \end{figure*}
    
    \begin{figure*}
    \centering
    \setlength\tabcolsep{0pt}
    \begin{tabular}{cccc}
	\includegraphics[height=4.6cm, clip=true, trim =8mm 0mm 48.5mm 5mm]{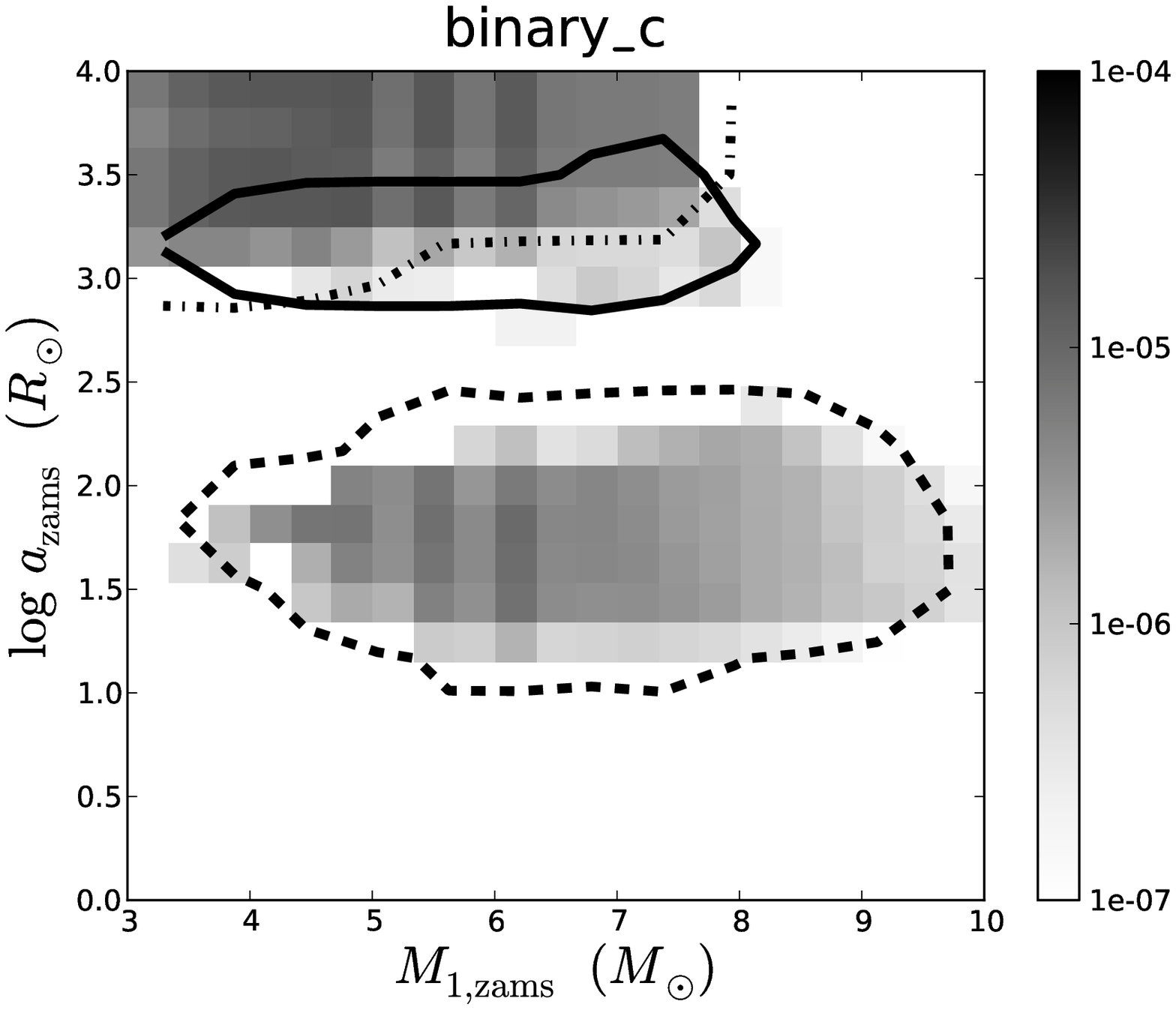} &
	\includegraphics[height=4.6cm, clip=true, trim =20mm 0mm 48.5mm 5mm]{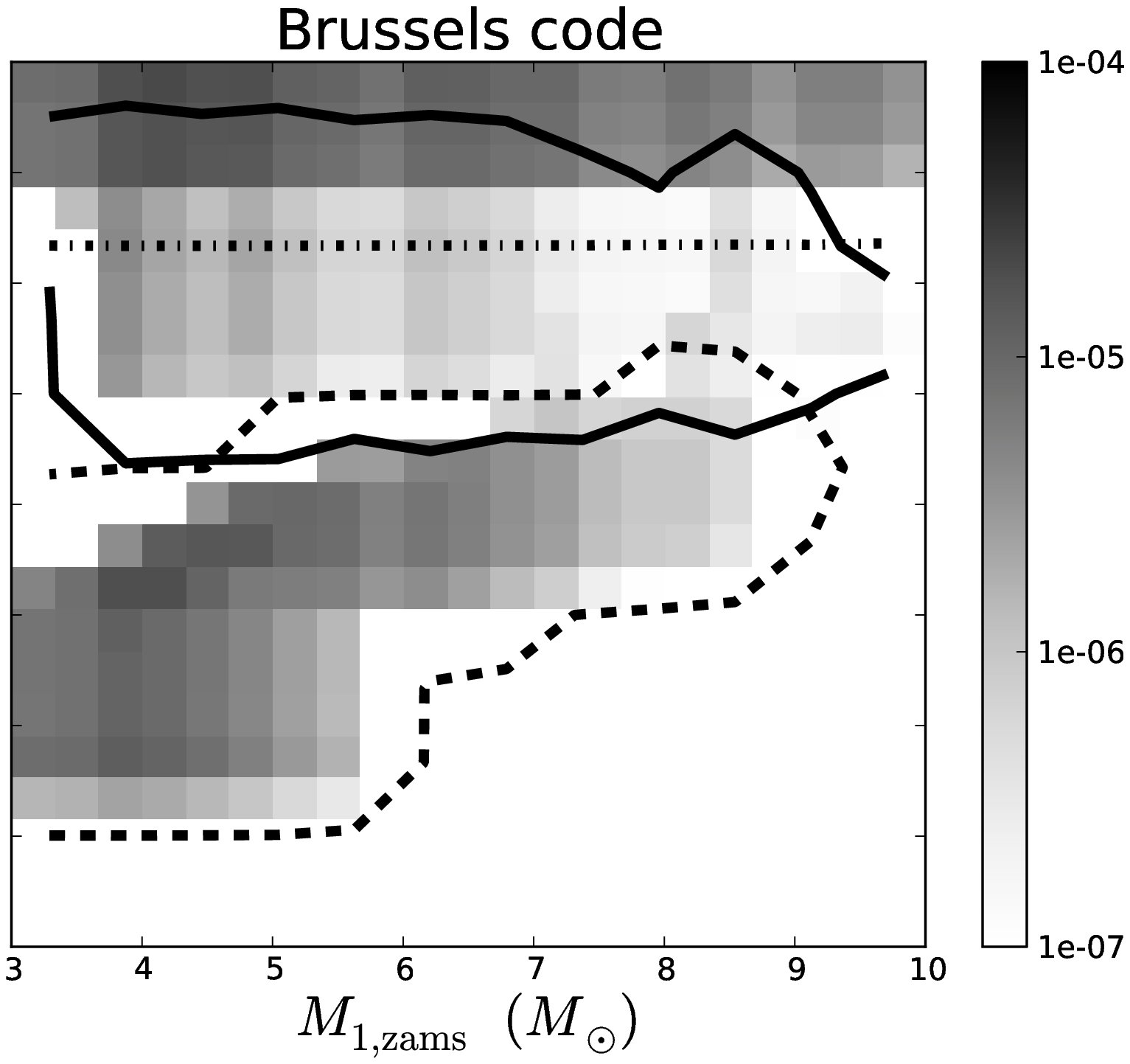} &
	\includegraphics[height=4.6cm, clip=true, trim =20mm 0mm 48.5mm 5mm]{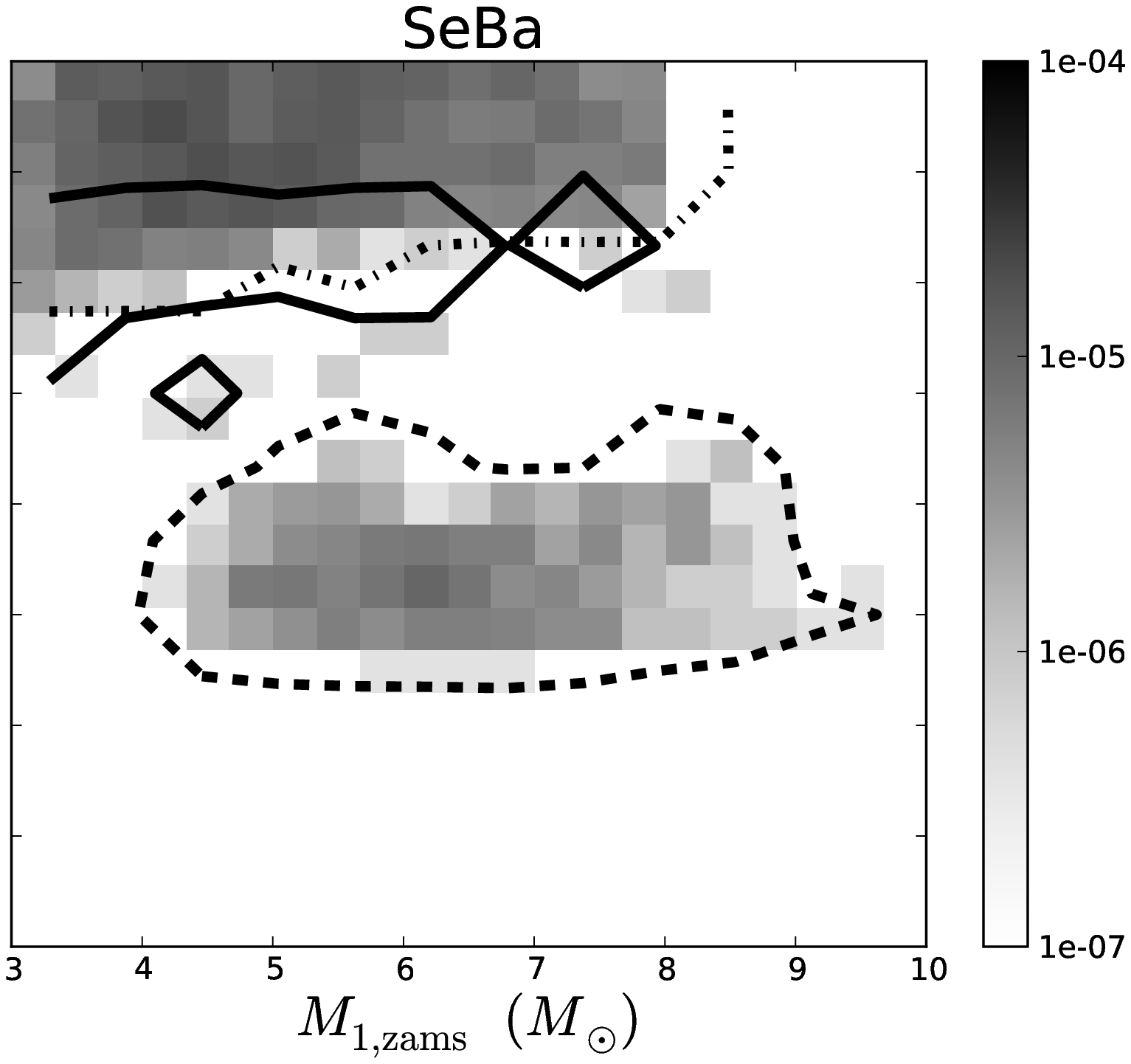} &
	\includegraphics[height=4.6cm, clip=true, trim =20mm 0mm 23mm 5mm]{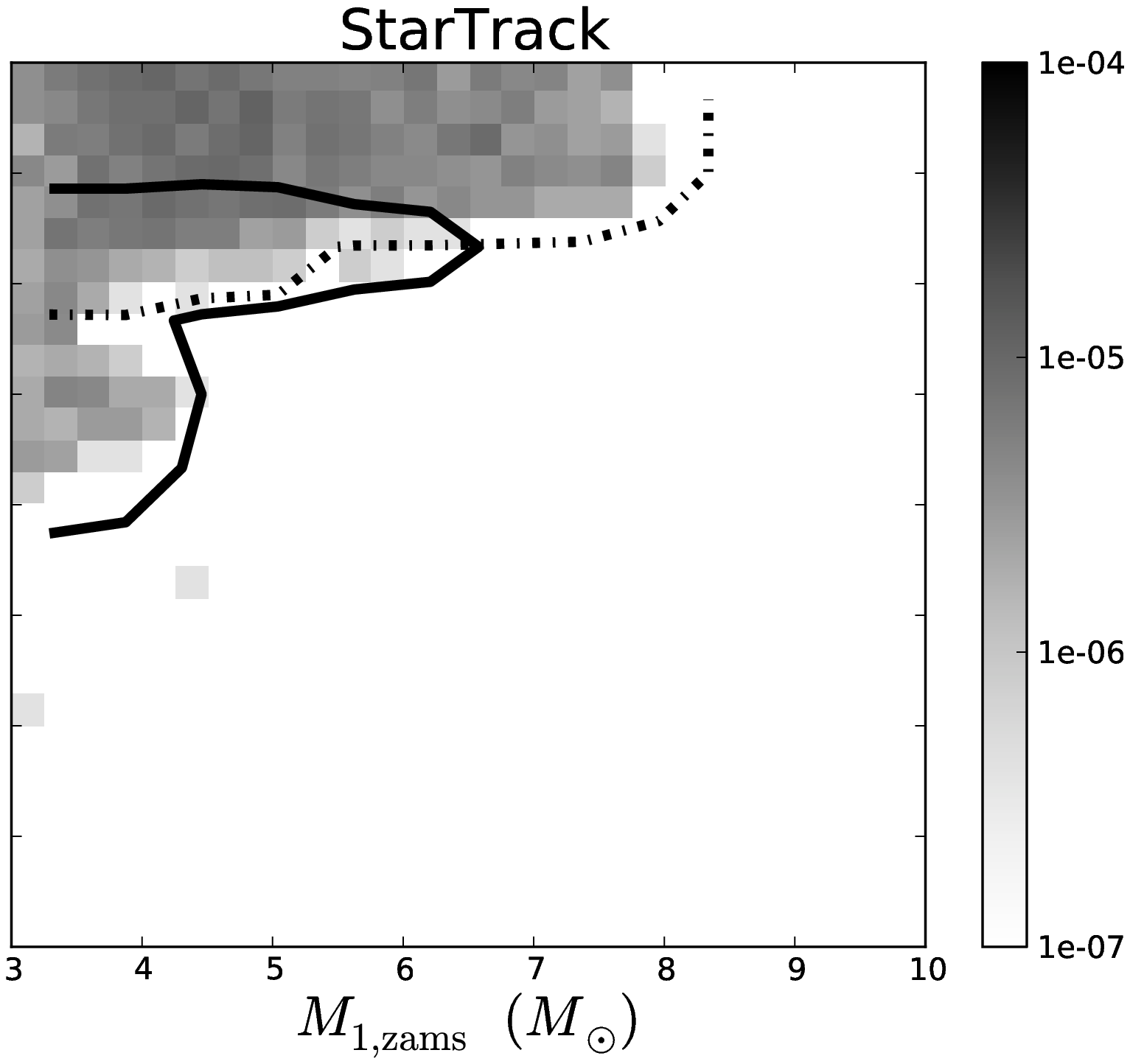} \\	
	\end{tabular}
    \caption{Initial orbital separation versus initial primary mass for all DWDs in the intermediate mass range. The contours represent the DWD population from a specific channel: channel~I (dash-dotted solid line), channel~II (solid line) and channel~III (dashed line). The contours of the DWD population from channel~III according to StarTrack and channel~IV according to all codes are not shown, as the birthrate from this channel is too low.} 
    \label{fig:dwd_zams_a_IM}
    \end{figure*}

\subsubsection{Channel I: detached evolution}
\label{sec:channelI}
\emph{Evolutionary path} Channel~I involves non-interacting
binaries. An example of a system was given for channel~1 in
Sect.\,\ref{sec:channel1}: a 5\Msolar~and 4\Msolar~star in a circular
orbit of $10^4$\Rsolar. When the first WD is born, the orbit has
increased to $[1.8, 1.8, 1.8, 1.8]\cdot10^4$\Rsolar. When the second WD
is born, the orbit has increased even more to $[4.9, 5.0, 4.9, 4.9]\cdot10^4$\Rsolar~with primary and secondary masses of
$[1.0, 0.94, 1.0, 1.0]$\Msolar~and $[0.87, 0.86, 0.87, 0.87]$\Msolar~respectively. 

\emph{Population}
There is a good agreement between the codes on the separations and masses of non-interacting DWDs, initially and at DWD formation. In the full mass range, initial separations are $a_{\rm zams}\approx (0.5-10)\cdot 10^3$\Rsolar. The codes binary\_c, SeBa, and StarTrack find non-interacting DWDs with WD masses between
0.5-1.4\Msolar~in wide orbits of $a_{\rm dwd}\approx (0.1-5.4)\cdot 10^4$\Rsolar. In the intermediate mass range, the initial separations are $a_{\rm zams}\gtrsim 1.5\cdot 10^3$\Rsolar~for binary\_c, SeBa and StarTrack.  
Both WD masses are $\gtrsim 0.75$\Msolar~and orbits are wide with separations $a_{\rm dwd}\approx (0.6-5.4)\cdot 10^4$\Rsolar~for binary\_c, SeBa and StarTrack. 
For the Brussels code, the separations are slightly higher at $a_{\rm zams}\gtrsim 2.8\cdot 10^3$\Rsolar~and $a_{\rm dwd}\approx (1.3-7.2)\cdot 10^4$\Rsolar, and both WD masses extend to slightly lower values of $\gtrsim 0.65$\Msolar.
Small differences between the populations are due to different MiMf-relations and different prescriptions for single stars (e.g. stellar radii), as for SWDs from channel~1. The birthrates in channel~I are very similar in the full mass range as well as in the intermediate mass range (Table\,\ref{tbl:birthrates_all}).

\subsubsection{Channel II: CE + CE} 
\label{sec:channelII}
\emph{Evolutionary path} The classical formation channel for close DWDs involves two CE-phases. First the primary star evolves into a WD via a phase of unstable mass transfer, i.e. via the evolutionary path described in Sect.\,\ref{sec:channel2}~and~\ref{sec:channel4} as channel~2~or~4 respectively. Subsequently the secondary initiates a CE-phase. It should be noted that for binary\_c, SeBa and StarTrack this channel includes, systems that evolve through one CE-phase in which both stars lose their (hydrogen) envelope, the so-called double CE-phase described in Sect.\,\ref{sec:BinEvol} and in eq.\,\ref{eq:ce_dspi}. Note that in the Brussels code, the double CE-phase is not considered. 

\emph{Population} In the full mass range there is a good agreement between the progenitors according to the binary\_c and SeBa code and a fair agreement with the StarTrack code. These three codes find that primaries of $M_{\rm 1, zams}\approx 1-8$\Msolar~contribute to this channel. For the majority the primaries have initial separations of $a_{\rm zams}\approx (0.1-2.5)\cdot 10^3$\Rsolar. The DWD populations as predicted by binary\_c and SeBa are similar, and comparable with the population of StarTrack. WD masses range from 0.35-1.4\Msolar~for primaries and 0.19-0.9\Msolar~for secondaries for binary\_c and SeBa, and slightly larger ranges for StarTrack of 0.2-1.4\Msolar~for primaries and 0.1-0.8\Msolar~for secondaries. The orbital separation of DWDs from channel~II is between a few tenths of solar radii to a few solar radii, however, the specific ranges of the three codes differ. The birthrates in channel~II are similar between the three codes (Table\,\ref{tbl:birthrates_all}).
In the intermediate mass range the codes agree that primaries and secondaries with initial mass between about 3 to 8\Msolar~can contribute to channel~II. In the Brussels code the mass range is slightly extended to higher masses of 10\Msolar~for primaries due to the MiMwd-relation.
There is an agreement on the initial separation of the majority of system, although the range of separations differs between the codes. For binary\_c and SeBa $a_{\rm zams}\approx (0.7-2.5)\cdot 10^3$\Rsolar, however, the range for StarTrack is extended to lower values of $a_{\rm zams}\approx (0.4-20)\cdot 10^2$\Rsolar~as noted above. Comparing with the Brussels code, the range is extended to lower as well as higher values ($(0.3-3.2)\cdot 10^3$\Rsolar). The higher maximum initial separation depends on the maximum radius in the single star prescriptions as discussed in channel~1. The difference in the lower minimum initial separation for the Brussels code has been noted for the SWDs in channel~2 as well. The Brussels code assumes that the primaries in these systems become WDs without a second interaction, where as in binary\_c, SeBa and StarTrack these systems merge in the second interaction of the primary star. The separations of DWDs are centred around 0.5\Rsolar, however, the distribution of separations is different between the codes: 0.17-10\Rsolar~for binary\_c, 0.06-1.18\Rsolar~for the Brussels code, 0.14-3.6\Rsolar~for SeBa and 0.05-11\Rsolar~for StarTrack. Primary WD masses are comparable between the codes, $[0.8-1.4,0.5-1.3,0.7-1.4,0.7-1.4]$\Msolar~where the ranges are the largest for the Brussels code. The maximum WD mass in the Brussels code is lower compared to the other codes due to the MiMwd-relation, see channel~1. The secondary WD masses at a given primary WD mass are lower in binary\_c, SeBa and StarTrack ($\lesssim 0.9$\Msolar) compared to the Brussels code ($\lesssim 1.3$\Msolar).

\emph{Effects}
Several effects influence the distribution of separations in Fig.\,\ref{fig:dwd_a_IM}. Even though the codes agree that the majority of DWDs from channel~II have separation around 0.5\Rsolar, the spread around this value varies between the codes. In the full mass range the maximum separation is 8\Rsolar~in the SeBa data, 22\Rsolar~in the StarTrack data and 31\Rsolar~for binary\_c. In the intermediate mass range it is 1\Rsolar~for the Brussels results, 4\Rsolar~in the SeBa data, 10\Rsolar~for binary\_c and 11\Rsolar~in the StarTrack data.  
The maximum separation is affected by the MiMwd-relation and winds. As seen in channel~2, the maximum orbital separation in the Brussels code is lower as winds are not taken into account and more mass is removed during the CE. The distribution of orbital separation in the Brussels data is also affected in a different way than in the others codes as this code assumes that AGB donors become WDs directly without a second phase of interaction (see also channel~2). In binary\_c, SeBa and StarTrack AGB donors can become helium stars, that fill their Roche lobes for a second time, resulting in lower average masses. This effect can be seen in Fig.\,\ref{fig:dwd_M2_IM} where the secondary mass in binary\_c, SeBa and StarTrack is $\lesssim 0.9$\Msolar~where as it is extended to $\lesssim 1.3$\Msolar~in the Brussels data. Mass loss in combination with the stability criteria, as also discussed for channel~2 causes high separations in the binary\_c data. However, the relatively high maximum separations found by the StarTrack code is not affected much by the difference in the MiMwd-relation and winds, but are affected by differences in the double CE-formalism (see below). 

All codes find that initially many DWD systems have high mass ratios, that in binary\_c, SeBa and StarTrack lead to a double CE-phase. As discussed for channel~2, there is a difference in the formalism of the double CE-phase between StarTrack on one hand, and binary\_c and SeBa on the other hand. As a result the separation after the double CE-phase is smaller according to the latter two codes, and a merger is more likely to happen. The birthrates of systems in the full (intermediate) mass range that evolve through a double CE-phase is $7.2\cdot 10^{-4}\peryr\ (7.9\cdot 10^{-5}\peryr)$ according to StarTrack, while it is $4.6\cdot 10^{-5}\peryr\ (2.5\cdot 10^{-5}\peryr)$ and $1.1\cdot 10^{-4}\peryr\ (3.2\cdot 10^{-5}\peryr)$ for binary\_c and SeBa respectively. An example of systems that merge according to binary\_c and SeBa, but form a DWD according to StarTrack are the systems at $a_{\rm zams}\lesssim 120$\Rsolar~in Fig.\,\ref{fig:dwd_zams_a} which lie at $a_{\rm dwd}\approx 0.07-1.2$\Rsolar~for $M_{\rm 1,dwd} \lesssim 0.35$\Msolar~in Fig.\,\ref{fig:dwd_a}. An example of systems that survive according to all codes, however, at smaller separations for binary\_c and SeBa compared to StarTrack, are systems with $M_{\rm 1, dwd} \gtrsim 0.7$\Msolar~and $a_{\rm dwd} \approx 4-10$\Rsolar~for StarTrack and $a_{\rm dwd} \lesssim 2$\Rsolar~for binary\_c and SeBa. 

An effect that plays a role in channel~II concerns the survival of a system during the mass-transfer event. As explained for channel~2, BPS codes compare the radius of the stripped star (i.e. exposed cores) to the corresponding Roche lobe to determine whether or not a merger takes place during the CE-event. For donor stars that become WDs directly after mass transfer ceases, i.e. without a hydrogen-poor helium burning phase, the Brussels and StarTrack code assume a zero-temperature WD where as binary\_c and SeBa assume the exposed core is expanded due to previous nuclear shell burning. The effect of this is that the radius of the stripped star is a factor of about 5 smaller in the Brussels and StarTrack code than in binary\_c and SeBa. Therefore a merger is less likely to take place. Therefore the minimum separation in the intermediate mass range is 0.06 and 0.05\Rsolar~in the Brussels code and the StarTrack code, respectively. While the minimum separation is about 0.15\Rsolar~in binary\_c and SeBa.

\subsubsection{Channel III: stable + CE}
\label{sec:channelIII}
\emph{Evolutionary path} In channel~III, the primary initiates stable mass transfer (alike channel~3~or~5 which are described in Sect.\,\ref{sec:channel3}~and~\ref{sec:channel5}). When the secondary fills its Roche lobe mass transfer is unstable\footnote{Note that there are two variations of this evolutionary path that are not included in this channel and Fig.\,\ref{fig:dwd_a}~to~\ref{fig:dwd_zams_a_IM}. First, systems in which the secondary becomes a WD before the primary are excluded in this channel. The birthrates of this evolutionary path are low ($[8.6,-,27,5.1]\cdot 10^{-5} \peryr$ in the full mass range. 
See also the discussion and footnote for channel~3 in Sect.\,\ref{sec:channel3} on this evolutionary path. Secondly, for systems with AGB donors that have suffered severe wind mass loss such that the mass ratio has reversed, the first phase of mass transfer can become stable as well. However, consequently the orbit widens to separations comparable to the separations of Channel~I~such that the secondary will not fill its Roche lobe. The birthrates of this evolutionary path are low as well ($[9.4, -, 6.6, 5.3]\cdot 10^{-4}\peryr$ in the full mass range.}

\emph{Population} There is an agreement between the codes about the main parameter space occupied by the DWDs from channel~III and their progenitors, however, the codes do not agree completely. The causes for differences in channel~III have been discussed previously in the context of SWDs (see the discussion on channel~3~and~5), but they lead to more pronounced differences in the DWD population than in the SWD population. 

In the intermediate mass range, the binary\_c, Brussels and SeBa code agree on the orbital characteristics of the main progenitors, $M_{\rm 1, zams}\approx 4-9$\Msolar~and $a_{\rm zams}\approx (0.2-2)\cdot 10^{2}$\Rsolar. There is also a rough agreement between these codes on the range of masses of both WDs. For primaries binary\_c and SeBa find $M_{\rm 1, dwd}\gtrsim 0.65$\Msolar~and the Brussels code $M_{\rm 1, dwd}\gtrsim 0.45$\Msolar~due to differences in the MiMwd-relation. For secondaries these three codes find $M_{\rm 2, dwd}\gtrsim 0.7$\Msolar. 
The maximum mass of the primary and secondary WDs varies between 1.2 and 1.4\Msolar.
The birthrates are high (a few times $10^{-4} \peryr$) in this channel according to binary\_c, the Brussels code and SeBa, however, the birthrate is a factor 1000 lower according to StarTrack. In the StarTrack simulation there are only two systems in channel~III in the intermediate mass range, and therefore we refrain from showing contours for this channel for the StarTrack data in Fig.\,\ref{fig:dwd_a_IM},~\ref{fig:dwd_M2_IM}~and~\ref{fig:dwd_zams_a_IM}.
Figure\,\ref{fig:dwd_zams_a_IM} shows an increase of progenitor systems at separations $a_{\rm zams}\lesssim 20$\Rsolar~and primary masses $M_{\rm 1,zams}\approx 3-5.5$\Msolar~in the Brussels simulation compared to those from the other codes. 
The effect carries through into the DWD population as the increase of systems in the data from the Brussels code with WD primary masses between 0.45-0.7\Msolar.
The orbital separation of DWDs in channel~III is very similar between binary\_c and SeBa, $a_{\rm dwd}\approx 0.1 -1.1 $\Rsolar, however, for the Brussels code $a_{\rm dwd}\approx 0.3-20$\Rsolar. The existence of wide systems in the Brussels code is not surprising, as this code also finds the widest SWDs from channel~3 in comparison with binary\_c and SeBa. As discussed previously in Sect.\,\ref{sec:channel3}, this is related to differences in the MiMwd-relation and angular momentum loss from winds. 
The gap at $M_{\rm 1, dwd}\approx 0.7-0.9$\Msolar~in Fig.\,\ref{fig:dwd_M2_IM} in the data from the Brussels code, is caused by the boundary between channel~3a~and~3b, as in Fig.\,\ref{fig:swd_final_m2_R3_IM} (Sect.\,\ref{sec:channel3}).

Regarding the populations of progenitors for low mass primaries, binary\_c, SeBa and StarTrack agree reasonably well. They both show that most DWDs in channel~III have initial separations of 5-20\Rsolar. However, the range of initial separations is extended to 25\Rsolar~in the population simulated by SeBa. SeBa and StarTrack also show an extra population compared to binary\_c ($a_{\rm zams} \approx140-270$\Rsolar~and $M_{\rm 1, zams} \lesssim 1.2$\Msolar). These two differences are due to differences in the stability of mass transfer for donors with shallow convective envelopes, as discussed for channel~3.   
Comparing the population of DWDs itself for low mass primaries, binary\_c, SeBa and StarTrack agree well. The codes show a population of DWDs with primary mass $M_{\rm 1, dwd}\approx 0.2-0.44$\Msolar~at a separation of $a_{\rm dwd}\approx 0.1-1.5$\Rsolar, with secondary masses $M_{\rm 2, dwd}$ around 0.6\Msolar. The extra population in the SeBa and StarTrack data lies at $a_{\rm dwd}\approx 10-50$\Rsolar~and $M_{\rm 1, dwd}\approx 0.4-0.47$\Msolar. 
The three codes show systems at $M_{\rm 2, dwd}$ about 0.3\Msolar, where in the binary\_c data this group is extended to higher primary WD masses of  $M_{\rm 1, dwd}\approx 0.2-0.7$\Msolar~in Fig.\,\ref{fig:dwd_M2}. These systems in binary\_c mainly evolve through a specific evolutionary path in which there is a phase of stable mass transfer between a He-MS and a WD, a so-called AM CVn-system. The birthrate of these systems is $5.0\cdot 10^{-4}\peryr$ according to binary\_c and negligible according to the other codes.

\emph{Effects} The extremely low birthrate of StarTrack in the
intermediate range is caused by a combination of effects discussed
previously. Firstly, stable mass transfer is more readily
realised in the other codes compared to StarTrack (see
channel~3). Only systems with $q_{\rm zams}\gtrsim 0.6$ undergo stable
mass transfer and become SWD systems according to StarTrack. For about
60\% of these systems, the secondary becomes massive enough
to collapse to a neutron star after nuclear burning ceases (in
accordance with the other codes). Secondly, the remaining systems merge when the
secondary star fills its Roche lobe. For AGB donors this is more
likely to happen in the StarTrack data, because of the difference in
the radii of stripped stars compared to binary\_c and SeBa (see
channel~2b).

The different methods of calculating mass transfer between the Brussels code and the other codes, cause an increase in systems in the data from the Brussels code, similar to channel~5. In particular for DWDs, it is important how the secondary responds to mass gain. The systems that survive in the Brussels code have $q_{\rm zams} > 0.85$, such that the orbit widens severely due to the mass transfer. However, according to binary\_c and SeBa, when the secondary accretes a significant amount of mass and is rejuvenated, its evolutionary timescale is reduced. As the secondary evolves, the system comes in contact and merges. The Brussels code assumes that the merger can be avoided for phases of shallow contact.  

The evolution of and mass transfer rates from evolved helium stars donors (see channel~3) are important for channel~III. It affects the DWD systems with high masses of the primary progenitor and primary WD, see Fig.\,\ref{fig:dwd_M2_IM}~and~\ref{fig:dwd_zams_a_IM}. The range of primary WD masses is extended to 1.2\Msolar~according to SeBa, and 1.3\Msolar~according to the Brussels code and 1.4\Msolar~according to binary\_c. Contrary to stable mass transfer from hydrogen rich donors, the core of evolved helium stars can grow significantly during stable mass transfer phases as the timescale for mass transfer can become comparable to the timescale of wind mass loss or nuclear evolution. If the mass transfer phase is relatively short, the core of the donor star does not have time to grow significantly and little mass is lost in the wind. With the assumption of conservative mass transfer, most of the envelope is then transferred to the secondary star which then is more likely to become a neutron star instead of a WD.

Differences in the radii of stripped stars causes a relative lack of close systems for the
Brussels code compared to the other codes. For channel~II this was discussed in the context of donor stars that
become a WD directly. However, in channel~III in the intermediate mass range many donor stars 
continue burning helium after the mass transfer event ceases. The radius of the stripped donor star depends on its mass, and for binary\_c, SeBa and StarTrack also on the evolutionary state of the donor stars (see also channel~2). 
When the donor star is stripped of its envelope before the AGB-phase, the core radius is a factor of about 4-5 larger in the Brussels code compared to binary\_c, SeBa and StarTrack. Therefore a merger is more likely to take place in the Brussels code.

\subsubsection{Channel IV: CE + stable}
\emph{Evolutionary path} 
In the final evolutionary channel for DWDs, when the primary fills its Roche lobe, mass transfer proceeds in an unstable manner (according to channel~2~or~4 which are described in Sect.\,\ref{sec:channel2}~and~\ref{sec:channel4}). However, when the secondary fills its Roche lobe mass transfer mass to the primary is stable. As a result the primary accretes mass.

\emph{Population}
The systems of channel~IV lie in a small and specific region of DWD parameter space (Fig.\,\ref{fig:dwd_a}~and~\ref{fig:dwd_M2}). The birthrates are low, 1.6$\cdot 10^{-4}\peryr$ and 5.5$\cdot 10^{-5} \peryr$ for binary\_c and SeBa respectively in the full mass range.   
We do not compare the population of this channel with the Brussels code as the progenitors according to binary\_c and SeBa are low mass binaries and the birthrate in the Brussels code is zero per definition (Sect.\,\ref{sec:standass}). We cannot compare the characteristics of the population of binary\_c and SeBa with that of the StarTrack code as in the simulations of the latter code there are no systems evolving through channel~IV indicating a birthrate of $<4\times 10^{-7}\peryr$. The birthrate is low according to StarTrack as unstable mass transfer is more readily realised in this code compared to binary\_c and SeBa (see also channel~3).
The binary\_c and SeBa code agree well on the binary parameters of the population of DWDs at DWD formation from this channel: separations of 10-30\Rsolar~and primary WD masses of 1.1-1.4\Msolar, and secondary WD masses of 0.15-0.20\Msolar. 
The progenitors systems in this channel are similar, $M_{\rm 1, zams}\approx 1-3$\Msolar~and $a_{\rm zams}\approx 50-400$\Rsolar. Differences in the population of DWD systems from this channel, their progenitors and the birthrates occur due to the uncertainty in the stability of mass transfer and the mass transfer rate (see also the discussion for channel~3). 
Note that in the current study we have assumed conservative mass transfer to all accretors, including WDs. This is not a physical picture, so a warning of caution needs to be given to trust the parameters of this population, nonetheless the similarities between the binary\_c and SeBa codes are striking.

\section{Backbones of the BPS codes}
\label{sec:TNS_inherent} 
\begin{table*}
\footnotesize
\caption{Numerical treatments in the different codes which are inherent to them. (Further explanation can be found in Sect.\,\ref{sec:TNS_inherent})}
\label{tbl:TNS}
\begin{center}
\begin{tabular}{|l|c|c|c|c|}
\hline
& \begin{tabular}{c}
binary\_c
\end{tabular}
& \begin{tabular}{c}
Brussels code
\end{tabular}
& \begin{tabular}{c}
SeBa
\end{tabular}
& \begin{tabular}{c}
StarTrack
\end{tabular}\\
\hline \hline
Single star prescriptions & HPT00 &  VB98 & HPT00  & HPT00 \\
Stability of RLOF & $q_{\rm crit}$ & $R_{\rm conv}$, $q_{\rm crit}$ & $\zeta$ & $\zeta$, q$_{\rm ddi}^{(1)}$ \\
Mass transfer rate & R$_{\rm d}$/R$_{\rm RL}^{(2)}$ & R$_{\rm d}$/R$_{\rm RL}^{(2)}$ & $\zeta$ $\rightarrow$ $\frac{M}{\tau_{*}}^{(3)}$ & $\zeta$ $\rightarrow$ $\frac{M}{\tau_{*}}^{(3)}$\\
Wind (AGB)
& R75, VW93, HPT00 & HG97 & R75, VW93, HPT00 & R75, VW93, HPT00 \\
AML (wind) & Donor (HTP02) & No & Donor & Donor \\
Helium star evolution & Yes & Not explicit & Yes & Yes\\
Population synthesis & Grid based & Grid based & Monte Carlo & Monte Carlo\\
\hline
\end{tabular}
\begin{flushleft}
\tablefoot{ References in the Table: HPT = \cite{HPT00}, VB98 = Vanbeveren et al. (1998), R75 = \cite{Reimers75}, VW93 = \cite{VW93}, HG97 = \cite{HoekGroenewegen97}, HTP02 =  \cite{HTP02}.}\\
{(1) Mass ratio threshold for delayed dynamical instability \citep{HW87}, dependent on evolutionary state of the donor.}\\
{(2) R$_{\rm RL}$ is the Roche radius of the donor star.} \\
{(3) $\tau_{*}$ = Characteristic timescale of mass transfer. Can be nuclear, Kelvin-Helmholtz, timescale of magnetic braking or of gravitational wave radiation.}
\end{flushleft}
\end{center}
\end{table*}

The structure of BPS codes can vary strongly, which complicates the process of comparing BPS codes. Some aspects of the code are relatively simple to adapt in order to let assumptions of different groups converge, where as other aspects are inherent to the code and are not straightforward to change. 
For example, where some codes use results from detailed single star evolution codes, written down in analytical formulae \citep[e.g.][]{Egg89,HPT00} to compute stellar parameters, others use the results of detailed binary evolution codes -- a grid over which one can interpolate -- and those results are integrated into the population code \citep[e.g.][]{DeD04}. The inherent differences will create differences between the results of the different groups. The main differences are summarised in Table\,\ref{tbl:TNS} and a more complete overview is given below. For most of the points the influence on a population it not immediately clear, therefore their effects are discussed in Sect.\,\ref{sec:sum}.

\subsection{Single star prescriptions}
The single star prescriptions, either given by analytical formulae or included in a grid of binary systems over which can be interpolated, determine which mass the WD will have when the star loses its envelope. Furthermore they determine the radii during the evolution of the star and therefore the moment at which the star fills its Roche lobe.
\begin{itemize}
\item \emph{binary\_c, SeBa, StarTrack:} the codes use analytical fitting formulae \citep{Hur00} from detailed single star evolution tracks, with an overshooting constant $\delta_{\rm ov}$ = 0.12 \cite[based on][]{Pols98}. In binary\_c different AGB models can be used, based on detailed models of \cite{Karakas02} for thermally pulsating AGB stars (TP-AGB). However, these are not used for this work. Prior to the work of \citet{Too12}, the single star prescriptions in SeBa were based on \citet{Egg89}. 
\item \emph{Brussels code:} intermediate mass single star prescriptions are taken from \citet{Scha92}. These tracks include convective overshooting by means of the following prescription: the overshooting distance $d_{over}$ is directly proportional to the pressure scale height $H_p$ according to $d_{over}=0.2H_p$. This corresponds to a slightly lower degree of overshooting than in the codes that use the overshooting constant $\delta_{ov}=0.12$ in the stability criteria, the latter corresponding to a $d_{over}/H_p$ between 0.22 and 0.4 depending on mass \citep[see][]{Hur00}. Stellar parameters which do not depend on whether the star is part of an interacting binary system are taken directly from this reference. Other stellar parameters, such as the remnant mass after RLOF, are taken from the detailed binary evolution code.
\end{itemize}

\subsection{Stability of mass transfer}
\label{sec:stability_A}
At the moment that one of the stars fills its Roche lobe mass transfer can
proceed in a stable manner or the system can evolve into a CE-phase
(Sect.\,\ref{sec:BinEvol}). In the simulation of the evolution of a binary system
the entire stellar structure is not explicitly followed in detail, and consequently, `stability checks' must be built-in to BPS codes to determine if RLOF will lead to a CE-phase. 
\begin{itemize}
\item \emph{binary\_c:} for every type of donor star and type of accretor star a critical mass ratio ($q_{\rm crit}$) is given. The mass ratio of the system during mass transfer is compared with the critical mass ratio for stable mass transfer and determines if mass transfer will proceed in a stable manner or not. An overview can be found in \citet{Cla13}. Note that in that paper two possibilities are described for the stability of Roche lobe overflowing helium stars to non-degenerate accretors. For this project the criterion as described in \cite{HTP02} is used.
\item \emph{Brussels code:} 
the boundary between stable and unstable RLOF is determined by whether the outer layers of the donor star are radiative or deeply convective respectively. For each stellar mass, the minimum stellar radius $R_{\rm conv}$ is given for which the envelope is convective. If the orbital period of the system under investigation is smaller than the theoretical orbital period at the time when $R_{\rm RL}=R_{\rm conv}$, mass transfer will proceed in a stable way.

If the mass ratio between the two stars is extreme ($q=M_{\rm a}/M_{\rm d}<0.2=q_{\rm crit}$) at the onset of mass transfer, this will result in an instability \citep[][]{Dar1879}. The donor star will be unable to extract sufficient angular momentum from the orbit to remain in synchronized rotation, resulting in the mass transfer episode quickly becoming dynamically unstable. Tidal interaction will cause the secondary to spiral into the donor's outer layers, a process that is treated identically to the CE-evolution (hence with $\beta=0$).

\item \emph{SeBa:} the stability and rate of mass transfer are dependent on the reaction
to mass change of the stellar radii and the corresponding Roche
lobes. 
The change in the Roche radius $R_{\rm RL}$ due to loss and transfer of mass M is given by
\begin{equation}
\zeta_{\rm RL}  \equiv \frac{d\ \rm  ln R_{\rm RL}}{d\ \rm ln M}, 
\label{eq:zeta_L}
\end{equation}
the adiabatic (i.e. immediate) response of the donor star's radius R is given by
\begin{equation}
\zeta_{\rm ad} \equiv \frac{d\ \rm  ln R}{d\ \rm ln M}.
\label{eq:zeta_ad}
\end{equation}
For every Roche lobe filling system, $\zeta_{\rm RL}$ and $\zeta_{\rm ad}$ are compared at every timestep. If $\zeta_{\rm RL} < \zeta_{\rm ad}$ we assume mass transfer proceeds in a stable manner \citep[e.g.][]{Web85, Pol94}. When $\zeta_{\rm RL} > \zeta_{\rm ad}$, mass transfer is dynamically unstable leading to a CE-phase. 

The value of $\zeta_{\rm RL}$ is calculated numerically by transferring a test mass of $10^{-5}$\Msolar. The advantage of this is that, because $\zeta_{\rm RL} = \zeta_{\rm RL}(M_{\rm d}, M_{\rm a}, a)$ and so $\zeta_{RL}$ is dependent on the mass accretion efficiency of the secondary, the (de)stabilising effect \citep[see][]{Sob97} of non-conservative stable mass transfer is taken into account automatically. Appropriate recipes of $\zeta_{\rm ad}$ are implemented in the code for every type of donor star. An overview can be found in \citet{Too12}, appendix A3 therein.

Furthermore, the orbital angular momentum is compared with the stellar spin angular momenta, to check whether a Darwin instability is encountered \citep{Dar1879}.
\item \emph{StarTrack}:
When a non-degenerate star fills its Roche lobe, $\zeta_{\rm ad}$ and $\zeta_{\rm RL}$ are calculated, similar to the case of SeBa. The value of $\zeta_{\rm ad}$ is determined by removing mass
from the star over a 1-year timestep \citep{Bel08a}. The value of $\zeta_{\rm RL}$ is determined by transferring a small amount (1\%) of the star's
mass toward the companion. 
In cases where the mass loss is so rapid such that the star loses thermal equilibrium, a `diagnostic diagram' is used to predict the stability of mass transfer \citep[see description in][sect. 5.2]{Bel08a}. 
The diagnostic diagram is a numerical tool
that was first calibrated using detailed stellar evolution calculations
of massive stars, and is currently being updated to include a range of
stellar models for low- and intermediate-mass stars.

In addition, there is also a check for a possible delayed dynamical instability. This occurs for stars with $M_{\rm d}/M_{\rm a} > q_{\rm ddi}$, with $q_{\rm ddi}$ based on \citet{HW87}, or when a Darwin instability is encountered, or when the trapping radius of the accretion stream \citep{king1999a} exceeds the Roche radius of the accreting star \citep[see][sect. 5.4]{ivanova2003a,Bel08a}. This latter point, however, is not considered for this work.
\end{itemize}

\subsection{Stable mass transfer}
\label{sec:stable}
To take into account various driving mechanism of stable RLOF, such as thermal readjustment or nuclear evolution of the donor, approximate prescriptions are used to determine
the mass transfer rate. Note that mass transfer rate refers to the
mass lost by the donor, which will always be equal to or greater than
the mass accretion rate, which refers to the mass gained by the companion.
\begin{itemize}
\item \emph{binary\_c:} the mass transfer rate is calculated as a function of the ratio of the stellar radius and the Roche radius \cite[based on][]{WhyteEggleton80}. A function is generated which follows the radius more closely during mass transfer on a thermal timescale and more loosely when the star is in thermally equilibrium. A smooth transition is build-in between the two. The formulation can be found in \citet{Cla13}. That paper also shows that the resulting mass transfer phases are comparable to that of the detailed binary stellar evolution code STARS \cite[based on][]{Egg71} in the duration of the mass transfer phases and the mass transfer rates for a set of models.
This method indirectly considers mass transfer on the nuclear and thermal timescale, but also on the timescale of gravitational wave radiation or magnetic braking are considered.
\item \emph{Brussels code:} the mass transfer rates are not explicitly calculated in the population code. It considers merely the initial and final masses. These are interpolated from the results of the detailed binary evolution code. The latter calculates the mass transfer rate during stable RLOF iteratively, by investigating how much mass needs to be lost during the current timestep for the donor star to remain confined by its Roche lobe (within the order of a few percent).
\item \emph{SeBa:} $\zeta_{\rm RL}$ is compared with appropriate values of $\zeta_{\rm eq}$ to determine if mass transfer is driven by the thermal readjustment or the nuclear evolution of the donor star. $\zeta_{\rm eq}$ represents the response of the donor star's radius R as is adjusts to the new thermal equilibrium:
\begin{equation}
\zeta_{\rm eq} = \left ( \frac{d\ \rm ln\ R}{d\ \rm ln\ M} \right )_{\rm th}.
\end{equation}
Appropriate recipes of $\zeta_{\rm eq}$ are implemented for every type of donor star. 
If $\zeta_{\rm RL} < \min(\zeta_{\rm ad}, \zeta_{\rm eq})$, mass transfer is driven by the nuclear evolution of the donor star and we assume mass transfer proceeds on the nuclear timescale of the donor star \citep[e.g.][]{Web85, Pol94}. If $\zeta_{\rm eq} < \zeta_{\rm RL} < \zeta_{\rm ad}$, RLOF is dynamically stable and driven by thermal readjustment of the donor, so that mass transfer proceeds on the thermal timescale of the donor star.  

In addition, stable mass transfer can be driven by angular momentum loss from magnetic braking or gravitational wave emission. When the timescale of angular momentum loss is shorter than the mass loss timescale determined above, we assume mass transfer is driven by angular momentum loss. For more detail see Appendix A.3 of \citet{Too12}.
\item \emph{StarTrack:}
For non-degenerate donors $\zeta_{\rm RL}$ and $\zeta_{\rm ad}$ are calculated, along with the thermal timescale $\tau_{\rm KH}$~\cite[based on][]{kalogera1996a}. Additionally, the equilibrium mass transfer timescale $\tau_{\rm eq}$ is calculated as a combination of RLOF both driven by angular momentum loss and the nuclear evolution of the star and/or the changes due to magnetic braking and gravitational wave radiation \cite[see][]{Bel08a}. If $\tau_{\rm eq} > \tau_{\rm KH}$ the mass losing star is in thermal equilibrium and mass transfer proceeds on $\dot{M}_{\rm eq} = M/\tau_{\rm eq}$. If $\tau_{\rm eq} \leq \tau_{\rm KH}$ mass transfer proceeds on a thermal timescale, given by $\dot{M}_{\rm KH} = M/\tau_{\rm KH}$. If $\dot{M}_{\rm eq}$ becomes positive the star falls out of equilibrium and the stability of mass transfer is determined by the \emph{diagnostic diagram} \cite[see][]{Bel08a}. In the case of WD donors, the mass transfer rate is always driven by gravitational radiation.
\end{itemize}

\subsection{Wind mass loss}\label{sec:wind}
The driving mechanisms of the wind and the explicit rate at which this material is lost are not yet completely understood. This results in different prescriptions to describe the rate of wind mass loss and the amount that can be lost \cite[e.g.][]{Wachter02}. 
We only discuss the wind-prescriptions that are relevant for low and intermediate mass stars. 
\begin{itemize}
\item \emph{binary\_c, SeBa, StarTrack:} for stars up to the early AGB the prescription of \cite{Reimers75} is adopted (with $\eta = 0.5$). To describe the wind mass loss of stars on the TP-AGB a prescription based on \cite{VW93} is implemented. Both prescriptions are defined in \cite{HPT00}. In binary\_c and StarTrack different prescriptions for the wind mass loss are available that used by different users of the respective codes. 
\item \emph{Brussels code:} For intermediate mass interacting binaries, the initial-final mass relation of WDs is determined by assuming the wind prescription of \cite{HoekGroenewegen97}. However, it should be noted that in the BPS code a star in an interacting binary does not have wind mass loss. For the most massive stars, wind mass loss is as is described in \citet{Van98}.
\end{itemize}

\subsection{Angular momentum loss from winds\label{sec:AMLwind}}
Sect.\,\ref{sec:mt_stable} and Appendix\,\ref{sec:AMLRLOF} describe the importance of angular momentum loss (AML) and the effect on the orbit. Not only mass lost during RLOF, but also wind carries angular momentum, which is lost when it leaves the system. The same prescriptions as described in Appendix.\,\ref{sec:AMLRLOF} can be applied to AML when material is lost through a wind and different prescriptions are used in the BPS codes.
\begin{itemize}
\item \emph{binary\_c:} different prescriptions of angular momentum loss through a stellar wind are available in binary\_c. In this study, as in \citet{Cla13}, wind angular momentum loss is as described in \cite{HTP02}. When no material is accreted by the companion star, the wind takes specific angular momentum of the donor. 
\item \emph{Brussels code:} Mass lost by a stellar wind in non-interacting systems is lost through the Jeans mode. Interacting systems do not have wind mass loss prior to interaction.

\item \emph{SeBa, StarTrack:} the material lost by a wind that is not accreted by the companion is lost from the system with specific angular momentum from the donor. 
\end{itemize}

\subsection{Evolution of helium stars}
A helium star is formed after a hydrogen-rich star with a helium core loses its hydrogen-rich envelope. When the core is not degenerate at that time, the evolution of the star continuous as a helium-burning star. Uncertainties in the evolution of helium stars encompasses the growth of this star, the wind mass loss and mass transfer phase, such as the stability and rate.
\begin{itemize}
\item \emph{binary\_c, SeBa, StarTrack:} the evolutionary tracks and wind
  prescription are based on \cite{HPT00}. The stability of mass transfer and the rate are described in previous sections.
\item \emph{Brussels code:} helium star evolution is not explicitly included in the code. It is assumed that the donor star always loses its entire H-rich envelope in one episode and becomes a WD afterwards, except in the case where a donor fills its Roche lobe for a second time as a helium star. In this case mass transfer is followed as described in Sect.\,\ref{sec:stable}, however, time-dependent evolutionary aspects of the helium star are not followed. This simplification is made because the intermediate step is not believed to have a large influence on the eventual masses and separation. However, this implicitly means that the most massive star will always become a WD first, which is not necessarily the case when helium star evolution is explicitly followed. For stars that lose mass during the planetary nebula phase, no resulting angular momentum loss is taken into account.
\end{itemize}

\subsection{Generating the initial stellar population}
The initial population can be chosen by a Monte Carlo method, or the choice can be made
grid-based. Nevertheless, if the method is well performed both methods should give the same results for a high enough resolution.
\begin{itemize}
\item \emph{binary\_c:} $N_{M_{1,\rm zams}}\times N_{M_{2,\rm zams}}\times N_{a_{\rm zams}}$ binaries are simulated, with $M_{\rm 1, zams}$, $M_{\rm 2, zams}$, $a_{\rm zams}$ chosen in logarithmic space. A probability is calculated for every system determined by the defined initial distributions.
\item \emph{Brussels code:} the code works with a three-dimensional grid of initial parameters: primary mass $M_{\rm 1, zams}$, mass ratio $q_{\rm zams}$ and orbital period $P_{\rm zams}$. According to the initial mass function, initial mass-ratio distribution and initial orbital period (or separation) distribution, each grid point is assigned a certain weight. Every system corresponding to such a grid point is then taken through its evolution.
\item \emph{SeBa, StarTrack:} initial parameters $M_{\rm 1, zams}$,
  $M_{\rm 2, zams}$, $a_{\rm zams}$ and the initial eccentricity $e_{\rm zams}$ are chosen randomly on a
  Monte-Carlo based-approach where the probability functions are given
  by the initial distributions. With this method, the resolution is
  highest in those regions of parameter space where most systems lie. 
\end{itemize}

\section{Typical variable assumptions in BPS codes}
\label{sec:TNS_equalized}
\begin{table*}
\footnotesize
\caption{Equalised assumptions for this research and the usual assumptions of the authors in the corresponding BPS codes.}
\label{tbl:TS_all}
\begin{center}
\begin{tabular}{|l|c|c|c|c||c|}
\hline
& \begin{tabular}{c}
binary\_c
\end{tabular}
& \begin{tabular}{c}
Brussels code
\end{tabular} 
& \begin{tabular}{c}
SeBa
\end{tabular}
& \begin{tabular}{c}
StarTrack
\end{tabular}
&\begin{tabular}{c}
This research
\end{tabular}\\
\hline \hline
$\beta$ (RLOF) & Variable & Conditional$^{(1)}$ & Variable & Conditional$^{(1)}$ & 1\\
AML (RLOF) & Isotropic re-emission & Ring$^{(2)}$ ($\eta=1.5\frac{(M_{\rm d}+M_{\rm a})^2}{M_{\rm d} M_{\rm a}}$) &  Orbit$^{(2)}$ ($\eta=2.5$) & Orbit$^{(2)}$ ($\eta=1$) & Orbit ($\eta=1$)\\
CE$^{(3)}$ & $\alpha$ (v2)  & $\alpha$ (v1) & $\gamma\alpha$ (T12) &
$\alpha$ (v1) & $\alpha$ (v1) \\
$\alpha_{\rm ce}\lambda_{\rm ce}/\gamma$ & Variable$^{(4)}$ & 1 & 2/1.75 & 1 & 1\\
Wind accretion & B-H$^{(5)}$ & No & B-H$^{(5)}$ & No$^{(6)}$ & No\\
Tides & Z77, H81, HTP02 & No & PZV96 & Z77, H81, HTP02, C07 & No\\
Magn. braking & RVJ83 & No & RVJ83 & IT03 & No\\
\hline
\end{tabular}
\tablefoot{
\footnotesize
References in the Table: T12 = \cite{Too12}, Z77 = \cite{Zah77}, H81 =
\cite{Hut81}, HTP02 = \cite{HTP02}, PZV96 = \cite{Por96}, C07 = \cite{Cla07}, RVJ83 = \cite{Rappaport83}, IT03 = \cite{IT03}.
\\(1) Constant for non-degenerate accretors, variable for accretion onto a WD. 
\\(2) Except during accretion onto a compact object, AML = istropic re-emission.
\\(3) v1 = prescription \cite{Web84}, v2 = prescription \cite{HTP02}.
\\(4) Based on detailed stellar structure models \citep{Izzard04Thesis, Cla13}.
\\(5) B-H = Prescription based on \cite{BondiHoyle44}.
\\(6) Wind accretion is taken into account for neutron star and black hole accretors assuming B-H-accretion}$^{(5)}$.
\end{center}
\end{table*}

\begin{table*}
\footnotesize
\caption{Equalised initial distribution and range of binary parameters and the usual distributions and ranges of the authors for the corresponding BPS codes.}
\label{tbl:grid_all}
\begin{center}
\begin{tabular}{|l|c|c|c|c||c|}
\hline
\begin{tabular}{c}
What?
\end{tabular}
& \begin{tabular}{c}
binary\_c
\end{tabular}
& \begin{tabular}{c}
Brussels code
\end{tabular}
& \begin{tabular}{c}
SeBa
\end{tabular}
& \begin{tabular}{c}
StarTrack
\end{tabular}
&\begin{tabular}{c}
This research
\end{tabular}\\
\hline \hline
$f(M_{\rm 1, zams})$ & KTG93 & KTG93 & KTG93 & KTG93 & KTG93\\
$M_{\rm 1, zams, min}$ (\Msolar) & 0.1 & 0.1 & 0.1 & 0.08 & 0.1\\
$M_{\rm 1, zams, max}$ (\Msolar) & 80 & 120 & 100 & 150 & 100\\
$f(a_{\rm zams})$ & $\propto a^{-1}$ & $\propto a^{-1}$ & $\propto a^{-1}$ & $\propto a^{-1}$ & $\propto a^{-1}$ (A83)\\
$a_{\rm zams, min}$ (\Rsolar) & max(5, $(R_{\rm a} + R_{\rm  b})/(1-e_{0})$) & $2-12^{(1)} (P=1$d) & $(R_{\rm a} + R_{\rm  b})/(1-e_{0})$ & $2(R_{\rm a} + R_{\rm  b})/(1-e_{0})$ & 5\\
$a_{\rm zams, max}$ (\Rsolar) & 5e6 & $5.8e2-2.2e3^{(1)}(P=3650$d) & 1e6 & 1e6 & 1e6\\
$f(q_{\rm zams})$ & Flat & Flat & Flat & Flat & Flat\\ 
$q_{\rm zams, min}$ & 0.01\Msolar/M$_{\rm 1,zams}$&  0.1\Msolar/M$_{\rm 1,zams}$ & 0 & 0.08\Msolar/M$_{\rm 1,zams}$ & 0.1\Msolar/M$_{\rm 1,zams}$\\
$q_{\rm zams, max}$ & 1 & 1 & 1 & 1 & 1 \\
$f(e_{\rm zams})$ & - & - & H75 &  H75 & -\\ 
$e_{\rm zams, min}$ & - & - & 0 & 0 & -\\
$e_{\rm zams,max}$ & - & - & 1 & 1 & - \\
Max time (Gyr) & 13.7 & 15 & 13.5 & 15 & 13.7 \\
Binary fraction (\%) & 100 & 100 & 50-100 & 50 & 100 \\
\hline
\end{tabular}
\tablefoot{$f(\xi)$ is the distribution of parameter $\xi$. '-' Indicates that no distribution of initial eccentricities is considered, instead $e_{\rm zams}=0$ a priori. Otherwise the distribution of initial eccentricities is $f(e_{\rm zams})$ with $e_{\rm zams}$ between $e_{\rm zams, min}$ and $e_{\rm zams, max}$.
\\References in the Table: KTG93 = \cite{KTG93}, A83 = \cite{Abt83}, H75 = \cite{Heggie75}
\\(1) Separations given for the binary masses under investigation.}
\end{center}
\end{table*}

Some aspects of the codes that are not straightforward to change have been discussed in the previous section. However, other aspects of the codes are relatively simple to adapt. These aspects are often contained in relatively isolated and parametrised functions. For this project we equalised these aspects in the codes as far as possible. However, we do not believe that all the assumptions made for this project are realistic. Previous publications of results from these BPS codes are based on different assumptions. Although we do not compare the effect of the different assumptions on stellar populations in this work, it is good to realise which assumptions are generally used. Therefore the usual assumptions made by the authors in their corresponding BPS code are summarised in Table\,\ref{tbl:TS_all} and are discussed in more detail below. Typical assumptions may vary between different users of the BPS codes.

\subsection{Accretion efficiency}
In this project mass transfer is assumed to be conservative to all types of stars. However, in general, the accretion efficiency depends on the type of accreting star and the mass transfer rate.

\begin{itemize}
\item \emph{binary\_c, SeBa:} in the case of non-degenerate accretors with radiative envelopes, the accretion efficiency mainly depends on the mass transfer rate and the thermal timescale of the accreting star. In the case of non-degenerate objects with convective envelopes, mass is transferred conservatively. In the case of a degenerate accretors, the accretion efficiency depends on the mass of the degenerate object and the mass transfer rate.
\item \emph{Brussels code:} the accretion efficiency onto a non-degenerate object is taken to be constant. If the mass ratio is below 0.2, mass transfer is unstable and the accretion efficiency is assumed to be zero (Sect.\,\ref{sec:stability_A}). To ensure continuity, between mass ratios 0.2 and 0.4 a linear interpolation is used for the accretion efficiency, between 0 and $\beta$ (usually 1). Note that for popcorn this transition was not implemented and the accretion efficiency is one between 0.2 and 0.4. 
In case of a degenerate accreting object, the regions in the (companion mass, orbital period)-parameter space from \citet{Hachisu08} are used to determine in which cases the WD can stably accrete up to 1.4 $M_{\odot}$. In all other cases, mass transfer towards WDs is assumed to become unstable, and is treated as a CE-phase.
\item \emph{StarTrack:} the accretion efficiency onto a non-degenerate object is taken to be constant. In the case of a degenerate accreting object, the accretion efficiency depends on the mass of the accreting object and the mass transfer rate \citep[see][sect. 5 therein]{Bel08a}.
\end{itemize}

\subsection{Angular momentum loss during RLOF}\label{sec:AMLRLOF}
\begin{figure*}
    \centering
	\includegraphics[width=10cm]{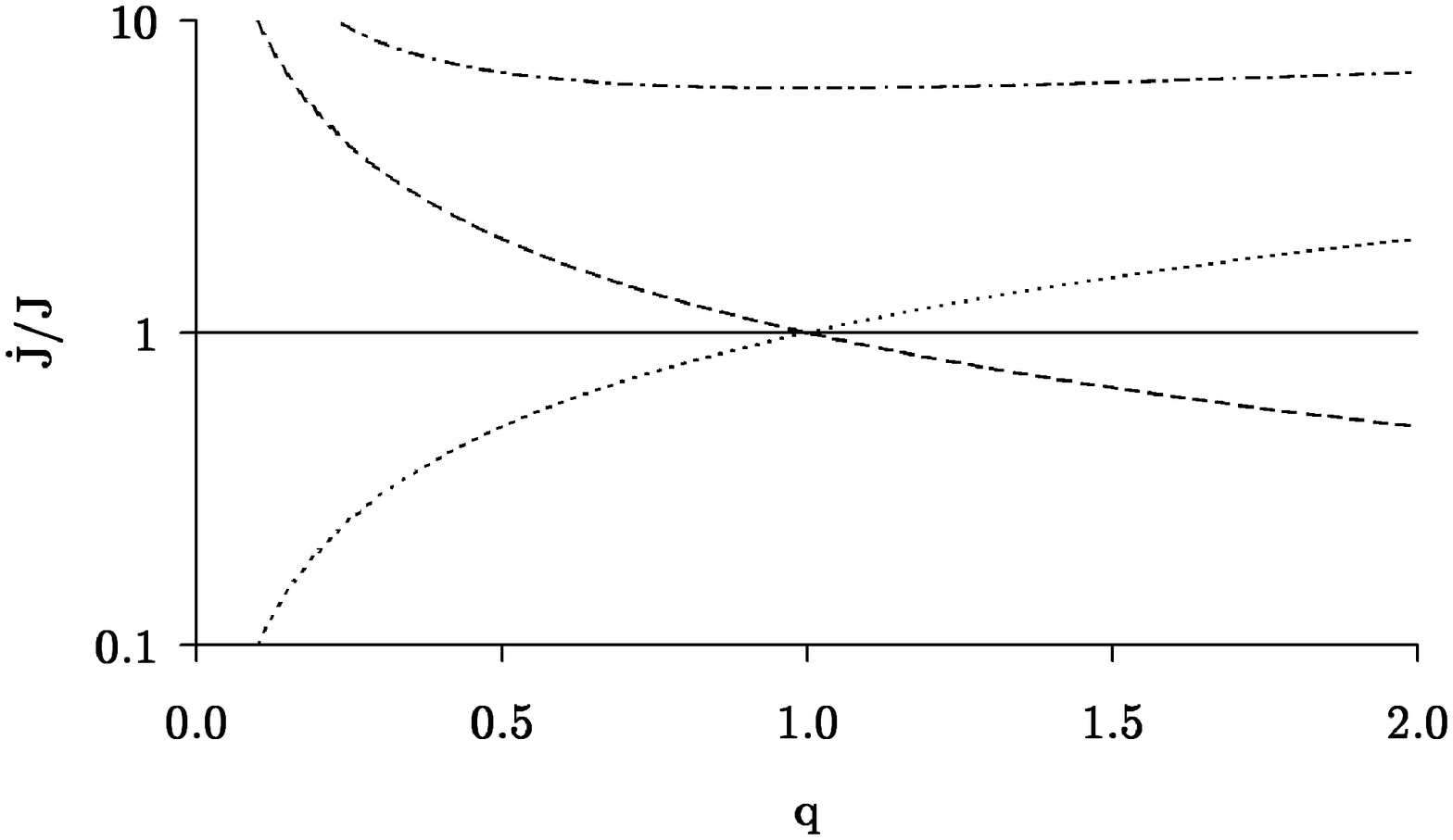} \\
    \caption{Angular momentum loss (in terms of $\frac{\dot{M}(1-\beta)}{M_{\rm d}+M_{\rm a}}$) as a function of mass ratio for four modes: specific angular momentum loss mode (solid, for $\eta=1$), Jeans mode (dotted), isotropic re-emission mode (dashed) and in the case of a circumbinary ring (dashed-dotted, for $a_{ring}/a=2.3$). See text for definition and explication of modes.} 
    \label{fig:AML}
    \end{figure*}

In BPS codes a wide range of prescriptions are used to describe angular momentum loss when material is lost in a phase of stable RLOF. They can be divided in four modes of AML or combinations of these modes \citep[see e.g.][for an overview of the effect of the different prescriptions on the stability of the system]{Sob97}. 

\begin{itemize}
\item Orbital angular momentum loss mode;\\
In this mode the mass is assumed to leave the binary system, with (a multiple of) the specific orbital angular momentum of the binary, i.e $\eta=constant$. 
\item Jeans mode;  \\
Mass is assumed to leave the system from the vicinity of the donor star in a fast spherically symmetric wind. In this mode, the wind matter does not interact with the system. It takes with it the specific orbital angular momentum of the donor in its relative orbit around the centre of mass. Making the assumption that the donor star can be approximated by a point mass, the specific angular momentum loss is as in eq.\,\ref{eq:Jloss} with: 
\begin{equation}
\eta = \frac{M_{\rm a}}{M_{\rm d}}.
\label{eq:Jloss_donor}
\end{equation}

\item Isotropic re-emission;\\
In this case mass is assumed to leave the system from the position of
the gainer in a spherically symmetric way (or at least symmetric with respect to the equatorial plane of the star). Possible scenarios are an enhanced stellar wind or bipolar jets. Further assumptions are as in the previous case, resulting in:
\begin{equation}
\eta = \frac{M_{\rm d}}{M_{\rm a}}.
\label{eq:Jloss_accretor}
\end{equation}

\item Circumbinary ring; \\
Finally, it is possible to assume that the matter will leave the binary through the formation of a non-corotating circumbinary ring, after passing through the second Lagrangian point $L_2$. The amount of angular momentum lost then depends on the radius of this ring $a_{\rm ring}$ compared to the orbital separation $a$:
\begin{equation}
\eta = \sqrt{\frac{a_{\rm ring}}{a}}\frac{(M_{\rm d}+M_{\rm a})^2}{M_{\rm d} M_{\rm a}}.
\label{eq:Jloss_ring}
\end{equation}
While an absolute minimum for $a_{ring}$ is the distance from the center of mass to $L_2$ (which can be shown to vary only very slightly during a mass transfer episode), it was shown by \citet{Sob97} that a more realistic value is 2.3 times the orbital separation.

\end{itemize}

Figure\,\ref{fig:AML} shows, for the four different AML modes, the angular momentum loss $\dot{J}/{J}$ as a function of mass ratio $q=M_{\rm a}/M_{\rm d}$. It is clear that the assumption of AML from a circumbinary ring always leads to the largest AML. The Jeans mode causes the least AML for systems with low mass ratios $q<1$, because the donor is then close to the center of mass of the system. As the mass ratio increases during mass transfer, the AML increases as well since the donor recedes from the center of mass. Conversely, the isotropic re-emission mode causes a large AML for low mass ratio systems, as the gainer is far from the center of mass. As the mass ratio rises, the gainer closes in on the center of mass and AML decreases. The orbital AML assumption results in an intermediate case between the two.

The choice of AML mode is critical for both the stability and the orbital evolution of mass transfer. To illustrate, in the case of the circumbinary ring mode (extracting the most angular momentum), a given amount of mass loss will lead to much more AML than in the case of Jeans mode AML. The former mode will thus result in a far greater number of systems that merge than the latter.

The typical assumptions in the BPS codes are:
\begin{itemize}
\item \emph{binary\_c:} in this work and the standard model in \citet{Cla13}, the material not accreted during the stable RLOF phase is lost as isotropic re-emission.
\item \emph{Brussels code:} the material is lost through the second Lagrangian point such that angular momentum is lost from a circumbinary ring with $a_{\rm ring}=2.3$. 
\item \emph{SeBa:} when the accretor is a non-degenerate star, the material lost carries 2.5 times the specific orbital angular momentum of the binary \citep{Por95, Nel01}. In the case of a degenerate accretor, the material lost carries specific orbital angular momentum of the accreting star.
\item \emph{StarTrack:} when the accretor is a non-degenerate star, the material lost carries one time the specific orbital angular momentum. In the case of a degenerate accretor, the material lost carries specific orbital angular momentum of the accreting star.
\end{itemize}

\subsection{Common envelope evolution}
There remain several uncertainties in the evolution of a CE-phase. For this reason, various BPS codes employ different CE-prescriptions (Sect.\,\ref{sec:mt_unstable}) and CE-efficiencies (such as $\alpha_{\rm ce}$) and both aspects are often varied within a BPS study for comparison. Here, we briefly describe the CE-parametrisations that are implemented most often by the authors in the four different codes.
 
\begin{itemize}
\item \emph{binary\_c:} to describe CE-evolution the prescription based on \cite{HTP02} is used. In the standard model of \citet{Cla13}, $\alpha_{\rm ce}$ is one, while $\lambda_{\rm ce}$ depends on the type of star, its mass and luminosity \citep[see][]{Izzard04Thesis, Cla13}. However, in the BPS code also the $\gamma$-prescription can be used.
\item \emph{Brussels code, StarTrack:} For standard calculations, the  prescription based on \cite{Web84} is used, where $\alpha_{\rm ce}$ and $\lambda_{\rm ce}$ are both one. In both codes different values for $\alpha_{\rm ce}$ and $\lambda_{\rm ce}$ can be implemented, as well as the $\gamma$-prescription \citep[for further information about the version of the $\gamma$-prescription implemented in StarTrack see][]{Bel08a,Rui11}. 
\item \emph{SeBa:} the standard model for simulating CE-evolution in SeBa is the $\gamma$-prescription, unless the binary contains a compact object or the CE is triggered by a Darwin instability \citep{Dar1879} for which the $\alpha$-formalism based on \citet{Web84} is used. The $\gamma$-formalism is introduced by \citet{Nel00} in order to better reproduce the mass ratio distribution of observed DWDs. 
The mass loss reduces the angular momentum of the system according to:
\begin{equation}
\frac{J_{\rm i}-J_{\rm f}}{J_{\rm i}} = \gamma \frac{M_{\rm d,env}}{M_{\rm d}+ M_{\rm a}},
\label{eq:ce_gamma}
\end{equation} 
where $J_{\rm i}$ and $J_{\rm f}$ are the angular momenta of the pre- and post-mass transfer binary respectively. 
The motivation for this formalism is the large amount of angular momentum available in binaries with similar mass objects that possibly can be used to expel the envelope. In SeBa $\gamma$ is taken to be equal to 1.5, and $\alpha_{\rm ce} \times \lambda_{\rm ce}$ is equal to two.
\end{itemize}

\subsection{Wind accretion}
Material that is lost in the form of a  stellar wind can be partly accreted by the companion star. The amount depends on properties of the wind (e.g. the velocity), the accreting star and the binary system (e.g. the separation). However, the exact amount accreted is ill-constrained.

\begin{itemize}
\item \emph{binary\_c, SeBa:} the accretion efficiency of wind material is determined by the Bondi-Hoyle prescription \cite[]{BondiHoyle44}. In binary\_c the accretion efficiency based on the wind Roche-lobe overflow model can be used \citep{Mohamed07, Mohamed12, Abate13}, however, is not used for this work.
\item \emph{Brussels code:} no material lost in the form of a stellar wind is accreted by the companion star.
\item \emph{StarTrack:}  material lost through a wind is in general not accreted by the companion star, except when the companion star is a neutron star or a black hole.
\end{itemize}

\subsection{Tides}
The general picture of tidal effects is clear, however, uncertainties remain due to missing knowledge about for example some dissipative processes.
\begin{itemize}
\item \emph{binary\_c:} tidal evolution is implemented as described by \cite{HTP02}, which is based on \citet{Hut81, Zah77}.
\item \emph{Brussels code:} tidal effects are not taken into account. In the Brussels code it is assumed that tidal effects will not influence the post-RLOF mass of the donor more than marginally.
\item \emph{SeBa:} tidal evolution is implemented as described by \cite{Por96}.
\item \emph{StarTrack:} tidal evolution is implemented as described by
  \cite{Cla07}, as well as \cite{HTP02}, which is based on
  \citet{Hut81, Zah77}.
\end{itemize}

\subsection{Magnetic braking}
Magnetic braking is important for low mass stars with convective envelopes. Nevertheless, this process is not fully understood and different prescriptions co-exist. 
\begin{itemize}
\item \emph{binary\_c, SeBa:} both codes use the prescription of \cite{Rappaport83}.
\item \emph{Brussels code:} the code is not used for the evolution of stellar objects with a mass lower than 3\Msolar, therefore magnetic braking is not considered.
\item \emph{StarTrack:}  the prescription of \cite{IT03} is used in
  standard calculations.
\end{itemize}

\subsection{Initial population}
\label{sec:init}
The choice for an initial distribution and the respective boundaries can severely affect the importance of a certain evolutionary channel through the normalisation of the simulation. 
For example, changing the upper boundary of the distribution of orbital separations from 10$^6$\Rsolar~to 10$^4$\Rsolar, increases the birthrates of interacting binaries by about 70\%. On the other hand, changing the upper boundary for the primary mass distribution within 
80\Msolar~to 150\Msolar~(see Table\,\ref{tbl:grid_all}), does not affect the normalisation. 

The assumptions made by the authors with their respective codes are summarised in Table\,\ref{tbl:grid_all}. 
Different aspects which need extra clarification are discussed below. Note that other users of the BPS codes under study here, other than the authors, may use different distribution functions and/or ranges. 

\begin{itemize}
\item \emph{binary\_c:} the initial eccentricity is zero, based on the work of \cite{HTP02}. The minimum initial separation is varied between 5\Rsolar~or the minimum separation at which a binary system with a certain mass is initially detached. 
The minimum and maximum separations and secondary masses are based on the work of \cite{Kou07}.
\item \emph{Brussels code:} the initial eccentricity is zero. No minimum and maximum separation is assumed for binaries, but a minimum and maximum initial orbital period, more specifically one day and 3650 days. In order to compare with the other codes, a conversion of orbital period to separations is given in Table\,\ref{tbl:grid_all}.
\item \emph{SeBa, StarTrack:} a distribution for the initial eccentricities is assumed (Table\,\ref{tbl:grid_all}). 
The initial semi-major axis is chosen between 10$^6$\Rsolar~and the minimum initial separation is the minimum separation at which a binary system with a certain mass is initially detached. 
\end{itemize}

\end{appendix}
\end{document}